\documentclass[tighten,times]{aastex631} %
\usepackage{amssymb}
\usepackage{amsmath}
\usepackage{amsfonts}
\usepackage{float}
\usepackage{relsize}
\usepackage{verbatim}
\usepackage{color}
\usepackage{comment}
\usepackage{graphicx}
\usepackage{bm}
\usepackage{enumitem}
\usepackage{xspace}
\usepackage{natbib}

\graphicspath{{./figures/}}
  
\newcommand{\DM}{\mathrm{DM}}
\newcommand{\tscatt}{\ensuremath{t_{\rm scatt}}}
\newcommand{\deltanu}{\ensuremath{\Delta\nu_{\rm d}}}
\newcommand{\deltat}{\ensuremath{\Delta t_{\rm d}}}

\newcommand{\mysec}[1]{Section~\ref{#1}}
\newcommand{\myfig}[1]{Figure~\ref{#1}}
\newcommand{\myfigs}[2]{Figures~\ref{#1}--\ref{#2}}
\newcommand{\mytab}[1]{Table~\ref{#1}}
\newcommand{\myeq}[1]{Equation~\ref{#1}}

\newcommand{\equad}{\rm EQUAD\xspace}
\newcommand{\efac}{\rm EFAC\xspace}
\newcommand{\ecorr}{\rm ECORR\xspace}
\newcommand{\eq}{\mathcal{Q}}
\newcommand{\ef}{\mathcal{F}}
\newcommand{\ec}{\mathcal{J}}
\newcommand{\enterprise}{{\tt Enterprise}\xspace}
\newcommand{\hasasia}{{\tt hasasia}\xspace}

\newcommand{\modelcurn}{\textsc{curn}\xspace}
\newcommand{\modelcurngamma}{\textsc{curn}${}^\gamma$\xspace}

\newcommand{\who}[1]{}  %

\newcommand{\resid}{\delta \mathbf{t}}
\newcommand{\delpar}{\boldsymbol{\epsilon}}%
\newcommand{\Tspan}{T_{\rm span}}

\defcitealias{abb+15}{NG9}
\defcitealias{abb+16}{NG9gwb}
\defcitealias{abb+18a}{NG11}
\defcitealias{abb+2018gwb}{NG11gwb}
\defcitealias{aab+20}{NG12}
\defcitealias{abb+2020gwb}{NG12gwb}
\defcitealias{NG15}{NG15}
\defcitealias{NG15detchar}{NG15detchar}
\defcitealias{NG15gwb}{NG15gwb}
\defcitealias{NG15interp}{NG15smbh}
\defcitealias{NG15cosmo}{NG15newphys}

\def\be{\begin{equation}}
\def\ee{\end{equation}}
\def\ba{\begin{eqnarray}}
\def\ea{\end{eqnarray}}

\shorttitle{NANOGrav 15-Yr Noise Budget}
\shortauthors{NANOGrav Collaboration}

\begin{document}

\title{
The NANOGrav 15-Year Data Set: Detector Characterization and Noise Budget
}

\author[0000-0001-5134-3925]{Gabriella Agazie}
\affiliation{Center for Gravitation, Cosmology and Astrophysics, Department of Physics, University of Wisconsin-Milwaukee,\\ P.O. Box 413, Milwaukee, WI 53201, USA}
\author[0000-0002-8935-9882]{Akash Anumarlapudi}
\affiliation{Center for Gravitation, Cosmology and Astrophysics, Department of Physics, University of Wisconsin-Milwaukee,\\ P.O. Box 413, Milwaukee, WI 53201, USA}
\author[0000-0003-0638-3340]{Anne M. Archibald}
\affiliation{Newcastle University, NE1 7RU, UK}
\author{Zaven Arzoumanian}
\affiliation{X-Ray Astrophysics Laboratory, NASA Goddard Space Flight Center, Code 662, Greenbelt, MD 20771, USA}
\author[0000-0003-2745-753X]{Paul T. Baker}
\affiliation{Department of Physics and Astronomy, Widener University, One University Place, Chester, PA 19013, USA}
\author[0000-0003-0909-5563]{Bence B\'{e}csy}
\affiliation{Department of Physics, Oregon State University, Corvallis, OR 97331, USA}
\author[0000-0002-2183-1087]{Laura Blecha}
\affiliation{Physics Department, University of Florida, Gainesville, FL 32611, USA}
\author[0000-0001-6341-7178]{Adam Brazier}
\affiliation{Cornell Center for Astrophysics and Planetary Science and Department of Astronomy, Cornell University, Ithaca, NY 14853, USA}
\affiliation{Cornell Center for Advanced Computing, Cornell University, Ithaca, NY 14853, USA}
\author[0000-0003-3053-6538]{Paul R. Brook}
\affiliation{Institute for Gravitational Wave Astronomy and School of Physics and Astronomy, University of Birmingham, Edgbaston, Birmingham B15 2TT, UK}
\author[0000-0003-4052-7838]{Sarah Burke-Spolaor}
\affiliation{Department of Physics and Astronomy, West Virginia University, P.O. Box 6315, Morgantown, WV 26506, USA}
\affiliation{Center for Gravitational Waves and Cosmology, West Virginia University, Chestnut Ridge Research Building, Morgantown, WV 26505, USA}
\author[0000-0003-3579-2522]{Maria Charisi}
\affiliation{Department of Physics and Astronomy, Vanderbilt University, 2301 Vanderbilt Place, Nashville, TN 37235, USA}
\author[0000-0002-2878-1502]{Shami Chatterjee}
\affiliation{Cornell Center for Astrophysics and Planetary Science and Department of Astronomy, Cornell University, Ithaca, NY 14853, USA}
\author[0000-0001-7587-5483]{Tyler Cohen}
\affiliation{Department of Physics, New Mexico Institute of Mining and Technology, 801 Leroy Place, Socorro, NM 87801, USA}
\author[0000-0002-4049-1882]{James M. Cordes}
\affiliation{Cornell Center for Astrophysics and Planetary Science and Department of Astronomy, Cornell University, Ithaca, NY 14853, USA}
\author[0000-0002-7435-0869]{Neil J. Cornish}
\affiliation{Department of Physics, Montana State University, Bozeman, MT 59717, USA}
\author[0000-0002-2578-0360]{Fronefield Crawford}
\affiliation{Department of Physics and Astronomy, Franklin \& Marshall College, P.O. Box 3003, Lancaster, PA 17604, USA}
\author[0000-0002-6039-692X]{H. Thankful Cromartie}
\altaffiliation{NASA Hubble Fellowship: Einstein Postdoctoral Fellow}
\affiliation{Cornell Center for Astrophysics and Planetary Science and Department of Astronomy, Cornell University, Ithaca, NY 14853, USA}
\author[0000-0002-1529-5169]{Kathryn Crowter}
\affiliation{Department of Physics and Astronomy, University of British Columbia, 6224 Agricultural Road, Vancouver, BC V6T 1Z1, Canada}
\author[0000-0002-2185-1790]{Megan E. DeCesar}
\affiliation{George Mason University, resident at the Naval Research Laboratory, Washington, DC 20375, USA}
\author[0000-0002-6664-965X]{Paul B. Demorest}
\affiliation{National Radio Astronomy Observatory, 1003 Lopezville Rd., Socorro, NM 87801, USA}
\author[0000-0001-8885-6388]{Timothy Dolch}
\affiliation{Department of Physics, Hillsdale College, 33 E. College Street, Hillsdale, MI 49242, USA}
\affiliation{Eureka Scientific, 2452 Delmer Street, Suite 100, Oakland, CA 94602-3017, USA}
\author{Brendan Drachler}
\affiliation{School of Physics and Astronomy, Rochester Institute of Technology, Rochester, NY 14623, USA}
\affiliation{Laboratory for Multiwavelength Astrophysics, Rochester Institute of Technology, Rochester, NY 14623, USA}
\author[0000-0001-7828-7708]{Elizabeth C. Ferrara}
\affiliation{Department of Astronomy, University of Maryland, College Park, MD 20742}
\affiliation{Center for Research and Exploration in Space Science and Technology, NASA/GSFC, Greenbelt, MD 20771}
\affiliation{NASA Goddard Space Flight Center, Greenbelt, MD 20771, USA}
\author[0000-0001-5645-5336]{William Fiore}
\affiliation{Department of Physics and Astronomy, West Virginia University, P.O. Box 6315, Morgantown, WV 26506, USA}
\affiliation{Center for Gravitational Waves and Cosmology, West Virginia University, Chestnut Ridge Research Building, Morgantown, WV 26505, USA}
\author[0000-0001-8384-5049]{Emmanuel Fonseca}
\affiliation{Department of Physics and Astronomy, West Virginia University, P.O. Box 6315, Morgantown, WV 26506, USA}
\affiliation{Center for Gravitational Waves and Cosmology, West Virginia University, Chestnut Ridge Research Building, Morgantown, WV 26505, USA}
\author[0000-0001-7624-4616]{Gabriel E. Freedman}
\affiliation{Center for Gravitation, Cosmology and Astrophysics, Department of Physics, University of Wisconsin-Milwaukee,\\ P.O. Box 413, Milwaukee, WI 53201, USA}
\author[0000-0001-6166-9646]{Nate Garver-Daniels}
\affiliation{Department of Physics and Astronomy, West Virginia University, P.O. Box 6315, Morgantown, WV 26506, USA}
\affiliation{Center for Gravitational Waves and Cosmology, West Virginia University, Chestnut Ridge Research Building, Morgantown, WV 26505, USA}
\author[0000-0001-8158-683X]{Peter A. Gentile}
\affiliation{Department of Physics and Astronomy, West Virginia University, P.O. Box 6315, Morgantown, WV 26506, USA}
\affiliation{Center for Gravitational Waves and Cosmology, West Virginia University, Chestnut Ridge Research Building, Morgantown, WV 26505, USA}
\author[0000-0003-4090-9780]{Joseph Glaser}
\affiliation{Department of Physics and Astronomy, West Virginia University, P.O. Box 6315, Morgantown, WV 26506, USA}
\affiliation{Center for Gravitational Waves and Cosmology, West Virginia University, Chestnut Ridge Research Building, Morgantown, WV 26505, USA}
\author[0000-0003-1884-348X]{Deborah C. Good}
\affiliation{Department of Physics, University of Connecticut, 196 Auditorium Road, U-3046, Storrs, CT 06269-3046, USA}
\affiliation{Center for Computational Astrophysics, Flatiron Institute, 162 5th Avenue, New York, NY 10010, USA}
\author[0000-0002-5070-7990]{Lydia Guertin}
\affiliation{Department of Physics and Astronomy, Haverford College, Haverford, PA 19041, USA}
\author[0000-0002-1146-0198]{Kayhan G\"{u}ltekin}
\affiliation{Department of Astronomy and Astrophysics, University of Michigan, Ann Arbor, MI 48109, USA}
\author[0000-0003-2742-3321]{Jeffrey S. Hazboun}
\affiliation{Department of Physics, Oregon State University, Corvallis, OR 97331, USA}
\author[0000-0003-1082-2342]{Ross J. Jennings}
\altaffiliation{NANOGrav Physics Frontiers Center Postdoctoral Fellow}
\affiliation{Department of Physics and Astronomy, West Virginia University, P.O. Box 6315, Morgantown, WV 26506, USA}
\affiliation{Center for Gravitational Waves and Cosmology, West Virginia University, Chestnut Ridge Research Building, Morgantown, WV 26505, USA}
\author[0000-0002-7445-8423]{Aaron D. Johnson}
\affiliation{Center for Gravitation, Cosmology and Astrophysics, Department of Physics, University of Wisconsin-Milwaukee,\\ P.O. Box 413, Milwaukee, WI 53201, USA}
\affiliation{Division of Physics, Mathematics, and Astronomy, California Institute of Technology, Pasadena, CA 91125, USA}
\author[0000-0001-6607-3710]{Megan L. Jones}
\affiliation{Center for Gravitation, Cosmology and Astrophysics, Department of Physics, University of Wisconsin-Milwaukee,\\ P.O. Box 413, Milwaukee, WI 53201, USA}
\author[0000-0002-3654-980X]{Andrew R. Kaiser}
\affiliation{Department of Physics and Astronomy, West Virginia University, P.O. Box 6315, Morgantown, WV 26506, USA}
\affiliation{Center for Gravitational Waves and Cosmology, West Virginia University, Chestnut Ridge Research Building, Morgantown, WV 26505, USA}
\author[0000-0001-6295-2881]{David L. Kaplan}
\affiliation{Center for Gravitation, Cosmology and Astrophysics, Department of Physics, University of Wisconsin-Milwaukee,\\ P.O. Box 413, Milwaukee, WI 53201, USA}
\author[0000-0002-6625-6450]{Luke Zoltan Kelley}
\affiliation{Department of Astronomy, University of California, Berkeley, 501 Campbell Hall \#3411, Berkeley, CA 94720, USA}
\author[0000-0002-0893-4073]{Matthew Kerr}
\affiliation{Space Science Division, Naval Research Laboratory, Washington, DC 20375-5352, USA}
\author[0000-0003-0123-7600]{Joey S. Key}
\affiliation{University of Washington Bothell, 18115 Campus Way NE, Bothell, WA 98011, USA}
\author[0000-0002-9197-7604]{Nima Laal}
\affiliation{Department of Physics, Oregon State University, Corvallis, OR 97331, USA}
\author[0000-0003-0721-651X]{Michael T. Lam}
\affiliation{School of Physics and Astronomy, Rochester Institute of Technology, Rochester, NY 14623, USA}
\affiliation{Laboratory for Multiwavelength Astrophysics, Rochester Institute of Technology, Rochester, NY 14623, USA}
\author[0000-0003-1096-4156]{William G. Lamb}
\affiliation{Department of Physics and Astronomy, Vanderbilt University, 2301 Vanderbilt Place, Nashville, TN 37235, USA}
\author{T. Joseph W. Lazio}
\affiliation{Jet Propulsion Laboratory, California Institute of Technology, 4800 Oak Grove Drive, Pasadena, CA 91109, USA}
\author[0000-0003-0771-6581]{Natalia Lewandowska}
\affiliation{Department of Physics, State University of New York at Oswego, Oswego, NY, 13126, USA}
\author[0000-0001-5766-4287]{Tingting Liu}
\affiliation{Department of Physics and Astronomy, West Virginia University, P.O. Box 6315, Morgantown, WV 26506, USA}
\affiliation{Center for Gravitational Waves and Cosmology, West Virginia University, Chestnut Ridge Research Building, Morgantown, WV 26505, USA}
\author[0000-0003-1301-966X]{Duncan R. Lorimer}
\affiliation{Department of Physics and Astronomy, West Virginia University, P.O. Box 6315, Morgantown, WV 26506, USA}
\affiliation{Center for Gravitational Waves and Cosmology, West Virginia University, Chestnut Ridge Research Building, Morgantown, WV 26505, USA}
\author[0000-0001-5373-5914]{Jing Luo}
\altaffiliation{Deceased}
\affiliation{Department of Astronomy \& Astrophysics, University of Toronto, 50 Saint George Street, Toronto, ON M5S 3H4, Canada}
\author[0000-0001-5229-7430]{Ryan S. Lynch}
\affiliation{Green Bank Observatory, P.O. Box 2, Green Bank, WV 24944, USA}
\author[0000-0002-4430-102X]{Chung-Pei Ma}
\affiliation{Department of Astronomy, University of California, Berkeley, 501 Campbell Hall \#3411, Berkeley, CA 94720, USA}
\affiliation{Department of Physics, University of California, Berkeley, CA 94720, USA}
\author[0000-0003-2285-0404]{Dustin R. Madison}
\affiliation{Department of Physics, University of the Pacific, 3601 Pacific Avenue, Stockton, CA 95211, USA}
\author[0000-0001-5481-7559]{Alexander McEwen}
\affiliation{Center for Gravitation, Cosmology and Astrophysics, Department of Physics, University of Wisconsin-Milwaukee,\\ P.O. Box 413, Milwaukee, WI 53201, USA}
\author[0000-0002-2885-8485]{James W. McKee}
\affiliation{E.A. Milne Centre for Astrophysics, University of Hull, Cottingham Road, Kingston-upon-Hull, HU6 7RX, UK}
\affiliation{Centre of Excellence for Data Science, Artificial Intelligence and Modelling (DAIM), University of Hull, Cottingham Road, Kingston-upon-Hull, HU6 7RX, UK}
\author[0000-0001-7697-7422]{Maura A. McLaughlin}
\affiliation{Department of Physics and Astronomy, West Virginia University, P.O. Box 6315, Morgantown, WV 26506, USA}
\affiliation{Center for Gravitational Waves and Cosmology, West Virginia University, Chestnut Ridge Research Building, Morgantown, WV 26505, USA}
\author[0000-0002-4642-1260]{Natasha McMann}
\affiliation{Department of Physics and Astronomy, Vanderbilt University, 2301 Vanderbilt Place, Nashville, TN 37235, USA}
\author[0000-0001-8845-1225]{Bradley W. Meyers}
\affiliation{Department of Physics and Astronomy, University of British Columbia, 6224 Agricultural Road, Vancouver, BC V6T 1Z1, Canada}
\affiliation{International Centre for Radio Astronomy Research, Curtin University, Bentley, WA 6102, Australia}
\author[0000-0002-4307-1322]{Chiara M. F. Mingarelli}
\affiliation{Center for Computational Astrophysics, Flatiron Institute, 162 5th Avenue, New York, NY 10010, USA}
\affiliation{Department of Physics, University of Connecticut, 196 Auditorium Road, U-3046, Storrs, CT 06269-3046, USA}
\affiliation{Department of Physics, Yale University, New Haven, CT 06520, USA}
\author[0000-0003-2898-5844]{Andrea Mitridate}
\affiliation{Deutsches Elektronen-Synchrotron DESY, Notkestr. 85, 22607 Hamburg, Germany}
\author[0000-0002-3616-5160]{Cherry Ng}
\affiliation{Dunlap Institute for Astronomy and Astrophysics, University of Toronto, 50 St. George St., Toronto, ON M5S 3H4, Canada}
\author[0000-0002-6709-2566]{David J. Nice}
\affiliation{Department of Physics, Lafayette College, Easton, PA 18042, USA}
\author[0000-0002-4941-5333]{Stella Koch Ocker}
\affiliation{Cornell Center for Astrophysics and Planetary Science and Department of Astronomy, Cornell University, Ithaca, NY 14853, USA}
\author[0000-0002-2027-3714]{Ken D. Olum}
\affiliation{Institute of Cosmology, Department of Physics and Astronomy, Tufts University, Medford, MA 02155, USA}
\author[0000-0001-5465-2889]{Timothy T. Pennucci}
\affiliation{Institute of Physics and Astronomy, E\"{o}tv\"{o}s Lor\'{a}nd University, P\'{a}zm\'{a}ny P. s. 1/A, 1117 Budapest, Hungary}
\author[0000-0002-8509-5947]{Benetge B. P. Perera}
\affiliation{Arecibo Observatory, HC3 Box 53995, Arecibo, PR 00612, USA}
\author[0000-0002-8826-1285]{Nihan S. Pol}
\affiliation{Department of Physics and Astronomy, Vanderbilt University, 2301 Vanderbilt Place, Nashville, TN 37235, USA}
\author[0000-0002-2074-4360]{Henri A. Radovan}
\affiliation{Department of Physics, University of Puerto Rico, Mayag\"{u}ez, PR 00681, USA}
\author[0000-0001-5799-9714]{Scott M. Ransom}
\affiliation{National Radio Astronomy Observatory, 520 Edgemont Road, Charlottesville, VA 22903, USA}
\author[0000-0002-5297-5278]{Paul S. Ray}
\affiliation{Space Science Division, Naval Research Laboratory, Washington, DC 20375-5352, USA}
\author[0000-0003-4915-3246]{Joseph D. Romano}
\affiliation{Department of Physics, Texas Tech University, Box 41051, Lubbock, TX 79409, USA}
\author[0009-0006-5476-3603]{Shashwat C. Sardesai}
\affiliation{Center for Gravitation, Cosmology and Astrophysics, Department of Physics, University of Wisconsin-Milwaukee,\\ P.O. Box 413, Milwaukee, WI 53201, USA}
\author[0000-0003-4391-936X]{Ann Schmiedekamp}
\affiliation{Department of Physics, Penn State Abington, Abington, PA 19001, USA}
\author[0000-0002-1283-2184]{Carl Schmiedekamp}
\affiliation{Department of Physics, Penn State Abington, Abington, PA 19001, USA}
\author[0000-0003-2807-6472]{Kai Schmitz}
\affiliation{Institute for Theoretical Physics, University of M\"{u}nster, 48149 M\"{u}nster, Germany}
\author[0000-0002-7283-1124]{Brent J. Shapiro-Albert}
\affiliation{Department of Physics and Astronomy, West Virginia University, P.O. Box 6315, Morgantown, WV 26506, USA}
\affiliation{Center for Gravitational Waves and Cosmology, West Virginia University, Chestnut Ridge Research Building, Morgantown, WV 26505, USA}
\affiliation{Giant Army, 915A 17th Ave, Seattle WA 98122}
\author[0000-0002-7778-2990]{Xavier Siemens}
\affiliation{Department of Physics, Oregon State University, Corvallis, OR 97331, USA}
\affiliation{Center for Gravitation, Cosmology and Astrophysics, Department of Physics, University of Wisconsin-Milwaukee,\\ P.O. Box 413, Milwaukee, WI 53201, USA}
\author[0000-0003-1407-6607]{Joseph Simon}
\altaffiliation{NSF Astronomy and Astrophysics Postdoctoral Fellow}
\affiliation{Department of Astrophysical and Planetary Sciences, University of Colorado, Boulder, CO 80309, USA}
\author[0000-0002-1530-9778]{Magdalena S. Siwek}
\affiliation{Center for Astrophysics, Harvard University, 60 Garden St, Cambridge, MA 02138}
\author[0000-0001-9784-8670]{Ingrid H. Stairs}
\affiliation{Department of Physics and Astronomy, University of British Columbia, 6224 Agricultural Road, Vancouver, BC V6T 1Z1, Canada}
\author[0000-0002-1797-3277]{Daniel R. Stinebring}
\affiliation{Department of Physics and Astronomy, Oberlin College, Oberlin, OH 44074, USA}
\author[0000-0002-7261-594X]{Kevin Stovall}
\affiliation{National Radio Astronomy Observatory, 1003 Lopezville Rd., Socorro, NM 87801, USA}
\author[0000-0002-2820-0931]{Abhimanyu Susobhanan}
\affiliation{Center for Gravitation, Cosmology and Astrophysics, Department of Physics, University of Wisconsin-Milwaukee,\\ P.O. Box 413, Milwaukee, WI 53201, USA}
\author[0000-0002-1075-3837]{Joseph K. Swiggum}
\altaffiliation{NANOGrav Physics Frontiers Center Postdoctoral Fellow}
\affiliation{Department of Physics, Lafayette College, Easton, PA 18042, USA}
\author[0000-0003-0264-1453]{Stephen R. Taylor}
\affiliation{Department of Physics and Astronomy, Vanderbilt University, 2301 Vanderbilt Place, Nashville, TN 37235, USA}
\author[0000-0002-2451-7288]{Jacob E. Turner}
\affiliation{Department of Physics and Astronomy, West Virginia University, P.O. Box 6315, Morgantown, WV 26506, USA}
\affiliation{Center for Gravitational Waves and Cosmology, West Virginia University, Chestnut Ridge Research Building, Morgantown, WV 26505, USA}
\author[0000-0001-8800-0192]{Caner Unal}
\affiliation{Department of Physics, Ben-Gurion University of the Negev, Be'er Sheva 84105, Israel}
\affiliation{Feza Gursey Institute, Bogazici University, Kandilli, 34684, Istanbul, Turkey}
\author[0000-0002-4162-0033]{Michele Vallisneri}
\affiliation{Jet Propulsion Laboratory, California Institute of Technology, 4800 Oak Grove Drive, Pasadena, CA 91109, USA}
\affiliation{Division of Physics, Mathematics, and Astronomy, California Institute of Technology, Pasadena, CA 91125, USA}
\author[0000-0003-4700-9072]{Sarah J. Vigeland}
\affiliation{Center for Gravitation, Cosmology and Astrophysics, Department of Physics, University of Wisconsin-Milwaukee,\\ P.O. Box 413, Milwaukee, WI 53201, USA}
\author[0000-0001-9678-0299]{Haley M. Wahl}
\affiliation{Department of Physics and Astronomy, West Virginia University, P.O. Box 6315, Morgantown, WV 26506, USA}
\affiliation{Center for Gravitational Waves and Cosmology, West Virginia University, Chestnut Ridge Research Building, Morgantown, WV 26505, USA}
\author[0000-0002-6020-9274]{Caitlin A. Witt}
\affiliation{Center for Interdisciplinary Exploration and Research in Astrophysics (CIERA), Northwestern University, Evanston, IL 60208}
\affiliation{Adler Planetarium, 1300 S. DuSable Lake Shore Dr., Chicago, IL 60605, USA}
\author[0000-0002-0883-0688]{Olivia Young}
\affiliation{School of Physics and Astronomy, Rochester Institute of Technology, Rochester, NY 14623, USA}
\affiliation{Laboratory for Multiwavelength Astrophysics, Rochester Institute of Technology, Rochester, NY 14623, USA}

\collaboration{1000}{The NANOGrav Collaboration}
\noaffiliation

\correspondingauthor{The NANOGrav Collaboration}
\email{comments@nanograv.org}

\begin{abstract}
Pulsar timing arrays (PTAs) are galactic-scale gravitational wave detectors. Each individual arm, composed of a millisecond pulsar, a radio telescope, and a kiloparsecs-long path, differs in its properties but, in aggregate, can be used to extract low-frequency gravitational wave (GW) signals. We present a noise and sensitivity analysis to accompany the NANOGrav 15-year data release and associated papers, along with an in-depth introduction to PTA noise models. As a first step in our analysis, we characterize each individual pulsar data set with three types of white noise parameters and two red noise parameters. These parameters, along with the timing model and, particularly, a piecewise-constant model for the time-variable dispersion measure, determine the  sensitivity curve over the low-frequency GW band we are searching. We tabulate information for all of the pulsars in this data release and present some representative sensitivity curves. We then combine the individual pulsar sensitivities using a signal-to-noise-ratio statistic to calculate the global sensitivity of the PTA to a stochastic background of GWs, obtaining a minimum noise characteristic strain of $7\times 10^{-15}$ at 5 nHz. A power law-integrated analysis shows rough agreement with the amplitudes recovered in NANOGrav's 15-year GW background analysis. While our phenomenological noise model does not model all known physical effects explicitly, it provides an accurate characterization of the noise in the data while preserving sensitivity to multiple classes of GW signals. 
  
\end{abstract}


\section{Introduction }

Millisecond pulsars (MSPs) are extremely stable rotators, with long-term  stability  comparable to that of accurate atomic clocks \citep{matsakis1997, hobbs2012, hobbs2020}. They are therefore uniquely well-suited astronomical objects to probe the low-frequency window of gravitational waves (GWs), from nanohertz to microhertz \citep{saz78,det79, hd83,fb90,psb+18}.

The most promising sources of low-frequency GWs are inspiralling supermassive black hole binaries (SMBHBs). The superposition of the GWs produced by the SMBHB population forms the stochastic GW background (GWB), and the most massive, closest binaries could be resolved as continuous GWs produced by individual sources \citep[see][for review]{tay21}.  The North American Nanohertz Observatory for Gravitational Waves \citep[NANOGrav;][]{aab+20nb} collaboration regularly monitors an array of highly stable MSPs spread across the sky to achieve the  sub-microsecond timing precision required for GW detection. In a GWB search, a common long-time-correlated (with more power at low frequency, or ``red'') signal should be detected among the pulsars in the array, combined with a quasi-quadrupolar signature in the angular correlations between pulsar pairs predicted by General Relativity, the so-called Hellings-Downs correlations \citep{hd83}. NANOGrav reported the detection of a common red signal in our 12.5-yr pulsar timing data set; however, the observations were insufficient to exhibit the Hellings-Downs correlations required to definitively associate this signature with GWs \citep{abb+2020gwb}. Similar GW-search results have subsequently been reported by other PTA collaborations, including  the European PTA \citep[EPTA;][]{chen+21}, the Parkes PTA \citep[PPTA;][]{goncharov2021}, and the International PTA \citep[IPTA;][]{antoniadis2022}, a collaboration between NANOGrav, the EPTA, the PPTA, and the recently formed Indian Pulsar  Timing Array  \citep[InPTA;][]{tarafdar2022}.

In this series of papers presenting the NANOGrav 15-year data release, we report statistically significant evidence for Hellings-Downs spatial correlations in the timing residuals for an ensemble of 67 pulsars (see \citealt{NG15gwb}, hereafter \citetalias{NG15gwb}).
 Because of the stochastic nature of the signal, a careful noise analysis of our experiment is  imperative.
As we have done in our previous papers \citep{abb+2016gwb,abb+2018gwb,abb+2020gwb}, we explicitly model both white (time-uncorrelated) and non-white (time-correlated) noise components present in the detector and incorporate those results in additional analysis steps.
Unlike in our previous data releases, in which this topic has been discussed in either the data presentation or GWB search papers, we devote a separate paper to the noise analysis of our 15-year data release (see \citealt{NG15}, hereafter \citetalias{NG15}). This  paper describes and employs a  noise analysis technique specifically designed for the data in this release. 
Our approach has been developed over our last four data releases and incorporates empirical expressions of noise sources discussed in \citet{cs10,cor13,sti13,dfg+13,abb+15,abb+2016gwb,lam+2016,lam+2017}.
Furthermore, we explicitly outline the structure of the covariance matrix we use since it incorporates our knowledge of noise source variances and their relation, if any, to each other.

The structure of the paper is as follows. 
In \S\ref{s:physical_noise} we present the known astrophysical sources of noise in our experiment. We present in \S\ref{s:phenom_noise} a phenomenological noise model that characterizes the white noise (WN) and its covariance as well as developing a simple spectral model for red noise (RN) present in the data, which includes possible contributions from the GWB and other sources.
The covariance matrices that take this knowledge of the noise and propagate it into later analyses are described in \S\ref{s:covmatrix}.
Noise characterization and its effect on our sensitivity to a GWB are summarized through transmission functions and sensitivity curves in \S\ref{s:PTA_nb}.
Finally, we consider possible future improvements to  noise characterization and reduction in \S\ref{s:future}. Notation used throughout the paper is listed in \mytab{t:notation}.

\section{Astrophysical Impacts on Pulse Arrival Times}\label{s:physical_noise}
Pulsar timing at the sub-microsecond level relies on using  MSPs with extremely stable rotation rates as high-precision cosmic clocks. We measure times of arrival (TOAs) of pulses from a pulsar by producing a summed pulse profile at every observation epoch and convolving with a pulse profile template using a matched-filter approach where the observed profile is a scaled and shifted version of a perfect template shape with additive noise \citep[e.g.,][]{taylor1992}. In this way, we are able to track a reference longitude on the pulsar, an arbitrarily chosen reference meridian. Measured TOAs are then subtracted from those predicted by a timing model to calculate pulsar timing residuals. Many known sources of noise both intrinsic and extrinsic to the pulsar impact the measured TOAs and must be carefully modeled in order to detect the  timing signatures expected due to GWs in the residuals. 

In order to keep this paper self-contained, we briefly discuss the numerous astrophysical perturbations to the TOAs, the fundamental measured quantities in our experiment. 
This is a deep, rich, and complicated subject. 
See \citet{cs10}, \cite{cor13},  \citet{sti13}, and \citet{vs2018} for more detail on all of these effects. Each subsection is followed by one or a few labels, such as ``chromatic,'' ``achromatic,'' etc., describing the noise. Chromatic (or chromaticity) refers to a dependence on the radio frequency of the observations. Achromatic features lack this dependence. Spatially-correlated noise processes show statistical cross-correlations that may depend on the sky position of the pulsars or possess the same correlation across all pulsars. These noise processes are particularly important to understand since PTAs search for the stochastic GWB which is expected to be spatially correlated according to the Hellings-Downs relationship.

\begin{figure}[h!tbp]
\centering
\includegraphics[width=0.95\textwidth]{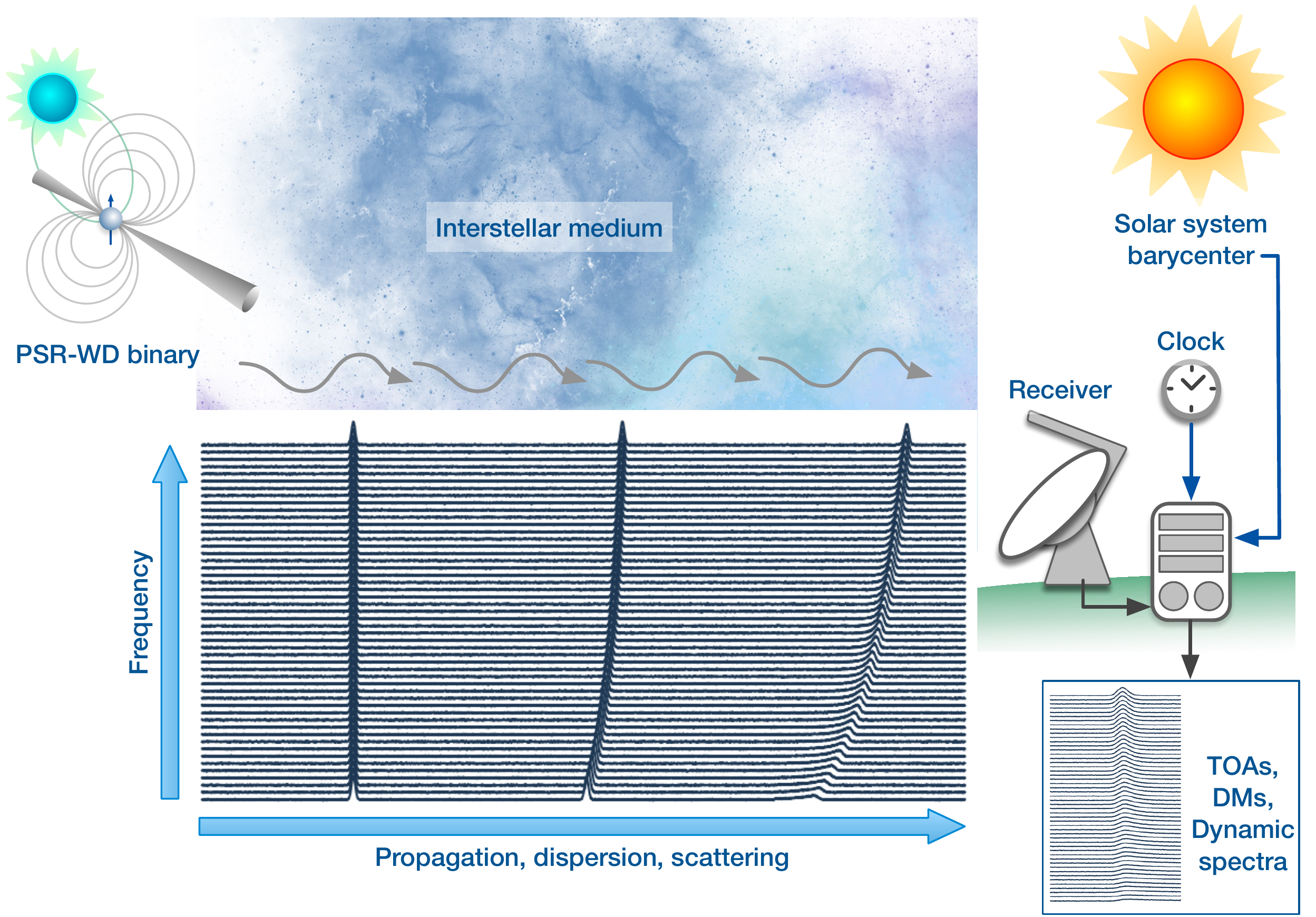}
\caption{An illustration of the signal path from the pulsar to the data products, highlighting a few relevant sources of noise. The pulsar emission itself is subject to jitter (and in rare cases, glitches), and the source model includes fits for spin and spindown, astrometric parameters, and possible binary orbit models. Propagation through the interstellar medium imposes pulse dispersion and scattering, which are both frequency- and time-dependent. At the receiver, along with thermal noise, there are effects related to clock corrections and solar system barycenter parameters, among others, that need to be modeled in order to produce a data set that includes pulse times of arrival and time-dependent dispersion measure estimates.}
\label{f:noise_sources}
\end{figure}

In the following subsections noise sources are discussed roughly in order from emission mechanisms at the pulsar, following the propagation order, through to observation of radio pulses at the telescopes. See \myfig{f:noise_sources} for a visual summary of these effects and their physical location relative to the pulsar-Earth lines of sight.

\subsection{TOA Errors from the Pulsar Emission Process, including Jitter --- Varying Chromaticity} \label{subsec:pulsar_clock}
Pheneomenologically,  {\it pulse jitter} comprises the stochastic phase variation of single pulses with respect to a mean phase that is locked to the rotation of a neutron star.   It appears to be a ubiquitous feature of radio pulsars and magnetars but is a much smaller (if not absent) effect in high-energy emission of pulsars. Pulse jitter limits the fundamental precision of our TOAs at the emission region itself. While the rotation of the neutron star represents the clock mechanism for our experiment, the TOAs we measure correspond to radio emission generated in the pulsar's magnetosphere. Pulse shapes averaged over many rotations are generally very stable, allowing for TOA measurements to precisions $\mathcal{O}(10^{-3})$ the duration of pulse widths. However, the shapes of individual pulses are not stable. This long-studied phenomenon is known as pulse jitter, in which single pulse components can vary in phase and amplitude, with phase variations of order the pulse width \citep{cordes1985, lam+2016,Parthasarathy+2021}. The net effect is that any finite sample of pulses will slightly differ in shape compared to another sample, creating a source of uncertainty in the measured phase of the spot of longitude that should be uncorrelated between observing epochs and  scale with the number of observed pulses  $N_{\rm p}$ as $\propto 1/\sqrt{N_{\rm p}}$. Jitter is strongly correlated across radio frequencies, though since pulse profiles vary in shape across frequency, the combined effect on timing due to jitter can have some frequency dependence, which has been measured for a large fraction of our MSPs \citep{lam+2019,Hebel2022}. However, jitter is entirely uncorrelated from pulse to pulse, i.e., it acts as WN in the residuals \citep{cordes1985}.

{\it Giant radio pulses} (GRPs) are individual pulses that have been observed to occur not periodically, show amplitudes many times larger than average and amplitude distributions that follow a power-law. They are another phenomenon that arises in a small subset of known pulsars. The millisecond pulsar B1937+21 was the second emitter of GRPs detected after the Crab pulsar \citep{staelin+1968,wolszczan_1984}. 
GRPs occur in specific ranges of pulse phase and are thus locked to the spin phase of the neutron star, but their large amplitudes occur stochastically without any distinct temporal pattern.
In the case of B1937+21 its GRPs have pulse widths ranging from micro- to nanoseconds with intensities that follow a power-law distribution and show brightness temperatures ranging up to 10$^{39}$\,K for a GRP of 15\,ns pulse width \citep{sallmen_backer_1995,cognard_1996,kinkhabwala_2000,soglasnov_2004}. 
GRPs from B1937+21 have been observed to have little effect on the timing precision of this pulsar. Extensive studies performed by \cite{mckee_2019} show that the GRPs from B1937+21 provide a timing solution negligibly different from the general timing solution of the pulsar.
Exhaustive studies of lower-level GRP emission have not been performed for the remaining NANOGrav pulsars though examination of several individual sources is underway. A frequently discussed pulsar characteristic for the occurrence of GRPs is a high magnetic field at the light cylinder \citep{cognard_1996}. Currently, it is not clear whether this characteristic is connected to the physics of GRPs. Further investigations need to be carried out before corresponding interpretations can be made. Another pulsar which also displays a high magnetic field at its light cylinder and was reported to be the third GRP emitter is PSR~ B1821-24A \citep{romani_2001}. Albeit not observed by NANOGrav, it is observed by other PTAs. No studies of the effect of its GRPs on its timing precision have been published yet. %
 Although caused by propagation effects in circum-pulsar environments rather than the emission process itself, highly magnified pulses were discovered in the MSP B1957+20 \citep{myc+18} and have been observed in a variety of eclipsing MSPs (see \citealt{lmj+23}). Their effect, if any, on MSP timing precision is largely unexplored.

{\it Mode changing} is a  phenomenon in which a pulsar's average profile switches between two or more discrete modes over a variety of sometimes-periodic timescales, ranging from seconds to days. The most dramatic, and first-identified, example of mode changing is the abrupt cessation of a pulsar's radio emission for several consecutive pulse periods, known as \textit{nulling} and  first identified by \cite{backer_1970}. Nulling has been observed in a large number of predominantly slow-period pulsars  \citep{huguenin_1970,page_1973,hesse_1974}. 
Nulling has not been observed in any MSP, but is  difficult to detect due to the high sensitivity required to detect MSP single pulses. 
 However, mode-changing has been observed  in  three MSPs so far -- PSRs~B1957+20 \citep{mahajan_2018}, J0621+1002 \citep{wang_2021}, and J1909$-$3744 \citep{miles_2022}.
 Of the three, only PSR J1909$-$3744 is observed by NANOGrav. 
 The mode changes detected from PSR~J1909--3744 are characterized by a weak and strong mode that are offset in phase and occur on a single-pulse timescale \citep{miles_2022}. Additionally, PSR~J1713+0747 shows evidence of distinct modes of low-amplitude {\it drifting subpulses} \citep{2016MNRAS.463.3239L}. For both pulsars, these variations should contribute similarly to WN as pulse-to-pulse jitter. We have shown that noise on timescales of several hours was shown to integrate down as WN \citep{2014ApJ...794...21D,2020ApJ...890..123S}. Mode-changing would likely be detectable in a  larger sample of the MSPs we observe with sufficient sensitivity. A detailed investigation of the impact of the nulling/mode changing over the timespans we observe is ongoing.

\subsection{Rotational Irregularities (Spin Noise) --- Achromatic} \label{subsec:spin_noise}

{\it Spin noise} results from rotational irregularities in the pulsars. The interiors of MSPs are thought to consist of a differentially-rotating superfluid core surrounded by an iron-rich crust \citep{lsc1998}.  
Rotational irregularities are thought to arise from torques in the pulsar magnetosphere \citep{c1987,klo+2006,cs2008,lhk+2010,gzy+2016} or coupling between the crust and superfluid core \citep{j1990,ml2014,vEl2018}. The result is a stochastic variation in the pulse period detectable over  timescales of many years.
Fortunately, spin noise is positively correlated with spin frequency derivative, with root-mean-square contribution of $\sigma_{\rm spin} \sim \dot{f}_{\rm s}^{1.0}$ over long  timescales \citep{sc2010,lam+2017,lower2020}. Since  MSPs have very small frequency derivatives, they typically show negligible spin noise. One counterexample is the NANOGrav pulsar B1937+21 \citep{sc2010}. 
Beyond rotational irregularities from neutron-star-specific origins, another proposed cause of the observed timing variations of this pulsar  is an asteroid belt of mass $\lesssim 0.05 M_{\oplus}$ \citep{shannon_2013}. 

{\it Glitches } represent a special case of spin noise where the spin of a pulsar experiences a step-change (see, e.g., \citealp{espinoza+11}). Although this usually results in a very small increase in spin frequency, examples of large increases \citep[``slow glitches''][]{Shabanova05} have been reported, or even \textit{decreases} (so-called ``anti-glitches") in X-ray pulsars and magnetars \citep{ray+19, archibald+13}. The glitch rate and typical fractional change of spin frequency show strong inverse correlations with pulsar age \citep{fuentes+17}. They are therefore rare in MSPs, although two examples   have been reported: first in the globular cluster pulsar B1821$-$24A \citep{cognard04}, and later in the NANOGrav pulsar J0613$-$0200 \citep{mckee+16}.  The fractional changes in spin frequency for these two glitches are only $8\pm1\times10^{-12}$ and $2.5\pm1\times10^{-12}$, respectively, making these the two smallest glitches listed in the Jodrell Bank Glitch Catalogue\footnote{\url{https://www.jb.man.ac.uk/pulsar/glitches.html}} \citep{espinoza+11}. The low glitch amplitudes make glitches very hard to detect in MSPs, with the PSR\,J0613$-$0200 glitch occurring in early 1998 but not being reported until 2016. Note that this glitch occurred prior to the start of our data set. However, other small and unrecognized glitches from this or other MSPs could prove problematic for detection of some proposed GW targets of PTAs, as a GW memory burst (affecting only the pulsar term) is predicted to induce an apparent spin-frequency step-change in the timing residuals of a single pulsar that would be indistinguishable from the signature of a small glitch \citep{cordes12}.

\subsection{Profile Changes in Frequency and Time --- Chromatic}

Pulse shapes intrinsically vary as a function of frequency and, for the most distant pulsars, due to multi-path propagation through the interstellar medium, or {\it scattering}; see \mysec{subsec:propagation}. 
We account for this effect in two different ways. Our traditional {\it narrowband} timing method uses a single template to fit pulses in a number of discrete frequency channels across  a given frequency band and then corrects for constant time offsets between channels in the timing models with a fitted functional form of the offsets versus frequency \citep{NG15}. Our ``wideband'' method \citep{pen14,2019ApJ...871...34P} uses a pulse portrait that contains information about the frequency-dependent pulse shape to fit for the TOA and the time-dependent dispersion measure, or DM (the integrated column density of free electrons along the line of sight), delay simultaneously per epoch.

The TOA creation process assumes that the pulse profile of the pulsar is constant and reproducible when integrating over a suitably long time (at least thousands of pulses). However, subtle pulse shape variations, likely attributable to either propagation effects  or incorrect polarization calibration, have been observed for several NANOGrav pulsars over long timescales \citep{brook18}.

In addition, 
in early 2021 (after the timespan of our data set), the NANOGrav pulsar J1713+0747 was found to have experienced a drastic change in its pulse shape on a timescale of less than one day \citep{lam21b,xu_2021, meyers_2021, singha_2021a}, before relaxing back towards its original pulse shape over the course of several months \citep{jennings_2022a}. As the signature in the timing residuals (generated using templates derived from pre-event data) was similar to that expected from interstellar medium (ISM) propagation effects \citep{lin+21}, this shape change was initially misinterpreted as being accompanied by a step-change in DM. It was later demonstrated that, as the pulse shape evolution over frequency was different before and after the event \citep{singha_2021b}, it is not trivial to measure a change in DM between epochs \citep{jennings_2022b}, and that the observed chromatic behavior is entirely consistent with a frequency-dependent pulse shape change unaccompanied by changes in ISM propagation. This event has necessitated re-examination of the previous chromatic timing features in PSR\,J1713+0747 data noted by \cite{lam21b} as well as one in PSR\,J1643$-$1224 \citep{shannon_2016}; it is not yet clear how much of an impact this phenomenon has on our ability to model pulsar timing behavior.

\subsection{Orbital Irregularities --- Achromatic}
\label{subsec:orbital}

The majority of pulsars observed by NANOGrav are in binary systems, typically with a white-dwarf companion \citep{Fonseca+2016,NG15}. A majority are in orbits well-described by Keplerian parameters, and also post-Keplerian parameters in many cases. However, several pulsars require different modeling of their binary systems, which we comment on here. Four (PSRs J0023+0923, J0636+5128, J1705$-$1903, and J2214+3000) are in low-eccentricity black-widow systems where tidal and wind effects can cause measurable orbital variations. Three of these pulsars (excluding PSR~J2214+3000) are modeled by higher-order orbital frequency derivatives. While PSR~J1705$-$1903 is newly added to the NANOGrav PTA and several new higher-order orbital-frequency derivatives have been measured in \citepalias{NG15} for the others, the orbital parameters for the other three measured in common with our 12.5-yr data set \citep{aab+20nb} remain stable with additional data \citep{NG15}, suggesting excess noise from parameter mis-estimation is low. Only PSR~J1705$-$1903 shows significant RN, with a very shallow spectral index, suggesting that mis-modeling does not significantly affect our GW analyses. Note that NANOGrav selects for ``well-behaved'' systems from a timing perspective, and so other such black-widow systems may indeed show significantly more noise due to irregular angular-momentum transfer, though since the orbital periods are all short compared to our GW signal ($<$1~day versus years) only a small amount of sensitivity to GWs is lost overall \citep{brd2015}.

Beyond these systems, PSR~J1024$-$0719 is a pulsar in a wide-binary system \citep{Kaplan+2016,Bassa+2016} such that we cannot feasibly measure a complete orbit; instead we model the orbital motion as a second derivative in the pulsar's spin frequency. In this case, the pulsar shows significant RN in individual noise and common noise analyses (see Table~\ref{t:rednoise}) and so unmodeled orbital variations may still be contributing to the noise in this pulsar. Lastly, there are two known pulsars in triple systems \citep{Thorsett+1999,Ransom+2014}, one of which (PSR J0337+1715) is currently being timed by NANOGrav due to its high-precision TOAs but is not included in \citetalias{NG15}. The timing of this pulsar requires higher-order effects from general relativity to be included in its timing model and so the modeling procedure involves more costly likelihood evaluations. Nonetheless, even though a complicated system, only a small amount of excess noise has been measured beyond the template-fitting uncertainties \citep{archibald+2018}, with no RN and a simplified DM fit compared to our procedure (see next subsection), making triple systems potentially significant contributors to future PTA efforts.

\subsection{ISM Propagation Effects --- Chromatic} \label{subsec:propagation}

{\it Dispersion} is the dominant frequency-dependent propagation effect in pulsar timing resulting from radio pulses traveling through the ionized ISM, interplanetary medium (IPM; or solar wind), and even Earth's ionosphere, with emission being temporally delayed as a function of frequency $\nu$ by an amount proportional to $\DM/\nu^2$. The relative motions of the Earth, pulsar, and ISM cause the sampled free electron content to vary, requiring us to estimate a DM for every observing epoch. These dispersive delays are also covariant with other propagation effects and the frequency-dependence of the pulse shape itself. While dispersive delays are accounted for in our timing models, other time-dependent propagation effects are not.

Multi-path propagation through the ISM causes extra time delays that may vary differently from the $\nu^{-2}$ dispersive delay along with distortions to the pulse shape that can cause mis-estimation of the TOAs in a variety of ways. The most visible effect is  scattering, manifesting as a broadening of the pulse shape mathematically described as a convolution with a pulse broadening function (PBF). The PBF has an approximately exponential tail, with the broadening scaling approximately as $\nu^{-4.4}$, though the value of the index is highly dependent on the physics and geometry of the medium. In the low-scattering regime, the primary impact of scattering is to delay the pulse with the roughly same frequency scaling \citep{hs08}. However, several of the NANOGrav pulsars show prominent scattering tails, especially those at high DM and/or those observed at our lowest radio frequencies.

Another frequency-dependent effect results from
the ray paths at different radio frequencies traversing different sets of electrons,  resulting in slightly different DMs at different  frequencies \citep{css16}. 
At sufficiently high levels of precision, this will require a frequency-dependent characterization of the DM at each epoch, currently outside the scope of our analysis.

{\it Refraction} also results from multi-path propagation and the deflection of the bulk set of rays through the ISM, causing the observed sky position of the pulsar  to vary. This results in an additional geometric delay to the pulses proportional to $\nu^{-4}$, with an additional delay caused by the variation in the angle-of-arrival proportional to $\nu^{-2}$ \citep{fc90}, the latter entirely covariant with the dispersive delay. 

{\it Scintillation} arises because pulsar images are extremely angularly compact, and the multiple ray paths interfere coherently with one another. The effect is most easily visualized observationally as heavy modulation of the dynamic spectrum, or the intensity of the pulsar as a function of frequency and time.
Normally referred to as diffractive interstellar scintillation (DISS), the process at a particular epoch can be characterized by a coherence bandwidth, \deltanu, and a coherence timescale, \deltat, with both quantities inversely proportional to pulsar distance.
DISS changes the PBF stochastically, resulting in a limit to the fundamental precision of our TOAs from propagation effects (versus intrinsically at the neutron star for jitter), with larger uncertainties at lower frequencies \citep[see e.g., in][]{lam+18optfreq}. When the total observation time, $T_{\rm obs}$, and total bandwidth, $B$, are very large with respect to \deltat\ and \deltanu , respectively, the effect of DISS on the timing is reduced. 
However, most NANOGrav MSPs are nearby, leading to  coherence  timescales of the same order or longer than our approximately 30-minute observation lengths and coherence bandwidths only factors of a few smaller than total observing bandwidths. This results in a ``finite scintle'' effect. This  is independent of pulse signal-to-noise (S/N) to first order \citep{1990ApJ...349..245C,lam+2016}, making the root-mean-square (RMS) timing noise due to scintillation  covariant with jitter in that respect. However, its  strong dependence on frequency allows it to be disentangled. In addition, measurements of the scintillation parameters can be used as priors on the analysis, though these are not applied to our current analyses. 

\subsection{Solar Wind Effects --- Chromatic (Spatially Correlated) } \label{subsec:solar_wind}

The solar wind is a stream of charged particles which escape from the solar corona due to their high ($\sim$1\,keV) energies \citep{marsch06}. This leads to an ionized IPM with an electron number density that decays with increasing distance from the Sun. Throughout the Earth's orbit the line of sight to a given pulsar cuts through different parts of the IPM,  resulting in an annual contribution to the pulsar's DM. The DM variations caused by the solar wind are subsumed in the generic DM variation model, DMX, used by NANOGrav; however, this excess DM can be modeled by assuming a spherically-symmetric electron number density as a function of Sun-pulsar separation angle $\theta$ \citep{edwards+06}
\begin{equation}
  \text{DM}_{\text{sw}}(\theta)=(4.85\times10^{-6} \textrm{pc})\, n_{0} \frac{\theta}{\sin\theta} ,
  \label{eq:DMsw}
\end{equation}
where $n_{0}$ is the nominal electron number density at a distance of 1~au, typically assumed to be several particles per cm$^{3}$ in pulsar timing codes and measured to be in the same range (e.g., \citealp{madison+2019,hazboun+2022}). The maximum amplitude (i.e., at the smallest angular separation from the Sun) of this periodic contribution to the DM depends on the pulsar's angle from the ecliptic plane, with pulsars very close to the ecliptic (latitude $| \beta | \lesssim 10^\circ$) showing very sharp peaks around the minimum solar separation \citep{jml+2017, donner+20}. The solar wind therefore represents both a chromatic and spatially-correlated signal \citep{thk+2016} among pulsars in a timing array. 

However, the solar wind is neither spherically symmetric nor static over the course of the 11-year solar cycle (e.g., \citealp{issautier01}), making the simple spherical model in Equation~\ref{eq:DMsw} inadequate to describe the solar wind contribution to DM for pulsars at our DM precisions \citep{madison+2019, tvs+19, hazboun+2022}. The model also  does not differentiate between the densities of the `fast' and `slow' solar wind \citep{tsb+21}, leading to large annual changes in amplitude. As shown in \citet{hazboun+2022} our current practice of measuring and removing the DM delay on a per-observation basis allows us to account for the changing DMs along each line of sight except in extreme circumstances where the Sun-pulsar separation angle is extremely small and dual-receiver observations are not available.
NANOGrav removes data for which the DM variation is likely to be too rapid to reliably represent it with DMX segments (see \S4.1 in \citetalias{NG15} for more details), but retains the raw data from  observing pulsars near the cusp of closest line of sight approach to the Sun for solar wind studies.

\subsection{Solar System Ephemeris --- Achromatic (Spatially Correlated)}
TOAs are measured at an observatory and then  must be  transformed to the quasi-inertial reference frame of the Solar System barycenter (SSB) (see e.g., \citealp{lorimer_and_kramer05} for full details). This  is done through the use of a planetary ephemeris, notably the Development Ephemeris (DE; \citealp{standish82}) maintained by NASA's Jet Propulsion Laboratory, most recently DE441 \citep{park+21}, and the Observatoire de Paris-maintained INPOP ephemerides \citep{fienga+09}, most recently INPOP19a \citep{fienga+19}. The dominant source of uncertainty in these ephemerides arises from inaccurate measurements of planetary masses, particularly the outer planets, each of which contributes a sinusoid to the timing residuals with period equal to the planetary orbit. This signature will be  spatially correlated among pulsars, roughly as a dipolar signal, in a PTA \citep{thk+2016}. The orbital periods of the giant planets range from 11.9\,--\,164.8\,yr, corresponding to frequencies of $\sim0.4$\,--\,2.7\,nHz, similar to the GWB frequency range that is probed by PTAs. The impact on PTA analyses is modeled by perturbing the orbital parameters in the ephemeris used to derive PTA noise limits and comparing with other models, as detailed in, e.g., \citet{vallisneri+20} and  \citet{chen+21}. The dependence of the GW statistics in NANOGrav's 11-yr data set on the version of the ephemeris used was the catalyst for the development of these methods \citep{abb+2018gwb}. These models allow the analyses to be bridged between different versions of the solar system ephemeris to obtain equivalent results. \citet{abb+2020gwb} and \citet{vallisneri+20} showed a diminishing dependency on the choice of ephemeris for PTA GWB results and all tests of different ephemeris versions on \citetalias{NG15}, including using a Bayesian ephemeris model \citep{vallisneri+20}, have shown insignificant differences in parameter recovery. In addition, PTA data sets have been used to identify non-GW common noise signals and have enabled limits to be placed on the masses of the outer planets, large asteroids, and undiscovered planets in the outer Solar System \citep{champion2010, caballero+18, guo+19}.

\subsection{Clock Errors --- Achromatic (Spatially Correlated)}

Pulsar data are referred to an observatory time standard when they are produced, but are transformed into Barycentric Dynamical Time (TDB) for timing calculations and comparison across data sets \citep{luo+2021,tempo2}. Observatory clocks are not perfect and so corrections are applied at various stages of this transformation. See \citet{luo+2021} for a detailed discussion of clock corrections and transformations used in pulsar timing. Any errors in these transformations, or drifts between time standards, can manifest in pulsar data as the same offset in all data sets with the clock error \citep{hobbs2012,miles+2023}. These offsets are spatially correlated in the sense that all affected pulsars will be impacted by the same shift at the same time. The spatial correlation function is therefore monopolar, i.e., the same for all pulsar sky positions. See \citet{thk+2016} for a complete discussion of clock errors and other spatially correlated noise processes in pulsar timing data.

\subsection{Measurements at the Observatories --- Varying Chromaticity} \label{subsec:obs_noise}

Radiometer noise from the observing systems used at our telescopes dominates our measurement uncertainty for our lowest S/N pulsars but becomes a less significant component at higher S/N; we are radiometer-noise-dominated for most of our pulsars at most epochs.  Random fluctuations due to radiometer noise should be time and frequency independent. However, the system temperature, and thus the S/N, is chromatic as it depends on frequency-dependent contributions from the receiver bandpass, Galactic background, etc. We have shown that the uncertainties from our template matching procedure follow the expectations from radiometer noise for lower S/N (see Appendix B in \citealt{abb+15} and note that we correct for this effect in our more recent data sets) and deviate due to jitter and diffractive interstellar scintillation at the higher S/Ns \citep{lam+2016}, regardless of the frequency. Therefore, even though the system temperature will be frequency-dependent, our per-channel modeling of the template-fitting error should contain no known biases. 

Incorrect polarization calibration could be a source of error in our data.
For every observation we correct for differential gain and phase variations in the two chiralities of polarization of the received radio waves as well as the changing parallactic angle. The latter effect is easily calculated based on the source's apparent position whereas the former is corrected for based on measurements of a noise diode taken prior to each observation and an unpolarized quasar roughly once per month by which we correct for variations in the noise source power. Mis-calibration causes alterations to the polarization-summed profiles used in timing \citep{stinebring+1984} which would then be a chromatic source of noise given the frequency dependence of the (polarization) profiles. If polarization calibration is incomplete, for example when a calibrator source is not observed at a particular epoch, we would expect some contribution to chromatic WN.

Strong radio-frequency interference (RFI) is removed in several stages of our pipeline as discussed in \citetalias{NG15}. Low-level RFI will still remain and perturb the TOA estimates in the template-matching procedure. RFI can take many forms -- remaining narrowband RFI will perturb individual frequency channels but broadband RFI can perturb all of the TOAs across the band. RFI can also be impulsive or periodic. Both can be removed from individual sub-integrations if strong enough to identify. If not, in either case RFI will be folded at the pulse period and diminished, but still affect the TOAs as a source of possibly-chromatic WN.

\section{Phenomenological Noise Model} \label{s:phenom_noise}

Independently fitting for each and every source of noise identified in \mysec{s:physical_noise} across all pulsars in the PTA would be extremely challenging. In order to efficiently characterize the noise of our detector, and allow noise modeling from unknown sources, we therefore use a phenomenological model that takes these effects into account while reducing the number of fit parameters dramatically. The model has two main parts, distinguished by the timescale and spectral characteristics of the noise being modeled. The first part models the white, or uncorrelated in time, noise that has a ``flat'' contribution to the  power spectra of timing residuals. We model this noise using three parameters which increase the uncertainties on the TOAs by accounting for WN unaccounted for in the template fitting process \citep{abb+15}. The second part uses Gaussian process regression \citep{rw06} to model low-frequency, time-correlated, or red, stochastic processes in the data. In theory, both of these models could be included in the covariance matrix for the TOA data, however, it is more numerically expeditious to separate out the Gaussian process formalism for real data analyses. See \mysec{s:covmatrix} for details about how this is accomplished.

\subsection{White Noise Model}\label{s:wn}

The uncertainties that are initially associated with pulsar TOAs $\sigma_{\rm S/N}$ are due to the finite S/N of the matched-filtering process used to calculate them \citep{taylor1992}. A template of the pulse shape, built from many observations, is convolved in the Fourier domain with the summed pulse profile from a single observing session, under the assumption that the data comprise a scaled and shifted version of the template added to WN \citep{2013CQGra..30v4001L}. However, there are other sources of WN (e.g., jitter, scattering, RFI, etc.; see Section~\ref{s:physical_noise}) that do not adhere to the assumptions of matched filtering (i.e., that the pulse shape is a copy of the template). These various noise effects cumulatively induce variance in the TOAs, and it is nontrivial to disentangle the distinct contributions of each noise source to the total TOA uncertainties. We therefore employ an empirical WN model for pulsar timing data that inflates the $\sigma_{\rm S/N}$ measured from the pulse template-matching process using three parameters.

Three WN parameters are used to adjust the TOA uncertainties in order to accurately reflect WN present in the data. This process yields a reduced-$\chi^2$ near unity for the fit to the timing residuals, if the timing model is complete and accurate. Various differences between pulsar timing backends and radio observatory receivers make it necessary to give different values of these WN parameters to each receiver/backend combination. These parameters thereby encode the trustworthiness of TOAs from each receiver/backend combination, down-weighting the TOAs from combinations where effects in addition to template matching reduce the reliability of the data. These three WN terms -- \efac ($\ef$), \equad ($\eq$) and \ecorr ($\ec$) --   come together with receiver/backend combination $re/be$ dependence as
\be\label{eq:sigmaTotal}
C_{ij} = \ef^2(re/be)[\sigma_{{\rm S/N},i}^2+\eq^2(re/be)]\; \delta_{ij}+ \ec^2(re/be)\;\mathcal{U}_{ij} 
\ee
where the $i,j$ denote TOA indices across all observing epochs,  $\delta_{ij}$ is the Kronecker delta and we omit the dependence on receiver and backend, $re/be$, from here on for simplicity. While \efac and \equad only add to the diagonal of $C$, where $C_{ij}$ are the elements of the covariance matrix to be discussed in \mysec{s:covmatrix}, the \ecorr terms are block diagonal for single observing epochs. \ecorr is modeled using a block diagonal matrix, $\mathcal{U}$, with values of $1$ for TOAs from the same observation and zeros for all other entries.

Historically, the \efac (error factor, $\ef$) parameter was the first WN parameter added to the pulsar timing covariance matrix. \efac is a scale factor on the $\sigma_{\rm S/N}$. The increase in uncertainties from EFAC attempts to account for underestimated template matching errors from low S/N ratio observing epochs  and template mismatches due to pulse profile variability. This parameter has tended towards $1$ as pulsar backends have improved, more high dynamic range (i.e., 8 or more bit sampling) digital systems have been implemented, and the treatment of low S/N TOAs has improved \citep{aab+20nb,aab+20wb}.

To include additional WN, the EQUAD ($\eq$) parameter\footnote{Note that there are two distinct definitions of \equad in the literature, depending on whether \efac multiplies only $\sigma_{\rm S/N}$, referred to as the \textsc{TempoNest} convention for the paper \citep{lah+14} where it was first used, or whether it multiplies the sum, in quadrature of $\sigma_{\rm S/N}$ and \equad. All three of the main pulsar timing packages and the \enterprise software stack use the latter convention, laid out in \myeq{eq:sigmaTotal}.} (first used in \citealt{nice+1995}) is added in quadrature following the usual rules of noise propagation. \equad  encompasses additional sources of WN not accounted for in the TOA uncertainties, and not modeled by \efac.

The most recent WN parameter added to the pulsar covariance matrix is \ecorr (error correlated in \emph{radio} frequency, $\ec$), which is also added in quadrature and specifically models noise correlated across radio frequencies within a single observing epoch\footnote{Here an observing epoch is effectively one observation with a single receiver. Multiple observations within a single MJD are not treated as correlated.}. \ecorr was first used in \citet{abb+15} to mitigate noise correlated across the narrowband TOAs obtained for a single observing epoch. In part this parameter became necessary due to NANOGrav's data acquisition strategy of using multiple TOAs across the full observing band for chromatic mitigation. While \ecorr is not strictly ``white noise'' from the perspective of being only diagonal in the covariance matrix, its spectral contribution is effectively white, i.e., constant in the frequency domain, for the frequencies important to GW searches. The correlation timescale modeled is only across pulses emitted from the pulsar within a relatively short ($\sim 30 \;{\rm min}$) observation. ECORR partially accounts for intra-band correlations caused by pulse jitter, but also includes other short-timescale correlations across radio frequency, which can be produced by, e.g., RFI and time-variable scattering \citep{lam+2019, shapiro-albert+2021}.

\subsection{Red Noise Model}\label{s:rn}

Various sources of noise in \mysec{s:physical_noise}, such as variations in the electron density of the ISM, instabilities in the spin of the pulsars, corrections to Earth-based clock systems, corrections to the solar system ephemeris model, and, of course, TOA shifts due to the stochastic GWB are time-correlated across long timescales. The characterization of these effects in the frequency domain, e.g., a power spectral density, shows that the effect has more power at lower (redder) frequencies, hence RN. In most cases the theoretical models for these sources of noise have a power spectral density $P(f)$ that follows a power law with frequency $f$  \citet{cs10}, $P(f)\sim A^2 f^{-\gamma}$, with amplitude $A$ and where the ``redness'' is dictated by the explicit minus sign in the exponent and $\gamma>0$. More complex models have been investigated, such as turnover models, but the straight power law is still considered an accurate and effective model for MSPs \citep{goncharov+2020}. 

There are two methods by which RN can be included in a pulsar timing analysis. One way is to use the Wiener-Khinchin theorem to write the correlations between different TOAs, $(t_i,t_j)$,
\be \label{eq:wiener-kinchin}
C^{\rm RN}_{ij}=\int_0^{f_{\rm Nyq}} {\rm d}\!f \cos[2\pi f (t_i-t_j)] P(f), 
\ee
where $C^{\rm RN}_{ij}$ are the elements of the covariance matrix as noted above and $f_{\rm Nyq}$ is the Nyquist frequency. Note that, strictly speaking, the integral in \myeq{eq:wiener-kinchin} does not converge for all forms of $P(f)$, in which case a low-frequency cutoff must be used, see \citet{vanhaasteren+2009} for an exhaustive discussion. The details of the RN model are discussed more in \mysec{s:covmatrix}. 

Alternatively, the perturbations of the TOAs due to RN can be modeled directly with a Fourier basis and a set of coefficients, $F\mathbf{a}$, where $F$ is an $(N_{\rm TOA} \times 2N_{\rm freq})$ Fourier design matrix, and $\mathbf{a}$ is a $2N_{\rm freq}$ vector with 2 coefficients for each of $N_{\rm freq}$ frequencies included. As will be discussed in \mysec{s:covmatrix}, the form of $P(f)$ can be dictated by imposing various functional forms on the $\mathbf{a}$ coefficients. This method is at the heart of PTA searches for the GWB \citep{Lentati:2016ygu,lentati+2013,vHv:2014} and achromatic (in radio frequency) RN models used for individual pulsars. In practice, the methods of Gaussian process regression are used in the analyses to accurately include the stochastic nature of these signals, and the explicit Fourier basis-modeling will eventually be incorporated into the covariance matrix, but in a different form, see \mysec{s:covmatrix}. It is these analyses that are used to fit for RN parameters used in the covariance matrix for pulsar timing packages. While NANOGrav uses the DMX model \citepalias{NG15} in this analysis, these same Gaussian methods are used by other PTA collaborations to model the variations in DM by including a dependence on the radio frequency \citep{Lentati:2016ygu}. 

\section{The NANOGrav Pulsar Covariance Matrix}\label{s:covmatrix}

PTA data analysis is done using the timing residuals $\delta \mathbf{t}$, produced by subtracting times of arrival predicted by a timing model $\mathbf{t}_M$ from the TOAs $\mathbf{t}$:
\be\label{eq:residuals}
\delta \mathbf{t} =\mathbf{t}-\mathbf{t}_M=\mathbf{t}_D-\mathbf{t}_M + F \mathbf{a} + \mathbf{n},
\ee
where the bolded symbols are column vectors. The difference between the actual underlying, deterministic, non-GW delays and the timing model, $\mathbf{t}_D-\mathbf{t}_M\equiv M \delpar$, is represented by a linear-order Taylor series in the (assumed small) perturbations to the model parameters, $\delta \mathbf{p}\equiv\delpar$, and the design matrix, $M$, where
\be\label{eq:design_matrix}
M_{ij}\equiv \left(\frac{\partial t_{M,i}}{\partial p_j}\right)_{\mathbf{p} = \mathbf{p}_0}
\ee
is evaluated (usually analytically, numerically otherwise) at the best-fit parameters, $\mathbf{p}_0$, from a linear least squares analysis. The use of the linearized timing models was introduced in \citet{esvh13} and is especially expeditious for full PTA analyses. However, full timing model fits can also be done as part of the Bayesian searches \citep[][Kaiser et al. in prep]{lah+14,vigeland+2014}. The $F \mathbf{a}$ term encodes the long-timescale correlated (red) noise, both for individual pulsar data sets and the GWB, see \mysec{s:phenom_noise}. Finally, $\mathbf{n}$ is the WN remaining in the residuals, assumed to have a multivariate Gaussian distribution,
\be\label{eq:noise_gaussian}
\mathcal{P}(\mathbf{n})=\frac{\exp\left(-\frac{1}{2}\mathbf{n}^\mathrm{T} C^{-1}\mathbf{n}\right)}{\sqrt{\det\left(2\pi C\right)}},
\ee
with covariance $C(\mathbf{E})\equiv \langle\mathbf{n},\mathbf{n}^\mathrm{T}\rangle$, which is a function of the set of WN parameters, e.g., $\mathbf{E}\equiv(\ef,\eq,\ec)$. We construct the likelihood function for a PTA by combining the timing residuals, $C$, and \myeq{eq:noise_gaussian}:
\begin{widetext}
\be \label{likelihood} 
\mathcal{P}(\resid|\delpar, \mathbf{a}, \mathbf{E}) = \frac{\exp\left(-\frac{1}{2}(\resid-M\delpar-F\mathbf{a})^\mathrm{T} C^{-1}(\resid-M\delpar-F\mathbf{a})\right)}{\sqrt{\det\left(2\pi C\right)}}\;.
\ee
\end{widetext}

When conducting a search for GWs in pulsar timing data we are not concerned with the values of the timing parameter perturbations included in $\delpar$. Using a Gaussian distribution for the prior probability distribution of the parameter perturbations, 
\be\label{eq:tm_gaussian}
\mathcal{P}(\delpar)=\frac{\exp\left(-\frac{1}{2}\delpar^\mathrm{T} X^{-1}\delpar\right)}{\sqrt{\det\left(2\pi X\right)}},
\ee
and defining $X\equiv \langle\delpar,\delpar^\mathrm{T}\rangle$, one can marginalize over the timing model parameter perturbations (see Appendix \ref{app_marginalize})
\ba\label{eq:tm_marginalize}
\mathcal{P}(\resid|\mathbf{a}, \mathbf{E}) &=& \int \mathcal{P}(\resid|\delpar, \mathbf{a}, \mathbf{E})\mathcal{P}(\delpar){\rm d}(\delpar)\\
&=& \frac{\exp\left(-\frac{1}{2}(\resid-F\mathbf{a})^T D^{-1}(\resid-F\mathbf{a})\right)}{\sqrt{\det\left(2\pi D\right)}},
\ea
where $D\equiv C+MXM^\mathrm{T}$ (Eq.~\ref{eq:D_matrix}).

In a similar fashion the Fourier basis coefficients are also marginalized over assuming that the RN, whether intrinsic to the pulsar or a GWB, is a Gaussian process,
\be\label{eq:a_gaussian}
\mathcal{P}(\mathbf{a})=\frac{\exp\left(-\frac{1}{2}\mathbf{a}^\mathrm{T} \varphi^{-1} \mathbf{a}\right)}{\sqrt{\det\left(2\pi \varphi\right)}},
\ee
where $\varphi\equiv \langle\mathbf{a}, \mathbf{a}^\mathrm{T}\rangle$ is a $(2N_{\rm freq}\times 2N_{\rm freq})$ diagonal matrix with dimensions dictated by the number of frequencies, $N_{\rm freq}$, used in the analysis and describing the variance of the RN coefficients. In principle these coefficients could be used as free parameters, however they are usually parameterized as a power law,
\be \label{eq:phi_red}
\varphi_i = \frac{A^2}{12\pi^2} \left(\frac{f_i}{f_{\rm yr}}\right)^{-\gamma}\;{\rm yr}^{3} {\rm d}\!f \sim \frac{A^2}{12\pi^2 \; T_{\rm span}} \left(\frac{f_i}{f_{\rm yr}}\right)^{-\gamma}\;{\rm yr}^{3},
\ee
or some other functional form for the power spectral density. Note that $f_{\rm yr} = 1~\mathrm{yr}^{-1}$, and the $\sim$ assumes the usual constant $\mathrm{d}f\sim1/\Tspan$, where $\Tspan$ is the total timespan (years) of the data set.  Marginalizing over the $\mathbf{a}$ coefficients gives the final form for a single pulsar likelihood,
\ba\label{eq:a_marginalize}
\mathcal{P}(\resid|\mathbf{E}) &=& \int \mathcal{P}(\resid|\mathbf{a}, \mathbf{E})\mathcal{P}(\mathbf{a}){\rm d}(\mathbf{a})\nonumber\\
&=& \frac{\exp\left(-\frac{1}{2}\resid^\mathrm{T} N^{-1}\resid\right)}{\sqrt{\det\left(2\pi N\right)}},
\ea
where $N\equiv D +F \varphi F^\mathrm{T}$, again see Appendix~\ref{app_marginalize} (Eq.~\ref{eq:N_matrix}).

In a single pulsar noise analysis using this likelihood, one searches over all of the parameters $(A,\gamma, \mathbf{E})$; however, the WN parameters are fixed to their maximum likelihood values during full PTA GW searches. The WN parameters are independent of the parameters varied in the GW search, and the necessary matrix inversion\footnote{For further mathematical discussion of this inversion, see Appendix \ref{app_marginalize}.} is prohibitively expensive from a computational standpoint. The marginalizations above were developed to reduce this computational expense to a single inversion of a $(2N_{\rm freq}\times 2N_{\rm freq})$ matrix \citep{vHv:2014}.

\subsection{Phenomenological Covariance Matrix} \label{subsec:phenommatrix}
As noted above, the phenomenological WN model is a function of the parameters $\mathbf{E}\equiv(\ef,\eq,\ec)$, which together account for the noise processes the NANOGrav-observed pulsars; see \mytab{table:noise}.
Collectively, the WN in each of our pulsar timing models can be expressed as a covariance matrix where each of $N_{\rm TOA}$ TOAs is specified with both its recorded time $t$ and its radio frequency $\nu$. This covariance matrix is then fit globally over all epochs and over all a pulsar's frontend-backend combinations. For a set of TOAs on a given epoch, we write the noise model in the form
\be \label{covmatrix} C^\mathrm{epoch} =
\begin{bmatrix}
\ef^2[\sigma_{\rm S/N}^2(\nu_1) + \eq^2] +\ec^2 & \ec^2 & \cdots & \ec^2\\ 
\ec^2 &\ef^2[\sigma_{\rm S/N}^2(\nu_2) + \eq^2] +\ec^2 & \cdots & \ec^2\\ 
\vdots & \vdots & \ddots & \vdots \\
\ec^2 & \ec^2 & \cdots & \ef^2[\sigma_{\rm S/N}^2(\nu_N) + \eq^2 ] +\ec^2
\end{bmatrix},
\ee
where $\ef$, $\eq$, and $\ec$ are the \efac, \equad, and \ecorr described in \mysec{s:phenom_noise}. This matrix is block diagonal, with the entries of $\ec^2$ connecting frequency channels from the same observations in blocks. The WN covariance matrix can then be written down as 
\be \label{eq:phenom_noise}
C =
 \begin{bmatrix}
C^\mathrm{epoch}(t_1) & 0 & \cdots & 0\\ 
0 & C^\mathrm{epoch}(t_2) & \cdots & 0 \\ 
\vdots & \vdots & \ddots & \vdots \\
0 & 0 & \cdots & C^\mathrm{epoch}(t_N)
\end{bmatrix},
\ee

\noindent
where the noise is independent between epochs, but the noise parameters are fit globally over all TOAs from a single pulsar. If one wants to include the RN in the covariance matrix the total noise would then be $C\rightarrow C + C^{\rm RN}$. This approach is used in pulsar timing software when  the best-fit amplitude and spectral index for RN in a given pulsar have already been fit for in an earlier analysis \citep{coles+2011}. 

It is tempting to try and map the phenomenological noise model to a signal covariance matrix made up of all of the effects from \mysec{s:physical_noise}. However, in practice, there are a few reasons for not doing so in a GW analysis. First of all, the complexity of these models makes this task unwieldy for GW searches, especially given that some of these effects are not directly measurable. Second, the noise measured through the phenomenological model is known to be larger than the sum of the various physical effects, so a signal based covariance matrix might not include all of the noise in our detector. In \mysec{app_signal_matrix} we sketch a signal covariance matrix to show how some of the terms from \mysec{s:physical_noise} would enter into the analysis.

\section{PTA Noise Budget}\label{s:PTA_nb}

\subsection{Comparison of Physical and Phenomenological Models}

The noise in \mysec{s:physical_noise} enters into the phenomenological noise model through a diverse set of physical processes and any particular physical effect may enter more than one part of the phenomenological model. From the standpoint of detecting GWs with PTAs the most salient aspect of this complex relationship is that the phenomenological model includes the known physical sources of noise and is pliable enough to include unknown sources as well. 

That pliability is demonstrated in \mytab{table:noise} where we provide a list of the various noise sources from \mysec{s:physical_noise}, and relationships between the phenomenological parameters in which they are included, along with a number of other descriptors that can be used to sort the sources of noise.

\begin{deluxetable}{lccccc}\label{table:noise}
\tablecaption{Connecting Noise Terms: Astrophysical noise sources, along with a summary of their attributes and connections to the phenomenological noise model, are listed. Short timescales are roughly the order of the observation length or less, while long timescales refer to periods between observation epochs.}
\tablehead{
\colhead{Noise Type} & 
\colhead{Symbol} & \colhead{Origin} & 
\colhead{Phenomenological Model Component} & \colhead{Timescale} &  \colhead{Spatially Correlated}
}
\startdata
Radiometer & $\sigma_{\rm S/N}$ & Telescope & $\sigma_{\rm S/N}$ & Short &  \\
Jitter & $\sigma_{\rm J}$ & Pulsar & $\ec$ & Short &  \\
Diffractive Interstellar Scintillation & $\sigma_{\rm DISS}$ & ISM & $\eq$, $\ec$ & Short &  \\
DM Mis-estimation & $\delta\DM$ & ISM & RN/DMX & Long &  \\
Solar Wind DM Mis-estimation & $\delta\DM_\mathrm{sw}$ & IPM & RN/DMX & Long & $\checkmark$  \\
Frequency-dependent DM & $\DM(\nu)$ & ISM & RN/DMX & Long &   \\
RFI & & Telescope & $\ec$ & Short &  \\
Polarization Mis-calibration & & Telescope & $\ec$ & Both &  \\
Scattering & & ISM & $\ef, \ec$ & Short & \\
Solar System Barycenter Mis-modeling & & Solar System & RN & Long & $\checkmark$ \\
Clock Errors & & Telescope & RN & Long & $\checkmark$ \\
\enddata

\end{deluxetable}

\subsubsection{DM Mis-estimation due to Asynchronous Measurements}

In order to estimate DM on each epoch, we make TOA measurements across a wide frequency range. We used two or more receiver bands at one or more telescopes to cover the total range. Data were recorded with each receiver independently, with   observations occurring within 30 minutes of each other at Arecibo but within several days at Green Bank due to efficiency constraints in changing the receivers. DMs can change rapidly (i.e., on timescale of a day or less) due to the passage of the line of sight through the inhomogeneous ISM, changes in the solar wind, etc. 
In the presence of this stochastic variation, \citet{lcc+15} estimated the effect of asynchronous measurement, finding an induced perturbation on the infinite-frequency arrival time with a shallow ($\gamma = 2/3$ for a Kolmogorov medium) RN spectral index.
The RMS error is given as (see Eq.~14 in \citealt{lcc+15})
\be 
\sigma_{\delta t_\infty} \approx 6.5~\mathrm{ns} \left(\frac{\nu}{\mathrm{GHz}}\right)^{-2} \left(\frac{r^2}{r^2-1}\right) \left(\frac{t_{\rm sep}}{\Delta t_{\rm d,1~GHz}/1000~\mathrm{s}}\right)^{5/6}.
\ee
This expression assumes a simplified model where pulsar observations are made at only two spot frequencies $\nu$ and $\nu'$,
where $r \equiv \nu/\nu'$ with $\nu > \nu'$, $t_{\rm sep}$ is the time spacing between the observations, and $\Delta t_{\rm d,1~GHz}$ is the scintillation timescale at a radio frequency of 1~GHz. 
For pulsars observed at Green Bank with $r = 1.5~\mathrm{GHz}/0.8~\mathrm{GHz} = 1.875$, a time between observations (in days) of $t_{\rm sep} \approx$ 2, and a typical scintillation timescale of approximately 1000~s at 1 GHz, we see that the RMS error from this effect is approximately 7 ns. These errors are therefore generally smaller than the intrinsic RN we measure for most pulsars.
For PSR J1713+0747, observed with all three telescopes, reducing the impact of rapidly-varying DM requires the window over which we fit a single DM to  be  shorter than typical (e.g., $\lesssim$1~day compared to six days) to avoid a larger DM mis-estimation error.

\subsubsection{DM Mis-estimation due to Additional Chromatic Effects}

There are a host of unmitigated chromatic effects in our timing model, though many or all are expected to be small except for specific lines of sight. 
Scattering is expected to be the next dominant effect.
It is highly frequency dependent, broadening the width of the PBF of the interstellar medium by an amount that is $\propto \nu^{-4.4}$ for a Kolmogorov medium.
When this scattering delay, $\tscatt \ll W$, where $W$ is the pulse width, the first-order effect is to delay the TOA by \tscatt.
However, when $\tscatt$ is comparable to $W$, there is distortion of the pulse shape and the time delay becomes a non-linear function of the scattering delay \citep{hs08}.
 We are likely only in this regime for one NANOGrav pulsar, J1903+0327, with a spin period of 2.2~ms and scattering delay of $\sim$150~$\mu$s at 1.5~GHz \citep{memo8}.

As discussed in \S\ref{subsec:propagation},
DM itself is a frequency-dependent quantity \citep{css16}.
Again considering a simplified model where pulsar observations are made at only two spot frequencies $\nu$ and $\nu'$,
we can write the TOA perturbation as 
\be 
\delta t_\infty = \frac{-r^2 t_{C,\nu} - r^2 \epsilon_\nu + t_{C,\nu'} + \epsilon_{\nu'}}{r^2 -1},
\ee
where $t_{C,\nu}$ and $t_{C,\nu'}$  are non-dispersive chromatic delays at frequencies $\nu$ and $\nu'$, and $\epsilon_{\nu}$ and $\epsilon_{\nu'}$ are additive, frequency-dependent errors at frequencies $\nu$ and $\nu'$. The additional error is uncorrelated between different frequencies for radiometer noise but is highly correlated for intrinsic pulse jitter, at least over modest frequency separations (see below).
The RMS timing error is found by calculating the square root of the variance.
Even when there is no WN, i.e., $\epsilon_\nu = \epsilon_{\nu'} = 0$, there is a non-zero perturbation that affects our timing. Frequency-dependent DM affects us at the 10s of ns level \citep{lam+18optfreq} for most pulsars at the frequencies we observe, though for a few at the highest DMs, that number can be substantially higher.

\subsubsection{Accounting for Jitter}
As noted in \mysec{s:phenom_noise}, the ECORR parameter was added to the pulsar signal covariance matrix to account for frequency-correlated WN.
 When examined over NANOGrav's 12.5-yr data set pulsars, we find that a global fiducial estimate for the RMS uncertainty from jitter is \citep{lam+2019}
\be
\sigma_{\rm J} \approx 60~\mathrm{ns}\left(\frac{P_{\rm s}}{5~\mathrm{ms}}\right)^{3/2} \left(\frac{T_{\rm obs}}{30~\mathrm{min}}\right)^{-1/2}
\ee
for a pulsar with a $P_{\rm s} = 5$-ms spin period and a $T_{\rm obs} = 30$~minute observing time, if we take the 1500 MHz estimate as a representative value. This uncertainty level makes up a non-negligible fraction of our known WN budget. In its general use, for narrowband timing, the ECORR estimation is higher than the physically-derived constraints in \citet{lam+2019} \citep[see also the comparison between ECORR and jitter in][]{lam+2016}, which prevents potential underestimation of the frequency-correlated noise . Likewise, the in-depth study of jitter in PTA pulsars revealed that jitter typically decorrelates given a sufficiently broad frequency range \citep{2014MNRAS.443.1463S}. This decorrelation bandwidth is currently not incorporated into the ECORR parameter because current receivers are narrow enough to avoid decorrelation; however, with the implementation of wideband receivers in the near future (see \mysec{s:future}), this decorrelation bandwidth may fall within the band of these new receivers. 

\subsubsection{The Impact of Polarization Mis-calibration}

For any given narrowband profile, polarization gain mis-calibration will cause a TOA uncertainty dependent on the fractional gain error $\varepsilon$ and the degree of circular polarization $\pi_{\rm V}$ for a pulsar with pulse width $W$ as \citep{2004NewAR..48.1413C}
\be 
\sigma_{\rm pol} \approx 1~\mu \mathrm{s}~\left(\frac{\varepsilon}{0.1}\right)\left(\frac{\pi_{\rm V}}{0.1}\right)\left(\frac{W}{100~\mu \mathrm{s}}\right).
\ee
The fiducial values given are representative of the NANOGrav pulsars and data \citep{lam+18optfreq} though the $\pi_{\rm V}$ and $W$ values can still vary quite widely \citep[e.g.,][]{2006ApJ...642.1004V,Gentile+2018}. Cross coupling induces a false circular polarization that adds to the measured circular polarization. While a seemingly large value, the uncertainty is on the narrowband TOAs and not the infinite-frequency arrival time (e.g., the epoch-averaged TOAs), and so the value should be reduced by the square root of the number of TOAs, nearly an order of magnitude reduction. Polarization mis-calibration uncertainties in the NANOGrav data set have not been fully explored but are expected to add a of order 100~ns uncertainty to our excess noise budget.

\subsection{Red Noise Analysis}\label{s:rednoise}
Since the strongest GW signal expected in PTAs manifests as a long-time correlated stochastic signal, understanding RN in pulsar timing data sets is of the utmost importance for mitigating false positives and/or parameter mis-estimation. The use of power-law RN models in pulsar timing analyses is ubiquitous and predates modern PTA data analysis techniques \citep{blandford+1984,cordes1985}. There are a number of more complex models for RN in PTA data sets \citep{hazboun:2020slice,leg+2018,Chalumeau+2022,goncharov+21a}, but we will demonstrate shortly that the power-law model is a reasonable and effective choice by carrying out an additional noise analysis.

\subsubsection{Accounting for Red Noise}\label{s:account_rn}
Searches for GWs in PTA data sets start with single-pulsar noise analyses and then move on to full analyses that include all of the pulsar data sets. In the case of the GWB this involves a correlated search across all pulsar data sets. The main reason for the individual noise analyses is to search for the WN parameters described in section \mysec{s:phenom_noise}. They involve full $N_{\rm TOA} \times N_{\rm TOA}$ ($\mathcal{O}(10^4)$ narrowband TOAs per pulsar, and a factor of $\sim$30 reduction on average for the wideband data) matrix inversions and upwards of 21 parameters per pulsar. Hence they are too costly to do across all pulsars at once in a full Bayesian search. It was shown in early, smaller data sets \citep{abb+15,dfg+13} that the WN parameters do not change substantially when the analysis is done across other pulsars, as long as there is a RN model included in both analyses, since this model can substantially change the amount of WN in some pulsars. The presence of two RN analyses, one for individual pulsars and one in the full GWB search, allows us to do an accounting of the RN in each pulsar and keep track for which pulsars the RN power moves from the individual noise channel into the common signal. 

\begin{figure}[h!tbp]
\centering
\includegraphics[width=0.43\textwidth]{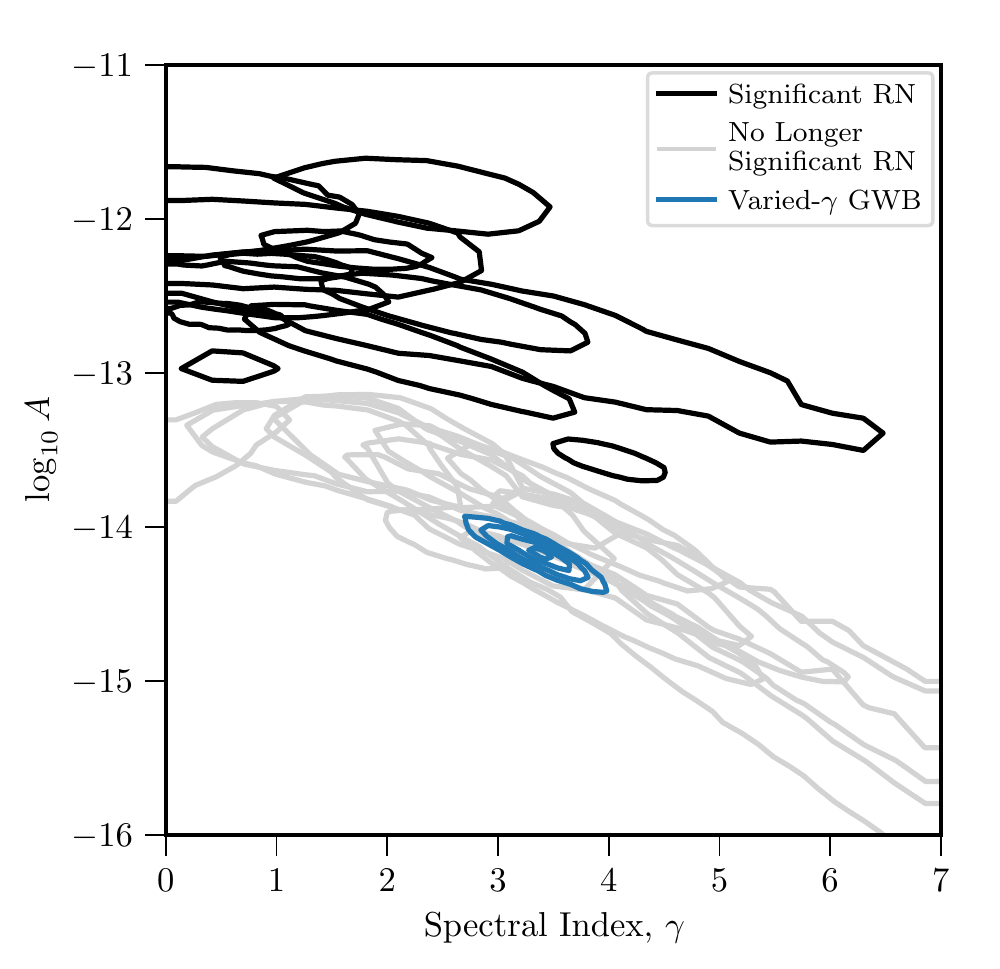}
\caption{Two-dimensional parameter posterior contours for the power-law RN models used in the pulsar noise models of the NG15 data set. The power spectral density is parameterized by a spectral index and amplitude. The black contours show the RN parameters of pulsars that have significant individual RN in both their single pulsar analyses and the full PTA analysis. The gray contours show RN parameters of pulsars that have significant  individual RN detections only in the single pulsar analyses, i.e., the power in the individual-pulsar model has moved into the common model. The blue concentric contours show the parameter recovery from a full PTA common process analysis \citepalias{NG15gwb}. Posteriors for pulsars without significant individual RN detections are not included for clarity.}
\label{f:rednoise_contours}
\end{figure}

In \mytab{t:rednoise} we show the pulsars that have significant RN detections in their individual noise analyses. Our criterion is the same as for \citetalias{NG15}; we consider the detection to be significant when the pulsar's RN has a Bayes factor $\mathcal{B}>100$, using the Savage-Dickey \citep{dickey1971} approximation, when possible. The italicized entries in \mytab{t:rednoise} denote the pulsars that no longer have significant detections of RN in the full PTA analysis, i.e., the power in the single pulsar RN model has moved into the common channel\footnote{Depending on the search this can either be a spatially-correlated process, as in searches for the GWB, or just a ``common'' process, described by only the power spectral density and not the spatial correlations. See \citetalias{NG15gwb}~for more details.}. The bolded entries denote the pulsars that continue to have significant RN detections.
\begin{table}[htp]
\caption{Individual RN model parameter values and 68\% credible intervals for pulsars with significant detections of RN. Pulsars with italicized entries only have significant detections in individual pulsar noise analyses while bolded entries show RN even when a common process is included. Note that $A_{\rm RN}$ is in strain amplitude to match the GW analyses, hence it is unit-less.}
\begin{center}
\begin{tabular}{|c|c|c|}
\hline
Pulsar 	& $\log_{10}A_{\rm RN}$  & $ \gamma_{\rm RN}$   \\ 
\hline
\textit{B1855+09} & $\mathit{-14.0_{-0.4}^{+0.3}}$ & $\mathit{3.9_{-0.8}^{+1.0}}$ \\
\textbf{B1937+21} & $\mathbf{-13.6_{-0.1}^{+0.1}}$ & $\mathbf{4.0_{-0.3}^{+0.4}}$ \\
\textbf{B1953+29} & $\mathbf{-12.8_{-0.3}^{+0.2}}$ & $\mathbf{1.8_{-0.7}^{+1.1}}$ \\
\textit{J0030+0451} & $\mathit{-14.4_{-0.5}^{+0.4}}$ & $\mathit{4.6_{-0.9}^{+1.1}}$ \\
\textit{J0437$-$4715} & $\mathit{-13.4_{-0.2}^{+0.2}}$ & $\mathit{0.5_{-0.4}^{+0.6}}$ \\
\textbf{J0610$-$2100} & $\mathbf{-12.9_{-0.5}^{+0.3}}$ & $\mathbf{4.1_{-1.9}^{+2.0}}$ \\
\textit{J0613$-$0200} & $\mathit{-13.8_{-0.3}^{+0.3}}$ & $\mathit{3.1_{-0.7}^{+0.9}}$ \\
\textbf{J1012+5307} & $\mathbf{-12.6_{-0.1}^{+0.1}}$ & $\mathbf{0.8_{-0.3}^{+0.3}}$ \\
\textit{J1600$-$3053} & $\mathit{-13.5_{-0.6}^{+0.2}}$ & $\mathit{1.6_{-0.7}^{+1.5}}$ \\
\textit{J1614$-$2230} & $\mathit{-14.9_{-0.8}^{+1.0}}$ & $\mathit{4.7_{-2.0}^{+1.6}}$ \\
\textbf{J1643$-$1224} & $\mathbf{-12.3_{-0.1}^{+0.1}}$ & $\mathbf{0.9_{-0.4}^{+0.4}}$ \\
\textbf{J1705$-$1903} & $\mathbf{-12.6_{-0.1}^{+0.1}}$ & $\mathbf{0.5_{-0.3}^{+0.4}}$ \\
\textit{J1713+0747} & $\mathit{-14.1_{-0.1}^{+0.1}}$ & $\mathit{2.6_{-0.4}^{+0.5}}$ \\
\textit{J1738+0333} & $\mathit{-14.6_{-0.6}^{+0.8}}$ & $\mathit{5.2_{-1.8}^{+1.3}}$ \\
\textit{J1744$-$1134} & $\mathit{-14.1_{-0.6}^{+0.4}}$ & $\mathit{3.6_{-1.2}^{+1.4}}$ \\
\textbf{J1745+1017} & $\mathbf{-11.9_{-0.1}^{+0.1}}$ & $\mathbf{2.4_{-0.5}^{+0.6}}$ \\
\textbf{J1747$-$4036} & $\mathbf{-12.6_{-0.2}^{+0.1}}$ & $\mathbf{2.4_{-0.7}^{+1.0}}$ \\
\textbf{J1802$-$2124} & $\mathbf{-12.2_{-0.2}^{+0.2}}$ & $\mathbf{1.8_{-0.6}^{+0.7}}$ \\
\textit{J1853+1303} & $\mathit{-13.5_{-0.4}^{+0.2}}$ & $\mathit{2.3_{-0.7}^{+1.1}}$ \\
\textbf{J1903+0327} & $\mathbf{-12.2_{-0.1}^{+0.1}}$ & $\mathbf{1.5_{-0.4}^{+0.4}}$ \\
\textit{J1909$-$3744} & $\mathit{-14.5_{-0.4}^{+0.3}}$ & $\mathit{4.1_{-0.9}^{+1.0}}$ \\
\textit{J1918$-$0642} & $\mathit{-13.8_{-0.7}^{+0.4}}$ & $\mathit{2.7_{-1.0}^{+1.5}}$ \\
\textbf{J1946+3417} & $\mathbf{-12.5_{-0.1}^{+0.1}}$ & $\mathbf{1.4_{-0.4}^{+0.5}}$ \\
\textbf{J2145$-$0750} & $\mathbf{-12.9_{-0.1}^{+0.1}}$ & $\mathbf{0.6_{-0.4}^{+0.5}}$ \\
\textit{J2234+0611} & $\mathit{-13.9_{-0.9}^{+0.5}}$ & $\mathit{3.2_{-1.9}^{+2.7}}$ \\
\hline
\end{tabular}
\end{center}
\label{t:rednoise}
\end{table}
\myfig{f:rednoise_contours} shows the RN power law posteriors for the same pulsars included in \mytab{t:rednoise}. Those italicized in \mytab{t:rednoise} are grayed in \myfig{f:rednoise_contours}, along with the overlaid posterior for a common process across all of the pulsars. Note that the gray contours follow a well known covariance trend between the spectral index and RN amplitude. See \citetalias{NG15gwb}~for more details, along with a way to ameliorate the covariance in parameter recoveries.
The results are broadly as one would expect. The pulsars that possess significant RN detections in their individual noise analysis, that lie along a similar slope of covariance to the common process recovery, lose their RN power to the common process. This is corroborated by a ``leave one out'' analysis in \citetalias{NG15gwb}, where individual pulsars are dropped out of a common uncorrelated red process (\modelcurn) analysis and an odds ratio is calculated between the analysis with $N$ pulsars versus $N-1$ pulsars. All of the grayed-out pulsars have positive dropout factors, i.e. the analysis with that pulsar is favored over the analysis without that pulsar, except for PSR~J0437$-$4715 and PSR~J1713+0747. These two pulsars are known to have challenging noise properties that are discussed below.

It is important to note that while there is obviously another intriguing cluster of posteriors that are a few orders of magnitude larger than the recovered \modelcurn, the RN in those pulsars does not seem to be common and would be excluded by the pulsars in the gray cluster that do not see this RN. Various Bayesian analyses find no \modelcurn in that part of parameter space \citepalias{NG15gwb}. Given the shallow spectral index this red noise is likely due to mis-modeled chromatic noise \citep{cs10}. While only the individual noise analysis results are shown in \mytab{t:rednoise} there is no change in the median RN values at the precision reported, and only slight changes in the credible intervals.

The individual RN models are working as expected. For pulsars where the RN seems to be a part of the \modelcurn the RN models are a stand-in for the common process in the individual noise analyses, but do not hold onto the RN during the full PTA analysis\footnote{The choice of priors is very important to ensure that this is the case \citep{hazboun+2020}.} where the common process instead models the RN. For pulsars where the RN seems to be truly intrinsic to only a single pulsar data set the RN models are successfully modeling that noise, which is \emph{not} able to contaminate the common signal. One interesting characteristic of the grayed-out posteriors is that they cover a large span in the spectral index, but all follow a roughly common slope. Pulsars with short time spans, e.g., PSR~J0437$-$4715 with only $4.8$ yrs of data and the gray posterior against the left side of \myfig{f:rednoise_contours}, may have shallow spectral index recoveries because they lack enough sensitivity at low frequencies where the GWB will rise above their WN floor. 

Next we present three separate noise analyses for four separate representative pulsars\footnote{Figures and analyses are included for all pulsars at the end of this manuscript.} in \myfig{f:excess_j1909} through \myfig{f:excess_b1937}. Each figure contains all of the components of the power-law model, a free spectral analysis, and what we refer to as an ``excess noise analysis,'' discussed below. The plots show RMS fluctuations of the timing residuals, i.e., the amplitude of the noise in the residuals across the frequencies shown, in units of $\log_{10}(\mathrm{s})$. We convert the parameter posterior probabilities from all noise models into amplitude for ease of comparison against the precision of the measurements. The most generic noise model uses a separate set of $\varphi$ coefficients (\myeq{eq:phi_red}), allowed to vary freely, without reference to any particular power spectral density model. This so-called \textit{free spectral model} \citep{Lentati:2016ygu,hazboun:2020slice} is a Bayesian spectrogram of the pulsar data. Here the spectra have been recovered from each pulsar's data across 30 frequencies, ranging from $1/\Tspan$ to $30/\Tspan$, where $\Tspan$ is the full time span of the PTA\footnote{The choice of these 30 frequencies is based on the usual Nyquist frequency sampling considerations and extends up to frequencies higher than where a GWB would be detectable due to the WN floor \citep{lentati+2013}. Unfortunately, the use of these models, with 30 frequency \emph{parameters} for each pulsar, is currently not feasible for full PTA analyses due to computational limitations.}. The solid orange ``violin'' plots show the posteriors, frequency by frequency, for the free spectral analysis. Significant posteriors are those where the violin plot is separated completely from the frequency axis, as in the second to lowest frequency free spectral (solid orange) posterior of \myfig{f:excess_j1909}. Insignificant posteriors have broad posteriors all the way down to the minimum values, while slightly significant detections have thin posteriors extending to the minimum.

The free spectral model posteriors represent a ``raw'' spectrum, which does not assume anything about the power in a pulsar data set. However, when we search for common uncorrelated processes across the pulsars a simpler power law model is used for both the RN and the \modelcurn. The recovered \modelcurn from \citetalias{NG15gwb}, plotted the same for all pulsars, is shown as the dashed red line. The horizontal gray dashed line shows the WN power spectral density of the residuals (subscript R), converted into these units as ${\rm WN}=\sqrt{P^{\rm (WN)}_R/\Tspan}$. The individual RN is shown with a dotted green line. This line may not be visible for pulsars where the power law amplitude is small or not very significant. Since an individual RN model is used for every pulsar, regardless of significance, we have included the maximum likelihood values of the RN for all pulsars. We do not include the RN from individual pulsar noise analyses in these figures for clarity. See \mytab{t:rednoise} below for a list of pulsars significant detections of RN. The solid black line shows the total of the common process RN, the WN, and any individual RN model and represents the total noise power (in these units) for the pulsar using these models. 

Sources of RN were studied in \citet{cs10}. Large RN with a shallower spectral index is thought to be due to modeling errors of time-correlated chromatic effects, while steeper spectral indices are thought to originate from achromatic processes either intrinsic to the pulsar system or due to a stochastic GWB. The pulsars can broadly be separated into four categories depending on what type of noise they are dominated by in their lowest frequencies.

\begin{quote}

\noindent \textbf{Common Process Dominated:} \myfig{f:excess_j1909} shows the noise budget for PSR J1909$-$3744, one of the most sensitive pulsars in the NANOGrav array. This pulsar has very low WN power spectral density, has a significant detection of power-law RN in its individual noise analysis, and shows significant recovery of power at a few low frequencies of the free spectral noise analysis. As denoted by the italicized text in \mytab{t:rednoise}, most of the RN power moves into the \modelcurn in a full PTA analysis, hence it is dominated by RN that appears to be a part of the common process. See \mysec{s:account_rn} for a full accounting of the RN across the PTA. 

Looking at the free spectral posteriors, one might ask whether a power-law model is sufficient for such a pulsar, but in fact the excess noise analysis shows no significant detections of additional noise. The standard model accounts for power in the 30 frequencies considered here. This is true across all of the pulsars in the array.
\begin{figure}[h!tbp]
\centering
\includegraphics[width=0.9\textwidth]{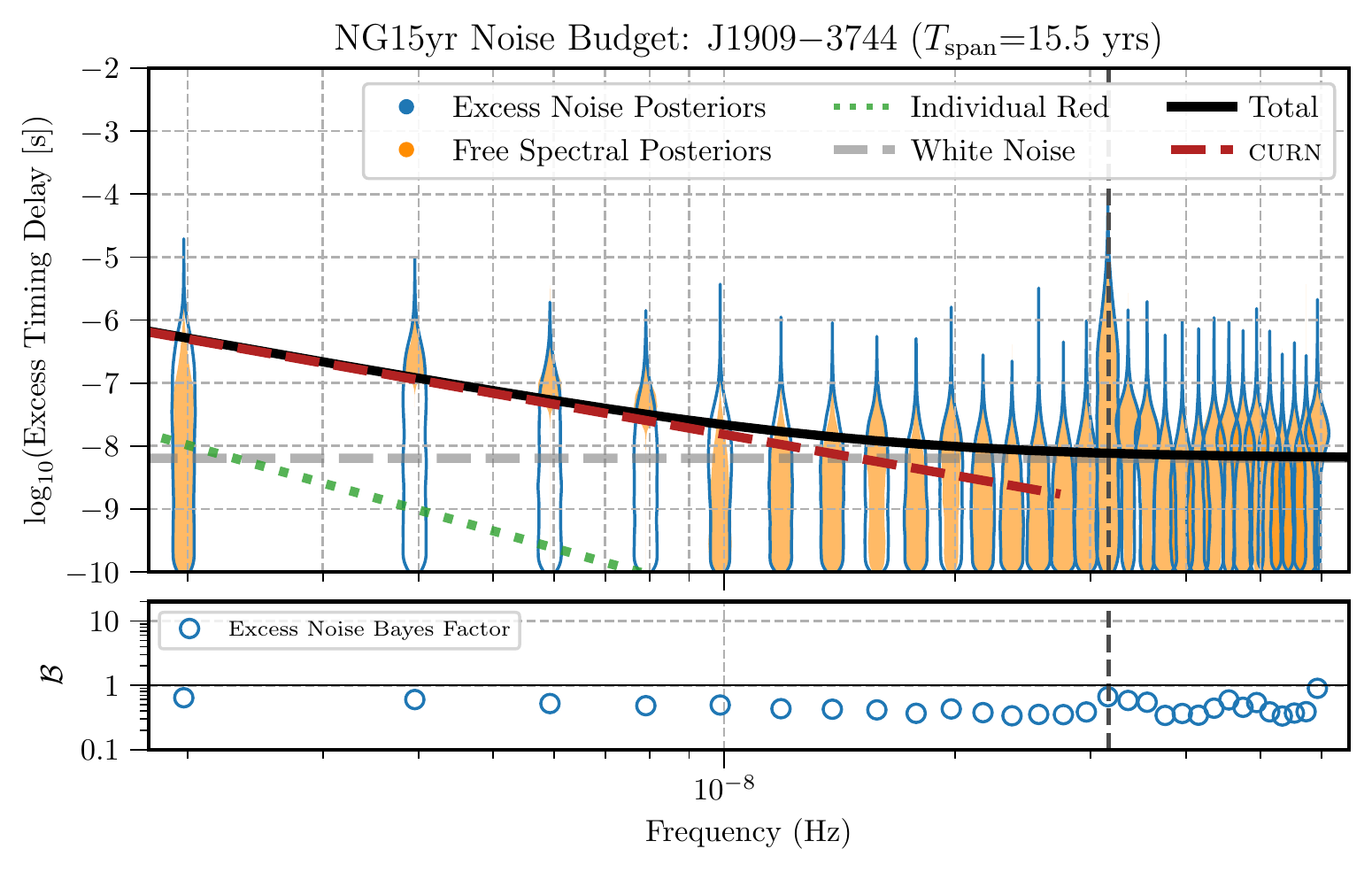}
\caption{The excess timing residual delay as a function of frequency for PSR~J1909$-$3744. The total noise (solid black line) includes WN (gray dashed line), common RN (red dashed line), and individual RN (green dashed line). The free spectral model (solid orange posteriors) does not dictate any relationship between the amplitude of the power at different frequencies. Holding the total noise model (solid black line) parameters constant and searching for additional noise using a free spectral model results in the excess noise shown in blue.  The vertical dashed line denotes  a frequency of 1~yr$^{-1}$. Bayes factors for these parameters, shown in the bottom panel, are fairly insignificant across all frequencies. 
}
\label{f:excess_j1909}
\end{figure}

\noindent \textbf{White Noise Dominated:} \myfig{f:excess_j0509} shows a representative of a second category of pulsar -- one dominated by WN at the lowest frequencies. PSR~J0509+0856 only has $3.6$ years of data, just making the cut for a pulsar included in NANOGrav gravitational analyses. One does not expect sensitivity to RN from this pulsar at the lowest frequencies considered here, since they are based on the longest time span covered by the full PTA data set. One can see that there are no significant detections of power at individual frequencies in either the free spectral model or the excess noise model. 
\begin{figure}[h!tbp]
\centering
\includegraphics[width=0.9\textwidth]{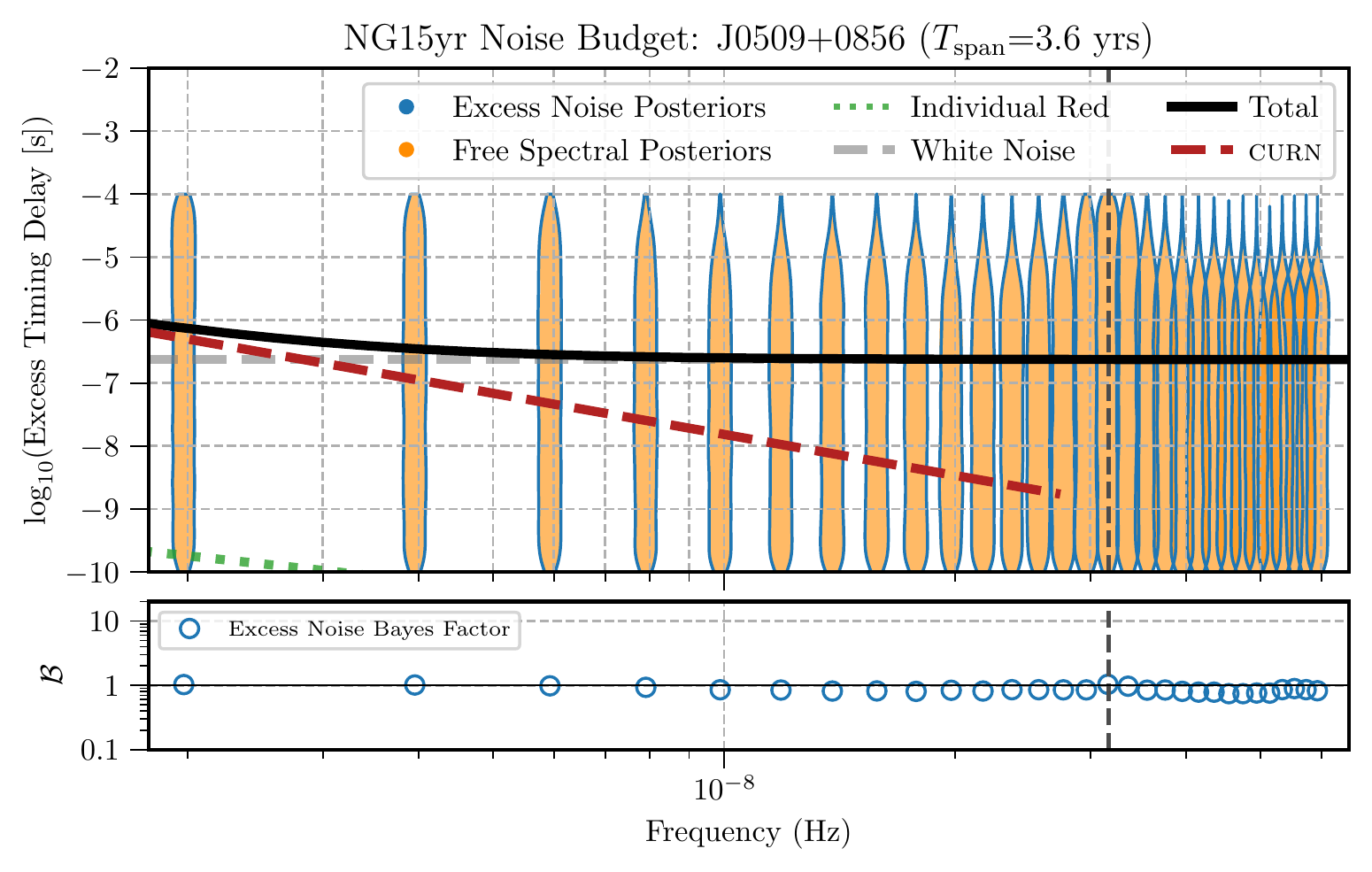}
\caption{The excess timing residual delay as a function of frequency for PSR~J0509+0856. See \myfig{f:excess_j1909} for details.
Bayes factors for these parameters,  shown in the bottom panel,  are insignificant across all frequencies. WN dominates this pulsar's short-timespan data set.}
\label{f:excess_j0509}
\end{figure}

\noindent \textbf{Shallow Red Noise Dominated:} \myfig{f:excess_j1903} shows a third type of pulsar, PSR~J1903+0327, also dominated by RN at the lowest frequencies, but which  does not move into the common channel during a full PTA analysis. PSR J1903+0327 is our highest DM pulsar and so significant unmodeled chromatic propagation effects are expected to impact the timing \citep[see e.g.,][]{memo8}. Here again, the power-law RN model is effective at modeling most of the noise picked up by the free spectral model, as seen by the insignificant Bayes factors.
\begin{figure}[h!tbp]
\centering
\includegraphics[width=0.9\textwidth]{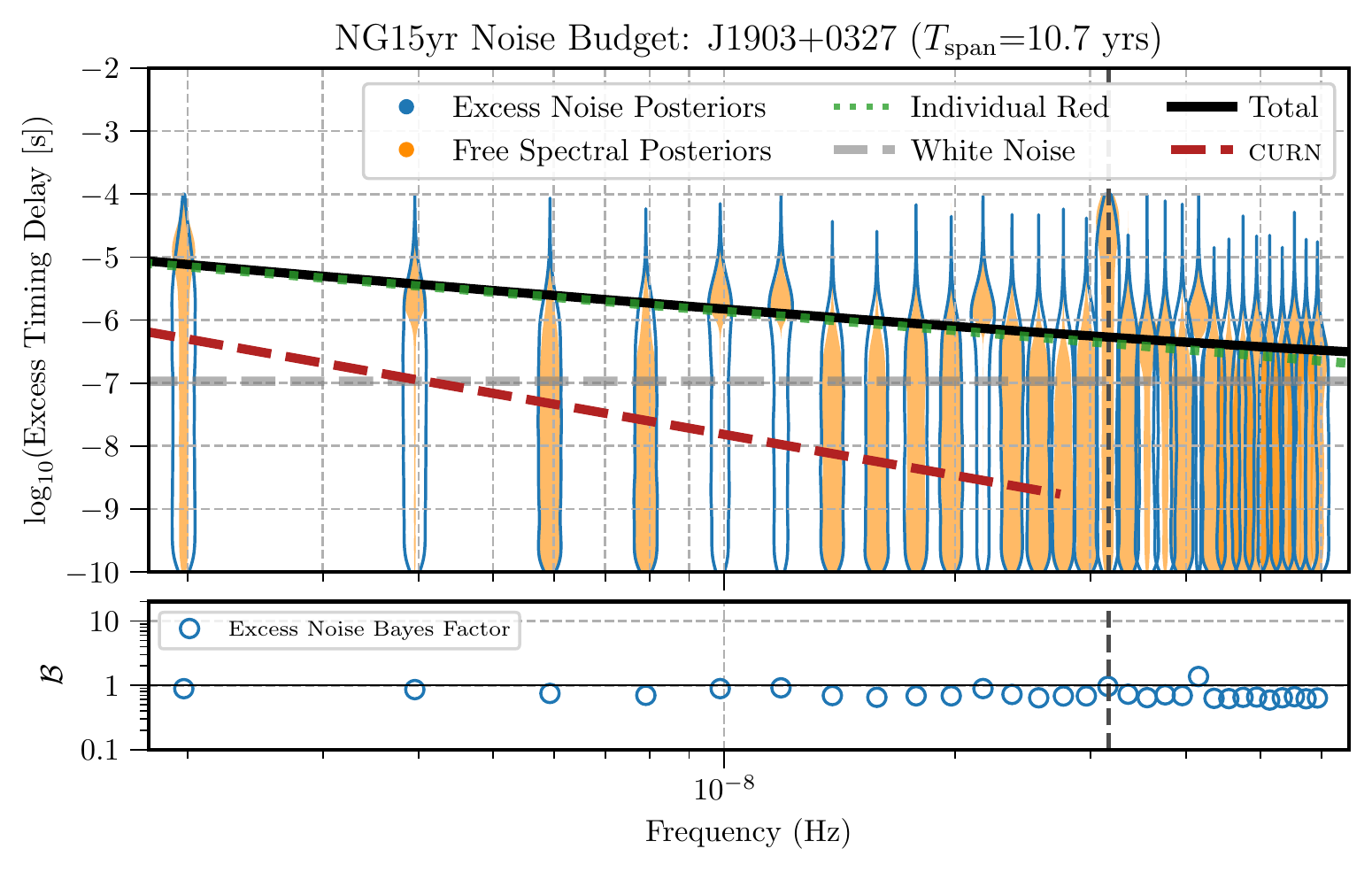}
\caption{The excess timing residual delay as a function of frequency for PSR~J1903+0327. See \myfig{f:excess_j1909} for details.
Bayes factors for these parameters,  shown in the bottom panel, are fairly insignificant across all frequencies. The lowest frequencies are dominated by RN intrinsic to the pulsar.  }
\label{f:excess_j1903}
\end{figure}

\noindent \textbf{Steep Red Noise Dominated:} \myfig{f:excess_b1937} shows a fourth type of pulsar, also dominated by RN at the lowest frequencies, that also does not move into the common channel during a full PTA analysis. PSR~B1937+21 is a well-known example of high amplitude, steep-spectral index RN which appears to be intrinsic to the pulsar system. Here again, the power-law RN model is effective at modeling most of the noise picked up by the free spectral model, as seen by the insignificant Bayes factors.
\begin{figure}[h!tbp]
\centering
\includegraphics[width=0.9\textwidth]{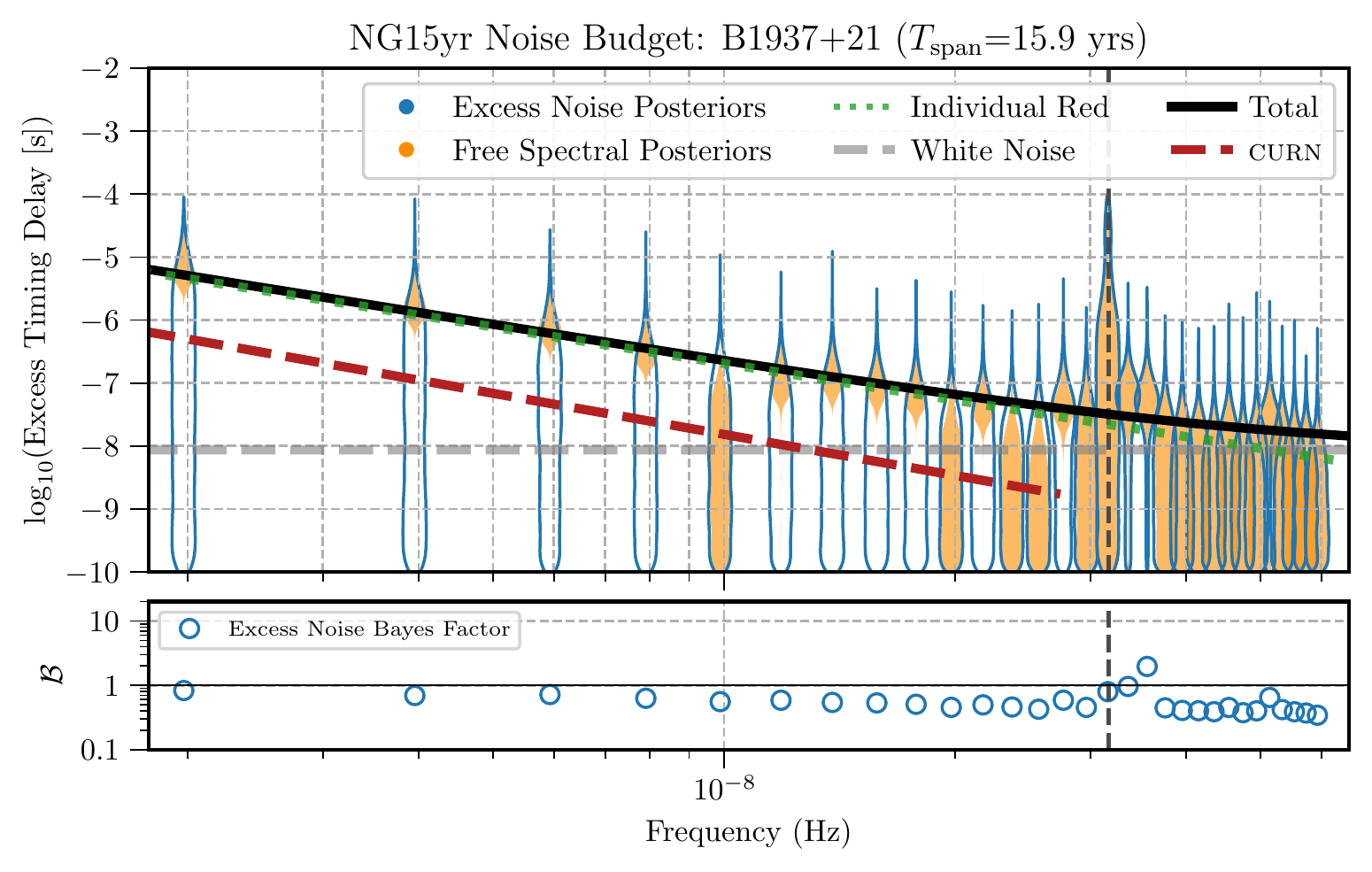}
\caption{The excess timing residual delay as a function of frequency for PSR~B1937+21. See \myfig{f:excess_j1909} for details.
Bayes factors for these parameters,  shown in the bottom panel, are fairly insignificant across all frequencies. The lowest frequencies are dominated by RN intrinsic to the pulsar.  }
\label{f:excess_b1937}
\end{figure}

\end{quote}

Noise budget figures for all of the pulsars used in \citet{NG15gwb} are included as \myfigs{f:budget_B1855+09}{f:budget_J2322+2057}.

\subsubsection{Comparison of Time-Correlated Noise Models}

In order to understand the quantities of noise that might be unaccounted for by the power-law model in our data set, we run an additional noise analysis on each pulsar. This analysis fixes the WN, power-law RN, and GWB parameters, i.e., the solid black line in \myfig{f:excess_j1909} through \myfig{f:excess_j0509}, and adds in an additional analysis, using a separate set of $\varphi$ coefficients, allowed to vary freely. This additional model is effectively the same as the free spectral analysis mentioned above, but this model now accounts for any excess noise unaccounted for in the usual noise model. As we will see, it is mostly superfluous when the power-law model is used. The WN parameters are from the individual pulsar noise analyses discussed in \mysec{s:wn}. In addition we use the 2-dimensional posterior power-law RN maximum likelihood values from a full PTA search for the \modelcurn along with the median value for the \modelcurn amplitude from that same analysis. The objective is to understand the noise unmodeled by the WN, RN and \modelcurn stochastic processes. As with the free spectral model the analyses are done over 30 frequencies ranging from $1/\Tspan$ to $30/\Tspan$. The new free spectral coefficients, referred to as ``excess noise,'' quantify how much power at various frequencies in each pulsar's data set is not modeled by the WN + RN + \modelcurn model. 

The results from this excess noise analysis are compared to the results of the other noise analyses for the four representative pulsars in \myfig{f:excess_j1909} to \myfig{f:excess_b1937}. The excess noise posteriors are shown as the hollow blue violins. The bottom panel shows the Bayes factor, calculated using a Savage-Dickey approximation \citep{dickey1971}, for the excess noise parameters.

\begin{figure}[h!tbp]
\centering
\includegraphics[width=0.95\textwidth]{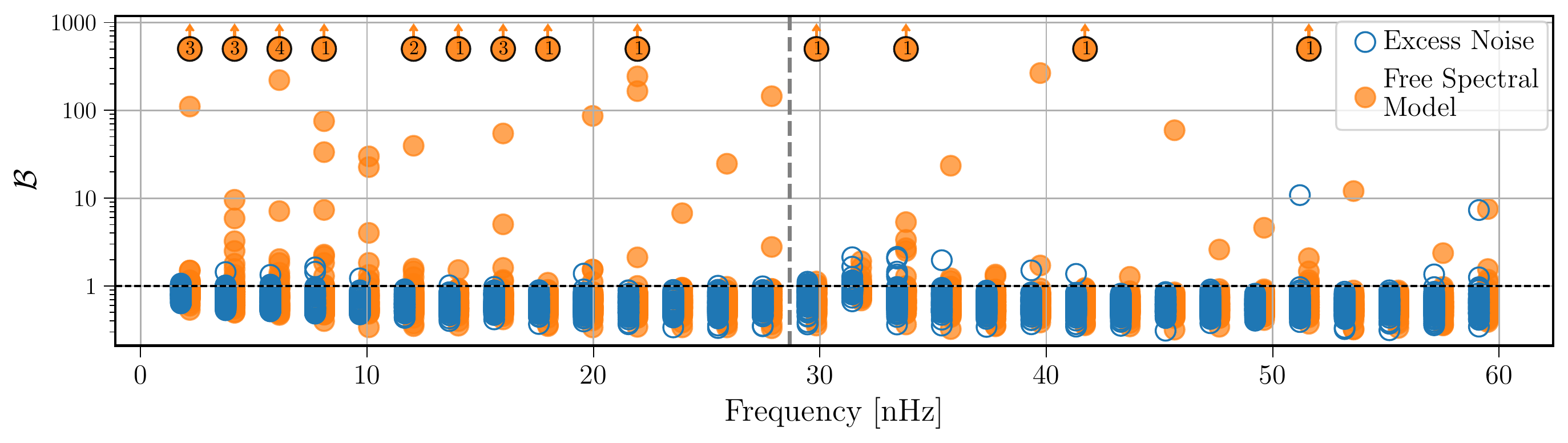}
\caption{Free spectral (orange) and excess noise (blue) model Bayes factors for all 67 pulsars used in the 15-yr GWB search.  The orange dots with arrows represent posteriors where the detection is too significant to use the Savage-Dickey approximation. The blue and orange circles represent the same frequencies, but the blue circles are offset slightly to the left of the orange for ease of viewing. The vertical dashed black line demarcates the 14th frequency. Only the lowest 14 frequencies were used in the GWB search. }
\label{f:excess_bfs}
\end{figure}

\myfig{f:excess_bfs} summarizes the excess noise results from all of the 67 pulsars in the NANOGrav PTA. We show  the individual Bayes factors for each pulsar across the 30 frequencies considered in our GWB search. Again, the solid orange dots show the Bayes factor for the free spectral parameters while the blue circles show the Bayes factors for the excess noise parameters. The orange dots with arrows at the top represent posteriors where the detection is too significant to use the Savage-Dickey approximation. The number of such detections is noted with a digit in the circle. The main takeaway from \myfig{f:excess_bfs} is that there is no case where a significant free spectral detection corresponds to a significant excess noise detection, revealing that the power-law model used across the PTA is a sufficient model, given the sensitivity of the data, for mitigating RN in individual pulsars and detecting RN from a common process. 

\subsection{Sensitivity Curves}

Detection sensitivity curves are commonly used in the GW community to summarize all aspects of detector characterization into a single ``figure of merit''. They are often used by the broader astrophysics community to assess the detectability of various GW sources. Here we use the Python package \hasasia \citep{hazboun:2019has}, based on the formalism developed in \citet{hazboun:2019sc}, to calculate sensitivity curves for the 67 pulsars in the NANOGrav PTA and then combine these into a sensitivity curve for the stochastic GWB. This combination is based on the signal-to-noise ratio of the GWB optimal statistic developed in \citet{abc+2009,ccs+2015,rsg2015}. 

\begin{figure}[h!tbp]
\centering
\fig{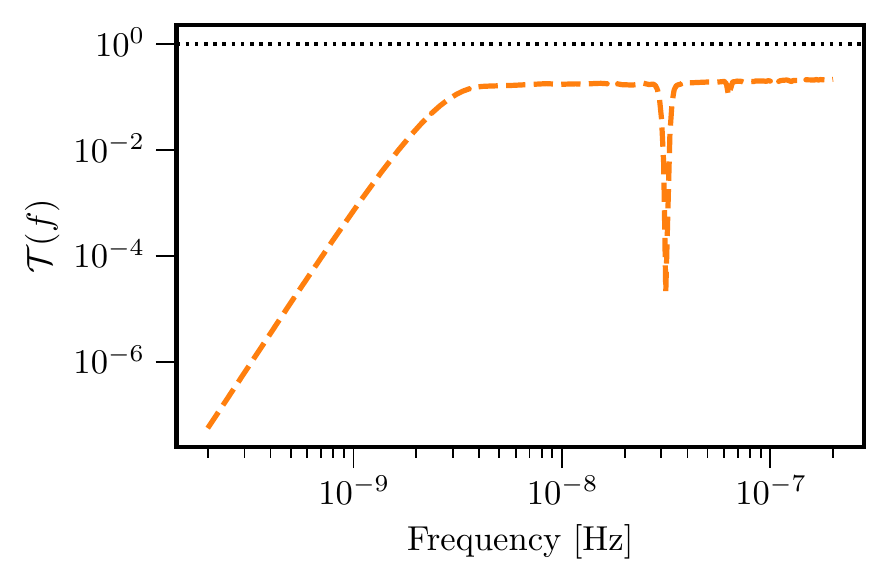}{0.5\textwidth}{(a)}

\fig{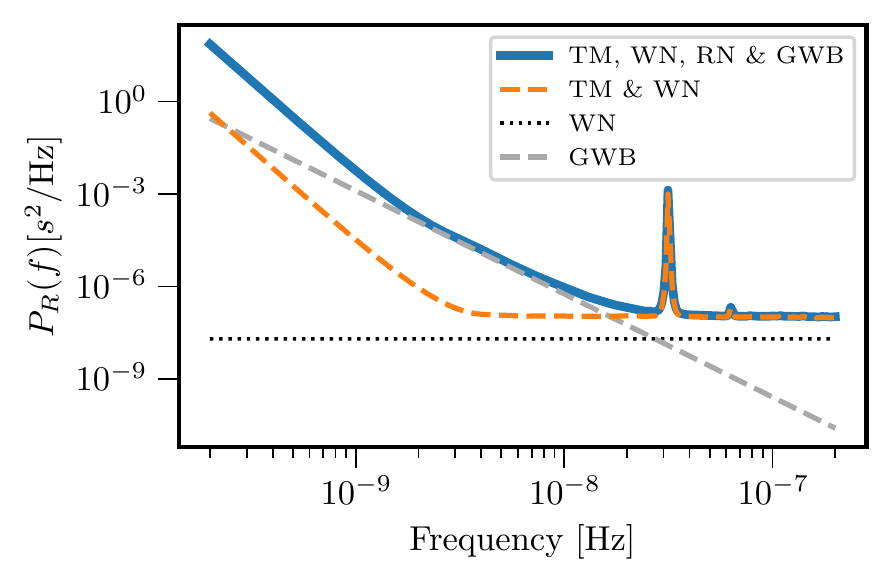}{0.5\textwidth}{(b)}

\fig{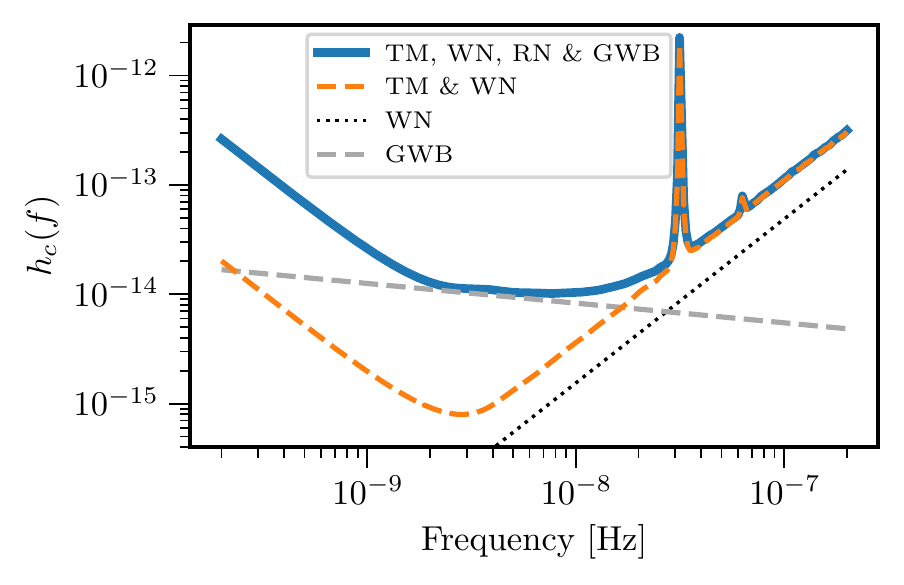}{0.5\textwidth}{(c)}

\caption{Various spectra for PSR J1909$-$3744. The transmission function 
is shown in (a) as a dashed orange line, while perfect transmission ($\mathcal{T}(f)=1$) is shown as the dotted black line. The various frequency-dependent features are discussed in the text. Various power spectral densities of the residuals, $P_R(f)$, are shown in (b).
The black dotted line shows the WN power spectral density $P^{\rm (WN)}_R$. The orange dashed line shows $P^{\rm (WN)}_R$  with the effects of the timing model (TM) included.
The gray dashed line shows the GWB reproduced using the median values of the \modelcurngamma model recovered in \citetalias{NG15gwb}.
The blue solid line shows the full power spectral density of the pulsar
and includes the effects of the timing model, WN, individual pulsar RN, if significant, and the GWB. In plot (c) $P_R(f)$ has been converted into units of characteristic strain, $h_c(f)$, for easy comparison with other GW detectors.}
\label{f:j1909_spectra}
\end{figure}

The effects of the timing model on searches for various astrophysically interesting signals, especially  noise correlated on long timescales, has been known for a long time \citep{blandford+1984}. These effects can be represented by a transmission function, $\mathcal{T}(f)$, that encodes the fraction of power transmitted through the timing model. In this way the transmission function acts as the transfer function for a pulsar \citep{blandford+1984,cordes1985}. A slightly different, but equivalent, method \citep{vhaasteren+2013} than described in \mysec{s:covmatrix} is used to marginalize over the timing model parameters and build the transmission function. Here the $G$-matrix, derived from the timing model design matrix, $M$, encodes the information about the timing model fit and acts to project the data and covariance matrix into a basis orthogonal to the timing model (see \citet{hazboun:2019sc} for details). The transmission function can be written in terms of $G$,
\be\label{eq:g_transmission}
\mathcal{T}(f)\equiv\frac{1}{N_{\rm TOA}}\sum_{i,j}\left(GG^\mathcal{T}\right)_{ij}e^{i2\pi f(t_i-t_j)}
\ee
and calculated for any pulsar in the array. In \myfig{f:j1909_spectra}$a$ the transmission function for PSR~J1909$-$3744 is shown. The transmission function has a few interesting features. First, the fit for the rotation frequency and its derivative (the spin and spindown) of the pulsar pulls a quadratic polynomial out of the data, which acts as a high pass filter, and limits sensitivity at the lowest frequencies. As the time span of the pulsar data set increases, the spindown parameters are fit more precisely and the frequency at which the transmission ``turns down'' moves to lower and lower frequencies. Second, the single frequency fits for the sky position/proper motion of the pulsar ($f= 1~{\rm yr}^{-1}$) and parallax ($f=2~ {\rm yr}^{-1}$) remove power in a narrow band around those frequencies. If the pulsar has a binary period within the frequency range, there will be another dip in the transmission function for that fit as well. The width of these dips is proportional to $1/T_{\rm span}$, and therefore narrows the longer a pulsar is timed. Lastly, the DM variation model used removes   power across the entire GW frequency band. The DMX model constitutes $>200$ parameters for a few of the pulsars timed and diminishes the power by a factor of $\sim3$ across the frequencies searched for GWs. Therefore the transmission function does not asymptote to $\mathcal{T}(f)\sim1$ at the highest frequencies. (Note that any DM variation model with the same frequency resolution would remove a similar amount of power.)

The next step in calculating a pulsar sensitivity curve is to calculate $\mathcal{N}^{-1}$, the noise-weighted transmission function. 
\be\label{eq:Ncal}
\mathcal{N}^{-1}(f)\equiv\frac{1}{2T_{\rm span}}\sum_{i,j}\left[G(G^T C G)^{-1}G^T\right]_{ij}e^{i2\pi f(t_i-t_j)},
\ee
which can also be thought of as the timing model-marginalized power spectral density, i.e., the Fourier domain equivalent of the timing model-marginalized covariance matrix. In fact, $\mathcal{N}=P_R(f)$ is the power spectral density of the pulsar's residuals, and hence the power output of a single arm of our Galactic-scale GW detector. The power spectral density can be transformed to units of GW strain by taking into account the response function  of TOAs to  GWs.

\myfig{f:j1909_spectra}$b$ shows three separate curves in residual power. The dotted line shows the WN power spectral density, $P^{\rm (WN)}_R$, where $P^{\rm (WN)}_R=\frac{1}{2T_{\rm span}}\sum_{i,j}C^{\rm (WN)}_{ij}e^{i2\pi(t_i-t_j)}$. This only includes the WN described in \mysec{s:wn} and shows what the WN power would look like without the effects of the transmission function. The orange dashed line shows the effects of the transmission function on the WN power. It is calculated by replacing $C$ in \myeq{eq:Ncal} with $C^{\rm (WN)}$. The solid blue line includes all of the modeled noise, modeling the same power as shown in the black solid line of \myfig{f:excess_j1909}.

The power spectral density in terms of pulsar residuals, $P_R$, can be converted into strain power spectral density,
\be\label{eq:response}
S(f)=\frac{P_R(f)}{\mathcal{R}}=12\pi^2f^2P_R(f),
\ee
by taking into account the pulsar response function to GWs. The sky-averaged response function useful for GWB characterization is simply 
\be
\mathcal{R}(f)=\frac{1}{12\pi^2 f^2},
\ee
where a factor of  three comes from the sky averaging and the $(2\pi f)^2$ stems from the fact that we are working with timing residuals instead of Doppler shifts in the pulse frequency, where $h_{\rm GWB}=\Delta f_{\rm s} / f_{\rm s}$. See \citet{hazboun:2019sc,shm+04} for more details about the pulsar residual response function. 

The strain power spectral density can then, in turn, be converted into units of characteristic strain, the units in which sensitivity curves are often plotted, via $h_c(f)\equiv\sqrt{fS(f)}=\pi f^\frac{3}{2}\sqrt{12P_R(f)}$. \myfig{f:j1909_spectra}$c$ shows the strain power spectral density for J1909$-$3744 converted into units of characteristic strain. In these units there is a distinctive positive slope in the curve at higher frequencies that goes like $\sim f^\frac{3}{2}$. It is also evident that the GWB acts as a noise floor for an individual pulsar at some frequencies. It is important to note that the correlated Earth-term power in the GWB as measured at the Earth (``Earth-term'') is the signal that we are searching for, but the power at the pulsar (``pulsar-term'') is uncorrelated across the pulsars \citep{cs2013}, and hence needs to be included in the noise budget and calculations of the sensitivity.

\begin{figure}[h!tbp]
\centering
\includegraphics[width=0.63\textwidth]{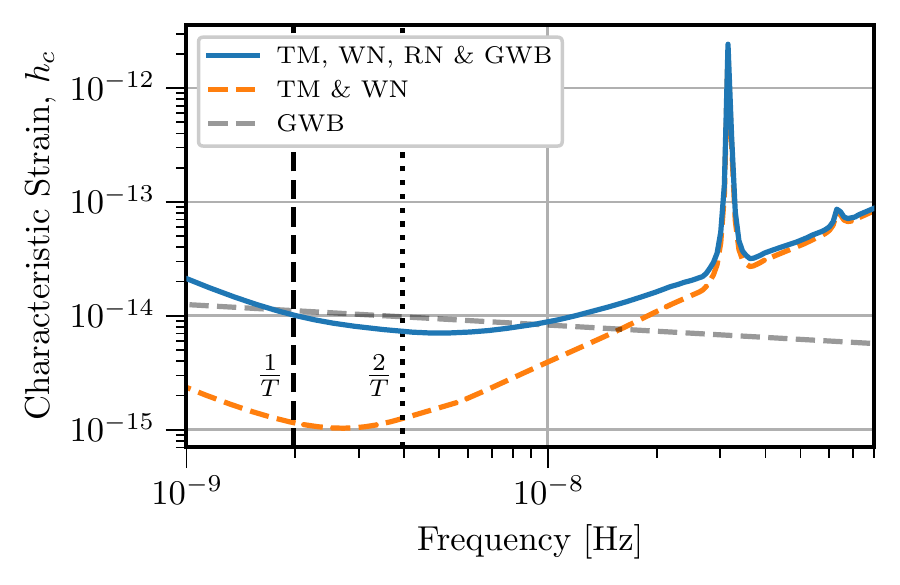}
\caption{Stochastic GWB sensitivity curves. The solid blue curve shows the characteristic strain sensitivity as a function of GW frequency for the NANOGrav 15-year data set and includes contributions from the WN, RN, and GWB self noise. The gray line shows the recovered GWB \citepalias{NG15gwb}, while the dashed orange curve is the sensitivity curve only including the contributions from WN. The importance of including time-correlated noise can be seen in the overly optimistic orange dashed curve. The vertical lines show frequencies at one and two times the inverse of the data set time span.}
\label{f:sensitivity_curve}
\end{figure}

\begin{figure}[h!tbp]
\centering
\includegraphics[width=0.63\textwidth]{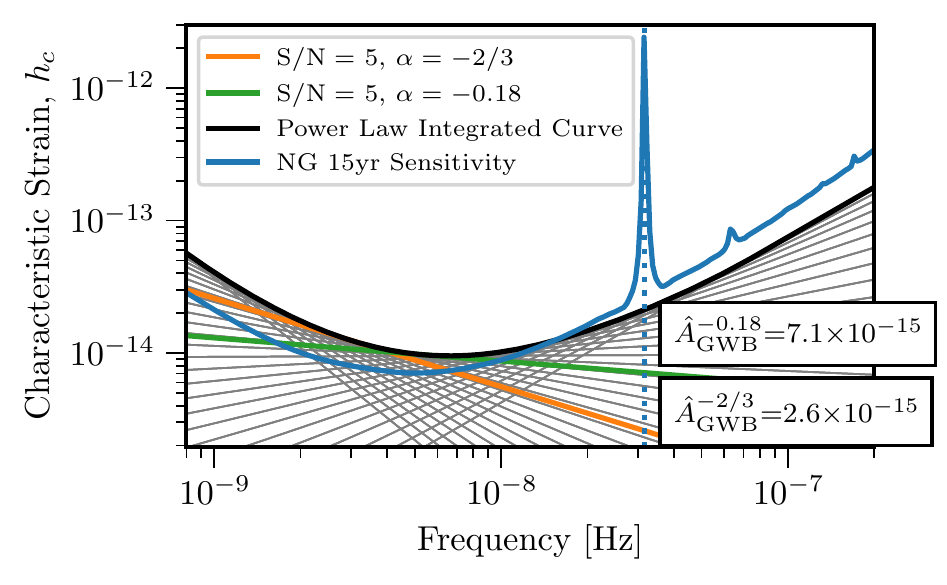}
\caption{Power-law integrated sensitivity curve. The full stochastic sensitivity curve for PTAs combines the individual pulsar sensitivities using the denominator of the optimal statistic \cite{ccs+2015,hazboun:2019sc}. As in \citet{thrane+2013} one can integrate the sensitivity curve against a set of power law stochastic backgrounds, here shown as the set of gray lines with $\alpha=-\frac{7}{4}$ to $\alpha=\frac{5}{4}$, and solve for the amplitude needed, $\hat{A}^\alpha_\mathrm{GWB}$, to meet a given threshold in S/N. The power-law integrated sensitivity curve, solid black, is the maximum of the set of those lines at each frequency. The blue line shows the GWB sensitivity, as described in Figure~\ref{f:sensitivity_curve}, and the orange and green lines show the power-law GWBs detectable with $\mathrm{S/N}=5$ for two discrete values of $\alpha$, described in the text.}
\label{f:pi_sensitivity_curve}
\end{figure}

The individual pulsar strain power spectral densities can be combined using the optimal statistic signal-to-noise ratio in the frequency domain to construct a detection sensitivity curve for the NANOGrav PTA\footnote{These curves are constructed using a statistic specific to a stochastic GWB. Curves for single, resolvable sources can also be constructed using a matched filter statistic \citep{hazboun:2019sc}.} \citep{hazboun:2019sc}, shown in \myfig{f:sensitivity_curve}. The blue curve is the fiducial curve to compare the stochastic gravitational signals against since it contains the self noise due to the background itself. The orange curve, which only includes WN, shows the importance of including the GWB and RN in calculating the sensitivity.
The sensitivity curve in \myfig{f:sensitivity_curve} is calculated for any possible spectral shape of the stochastic GWB. In order to understand the detectability of that background one needs to integrate the GWB power spectral density against the full ``effective'' power spectral density in the frequency domain. Following \citet{thrane+2013} we determine the power law-integrated sensitivity curve by calculating the amplitude of a power-law GWB we would detect with an $\mathrm{S/N} = 5$, as reported by the optimal statistic in \citetalias{NG15gwb}. In \myfig{f:pi_sensitivity_curve} we show the detection sensitivity curve in blue and the power law-integrated curve in black\footnote{In order to make the full curve aesthetically complete a wide range of spectral indices are used.}. Two particular spectral index values are highlighted: $\alpha=-2/3$ ($\gamma=13/3$), the value theoretically expected for a population of binaries where the loss of energy to GWs dominates their evolution and $\alpha=-0.18$ ($\gamma=3.35$), the median value for $\alpha$ obtained from the GWB search in \citetalias{NG15gwb}.
The power law-integrated calculation gives us another important accounting of the noise budget for the NANOGrav detector. This analysis demonstrates that by folding together our knowledge of the noise characterization of the individual pulsars and the parameters for the common process, we can estimate the recovered amplitude, $\hat{A}^\alpha_\mathrm{GWB}$, for the two values of the GWB spectral index used in \citetalias{NG15gwb}.

\section{Improvements and Future Outlook}
\label{s:future}

\subsection{Near-Future Wideband Noise Mitigation Strategies}

Over the last several decades, the bandwidths of the radio telescope receivers used for pulsar observations have steadily increased, with many  receivers  now capable of observing  over more than one or even several octaves of bandwidth. Some examples of receivers used for pulsar timing array science are the Parkes-Murriyang Ultra-Wideband Low Receiver (UWL; 0.7-‐4.2 GHz, \citealt{Hobbs2020uwb}), the Effelsberg Ultra-Broadband Receiver (UBB; 1.3--6 GHz\footnote{\url{https://www.mpifr-bonn.mpg.de/research/fundamental/ubb}}), and, for NANOGrav, a 0.7‐-4.0 GHz Ultra Wideband Receiver (UWBR; \citealt{2020AAS...23517517B}) being commissioned at the GBT, as well as the octave capability of CHIME (400--800 MHz, \citealt{chime/pulsar2021}). Processing the data streams remains a ``big data’’ problem in which raw voltage data is not desirable. 

For pulsar timing applications, data is first coherently dedispersed to remove DM delays in real time. Traditionally,  TOAs were then calculated over a set of discrete frequency channels. Comparing TOAs over wide frequency ranges then yielded deviations of DM from the fiducial value used in the dedispersion kernel. However, this traditional procedure ignored the pulse profile evolution apparent across wide frequency ranges which is covariant with the DM. This has long been accounted for in NANOGrav timing models through fitting frequency-dependent (FD) pulse profile parameters to the channelized TOAs. The move toward even wider and continuously sampled bandwidths, however, favors a newer method of timing TOAs, by representing the profile as a 2D function that takes into account a fiducial DM, pulse profile frequency evolution, and scattering in a procedure known as ``pulse portraiture’’ (\citealt{pen14}, \citealt{liu14}) which reduces covariances. Starting with the 12.5-year data release, NANOGrav has been creating both narrowband and wideband data sets (\citealt{aab+20nb}, \citealt{aab+20wb}). With each new data release, we expect the wideband data set to become more and more important, as data from the newer wideband receivers will become a larger fraction of each existing data set. Additionally, the portraiture framework will be critical for probing frequency-dependent DM behavior, expected for some pulsars across very wide bandwidths \citep{css16}.

Not all near-future noise mitigation strategies depend on new hardware. We highlight a signal processing technique known as cyclic spectroscopy (CS). \citet{2011MNRAS.416.2821D} and \citet{2013ApJ...779...99W} used CS as a deconvolution method to analyze data from PSR~B1937+21, both resolving an intrinsic profile as well as revealing scintillation features with better resolution than that derived from filterbank techniques. Additional work since has shown how we can improve pulsar scintillation and TOA measurements \citep{2014ApJ...790L..22A,2015ApJ...815...89P,2021ApJ...913...98D,turner23}, where CS is most effective for pulsars with high profile S/N and/or high scattering timescales, as is the case for PSR~B1937+21. While PSR~B1937+21 is more highly scattered than most current PTA pulsars, future telescopes and upcoming receivers on existing telescopes will likely discover new, distant pulsars for which CS deconvolution is applicable. In addition, some currently known pulsars may become PTA-quality.

We expect significant S/N improvement both from the highly resolved pulse scintillation structure enabled by CS and from improved RFI mitigation due to frequency binning. This will be especially important for the heavy RFI-contamination expected in the UWBR context. The computational limit, however, is that CS requires either saved baseband data or cyclic spectra written in real time, much like pulse profiles are currently written in real time after the application of coherent dedispersion. For the second (preferable) case, a real-time CS-backend on the GBT's UWBR is currently under development.

\subsection{Noise Mitigation for Different GW Sources}
PTA collaborations continually strive to understand and mitigate the noise in pulsar data sets. A deep understanding of the noise allows PTA collaborations to forecast the sensitivity of our detector to various GW signals and even tune it towards specific sources in some cases \citep{liu+2023,speri+2023}.  \myfig{f:sensitivity_curve} clearly demonstrates  the strong effect of the stochastic GWB on the sensitivity of our detector. As PTAs enter the post-GWB detection era, the GWB will in fact be a foreground source of noise, relative to the many individual GW sources searched for by PTAs. While continued resolution of the background \citep{pol+2021} will  allow us to mitigate up to half of the power in the GWB, half of the power in any given pulsar is uncorrelated across the pulsars, i.e., the pulsar term. Flagship GW analyses, e.g. for resolvable supermassive binary black holes \citep{ng12p5cw}, already necessitate joint searches for a GWB and the deterministic signal to mitigate false positives. Searches for steeper spectral index stochastic backgrounds will require continued long time span observations \citep{kaiser+2022}, while searches for resolvable binary signals will be served by better sensitivity at higher frequencies \citep{liu+2023}. 

In recent years a number of new strategies for noise modeling of pulsar timing data sets have been developed \citep{leg+2018,Chalumeau+2022,goncharov+21a} to mitigate against higher-order chromatic propagation effects like scattering and telescope issues (i.e., ``band noise'' and ``system noise''). The latter noise types are characterized using power law RN models exclusively in a particular frequency band or for a specific pulsar backend. Largely these advances come in the form of Gaussian process models, built into the Bayesian noise analyses and also used in the full PTA GW searches. In \citet{falxa+2023} it was shown that including a large number of frequencies in the DM modeling was necessary to mitigate against spurious high-frequency single-source detections. These efforts reveal the need for and  potential of these noise mitigation strategies for increasing sensitivity to high-frequency, deterministic GW sources. 

\subsection{The Future of PTA Noise Mitigation}

New noise mitigation strategies for PTAs offer a two-fold benefit for the low-frequency search for GWs. Mitigation techniques that involve new data will allow for better and more sensitive searches of GWs from sources in the higher end of our frequency band. Strategies that allow for mitigation of noise across the $15+$ year data set allow us to extract more sensitivity out of the data that we already possess. These techniques, as well as the in-depth characterization of PTAs as Galactic-scale detectors, will allow PTAs  to reach their full potential as  sensitive nanohertz GW observatories. 

\section{Conclusion}

The North American Nanohertz Observatory for Gravitational Waves is a Galactic-scale GW detector made up of some of the most sensitive radio telescopes in the world and nature's most stable clocks. The astrophysical makeup of our detector requires careful noise analysis, starting from the underlying physical processes adding uncertainty to our measurements. It also requires that we use a well-motivated phenomenological model in order to incorporate all sources of noise, whether explicitly accounted for or not, in order to be as conservative as possible in our characterization. 

We have carried out a full characterization of the noise in our nanohertz GW detector, including a full accounting of RN in the pulsars and new analyses to corroborate the use of power-law RN models. These noise analyses allow us to construct detection sensitivity curves for the stochastic GWB which are in agreement with the results of \citetalias{NG15gwb}. We find that our phenomenological model faithfully encapsulates the noise in our detector.

\begin{acknowledgments}

The NANOGrav collaboration receives support from National Science Foundation (NSF) Physics Frontiers Center award numbers 1430284 and 2020265, the Gordon and Betty Moore Foundation, an NSERC Discovery Grant, and CIFAR. The Arecibo Observatory is a facility of the NSF operated under cooperative agreement (AST-1744119) by the University of Central Florida (UCF) in alliance with Universidad Ana G. M{\'e}ndez (UAGM) and Yang Enterprises (YEI), Inc. The Green Bank Observatory is a facility of the NSF operated under cooperative agreement by Associated Universities, Inc. The National Radio Astronomy Observatory is a facility of the NSF operated under cooperative agreement by Associated Universities, Inc. 
NANOGrav is part of the International Pulsar Timing Array (IPTA); we would like to thank our IPTA colleagues for their help with this paper. Particular thanks to Joris Verbiest for extensive conversations and comments concerning early drafts of this manuscript.

This work was conducted in part using the resources of the Advanced Computing Center for Research and Education (ACCRE) at Vanderbilt University, Nashville, TN. This work was facilitated through the use of advanced computational, storage, and networking infrastructure provided by the Hyak supercomputer system at the University of Washington.

L.B. acknowledges support from the National Science Foundation under award AST-1909933 and from the Research Corporation for Science Advancement under Cottrell Scholar Award No. 27553.
P.R.B. is supported by the Science and Technology Facilities Council, grant number ST/W000946/1.
S.B. gratefully acknowledges the support of a Sloan Fellowship, and the support of NSF under award \#1815664.
M.C. and S.R.T. acknowledge support from NSF AST-2007993.
M.C. and N.S.P. were supported by the Vanderbilt Initiative in Data Intensive Astrophysics (VIDA) Fellowship.
Support for this work was provided by the NSF through the Grote Reber Fellowship Program administered by Associated Universities, Inc./National Radio Astronomy Observatory.
Support for H.T.C. is provided by NASA through the NASA Hubble Fellowship Program grant \#HST-HF2-51453.001 awarded by the Space Telescope Science Institute, which is operated by the Association of Universities for Research in Astronomy, Inc., for NASA, under contract NAS5-26555.
K.C. is supported by a UBC Four Year Fellowship (6456).
M.E.D. acknowledges support from the Naval Research Laboratory by NASA under contract S-15633Y.
T.D. and M.T.L. are supported by an NSF Astronomy and Astrophysics Grant (AAG) award number 2009468.
E.C.F. is supported by NASA under award number 80GSFC21M0002.
G.E.F., S.C.S., and S.J.V. are supported by NSF award PHY-2011772.
The Flatiron Institute is supported by the Simons Foundation.
A.D.J. and M.V. acknowledge support from the Caltech and Jet Propulsion Laboratory President's and Director's Research and Development Fund.
A.D.J. acknowledges support from the Sloan Foundation.
The work of N.La. and X.S. is partly supported by the George and Hannah Bolinger Memorial Fund in the College of Science at Oregon State University.
N.La. acknowledges the support from Larry W. Martin and Joyce B. O'Neill Endowed Fellowship in the College of Science at Oregon State University.
Part of this research was carried out at the Jet Propulsion Laboratory, California Institute of Technology, under a contract with the National Aeronautics and Space Administration (80NM0018D0004).
D.R.L. and M.A.M. are supported by NSF \#1458952.
M.A.M. is supported by NSF \#2009425.
C.M.F.M. was supported in part by the National Science Foundation under Grants No. NSF PHY-1748958 and AST-2106552.
A.Mi. is supported by the Deutsche Forschungsgemeinschaft under Germany's Excellence Strategy - EXC 2121 Quantum Universe - 390833306.
The Dunlap Institute is funded by an endowment established by the David Dunlap family and the University of Toronto.
K.D.O. was supported in part by NSF Grant No. 2207267.
T.T.P. acknowledges support from the Extragalactic Astrophysics Research Group at E\"{o}tv\"{o}s Lor\'{a}nd University, funded by the E\"{o}tv\"{o}s Lor\'{a}nd Research Network (ELKH), which was used during the development of this research.
S.M.R. and I.H.S. are CIFAR Fellows.
Portions of this work performed at NRL were supported by ONR 6.1 basic research funding.
J.D.R. also acknowledges support from start-up funds from Texas Tech University.
J.S. is supported by an NSF Astronomy and Astrophysics Postdoctoral Fellowship under award AST-2202388, and acknowledges previous support by the NSF under award 1847938.
S.R.T. acknowledges support from an NSF CAREER award \#2146016.
C.U. acknowledges support from BGU (Kreitman fellowship), and the Council for Higher Education and Israel Academy of Sciences and Humanities (Excellence fellowship).
C.A.W. acknowledges support from CIERA, the Adler Planetarium, and the Brinson Foundation through a CIERA-Adler postdoctoral fellowship.
O.Y. is supported by the National Science Foundation Graduate Research Fellowship under Grant No. DGE-2139292.
\end{acknowledgments}

\section*{Author Contributions}

An alphabetical-order author list was used for this paper in recognition of the fact that a large, decade-timescale project such as NANOGrav is necessarily the result of the work of many people. All authors contributed to the activities of the NANOGrav collaboration leading to the work presented here and reviewed the manuscript, text, and figures prior to the paper's submission. Additional specific contributions to this paper are as follows. G.A., A.A., A.M.A., Z.A., P.T.B., P.R.B., H.T.C., K.C., M.E.D., P.B.D., T.D., E.C.F., W.F., E.F., G.E.F., N.G., P.A.G., J.G., D.C.G., J.S.H., R.J.J., M.L.J., D.L.K., M.K., M.T.L., D.R.L., J.L., R.S.L., A.M., M.A.M., N.M., B.W.M., C.N., D.J.N., T.T.P., B.B.P.P., N.S.P., H.A.R., S.M.R., P.S.R., A.S., C.S., B.J.S., I.H.S., K.S., A.S., J.K.S., and H.M.W.~developed the 15-yr data set through a combination of observations, arrival time calculations, data checks and refinements, and timing-model development and analysis; additional specific contributions to the data set are summarized in \citetalias{NG15}. J.S.H. coordinated the writing of this paper. S.C., T.D., L.G., J.S.H., M.T.L., N.Le., J.W.M., M.A.M., S.K.O, B.B.B.P., and D.R.S. wrote the paper and collected the bibliography. J.S.H., N.La., J.S., and X.S carried out the noise analyses. J.S.H carried out the sensitivity analysis. 

\facilities{Arecibo, GBT, VLA}

\software{\texttt{ENTERPRISE} \citep{enterprise}, \texttt{enterprise\_extensions} \citep{e_e}, \texttt{hasasia} \citep{hazboun:2019has}, \texttt{libstempo} \citep{libstempo}, \texttt{matplotlib} \citep{matplotlib}, \texttt{PTMCMC} \citep{ptmcmc}} 

\bibliographystyle{aasjournal} 
\bibliography{noise}

\begin{thebibliography}{}
\expandafter\ifx\csname natexlab\endcsname\relax\def\natexlab#1{#1}\fi
\providecommand{\url}[1]{\href{#1}{#1}}
\providecommand{\dodoi}[1]{doi:~\href{http://doi.org/#1}{\nolinkurl{#1}}}
\providecommand{\doeprint}[1]{\href{http://ascl.net/#1}{\nolinkurl{http://ascl.net/#1}}}
\providecommand{\doarXiv}[1]{\href{https://arxiv.org/abs/#1}{\nolinkurl{https://arxiv.org/abs/#1}}}

\bibitem[{{Agazie} {et~al.}(2023{\natexlab{a}}){Agazie}, {Alam}, Anumarlapudi,
  Archibald, Arzoumanian, \& Others}]{NG15}
{Agazie}, G., {Alam}, M.~F., Anumarlapudi, A., {et~al.} 2023{\natexlab{a}},
  \apjl, \dodoi{10.3847/2041-8213/acda9a}

\bibitem[{{Agazie} {et~al.}(2023{\natexlab{b}}){Agazie}, Anumarlapudi,
  Archibald, Arzoumanian, {Baker}, \& Others}]{NG15gwb}
{Agazie}, G., Anumarlapudi, A., Archibald, A.~M., {et~al.} 2023{\natexlab{b}},
  \apjl, \dodoi{10.3847/2041-8213/acdac6}

\bibitem[{{Alam} {et~al.}(2021{\natexlab{a}}){Alam}, {Arzoumanian}, {Baker},
  {Blumer}, {Bohler}, {Brazier}, {Brook}, {Burke-Spolaor}, {Caballero},
  {Camuccio}, {Chamberlain}, {Chatterjee}, {Cordes}, {Cornish}, {Crawford},
  {Cromartie}, {Decesar}, {Demorest}, {Dolch}, {Ellis}, {Ferdman}, {Ferrara},
  {Fiore}, {Fonseca}, {Garcia}, {Garver-Daniels}, {Gentile}, {Good},
  {Gusdorff}, {Halmrast}, {Hazboun}, {Islo}, {Jennings}, {Jessup}, {Jones},
  {Kaiser}, {Kaplan}, {Kelley}, {Key}, {Lam}, {Lazio}, {Lorimer}, {Luo},
  {Lynch}, {Madison}, {Maraccini}, {McLaughlin}, {Mingarelli}, {Ng}, {Nguyen},
  {Nice}, {Pennucci}, {Pol}, {Ramette}, {Ransom}, {Ray}, {Shapiro-Albert},
  {Siemens}, {Simon}, {Spiewak}, {Stairs}, {Stinebring}, {Stovall}, {Swiggum},
  {Taylor}, {Tripepi}, {Vallisneri}, {Vigeland}, {Witt}, {Zhu}, \& {Nanograv
  Collaboration}}]{aab+20nb}
{Alam}, M.~F., {Arzoumanian}, Z., {Baker}, P.~T., {et~al.} 2021{\natexlab{a}},
  \apjs, 252, 4, \dodoi{10.3847/1538-4365/abc6a0}

\bibitem[{{Alam} {et~al.}(2021{\natexlab{b}}){Alam}, {Arzoumanian}, {Baker},
  {Blumer}, {Bohler}, {Brazier}, {Brook}, {Burke-Spolaor}, {Caballero},
  {Camuccio}, {Chamberlain}, {Chatterjee}, {Cordes}, {Cornish}, {Crawford},
  {Cromartie}, {Decesar}, {Demorest}, {Dolch}, {Ellis}, {Ferdman}, {Ferrara},
  {Fiore}, {Fonseca}, {Garcia}, {Garver-Daniels}, {Gentile}, {Good},
  {Gusdorff}, {Halmrast}, {Hazboun}, {Islo}, {Jennings}, {Jessup}, {Jones},
  {Kaiser}, {Kaplan}, {Kelley}, {Key}, {Lam}, {Lazio}, {Lorimer}, {Luo},
  {Lynch}, {Madison}, {Maraccini}, {McLaughlin}, {Mingarelli}, {Ng}, {Nguyen},
  {Nice}, {Pennucci}, {Pol}, {Ramette}, {Ransom}, {Ray}, {Shapiro-Albert},
  {Siemens}, {Simon}, {Spiewak}, {Stairs}, {Stinebring}, {Stovall}, {Swiggum},
  {Taylor}, {Tripepi}, {Vallisneri}, {Vigeland}, {Witt}, {Zhu}, \& {Nanograv
  Collaboration}}]{aab+20wb}
---. 2021{\natexlab{b}}, \apjs, 252, 5, \dodoi{10.3847/1538-4365/abc6a1}

\bibitem[{{Anholm} {et~al.}(2009){Anholm}, {Ballmer}, {Creighton}, {Price}, \&
  {Siemens}}]{abc+2009}
{Anholm}, M., {Ballmer}, S., {Creighton}, J.~D.~E., {Price}, L.~R., \&
  {Siemens}, X. 2009, \prd, 79, 084030, \dodoi{10.1103/PhysRevD.79.084030}

\bibitem[{{Antoniadis} {et~al.}(2022){Antoniadis}, {Arzoumanian}, {Babak},
  {Bailes}, {Bak Nielsen}, {Baker}, {Bassa}, {B{\'e}csy}, {Berthereau},
  {Bonetti}, {Brazier}, {Brook}, {Burgay}, {Burke-Spolaor}, {Caballero},
  {Casey-Clyde}, {Chalumeau}, {Champion}, {Charisi}, {Chatterjee}, {Chen},
  {Cognard}, {Cordes}, {Cornish}, {Crawford}, {Cromartie}, {Crowter}, {Dai},
  {DeCesar}, {Demorest}, {Desvignes}, {Dolch}, {Drachler}, {Falxa}, {Ferrara},
  {Fiore}, {Fonseca}, {Gair}, {Garver-Daniels}, {Goncharov}, {Good}, {Graikou},
  {Guillemot}, {Guo}, {Hazboun}, {Hobbs}, {Hu}, {Islo}, {Janssen}, {Jennings},
  {Johnson}, {Jones}, {Kaiser}, {Kaplan}, {Karuppusamy}, {Keith}, {Kelley},
  {Kerr}, {Key}, {Kramer}, {Lam}, {Lamb}, {Lazio}, {Lee}, {Lentati}, {Liu},
  {Luo}, {Lynch}, {Lyne}, {Madison}, {Main}, {Manchester}, {McEwen}, {McKee},
  {McLaughlin}, {Mickaliger}, {Mingarelli}, {Ng}, {Nice}, {Os{\l}owski},
  {Parthasarathy}, {Pennucci}, {Perera}, {Perrodin}, {Petiteau}, {Pol},
  {Porayko}, {Possenti}, {Ransom}, {Ray}, {Reardon}, {Russell}, {Samajdar},
  {Sampson}, {Sanidas}, {Sarkissian}, {Schmitz}, {Schult}, {Sesana},
  {Shaifullah}, {Shannon}, {Shapiro-Albert}, {Siemens}, {Simon}, {Smith},
  {Speri}, {Spiewak}, {Stairs}, {Stappers}, {Stinebring}, {Swiggum}, {Taylor},
  {Theureau}, {Tiburzi}, {Vallisneri}, {van der Wateren}, {Vecchio},
  {Verbiest}, {Vigeland}, {Wahl}, {Wang}, {Wang}, {Wang}, {Witt}, {Zhang}, \&
  {Zhu}}]{antoniadis2022}
{Antoniadis}, J., {Arzoumanian}, Z., {Babak}, S., {et~al.} 2022, \mnras, 510,
  4873, \dodoi{10.1093/mnras/stab3418}

\bibitem[{{Archibald} {et~al.}(2014){Archibald}, {Kondratiev}, {Hessels}, \&
  {Stinebring}}]{2014ApJ...790L..22A}
{Archibald}, A.~M., {Kondratiev}, V.~I., {Hessels}, J.~W.~T., \& {Stinebring},
  D.~R. 2014, \apjl, 790, L22, \dodoi{10.1088/2041-8205/790/2/L22}

\bibitem[{{Archibald} {et~al.}(2018){Archibald}, {Gusinskaia}, {Hessels},
  {Deller}, {Kaplan}, {Lorimer}, {Lynch}, {Ransom}, \&
  {Stairs}}]{archibald+2018}
{Archibald}, A.~M., {Gusinskaia}, N.~V., {Hessels}, J. W.~T., {et~al.} 2018,
  \nat, 559, 73, \dodoi{10.1038/s41586-018-0265-1}

\bibitem[{{Archibald} {et~al.}(2013){Archibald}, {Kaspi}, {Ng},
  {Gourgouliatos}, {Tsang}, {Scholz}, {Beardmore}, {Gehrels}, \&
  {Kennea}}]{archibald+13}
{Archibald}, R.~F., {Kaspi}, V.~M., {Ng}, C.-Y., {et~al.} 2013, Nature, 497,
  591, \dodoi{10.1038/nature12159}

\bibitem[{Arzoumanian {et~al.}(2015)Arzoumanian, Brazier, Burke-Spolaor,
  Chamberlin, Chatterjee, Christy, Cordes, Cornish, Crowter, Demorest, Dolch,
  Ellis, Ferdman, Fonseca, Garver-Daniels, Gonzalez, Jenet, Jones, Jones,
  Kaspi, Koop, Lam, Lazio, Levin, Lommen, Lorimer, Luo, Lynch, Madison,
  McLaughlin, Mcwilliams, Nice, Palliyaguru, Pennucci, Ransom, Siemens, Stairs,
  Stinebring, Stovall, Swiggum, Vallisneri, van Haasteren, Wang, \&
  Zhu}]{abb+15}
Arzoumanian, Z., Brazier, A., Burke-Spolaor, S., {et~al.} 2015, \apj, 813, 65,
  \dodoi{10.1088/0004-637X/813/1/65}

\bibitem[{{Arzoumanian} {et~al.}(2016){Arzoumanian}, {Brazier},
  {Burke-Spolaor}, {Chamberlin}, {Chatterjee}, {Christy}, {Cordes}, {Cornish},
  {Crowter}, {Demorest}, {Deng}, {Dolch}, {Ellis}, {Ferdman}, {Fonseca},
  {Garver-Daniels}, {Gonzalez}, {Jenet}, {Jones}, {Jones}, {Kaspi}, {Koop},
  {Lam}, {Lazio}, {Levin}, {Lommen}, {Lorimer}, {Luo}, {Lynch}, {Madison},
  {McLaughlin}, {McWilliams}, {Mingarelli}, {Nice}, {Palliyaguru}, {Pennucci},
  {Ransom}, {Sampson}, {Sanidas}, {Sesana}, {Siemens}, {Simon}, {Stairs},
  {Stinebring}, {Stovall}, {Swiggum}, {Taylor}, {Vallisneri}, {van Haasteren},
  {Wang}, {Zhu}, \& {NANOGrav Collaboration}}]{abb+2016gwb}
{Arzoumanian}, Z., {Brazier}, A., {Burke-Spolaor}, S., {et~al.} 2016, \apj,
  821, 13, \dodoi{10.3847/0004-637X/821/1/13}

\bibitem[{{Arzoumanian} {et~al.}(2018){Arzoumanian}, {Baker}, {Brazier},
  {Burke-Spolaor}, {Chamberlin}, {Chatterjee}, {Christy}, {Cordes}, {Cornish},
  {Crawford}, {Thankful Cromartie}, {Crowter}, {DeCesar}, {Demorest}, {Dolch},
  {Ellis}, {Ferdman}, {Ferrara}, {Folkner}, {Fonseca}, {Garver-Daniels},
  {Gentile}, {Haas}, {Hazboun}, {Huerta}, {Islo}, {Jones}, {Jones}, {Kaplan},
  {Kaspi}, {Lam}, {Lazio}, {Levin}, {Lommen}, {Lorimer}, {Luo}, {Lynch},
  {Madison}, {McLaughlin}, {McWilliams}, {Mingarelli}, {Ng}, {Nice}, {Park},
  {Pennucci}, {Pol}, {Ransom}, {Ray}, {Rasskazov}, {Siemens}, {Simon},
  {Spiewak}, {Stairs}, {Stinebring}, {Stovall}, {Swiggum}, {Taylor},
  {Vallisneri}, {van Haasteren}, {Vigeland}, {Zhu}, \& {NANOGrav
  Collaboration}}]{abb+2018gwb}
{Arzoumanian}, Z., {Baker}, P.~T., {Brazier}, A., {et~al.} 2018, \apj, 859, 47,
  \dodoi{10.3847/1538-4357/aabd3b}

\bibitem[{{Arzoumanian} {et~al.}(2020){Arzoumanian}, {Baker}, {Blumer},
  {B{\'e}csy}, {Brazier}, {Brook}, {Burke-Spolaor}, {Chatterjee}, {Chen},
  {Cordes}, {Cornish}, {Crawford}, {Cromartie}, {Decesar}, {Demorest}, {Dolch},
  {Ellis}, {Ferrara}, {Fiore}, {Fonseca}, {Garver-Daniels}, {Gentile}, {Good},
  {Hazboun}, {Holgado}, {Islo}, {Jennings}, {Jones}, {Kaiser}, {Kaplan},
  {Kelley}, {Key}, {Laal}, {Lam}, {Lazio}, {Lorimer}, {Luo}, {Lynch},
  {Madison}, {McLaughlin}, {Mingarelli}, {Ng}, {Nice}, {Pennucci}, {Pol},
  {Ransom}, {Ray}, {Shapiro-Albert}, {Siemens}, {Simon}, {Spiewak}, {Stairs},
  {Stinebring}, {Stovall}, {Sun}, {Swiggum}, {Taylor}, {Turner}, {Vallisneri},
  {Vigeland}, {Witt}, \& {Nanograv Collaboration}}]{abb+2020gwb}
{Arzoumanian}, Z., {Baker}, P.~T., {Blumer}, H., {et~al.} 2020, \apjl, 905,
  L34, \dodoi{10.3847/2041-8213/abd401}

\bibitem[{{Arzoumanian} {et~al.}(2023){Arzoumanian}, {Baker}, {Blecha},
  {Blumer}, {Brazier}, {Brook}, {Burke-Spolaor}, {B{\'e}csy}, {Casey-Clyde},
  {Charisi}, {Chatterjee}, {Chen}, {Cordes}, {Cornish}, {Crawford},
  {Cromartie}, {DeCesar}, {Demorest}, {Dolch}, {Drachler}, {Ellis}, {Ferrara},
  {Fiore}, {Fonseca}, {Freedman}, {Garver-Daniels}, {Gentile}, {Glaser},
  {Good}, {G{\"u}ltekin}, {Hazboun}, {Jennings}, {Johnson}, {Jones}, {Kaiser},
  {Kaplan}, {Kelley}, {Shapiro Key}, {Laal}, {Lam}, {Lamb}, {Lazio},
  {Lewandowska}, {Liu}, {Lorimer}, {Luo}, {Lynch}, {Madison}, {McEwen},
  {McLaughlin}, {Mingarelli}, {Ng}, {Nice}, {Ocker}, {Olum}, {Pennucci}, {Pol},
  {Ransom}, {Ray}, {Romano}, {Shapiro-Albert}, {Siemens}, {Simon}, {Siwek},
  {Spiewak}, {Stairs}, {Stinebring}, {Stovall}, {Swiggum}, {Sydnor}, {Taylor},
  {Turner}, {Vallisneri}, {Vigeland}, {Wahl}, {Walsh}, {Witt}, \&
  {Young}}]{ng12p5cw}
{Arzoumanian}, Z., {Baker}, P.~T., {Blecha}, L., {et~al.} 2023, arXiv e-prints,
  arXiv:2301.03608, \dodoi{10.48550/arXiv.2301.03608}

\bibitem[{{Backer}(1970)}]{backer_1970}
{Backer}, D.~C. 1970, \nat, 228, 42, \dodoi{10.1038/228042a0}

\bibitem[{{Bassa} {et~al.}(2016){Bassa}, {Janssen}, {Stappers}, {Tauris},
  {Wevers}, {Jonker}, {Lentati}, {Verbiest}, {Desvignes}, {Graikou},
  {Guillemot}, {Freire}, {Lazarus}, {Caballero}, {Champion}, {Cognard},
  {Jessner}, {Jordan}, {Karuppusamy}, {Kramer}, {Lazaridis}, {Lee}, {Liu},
  {Lyne}, {McKee}, {Os{\l}owski}, {Perrodin}, {Sanidas}, {Shaifullah}, {Smits},
  {Theureau}, {Tiburzi}, \& {Zhu}}]{Bassa+2016}
{Bassa}, C.~G., {Janssen}, G.~H., {Stappers}, B.~W., {et~al.} 2016, \mnras,
  460, 2207, \dodoi{10.1093/mnras/stw1134}

\bibitem[{{Blandford} {et~al.}(1984){Blandford}, {Narayan}, \&
  {Romani}}]{blandford+1984}
{Blandford}, R., {Narayan}, R., \& {Romani}, R.~W. 1984, Journal of
  Astrophysics and Astronomy, 5, 369, \dodoi{10.1007/BF02714466}

\bibitem[{{Bochenek} {et~al.}(2015){Bochenek}, {Ransom}, \&
  {Demorest}}]{brd2015}
{Bochenek}, C., {Ransom}, S., \& {Demorest}, P. 2015, \apjl, 813, L4,
  \dodoi{10.1088/2041-8205/813/1/L4}

\bibitem[{{Brook} {et~al.}(2018){Brook}, {Karastergiou}, {McLaughlin}, {Lam},
  {Arzoumanian}, {Chatterjee}, {Cordes}, {Crowter}, {DeCesar}, {Demorest},
  {Dolch}, {Ellis}, {Ferdman}, {Ferrara}, {Fonseca}, {Gentile}, {Jones},
  {Jones}, {Lazio}, {Levin}, {Lorimer}, {Lynch}, {Ng}, {Nice}, {Pennucci},
  {Ransom}, {Ray}, {Spiewak}, {Stairs}, {Stinebring}, {Stovall}, {Swiggum}, \&
  {Zhu}}]{brook18}
{Brook}, P.~R., {Karastergiou}, A., {McLaughlin}, M.~A., {et~al.} 2018, \apj,
  868, 122, \dodoi{10.3847/1538-4357/aae9e3}

\bibitem[{{Bulatek} \& {White}(2020)}]{2020AAS...23517517B}
{Bulatek}, A., \& {White}, S. 2020, in American Astronomical Society Meeting
  Abstracts, Vol. 235, American Astronomical Society Meeting Abstracts \#235,
  175.17

\bibitem[{{Caballero} {et~al.}(2018){Caballero}, {Guo}, {Lee}, {Lazarus},
  {Champion}, {Desvignes}, {Kramer}, {Plant}, {Arzoumanian}, {Bailes}, {Bassa},
  {Bhat}, {Brazier}, {Burgay}, {Burke-Spolaor}, {Chamberlin}, {Chatterjee},
  {Cognard}, {Cordes}, {Dai}, {Demorest}, {Dolch}, {Ferdman}, {Fonseca},
  {Gair}, {Garver-Daniels}, {Gentile}, {Gonzalez}, {Graikou}, {Guillemot},
  {Hobbs}, {Janssen}, {Karuppusamy}, {Keith}, {Kerr}, {Lam}, {Lasky}, {Lazio},
  {Levin}, {Liu}, {Lommen}, {Lorimer}, {Lynch}, {Madison}, {Manchester},
  {McKee}, {McLaughlin}, {McWilliams}, {Mingarelli}, {Nice}, {Os{\l}owski},
  {Palliyaguru}, {Pennucci}, {Perera}, {Perrodin}, {Possenti}, {Ransom},
  {Reardon}, {Sanidas}, {Sesana}, {Shaifullah}, {Shannon}, {Siemens}, {Simon},
  {Spiewak}, {Stairs}, {Stappers}, {Stinebring}, {Stovall}, {Swiggum},
  {Taylor}, {Theureau}, {Tiburzi}, {Toomey}, {van Haasteren}, {van Straten},
  {Verbiest}, {Wang}, {Zhu}, \& {Zhu}}]{caballero+18}
{Caballero}, R.~N., {Guo}, Y.~J., {Lee}, K.~J., {et~al.} 2018, \mnras, 481,
  5501, \dodoi{10.1093/mnras/sty2632}

\bibitem[{{Chalumeau} {et~al.}(2022){Chalumeau}, {Babak}, {Petiteau}, {Chen},
  {Samajdar}, {Caballero}, {Theureau}, {Guillemot}, {Desvignes},
  {Parthasarathy}, {Liu}, {Shaifullah}, {Hu}, {van der Wateren}, {Antoniadis},
  {Bak Nielsen}, {Bassa}, {Berthereau}, {Burgay}, {Champion}, {Cognard},
  {Falxa}, {Ferdman}, {Freire}, {Gair}, {Graikou}, {Guo}, {Jang}, {Janssen},
  {Karuppusamy}, {Keith}, {Kramer}, {Lee}, {Liu}, {Lyne}, {Main}, {McKee},
  {Mickaliger}, {Perera}, {Perrodin}, {Porayko}, {Possenti}, {Sanidas},
  {Sesana}, {Speri}, {Stappers}, {Tiburzi}, {Vecchio}, {Verbiest}, {Wang},
  {Wang}, \& {Xu}}]{Chalumeau+2022}
{Chalumeau}, A., {Babak}, S., {Petiteau}, A., {et~al.} 2022, \mnras, 509, 5538,
  \dodoi{10.1093/mnras/stab3283}

\bibitem[{{Chamberlin} {et~al.}(2015){Chamberlin}, {Creighton}, {Siemens},
  {Demorest}, {Ellis}, {Price}, \& {Romano}}]{ccs+2015}
{Chamberlin}, S.~J., {Creighton}, J.~D.~E., {Siemens}, X., {et~al.} 2015, \prd,
  91, 044048, \dodoi{10.1103/PhysRevD.91.044048}

\bibitem[{{Champion} {et~al.}(2010){Champion}, {Hobbs}, {Manchester},
  {Edwards}, {Backer}, {Bailes}, {Bhat}, {Burke-Spolaor}, {Coles}, {Demorest},
  {Ferdman}, {Folkner}, {Hotan}, {Kramer}, {Lommen}, {Nice}, {Purver},
  {Sarkissian}, {Stairs}, {van Straten}, {Verbiest}, \&
  {Yardley}}]{champion2010}
{Champion}, D.~J., {Hobbs}, G.~B., {Manchester}, R.~N., {et~al.} 2010, \apjl,
  720, L201, \dodoi{10.1088/2041-8205/720/2/L201}

\bibitem[{{Chen} {et~al.}(2021){Chen}, {Caballero}, {Guo}, {Chalumeau}, {Liu},
  {Shaifullah}, {Lee}, {Babak}, {Desvignes}, {Parthasarathy}, {Hu}, {van der
  Wateren}, {Antoniadis}, {Bak Nielsen}, {Bassa}, {Berthereau}, {Burgay},
  {Champion}, {Cognard}, {Falxa}, {Ferdman}, {Freire}, {Gair}, {Graikou},
  {Guillemot}, {Jang}, {Janssen}, {Karuppusamy}, {Keith}, {Kramer}, {Liu},
  {Lyne}, {Main}, {McKee}, {Mickaliger}, {Perera}, {Perrodin}, {Petiteau},
  {Porayko}, {Possenti}, {Samajdar}, {Sanidas}, {Sesana}, {Speri}, {Stappers},
  {Theureau}, {Tiburzi}, {Vecchio}, {Verbiest}, {Wang}, {Wang}, \&
  {Xu}}]{chen+21}
{Chen}, S., {Caballero}, R.~N., {Guo}, Y.~J., {et~al.} 2021, \mnras, 508, 4970,
  \dodoi{10.1093/mnras/stab2833}

\bibitem[{{Cheng}(1987)}]{c1987}
{Cheng}, K.~S. 1987, \apj, 321, 805, \dodoi{10.1086/165673}

\bibitem[{{CHIME/Pulsar Collaboration} {et~al.}(2021){CHIME/Pulsar
  Collaboration}, {Amiri}, {Bandura}, {Boyle}, {Brar}, {Cliche}, {Crowter},
  {Cubranic}, {Demorest}, {Denman}, {Dobbs}, {Dong}, {Fandino}, {Fonseca},
  {Good}, {Halpern}, {Hill}, {H{\"o}fer}, {Kaspi}, {Landecker}, {Leung}, {Lin},
  {Luo}, {Masui}, {McKee}, {Mena-Parra}, {Meyers}, {Michilli}, {Naidu},
  {Newburgh}, {Ng}, {Patel}, {Pinsonneault-Marotte}, {Ransom}, {Renard},
  {Scholz}, {Shaw}, {Sikora}, {Stairs}, {Tan}, {Tendulkar}, {Tretyakov},
  {Vanderlinde}, {Wang}, \& {Wang}}]{chime/pulsar2021}
{CHIME/Pulsar Collaboration}, {Amiri}, M., {Bandura}, K.~M., {et~al.} 2021,
  \apjs, 255, 5, \dodoi{10.3847/1538-4365/abfdcb}

\bibitem[{{Cognard} \& {Backer}(2004)}]{cognard04}
{Cognard}, I., \& {Backer}, D.~C. 2004, \apjl, 612, L125,
  \dodoi{10.1086/424692}

\bibitem[{{Cognard} {et~al.}(1996){Cognard}, {Shrauner}, {Taylor}, \&
  {Thorsett}}]{cognard_1996}
{Cognard}, I., {Shrauner}, J.~A., {Taylor}, J.~H., \& {Thorsett}, S.~E. 1996,
  \apjl, 457, L81, \dodoi{10.1086/309894}

\bibitem[{{Coles} {et~al.}(2011){Coles}, {Hobbs}, {Champion}, {Manchester}, \&
  {Verbiest}}]{coles+2011}
{Coles}, W., {Hobbs}, G., {Champion}, D.~J., {Manchester}, R.~N., \&
  {Verbiest}, J.~P.~W. 2011, \mnras, 418, 561,
  \dodoi{10.1111/j.1365-2966.2011.19505.x}

\bibitem[{{Cordes}(2013)}]{cor13}
{Cordes}, J.~M. 2013, Classical and Quantum Gravity, 30, 224002,
  \dodoi{10.1088/0264-9381/30/22/224002}

\bibitem[{{Cordes} \& {Downs}(1985)}]{cordes1985}
{Cordes}, J.~M., \& {Downs}, G.~S. 1985, \apjs, 59, 343, \dodoi{10.1086/191076}

\bibitem[{{Cordes} \& {Jenet}(2012)}]{cordes12}
{Cordes}, J.~M., \& {Jenet}, F.~A. 2012, ApJ, 752, 54,
  \dodoi{10.1088/0004-637X/752/1/54}

\bibitem[{{Cordes} {et~al.}(2004){Cordes}, {Kramer}, {Lazio}, {Stappers},
  {Backer}, \& {Johnston}}]{2004NewAR..48.1413C}
{Cordes}, J.~M., {Kramer}, M., {Lazio}, T.~J.~W., {et~al.} 2004, \nar, 48,
  1413, \dodoi{10.1016/j.newar.2004.09.040}

\bibitem[{{Cordes} \& {Shannon}(2008)}]{cs2008}
{Cordes}, J.~M., \& {Shannon}, R.~M. 2008, \apj, 682, 1152,
  \dodoi{10.1086/589425}

\bibitem[{{Cordes} \& {Shannon}(2010)}]{cs10}
---. 2010, arXiv e-prints, arXiv:1010.3785.
\newblock \doarXiv{1010.3785}

\bibitem[{{Cordes} {et~al.}(2016){Cordes}, {Shannon}, \& {Stinebring}}]{css16}
{Cordes}, J.~M., {Shannon}, R.~M., \& {Stinebring}, D.~R. 2016, \apj, 817, 16,
  \dodoi{10.3847/0004-637X/817/1/16}

\bibitem[{{Cordes} {et~al.}(1990){Cordes}, {Wolszczan}, {Dewey}, {Blaskiewicz},
  \& {Stinebring}}]{1990ApJ...349..245C}
{Cordes}, J.~M., {Wolszczan}, A., {Dewey}, R.~J., {Blaskiewicz}, M., \&
  {Stinebring}, D.~R. 1990, \apj, 349, 245, \dodoi{10.1086/168310}

\bibitem[{{Cornish} \& {Sesana}(2013)}]{cs2013}
{Cornish}, N.~J., \& {Sesana}, A. 2013, Classical and Quantum Gravity, 30,
  224005, \dodoi{10.1088/0264-9381/30/22/224005}

\bibitem[{{Demorest}(2011)}]{2011MNRAS.416.2821D}
{Demorest}, P.~B. 2011, \mnras, 416, 2821,
  \dodoi{10.1111/j.1365-2966.2011.19230.x}

\bibitem[{{Demorest} {et~al.}(2013){Demorest}, {Ferdman}, {Gonzalez}, {Nice},
  {Ransom}, {Stairs}, {Arzoumanian}, {Brazier}, {Burke-Spolaor}, {Chamberlin},
  {Cordes}, {Ellis}, {Finn}, {Freire}, {Giampanis}, {Jenet}, {Kaspi}, {Lazio},
  {Lommen}, {McLaughlin}, {Palliyaguru}, {Perrodin}, {Shannon}, {Siemens},
  {Stinebring}, {Swiggum}, \& {Zhu}}]{dfg+13}
{Demorest}, P.~B., {Ferdman}, R.~D., {Gonzalez}, M.~E., {et~al.} 2013, \apj,
  762, 94, \dodoi{10.1088/0004-637X/762/2/94}

\bibitem[{{Detweiler}(1979)}]{det79}
{Detweiler}, S. 1979, \apj, 234, 1100, \dodoi{10.1086/157593}

\bibitem[{Dickey(1971)}]{dickey1971}
Dickey, J.~M. 1971, The Annals of Mathematical Statistics, 42, 204.
\newblock \url{http://www.jstor.org/stable/2958475}

\bibitem[{{Dolch} {et~al.}(2014){Dolch}, {Lam}, {Cordes}, {Chatterjee},
  {Bassa}, {Bhattacharyya}, {Champion}, {Cognard}, {Crowter}, {Demorest},
  {Hessels}, {Janssen}, {Jenet}, {Jones}, {Jordan}, {Karuppusamy}, {Keith},
  {Kondratiev}, {Kramer}, {Lazarus}, {Lazio}, {Lee}, {McLaughlin}, {Roy},
  {Shannon}, {Stairs}, {Stovall}, {Verbiest}, {Madison}, {Palliyaguru},
  {Perrodin}, {Ransom}, {Stappers}, {Zhu}, {Dai}, {Desvignes}, {Guillemot},
  {Liu}, {Lyne}, {Perera}, {Petroff}, {Rankin}, \&
  {Smits}}]{2014ApJ...794...21D}
{Dolch}, T., {Lam}, M.~T., {Cordes}, J., {et~al.} 2014, \apj, 794, 21,
  \dodoi{10.1088/0004-637X/794/1/21}

\bibitem[{{Dolch} {et~al.}(2021){Dolch}, {Stinebring}, {Jones}, {Zhu}, {Lynch},
  {Cohen}, {Demorest}, {Lam}, {Levin}, {McLaughlin}, \&
  {Palliyaguru}}]{2021ApJ...913...98D}
{Dolch}, T., {Stinebring}, D.~R., {Jones}, G., {et~al.} 2021, \apj, 913, 98,
  \dodoi{10.3847/1538-4357/abf48b}

\bibitem[{{Donner} {et~al.}(2020){Donner}, {Verbiest}, {Tiburzi},
  {Os{\l}owski}, {K{\"u}nsem{\"o}ller}, {Bak Nielsen}, {Grie{\ss}meier},
  {Serylak}, {Kramer}, {Anderson}, {Wucknitz}, {Keane}, {Kondratiev}, {Sobey},
  {McKee}, {Bilous}, {Breton}, {Br{\"u}ggen}, {Ciardi}, {Hoeft}, {van Leeuwen},
  \& {Vocks}}]{donner+20}
{Donner}, J.~Y., {Verbiest}, J.~P.~W., {Tiburzi}, C., {et~al.} 2020, \aap, 644,
  A153, \dodoi{10.1051/0004-6361/202039517}

\bibitem[{{Edwards} {et~al.}(2006){Edwards}, {Hobbs}, \&
  {Manchester}}]{edwards+06}
{Edwards}, R.~T., {Hobbs}, G.~B., \& {Manchester}, R.~N. 2006, \mnras, 372,
  1549, \dodoi{10.1111/j.1365-2966.2006.10870.x}

\bibitem[{Ellis \& van Haasteren(2017)}]{ptmcmc}
Ellis, J., \& van Haasteren, R. 2017, jellis18/PTMCMCSampler: Official Release,
  \dodoi{10.5281/zenodo.1037579}

\bibitem[{{Ellis} {et~al.}(2013){Ellis}, {Siemens}, \& {van
  Haasteren}}]{esvh13}
{Ellis}, J.~A., {Siemens}, X., \& {van Haasteren}, R. 2013, \apj, 769, 63,
  \dodoi{10.1088/0004-637X/769/1/63}

\bibitem[{{Ellis} {et~al.}(2019){Ellis}, {Vallisneri}, {Taylor}, \&
  {Baker}}]{enterprise}
{Ellis}, J.~A., {Vallisneri}, M., {Taylor}, S.~R., \& {Baker}, P.~T. 2019,
  {ENTERPRISE: Enhanced Numerical Toolbox Enabling a Robust PulsaR Inference
  SuitE}.
\newblock \doeprint{1912.015}

\bibitem[{{Espinoza} {et~al.}(2011){Espinoza}, {Lyne}, {Stappers}, \&
  {Kramer}}]{espinoza+11}
{Espinoza}, C.~M., {Lyne}, A.~G., {Stappers}, B.~W., \& {Kramer}, M. 2011,
  MNRAS, 414, 1679, \dodoi{10.1111/j.1365-2966.2011.18503.x}

\bibitem[{{Falxa} {et~al.}(2023){Falxa}, {Babak}, {Baker}, {B{\'e}csy},
  {Chalumeau}, {Chen}, {Chen}, {Cornish}, {Guillemot}, {Hazboun}, {Mingarelli},
  {Parthasarathy}, {Petiteau}, {Pol}, {Sesana}, {Spolaor}, {Taylor},
  {Theureau}, {Vallisneri}, {Vigeland}, {Witt}, {Zhu}, {Antoniadis},
  {Arzoumanian}, {Bailes}, {Bhat}, {Blecha}, {Brazier}, {Brook}, {Caballero},
  {Cameron}, {Casey-Clyde}, {Champion}, {Charisi}, {Chatterjee}, {Cognard},
  {Cordes}, {Crawford}, {Cromartie}, {Crowter}, {Dai}, {DeCesar}, {Demorest},
  {Desvignes}, {Dolch}, {Drachler}, {Feng}, {Ferrara}, {Fiore}, {Fonseca},
  {Garver-Daniels}, {Glaser}, {Goncharov}, {Good}, {Griessmeier}, {Guo},
  {G{\"u}ltekin}, {Hobbs}, {Hu}, {Islo}, {Jang}, {Jennings}, {Johnson},
  {Jones}, {Kaczmarek}, {Kaiser}, {Kaplan}, {Keith}, {Kelley}, {Kerr}, {Key},
  {Laal}, {Lam}, {Lamb}, {Lazio}, {Liu}, {Liu}, {Luo}, {Lynch}, {Madison},
  {Main}, {Manchester}, {McEwen}, {McKee}, {McLaughlin}, {Ng}, {Nice}, {Ocker},
  {Olum}, {Os{\l}owski}, {Pennucci}, {Perera}, {Perrodin}, {Porayko},
  {Possenti}, {Quelquejay-Leclere}, {Ransom}, {Ray}, {Reardon}, {Russell},
  {Samajdar}, {Sarkissian}, {Schult}, {Shaifullah}, {Shannon},
  {Shapiro-Albert}, {Siemens}, {Simon}, {Siwek}, {Smith}, {Speri}, {Spiewak},
  {Stairs}, {Stappers}, {Stinebring}, {Swiggum}, {Tiburzi}, {Turner},
  {Vecchio}, {Verbiest}, {Wahl}, {Wang}, {Wang}, {Wang}, {Wu}, {Zhang},
  {Zhang}, \& {IPTA Collaboration}}]{falxa+2023}
{Falxa}, M., {Babak}, S., {Baker}, P.~T., {et~al.} 2023, \mnras, 521, 5077,
  \dodoi{10.1093/mnras/stad812}

\bibitem[{{Fienga} {et~al.}(2019){Fienga}, {Deram}, {Viswanathan}, {Di Ruscio},
  {Bernus}, {Durante}, {Gastineau}, \& {Laskar}}]{fienga+19}
{Fienga}, A., {Deram}, P., {Viswanathan}, V., {et~al.} 2019, Notes
  Scientifiques et Techniques de l'Institut de Mecanique Celeste, 109

\bibitem[{{Fienga} {et~al.}(2009){Fienga}, {Laskar}, {Morley}, {Manche},
  {Kuchynka}, {Le Poncin-Lafitte}, {Budnik}, {Gastineau}, \&
  {Somenzi}}]{fienga+09}
{Fienga}, A., {Laskar}, J., {Morley}, T., {et~al.} 2009, \aap, 507, 1675,
  \dodoi{10.1051/0004-6361/200911755}

\bibitem[{{Fonseca} {et~al.}(2016){Fonseca}, {Pennucci}, {Ellis}, {Stairs},
  {Nice}, {Ransom}, {Demorest}, {Arzoumanian}, {Crowter}, {Dolch}, {Ferdman},
  {Gonzalez}, {Jones}, {Jones}, {Lam}, {Levin}, {McLaughlin}, {Stovall},
  {Swiggum}, \& {Zhu}}]{Fonseca+2016}
{Fonseca}, E., {Pennucci}, T.~T., {Ellis}, J.~A., {et~al.} 2016, \apj, 832,
  167, \dodoi{10.3847/0004-637X/832/2/167}

\bibitem[{{Foster} \& {Backer}(1990)}]{fb90}
{Foster}, R.~S., \& {Backer}, D.~C. 1990, \apj, 361, 300,
  \dodoi{10.1086/169195}

\bibitem[{{Foster} \& {Cordes}(1990)}]{fc90}
{Foster}, R.~S., \& {Cordes}, J.~M. 1990, \apj, 364, 123,
  \dodoi{10.1086/169393}

\bibitem[{{Fuentes} {et~al.}(2017){Fuentes}, {Espinoza}, {Reisenegger}, {Shaw},
  {Stappers}, \& {Lyne}}]{fuentes+17}
{Fuentes}, J.~R., {Espinoza}, C.~M., {Reisenegger}, A., {et~al.} 2017, \aap,
  608, A131, \dodoi{10.1051/0004-6361/201731519}

\bibitem[{{Gao} {et~al.}(2016){Gao}, {Zhang}, {Yi}, {Xie}, \& {Fu}}]{gzy+2016}
{Gao}, X.-D., {Zhang}, S.-N., {Yi}, S.-X., {Xie}, Y., \& {Fu}, J.-N. 2016,
  \mnras, 459, 402, \dodoi{10.1093/mnras/stw631}

\bibitem[{Geiger \& Lam(2022)}]{memo8}
Geiger, A., \& Lam, M.~T. 2022, NANOGrav Memo Series No. 8, The
  Frequency-Dependent Scattering of Pulsar J1903+0327, Tech. rep., NANOGrav

\bibitem[{{Gentile} {et~al.}(2018){Gentile}, {McLaughlin}, {Demorest},
  {Stairs}, {Arzoumanian}, {Crowter}, {Dolch}, {DeCesar}, {Ellis}, {Ferdman},
  {Ferrara}, {Fonseca}, {Gonzalez}, {Jones}, {Jones}, {Lam}, {Levin},
  {Lorimer}, {Lynch}, {Ng}, {Nice}, {Pennucci}, {Ransom}, {Ray}, {Spiewak},
  {Stovall}, {Swiggum}, \& {Zhu}}]{Gentile+2018}
{Gentile}, P.~A., {McLaughlin}, M.~A., {Demorest}, P.~B., {et~al.} 2018, \apj,
  862, 47, \dodoi{10.3847/1538-4357/aac9c9}

\bibitem[{{Goncharov} {et~al.}(2020){Goncharov}, {Zhu}, \&
  {Thrane}}]{goncharov+2020}
{Goncharov}, B., {Zhu}, X.-J., \& {Thrane}, E. 2020, \mnras, 497, 3264,
  \dodoi{10.1093/mnras/staa2081}

\bibitem[{{Goncharov} {et~al.}(2021{\natexlab{a}}){Goncharov}, {Shannon},
  {Reardon}, {Hobbs}, {Zic}, {Bailes}, {Cury{\l}o}, {Dai}, {Kerr}, {Lower},
  {Manchester}, {Mandow}, {Middleton}, {Miles}, {Parthasarathy}, {Thrane},
  {Thyagarajan}, {Xue}, {Zhu}, {Cameron}, {Feng}, {Luo}, {Russell},
  {Sarkissian}, {Spiewak}, {Wang}, {Wang}, {Zhang}, \& {Zhang}}]{goncharov2021}
{Goncharov}, B., {Shannon}, R.~M., {Reardon}, D.~J., {et~al.}
  2021{\natexlab{a}}, \apjl, 917, L19, \dodoi{10.3847/2041-8213/ac17f4}

\bibitem[{{Goncharov} {et~al.}(2021{\natexlab{b}}){Goncharov}, {Reardon},
  {Shannon}, {Zhu}, {Thrane}, {Bailes}, {Bhat}, {Dai}, {Hobbs}, {Kerr},
  {Manchester}, {Os{\l}owski}, {Parthasarathy}, {Russell}, {Spiewak},
  {Thyagarajan}, \& {Wang}}]{goncharov+21a}
{Goncharov}, B., {Reardon}, D.~J., {Shannon}, R.~M., {et~al.}
  2021{\natexlab{b}}, \mnras, 502, 478, \dodoi{10.1093/mnras/staa3411}

\bibitem[{{Guo} {et~al.}(2019){Guo}, {Li}, {Lee}, \& {Caballero}}]{guo+19}
{Guo}, Y.~J., {Li}, G.~Y., {Lee}, K.~J., \& {Caballero}, R.~N. 2019, \mnras,
  489, 5573, \dodoi{10.1093/mnras/stz2515}

\bibitem[{Hazboun {et~al.}(2019{\natexlab{a}})Hazboun, Romano, \&
  Smith}]{hazboun:2019has}
Hazboun, J., Romano, J., \& Smith, T. 2019{\natexlab{a}}, J. Open Source
  Softw., 4, 1775, \dodoi{10.21105/joss.01775}

\bibitem[{Hazboun {et~al.}(2019{\natexlab{b}})Hazboun, Romano, \&
  Smith}]{hazboun:2019sc}
Hazboun, J.~S., Romano, J.~D., \& Smith, T.~L. 2019{\natexlab{b}}, Phys. Rev.
  D, 100, 104028, \dodoi{10.1103/PhysRevD.100.104028}

\bibitem[{{Hazboun} {et~al.}(2020{\natexlab{a}}){Hazboun}, {Simon}, {Siemens},
  \& {Romano}}]{hazboun+2020}
{Hazboun}, J.~S., {Simon}, J., {Siemens}, X., \& {Romano}, J.~D.
  2020{\natexlab{a}}, \apjl, 905, L6, \dodoi{10.3847/2041-8213/abca92}

\bibitem[{{Hazboun} {et~al.}(2020{\natexlab{b}}){Hazboun}, {Simon}, {Taylor},
  {Lam}, {Vigeland}, {Islo}, {Key}, {Arzoumanian}, {Baker}, {Brazier}, {Brook},
  {Burke-Spolaor}, {Chatterjee}, {Cordes}, {Cornish}, {Crawford}, {Crowter},
  {Cromartie}, {DeCesar}, {Demorest}, {Dolch}, {Ellis}, {Ferdman}, {Ferrara},
  {Fonseca}, {Garver-Daniels}, {Gentile}, {Good}, {Holgado}, {Huerta},
  {Jennings}, {Jones}, {Jones}, {Kaiser}, {Kaplan}, {Kelley}, {Lazio}, {Levin},
  {Lommen}, {Lorimer}, {Luo}, {Lynch}, {Madison}, {McLaughlin}, {McWilliams},
  {Mingarelli}, {Ng}, {Nice}, {Pennucci}, {Pol}, {Ransom}, {Ray}, {Siemens},
  {Spiewak}, {Stairs}, {Stinebring}, {Stovall}, {Swiggum}, {Turner},
  {Vallisneri}, {van Haasteren}, {Witt}, \& {Zhu}}]{hazboun:2020slice}
{Hazboun}, J.~S., {Simon}, J., {Taylor}, S.~R., {et~al.} 2020{\natexlab{b}},
  \apj, 890, 108, \dodoi{10.3847/1538-4357/ab68db}

\bibitem[{{Hazboun} {et~al.}(2022){Hazboun}, {Simon}, {Madison}, {Arzoumanian},
  {Cromartie}, {Crowter}, {Decesar}, {Demorest}, {Dolch}, {Ellis}, {Ferdman},
  {Ferrara}, {Fonseca}, {Gentile}, {Jones}, {Jones}, {Lam}, {Levin}, {Lorimer},
  {Lynch}, {McLaughlin}, {Ng}, {Nice}, {Pennucci}, {Ransom}, {Ray}, {Spiewak},
  {Stairs}, {Stovall}, {Swiggum}, {Zhu}, \& {The Nanograv
  Collaboration}}]{hazboun+2022}
{Hazboun}, J.~S., {Simon}, J., {Madison}, D.~R., {et~al.} 2022, \apj, 929, 39,
  \dodoi{10.3847/1538-4357/ac5829}

\bibitem[{{Hebel}(2022)}]{Hebel2022}
{Hebel}, J. 2022, Master's thesis, Rochester Institute of Technology

\bibitem[{{Hellings} \& {Downs}(1983)}]{hd83}
{Hellings}, R.~W., \& {Downs}, G.~S. 1983, \apjl, 265, L39,
  \dodoi{10.1086/183954}

\bibitem[{{Hemberger} \& {Stinebring}(2008)}]{hs08}
{Hemberger}, D.~A., \& {Stinebring}, D.~R. 2008, \apjl, 674, L37,
  \dodoi{10.1086/528985}

\bibitem[{{Hesse} \& {Wielebinski}(1974)}]{hesse_1974}
{Hesse}, K.~H., \& {Wielebinski}, R. 1974, \aap, 31, 409

\bibitem[{{Hobbs} \& {Edwards}(2012)}]{tempo2}
{Hobbs}, G., \& {Edwards}, R. 2012, {Tempo2: Pulsar Timing Package}.
\newblock \doeprint{1210.015}

\bibitem[{{Hobbs} {et~al.}(2012){Hobbs}, {Coles}, {Manchester}, {Keith},
  {Shannon}, {Chen}, {Bailes}, {Bhat}, {Burke-Spolaor}, {Champion},
  {Chaudhary}, {Hotan}, {Khoo}, {Kocz}, {Levin}, {Oslowski}, {Preisig}, {Ravi},
  {Reynolds}, {Sarkissian}, {van Straten}, {Verbiest}, {Yardley}, \&
  {You}}]{hobbs2012}
{Hobbs}, G., {Coles}, W., {Manchester}, R.~N., {et~al.} 2012, \mnras, 427,
  2780, \dodoi{10.1111/j.1365-2966.2012.21946.x}

\bibitem[{{Hobbs} {et~al.}(2020{\natexlab{a}}){Hobbs}, {Guo}, {Caballero},
  {Coles}, {Lee}, {Manchester}, {Reardon}, {Matsakis}, {Tong}, {Arzoumanian},
  {Bailes}, {Bassa}, {Bhat}, {Brazier}, {Burke-Spolaor}, {Champion},
  {Chatterjee}, {Cognard}, {Dai}, {Desvignes}, {Dolch}, {Ferdman}, {Graikou},
  {Guillemot}, {Janssen}, {Keith}, {Kerr}, {Kramer}, {Lam}, {Liu}, {Lyne},
  {Lazio}, {Lynch}, {McKee}, {McLaughlin}, {Mingarelli}, {Nice}, {Os{\l}owski},
  {Pennucci}, {Perera}, {Perrodin}, {Possenti}, {Russell}, {Sanidas}, {Sesana},
  {Shaifullah}, {Shannon}, {Simon}, {Spiewak}, {Stairs}, {Stappers}, {Swiggum},
  {Taylor}, {Theureau}, {Toomey}, {van Haasteren}, {Wang}, {Wang}, \&
  {Zhu}}]{hobbs2020}
{Hobbs}, G., {Guo}, L., {Caballero}, R.~N., {et~al.} 2020{\natexlab{a}},
  \mnras, 491, 5951, \dodoi{10.1093/mnras/stz3071}

\bibitem[{{Hobbs} {et~al.}(2020{\natexlab{b}}){Hobbs}, {Manchester}, {Dunning},
  {Jameson}, {Roberts}, {George}, {Green}, {Tuthill}, {Toomey}, {Kaczmarek},
  {Mader}, {Marquarding}, {Ahmed}, {Amy}, {Bailes}, {Beresford}, {Bhat},
  {Bock}, {Bourne}, {Bowen}, {Brothers}, {Cameron}, {Carretti}, {Carter},
  {Castillo}, {Chekkala}, {Cheng}, {Chung}, {Craig}, {Dai}, {Dawson},
  {Dempsey}, {Doherty}, {Dong}, {Edwards}, {Ergesh}, {Gao}, {Han}, {Hayman},
  {Indermuehle}, {Jeganathan}, {Johnston}, {Kanoniuk}, {Kesteven}, {Kramer},
  {Leach}, {Mcintyre}, {Moss}, {Os{\l}owski}, {Phillips}, {Pope}, {Preisig},
  {Price}, {Reeves}, {Reilly}, {Reynolds}, {Robishaw}, {Roush}, {Ruckley},
  {Sadler}, {Sarkissian}, {Severs}, {Shannon}, {Smart}, {Smith}, {Smith},
  {Sobey}, {Staveley-Smith}, {Tzioumis}, {van Straten}, {Wang}, {Wen}, \&
  {Whiting}}]{Hobbs2020uwb}
{Hobbs}, G., {Manchester}, R.~N., {Dunning}, A., {et~al.} 2020{\natexlab{b}},
  \pasa, 37, e012, \dodoi{10.1017/pasa.2020.2}

\bibitem[{{Huguenin} {et~al.}(1970){Huguenin}, {Taylor}, \&
  {Troland}}]{huguenin_1970}
{Huguenin}, G.~R., {Taylor}, J.~H., \& {Troland}, T.~H. 1970, \apj, 162, 727,
  \dodoi{10.1086/150704}

\bibitem[{Hunter(2007)}]{matplotlib}
Hunter, J.~D. 2007, Computing in Science Engineering, 9, 90,
  \dodoi{10.1109/MCSE.2007.55}

\bibitem[{{Issautier} {et~al.}(2001){Issautier}, {Hoang}, {Moncuquet}, \&
  {Meyer-Vernet}}]{issautier01}
{Issautier}, K., {Hoang}, S., {Moncuquet}, M., \& {Meyer-Vernet}, N. 2001,
  \ssr, 97, 105, \dodoi{10.1023/A:1011878228168}

\bibitem[{{Jennings} {et~al.}(2022{\natexlab{a}}){Jennings}, {Nanograv
  Collaboration}, \& {Chime/Pulsar Collaboration}}]{jennings_2022a}
{Jennings}, R.~J., {Nanograv Collaboration}, \& {Chime/Pulsar Collaboration}.
  2022{\natexlab{a}}, The Astronomer's Telegram, 15223, 1

\bibitem[{{Jennings} {et~al.}(2022{\natexlab{b}}){Jennings}, {Cordes},
  {Chatterjee}, {McLaughlin}, {Demorest}, {Arzoumanian}, {Baker}, {Blumer},
  {Brook}, {Cohen}, {Crawford}, {Cromartie}, {DeCesar}, {Dolch}, {Ferrara},
  {Fonseca}, {Good}, {Hazboun}, {Jones}, {Kaplan}, {Lam}, {Lazio}, {Lorimer},
  {Luo}, {Lynch}, {McKee}, {Madison}, {Meyers}, {Mingarelli}, {Nice},
  {Pennucci}, {Perera}, {Pol}, {Ransom}, {Ray}, {Shapiro-Albert}, {Siemens},
  {Stairs}, {Stinebring}, {Swiggum}, {Tan}, {Taylor}, {Vigeland}, \&
  {Witt}}]{jennings_2022b}
{Jennings}, R.~J., {Cordes}, J.~M., {Chatterjee}, S., {et~al.}
  2022{\natexlab{b}}, arXiv e-prints, arXiv:2210.12266.
\newblock \doarXiv{2210.12266}

\bibitem[{{Jones} {et~al.}(2017){Jones}, {McLaughlin}, {Lam}, {Cordes},
  {Levin}, {Chatterjee}, {Arzoumanian}, {Crowter}, {Demorest}, {Dolch},
  {Ellis}, {Ferdman}, {Fonseca}, {Gonzalez}, {Jones}, {Lazio}, {Nice},
  {Pennucci}, {Ransom}, {Stinebring}, {Stairs}, {Stovall}, {Swiggum}, \&
  {Zhu}}]{jml+2017}
{Jones}, M.~L., {McLaughlin}, M.~A., {Lam}, M.~T., {et~al.} 2017, \apj, 841,
  125, \dodoi{10.3847/1538-4357/aa73df}

\bibitem[{{Jones}(1990)}]{j1990}
{Jones}, P.~B. 1990, \mnras, 246, 364

\bibitem[{{Kaiser} {et~al.}(2022){Kaiser}, {Pol}, {McLaughlin}, {Chen},
  {Hazboun}, {Kelley}, {Simon}, {Taylor}, {Vigeland}, \& {Witt}}]{kaiser+2022}
{Kaiser}, A.~R., {Pol}, N.~S., {McLaughlin}, M.~A., {et~al.} 2022, \apj, 938,
  115, \dodoi{10.3847/1538-4357/ac86cc}

\bibitem[{{Kaplan} {et~al.}(2016){Kaplan}, {Kupfer}, {Nice}, {Irrgang},
  {Heber}, {Arzoumanian}, {Beklen}, {Crowter}, {DeCesar}, {Demorest}, {Dolch},
  {Ellis}, {Ferdman}, {Ferrara}, {Fonseca}, {Gentile}, {Jones}, {Jones},
  {Kreuzer}, {Lam}, {Levin}, {Lorimer}, {Lynch}, {McLaughlin}, {Miller}, {Ng},
  {Pennucci}, {Prince}, {Ransom}, {Ray}, {Spiewak}, {Stairs}, {Stovall},
  {Swiggum}, \& {Zhu}}]{Kaplan+2016}
{Kaplan}, D.~L., {Kupfer}, T., {Nice}, D.~J., {et~al.} 2016, \apj, 826, 86,
  \dodoi{10.3847/0004-637X/826/1/86}

\bibitem[{{Kinkhabwala} \& {Thorsett}(2000)}]{kinkhabwala_2000}
{Kinkhabwala}, A., \& {Thorsett}, S.~E. 2000, \apj, 535, 365,
  \dodoi{10.1086/308844}

\bibitem[{{Kramer} {et~al.}(2006){Kramer}, {Lyne}, {O'Brien}, {Jordan}, \&
  {Lorimer}}]{klo+2006}
{Kramer}, M., {Lyne}, A.~G., {O'Brien}, J.~T., {Jordan}, C.~A., \& {Lorimer},
  D.~R. 2006, Science, 312, 549, \dodoi{10.1126/science.1124060}

\bibitem[{{Lam}(2021)}]{lam21b}
{Lam}, M.~T. 2021, Research Notes of the American Astronomical Society, 5, 167,
  \dodoi{10.3847/2515-5172/ac1670}

\bibitem[{{Lam} {et~al.}(2015){Lam}, {Cordes}, {Chatterjee}, \&
  {Dolch}}]{lcc+15}
{Lam}, M.~T., {Cordes}, J.~M., {Chatterjee}, S., \& {Dolch}, T. 2015, \apj,
  801, 130, \dodoi{10.1088/0004-637X/801/2/130}

\bibitem[{{Lam} {et~al.}(2018{\natexlab{a}}){Lam}, {McLaughlin}, {Cordes},
  {Chatterjee}, \& {Lazio}}]{lam+18optfreq}
{Lam}, M.~T., {McLaughlin}, M.~A., {Cordes}, J.~M., {Chatterjee}, S., \&
  {Lazio}, T.~J.~W. 2018{\natexlab{a}}, \apj, 861, 12,
  \dodoi{10.3847/1538-4357/aac48d}

\bibitem[{{Lam} {et~al.}(2016){Lam}, {Cordes}, {Chatterjee}, {Arzoumanian},
  {Crowter}, {Demorest}, {Dolch}, {Ellis}, {Ferdman}, {Fonseca}, {Gonzalez},
  {Jones}, {Jones}, {Levin}, {Madison}, {McLaughlin}, {Nice}, {Pennucci},
  {Ransom}, {Siemens}, {Stairs}, {Stovall}, {Swiggum}, \& {Zhu}}]{lam+2016}
{Lam}, M.~T., {Cordes}, J.~M., {Chatterjee}, S., {et~al.} 2016, \apj, 819, 155,
  \dodoi{10.3847/0004-637X/819/2/155}

\bibitem[{{Lam} {et~al.}(2017){Lam}, {Cordes}, {Chatterjee}, {Arzoumanian},
  {Crowter}, {Demorest}, {Dolch}, {Ellis}, {Ferdman}, {Fonseca}, {Gonzalez},
  {Jones}, {Jones}, {Levin}, {Madison}, {McLaughlin}, {Nice}, {Pennucci},
  {Ransom}, {Shannon}, {Siemens}, {Stairs}, {Stovall}, {Swiggum}, \&
  {Zhu}}]{lam+2017}
---. 2017, \apj, 834, 35, \dodoi{10.3847/1538-4357/834/1/35}

\bibitem[{{Lam} {et~al.}(2018{\natexlab{b}}){Lam}, {Ellis}, {Grillo}, {Jones},
  {Hazboun}, {Brook}, {Turner}, {Chatterjee}, {Cordes}, {Lazio}, {DeCesar},
  {Arzoumanian}, {Blumer}, {Cromartie}, {Demorest}, {Dolch}, {Ferdman},
  {Ferrara}, {Fonseca}, {Garver-Daniels}, {Gentile}, {Gupta}, {Lorimer},
  {Lynch}, {Madison}, {McLaughlin}, {Ng}, {Nice}, {Pennucci}, {Ransom},
  {Spiewak}, {Stairs}, {Stinebring}, {Stovall}, {Swiggum}, {Vigeland}, \&
  {Zhu}}]{leg+2018}
{Lam}, M.~T., {Ellis}, J.~A., {Grillo}, G., {et~al.} 2018{\natexlab{b}}, \apj,
  861, 132, \dodoi{10.3847/1538-4357/aac770}

\bibitem[{{Lam} {et~al.}(2019){Lam}, {McLaughlin}, {Arzoumanian}, {Blumer},
  {Brook}, {Cromartie}, {Demorest}, {DeCesar}, {Dolch}, {Ellis}, {Ferdman},
  {Ferrara}, {Fonseca}, {Garver-Daniels}, {Gentile}, {Jones}, {Lorimer},
  {Lynch}, {Ng}, {Nice}, {Pennucci}, {Ransom}, {Spiewak}, {Stairs}, {Stovall},
  {Swiggum}, {Vigeland}, \& {Zhu}}]{lam+2019}
{Lam}, M.~T., {McLaughlin}, M.~A., {Arzoumanian}, Z., {et~al.} 2019, \apj, 872,
  193, \dodoi{10.3847/1538-4357/ab01cd}

\bibitem[{{Langlois} {et~al.}(1998){Langlois}, {Sedrakian}, \&
  {Carter}}]{lsc1998}
{Langlois}, D., {Sedrakian}, D.~M., \& {Carter}, B. 1998, \mnras, 297, 1189,
  \dodoi{10.1046/j.1365-8711.1998.01575.x}

\bibitem[{{Lentati} {et~al.}(2014){Lentati}, {Alexander}, {Hobson}, {Feroz},
  {van Haasteren}, {Lee}, \& {Shannon}}]{lah+14}
{Lentati}, L., {Alexander}, P., {Hobson}, M.~P., {et~al.} 2014, \mnras, 437,
  3004, \dodoi{10.1093/mnras/stt2122}

\bibitem[{{Lentati} {et~al.}(2013){Lentati}, {Alexander}, {Hobson}, {Taylor},
  {Gair}, {Balan}, \& {van Haasteren}}]{lentati+2013}
---. 2013, \prd, 87, 104021, \dodoi{10.1103/PhysRevD.87.104021}

\bibitem[{Lentati {et~al.}(2016)}]{Lentati:2016ygu}
Lentati, L., {et~al.} 2016, Mon. Not. Roy. Astron. Soc., 458, 2161,
  \dodoi{10.1093/mnras/stw395}

\bibitem[{{Lin} {et~al.}(2021){Lin}, {Lin}, {Luo}, {Main}, {McKee}, {Pen},
  {Simard}, \& {van Kerkwijk}}]{lin+21}
{Lin}, F.~X., {Lin}, H.-H., {Luo}, J., {et~al.} 2021, \mnras, 508, 1115,
  \dodoi{10.1093/mnras/stab2529}

\bibitem[{{Lin} {et~al.}(2023){Lin}, {Main}, {Jow}, {Li}, {Pen}, \& {van
  Kerkwijk}}]{lmj+23}
{Lin}, F.~X., {Main}, R.~A., {Jow}, D., {et~al.} 2023, \mnras, 519, 121,
  \dodoi{10.1093/mnras/stac3456}

\bibitem[{{Liu} {et~al.}(2014){Liu}, {Desvignes}, {Cognard}, {Stappers},
  {Verbiest}, {Lee}, {Champion}, {Kramer}, {Freire}, \& {Karuppusamy}}]{liu14}
{Liu}, K., {Desvignes}, G., {Cognard}, I., {et~al.} 2014, \mnras, 443, 3752,
  \dodoi{10.1093/mnras/stu1420}

\bibitem[{{Liu} {et~al.}(2016){Liu}, {Bassa}, {Janssen}, {Karuppusamy},
  {McKee}, {Kramer}, {Lee}, {Perrodin}, {Purver}, {Sanidas}, {Smits},
  {Stappers}, {Weltevrede}, \& {Zhu}}]{2016MNRAS.463.3239L}
{Liu}, K., {Bassa}, C.~G., {Janssen}, G.~H., {et~al.} 2016, \mnras, 463, 3239,
  \dodoi{10.1093/mnras/stw2223}

\bibitem[{{Liu} {et~al.}(2023){Liu}, {Cohen}, {McGrath}, {Demorest}, \&
  {Vigeland}}]{liu+2023}
{Liu}, T., {Cohen}, T., {McGrath}, C., {Demorest}, P., \& {Vigeland}, S. 2023,
  arXiv e-prints, arXiv:2301.07135, \dodoi{10.48550/arXiv.2301.07135}

\bibitem[{{Lommen} \& {Demorest}(2013)}]{2013CQGra..30v4001L}
{Lommen}, A.~N., \& {Demorest}, P. 2013, Classical and Quantum Gravity, 30,
  224001, \dodoi{10.1088/0264-9381/30/22/224001}

\bibitem[{Lorimer \& Kramer(2005)}]{lorimer_and_kramer05}
Lorimer, D.~R., \& Kramer, M. 2005, Handbook of Pulsar Astronomy (Cambridge
  University Press)

\bibitem[{{Lower} {et~al.}(2020){Lower}, {Bailes}, {Shannon}, {Johnston},
  {Flynn}, {Os{\l}owski}, {Gupta}, {Farah}, {Bateman}, {Green}, {Hunstead},
  {Jameson}, {Jankowski}, {Parthasarathy}, {Price}, {Sutherland}, {Temby}, \&
  {Venkatraman Krishnan}}]{lower2020}
{Lower}, M.~E., {Bailes}, M., {Shannon}, R.~M., {et~al.} 2020, \mnras, 494,
  228, \dodoi{10.1093/mnras/staa615}

\bibitem[{{Luo} {et~al.}(2021){Luo}, {Ransom}, {Demorest}, {Ray}, {Archibald},
  {Kerr}, {Jennings}, {Bachetti}, {van Haasteren}, {Champagne}, {Colen},
  {Phillips}, {Zimmerman}, {Stovall}, {Lam}, \& {Jenet}}]{luo+2021}
{Luo}, J., {Ransom}, S., {Demorest}, P., {et~al.} 2021, \apj, 911, 45,
  \dodoi{10.3847/1538-4357/abe62f}

\bibitem[{{Lyne} {et~al.}(2010){Lyne}, {Hobbs}, {Kramer}, {Stairs}, \&
  {Stappers}}]{lhk+2010}
{Lyne}, A., {Hobbs}, G., {Kramer}, M., {Stairs}, I., \& {Stappers}, B. 2010,
  Science, 329, 408, \dodoi{10.1126/science.1186683}

\bibitem[{{Madison} {et~al.}(2019){Madison}, {Cordes}, {Arzoumanian},
  {Chatterjee}, {Crowter}, {DeCesar}, {Demorest}, {Dolch}, {Ellis}, {Ferdman},
  {Ferrara}, {Fonseca}, {Gentile}, {Jones}, {Jones}, {Lam}, {Levin}, {Lorimer},
  {Lynch}, {McLaughlin}, {Mingarelli}, {Ng}, {Nice}, {Pennucci}, {Ransom},
  {Ray}, {Spiewak}, {Stairs}, {Stovall}, {Swiggum}, \& {Zhu}}]{madison+2019}
{Madison}, D.~R., {Cordes}, J.~M., {Arzoumanian}, Z., {et~al.} 2019, \apj, 872,
  150, \dodoi{10.3847/1538-4357/ab01fd}

\bibitem[{{Mahajan} {et~al.}(2018){Mahajan}, {van Kerkwijk}, {Main}, \&
  {Pen}}]{mahajan_2018}
{Mahajan}, N., {van Kerkwijk}, M.~H., {Main}, R., \& {Pen}, U.-L. 2018, \apjl,
  867, L2, \dodoi{10.3847/2041-8213/aae713}

\bibitem[{{Main} {et~al.}(2018){Main}, {Yang}, {Chan}, {Li}, {Lin}, {Mahajan},
  {Pen}, {Vanderlinde}, \& {van Kerkwijk}}]{myc+18}
{Main}, R., {Yang}, I.~S., {Chan}, V., {et~al.} 2018, \nat, 557, 522,
  \dodoi{10.1038/s41586-018-0133-z}

\bibitem[{{Marsch}(2006)}]{marsch06}
{Marsch}, E. 2006, Living Reviews in Solar Physics, 3, 1,
  \dodoi{10.12942/lrsp-2006-1}

\bibitem[{{Matsakis} {et~al.}(1997){Matsakis}, {Taylor}, \&
  {Eubanks}}]{matsakis1997}
{Matsakis}, D.~N., {Taylor}, J.~H., \& {Eubanks}, T.~M. 1997, \aap, 326, 924

\bibitem[{{McKee} {et~al.}(2016){McKee}, {Janssen}, {Stappers}, {Lyne},
  {Caballero}, {Lentati}, {Desvignes}, {Jessner}, {Jordan}, {Karuppusamy},
  {Kramer}, {Cognard}, {Champion}, {Graikou}, {Lazarus}, {Os{\l}owski},
  {Perrodin}, {Shaifullah}, {Tiburzi}, \& {Verbiest}}]{mckee+16}
{McKee}, J.~W., {Janssen}, G.~H., {Stappers}, B.~W., {et~al.} 2016, MNRAS, 461,
  2809, \dodoi{10.1093/mnras/stw1442}

\bibitem[{{McKee} {et~al.}(2019){McKee}, {Stappers}, {Bassa}, {Chen},
  {Cognard}, {Gaikwad}, {Janssen}, {Karuppusamy}, {Kramer}, {Lee}, {Liu},
  {Perrodin}, {Sanidas}, {Smits}, {Wang}, \& {Zhu}}]{mckee_2019}
{McKee}, J.~W., {Stappers}, B.~W., {Bassa}, C.~G., {et~al.} 2019, \mnras, 483,
  4784, \dodoi{10.1093/mnras/sty3058}

\bibitem[{{Melatos} \& {Link}(2014)}]{ml2014}
{Melatos}, A., \& {Link}, B. 2014, \mnras, 437, 21,
  \dodoi{10.1093/mnras/stt1828}

\bibitem[{{Meyers} \& {Chime/Pulsar Collaboration}(2021)}]{meyers_2021}
{Meyers}, B., \& {Chime/Pulsar Collaboration}. 2021, The Astronomer's Telegram,
  14652, 1

\bibitem[{{Miles} {et~al.}(2022){Miles}, {Shannon}, {Bailes}, {Reardon},
  {Buchner}, {Middleton}, \& {Spiewak}}]{miles_2022}
{Miles}, M.~T., {Shannon}, R.~M., {Bailes}, M., {et~al.} 2022, \mnras, 510,
  5908, \dodoi{10.1093/mnras/stab3549}

\bibitem[{{Miles} {et~al.}(2023){Miles}, {Shannon}, {Bailes}, {Reardon},
  {Keith}, {Cameron}, {Parthasarathy}, {Shamohammadi}, {Spiewak}, {van
  Straten}, {Buchner}, {Camilo}, {Geyer}, {Karastergiou}, {Kramer}, {Serylak},
  {Theureau}, \& {Venkatraman Krishnan}}]{miles+2023}
---. 2023, \mnras, 519, 3976, \dodoi{10.1093/mnras/stac3644}

\bibitem[{{Nice} \& {Taylor}(1995)}]{nice+1995}
{Nice}, D.~J., \& {Taylor}, J.~H. 1995, \apj, 441, 429, \dodoi{10.1086/175367}

\bibitem[{{Page}(1973)}]{page_1973}
{Page}, C.~G. 1973, \mnras, 163, 29, \dodoi{10.1093/mnras/163.1.29}

\bibitem[{{Palliyaguru} {et~al.}(2015){Palliyaguru}, {Stinebring},
  {McLaughlin}, {Demorest}, \& {Jones}}]{2015ApJ...815...89P}
{Palliyaguru}, N., {Stinebring}, D., {McLaughlin}, M., {Demorest}, P., \&
  {Jones}, G. 2015, \apj, 815, 89, \dodoi{10.1088/0004-637X/815/2/89}

\bibitem[{{Park} {et~al.}(2021){Park}, {Folkner}, {Williams}, \&
  {Boggs}}]{park+21}
{Park}, R.~S., {Folkner}, W.~M., {Williams}, J.~G., \& {Boggs}, D.~H. 2021,
  \aj, 161, 105, \dodoi{10.3847/1538-3881/abd414}

\bibitem[{{Parthasarathy} {et~al.}(2021){Parthasarathy}, {Bailes}, {Shannon},
  {van Straten}, {Os{\l}owski}, {Johnston}, {Spiewak}, {Reardon}, {Kramer},
  {Venkatraman Krishnan}, {Pennucci}, {Abbate}, {Buchner}, {Camilo},
  {Champion}, {Geyer}, {Hugo}, {Jameson}, {Karastergiou}, {Keith}, \&
  {Serylak}}]{Parthasarathy+2021}
{Parthasarathy}, A., {Bailes}, M., {Shannon}, R.~M., {et~al.} 2021, \mnras,
  502, 407, \dodoi{10.1093/mnras/stab037}

\bibitem[{{Pennucci}(2019)}]{2019ApJ...871...34P}
{Pennucci}, T.~T. 2019, \apj, 871, 34, \dodoi{10.3847/1538-4357/aaf6ef}

\bibitem[{{Pennucci} {et~al.}(2014){Pennucci}, {Demorest}, \& {Ransom}}]{pen14}
{Pennucci}, T.~T., {Demorest}, P.~B., \& {Ransom}, S.~M. 2014, \apj, 790, 93,
  \dodoi{10.1088/0004-637X/790/2/93}

\bibitem[{{Perera} {et~al.}(2018){Perera}, {Stappers}, {Babak}, {Keith},
  {Antoniadis}, {Bassa}, {Caballero}, {Champion}, {Cognard}, {Desvignes},
  {Graikou}, {Guillemot}, {Janssen}, {Karuppusamy}, {Kramer}, {Lazarus},
  {Lentati}, {Liu}, {Lyne}, {McKee}, {Os{\l}owski}, {Perrodin}, {Sanidas},
  {Sesana}, {Shaifullah}, {Theureau}, {Verbiest}, \& {Taylor}}]{psb+18}
{Perera}, B.~B.~P., {Stappers}, B.~W., {Babak}, S., {et~al.} 2018, \mnras, 478,
  218, \dodoi{10.1093/mnras/sty1116}

\bibitem[{{Pol} {et~al.}(2021){Pol}, {Taylor}, {Kelley}, {Vigeland}, {Simon},
  {Chen}, {Arzoumanian}, {Baker}, {B{\'e}csy}, {Brazier}, {Brook},
  {Burke-Spolaor}, {Chatterjee}, {Cordes}, {Cornish}, {Crawford}, {Thankful
  Cromartie}, {Decesar}, {Demorest}, {Dolch}, {Ferrara}, {Fiore}, {Fonseca},
  {Garver-Daniels}, {Good}, {Hazboun}, {Jennings}, {Jones}, {Kaiser}, {Kaplan},
  {Shapiro Key}, {Lam}, {Lazio}, {Luo}, {Lynch}, {Madison}, {McEwen},
  {McLaughlin}, {Mingarelli}, {Ng}, {Nice}, {Pennucci}, {Ransom}, {Ray},
  {Shapiro-Albert}, {Siemens}, {Stairs}, {Stinebring}, {Swiggum}, {Vallisneri},
  {Wahl}, {Witt}, \& {Nanograv Collaboration}}]{pol+2021}
{Pol}, N.~S., {Taylor}, S.~R., {Kelley}, L.~Z., {et~al.} 2021, \apjl, 911, L34,
  \dodoi{10.3847/2041-8213/abf2c9}

\bibitem[{{Ransom} {et~al.}(202314){Ransom}, {Stairs}, {Archibald}, {Hessels},
  {Kaplan}, {van Kerkwijk}, {Boyles}, {Deller}, {Chatterjee},
  {Schechtman-Rook}, {Berndsen}, {Lynch}, {Lorimer}, {Karako-Argaman}, {Kaspi},
  {Kondratiev}, {McLaughlin}, {van Leeuwen}, {Rosen}, {Roberts}, \&
  {Stovall}}]{Ransom+2014}
{Ransom}, S.~M., {Stairs}, I.~H., {Archibald}, A.~M., {et~al.} 202314, \nat,
  505, 520, \dodoi{10.1038/nature12917}

\bibitem[{{Ray} {et~al.}(2019){Ray}, {Guillot}, {Ho}, {Kerr}, {Enoto},
  {Gendreau}, {Arzoumanian}, {Altamirano}, {Bogdanov}, {Campion},
  {Chakrabarty}, {Deneva}, {Jaisawal}, {Kozon}, {Malacaria}, {Strohmayer}, \&
  {Wolff}}]{ray+19}
{Ray}, P.~S., {Guillot}, S., {Ho}, W. C.~G., {et~al.} 2019, \apj, 879, 130,
  \dodoi{10.3847/1538-4357/ab24d8}

\bibitem[{{Romani} \& {Johnston}(2001)}]{romani_2001}
{Romani}, R.~W., \& {Johnston}, S. 2001, \apjl, 557, L93,
  \dodoi{10.1086/323415}

\bibitem[{{Rosado} {et~al.}(2015){Rosado}, {Sesana}, \& {Gair}}]{rsg2015}
{Rosado}, P.~A., {Sesana}, A., \& {Gair}, J. 2015, \mnras, 451, 2417,
  \dodoi{10.1093/mnras/stv1098}

\bibitem[{{Sallmen} \& {Backer}(1995)}]{sallmen_backer_1995}
{Sallmen}, S., \& {Backer}, D.~C. 1995, in Astronomical Society of the Pacific
  Conference Series, Vol.~72, Millisecond Pulsars. A Decade of Surprise, ed.
  A.~S. {Fruchter}, M.~{Tavani}, \& D.~C. {Backer}, 340

\bibitem[{{Sazhin}(1978)}]{saz78}
{Sazhin}, M.~V. 1978, \sovast, 22, 36

\bibitem[{{Sesana} {et~al.}(2004){Sesana}, {Haardt}, {Madau}, \&
  {Volonteri}}]{shm+04}
{Sesana}, A., {Haardt}, F., {Madau}, P., \& {Volonteri}, M. 2004, \apj, 611,
  623, \dodoi{10.1086/422185}

\bibitem[{{Shabanova}(2005)}]{Shabanova05}
{Shabanova}, T.~V. 2005, \mnras, 356, 1435,
  \dodoi{10.1111/j.1365-2966.2004.08580.x}

\bibitem[{{Shannon} \& {Cordes}(2010)}]{sc2010}
{Shannon}, R.~M., \& {Cordes}, J.~M. 2010, \apj, 725, 1607,
  \dodoi{10.1088/0004-637X/725/2/1607}

\bibitem[{{Shannon} {et~al.}(2013){Shannon}, {Cordes}, {Metcalfe}, {Lazio},
  {Cognard}, {Desvignes}, {Janssen}, {Jessner}, {Kramer}, {Lazaridis},
  {Purver}, {Stappers}, \& {Theureau}}]{shannon_2013}
{Shannon}, R.~M., {Cordes}, J.~M., {Metcalfe}, T.~S., {et~al.} 2013, \apj, 766,
  5, \dodoi{10.1088/0004-637X/766/1/5}

\bibitem[{{Shannon} {et~al.}(2014){Shannon}, {Os{\l}owski}, {Dai}, {Bailes},
  {Hobbs}, {Manchester}, {van Straten}, {Raithel}, {Ravi}, {Toomey}, {Bhat},
  {Burke-Spolaor}, {Coles}, {Keith}, {Kerr}, {Levin}, {Sarkissian}, {Wang},
  {Wen}, \& {Zhu}}]{2014MNRAS.443.1463S}
{Shannon}, R.~M., {Os{\l}owski}, S., {Dai}, S., {et~al.} 2014, \mnras, 443,
  1463, \dodoi{10.1093/mnras/stu1213}

\bibitem[{{Shannon} {et~al.}(2016){Shannon}, {Lentati}, {Kerr}, {Bailes},
  {Bhat}, {Coles}, {Dai}, {Dempsey}, {Hobbs}, {Keith}, {Lasky}, {Levin},
  {Manchester}, {Os{\l}owski}, {Ravi}, {Reardon}, {Rosado}, {Spiewak}, {van
  Straten}, {Toomey}, {Wang}, {Wen}, {You}, \& {Zhu}}]{shannon_2016}
{Shannon}, R.~M., {Lentati}, L.~T., {Kerr}, M., {et~al.} 2016, \apjl, 828, L1,
  \dodoi{10.3847/2041-8205/828/1/L1}

\bibitem[{{Shapiro-Albert} {et~al.}(2021){Shapiro-Albert}, {Hazboun},
  {McLaughlin}, \& {Lam}}]{shapiro-albert+2021}
{Shapiro-Albert}, B.~J., {Hazboun}, J.~S., {McLaughlin}, M.~A., \& {Lam}, M.~T.
  2021, \apj, 909, 219, \dodoi{10.3847/1538-4357/abdc29}

\bibitem[{{Shapiro-Albert} {et~al.}(2020){Shapiro-Albert}, {McLaughlin}, {Lam},
  {Cordes}, \& {Swiggum}}]{2020ApJ...890..123S}
{Shapiro-Albert}, B.~J., {McLaughlin}, M.~A., {Lam}, M.~T., {Cordes}, J.~M., \&
  {Swiggum}, J.~K. 2020, \apj, 890, 123, \dodoi{10.3847/1538-4357/ab65f8}

\bibitem[{{Singha} {et~al.}(2021{\natexlab{a}}){Singha}, {Joshi}, {Maan},
  {Gupta}, {Agarwal}, {Bagchi}, {Batra}, {Basu}, {Bethapudi}, {Choudhary},
  {Dandapat}, {Desai}, {De}, {Dey}, {Gopakumar}, {Girgaonkar}, {Krishnakumar},
  {Manoharan}, {Naidu}, {Nobleson}, {Pandey}, {Pathak}, {Prabu}, {Surnis},
  {Susarla}, \& {Susobhanan}}]{singha_2021a}
{Singha}, J., {Joshi}, B.~C., {Maan}, Y., {et~al.} 2021{\natexlab{a}}, The
  Astronomer's Telegram, 14667, 1

\bibitem[{{Singha} {et~al.}(2021{\natexlab{b}}){Singha}, {Surnis}, {Joshi},
  {Tarafdar}, {Rana}, {Susobhanan}, {Girgaonkar}, {Kolhe}, {Agarwal}, {Desai},
  {Prabu}, {Bathula}, {Dandapat}, {Dey}, {Hisano}, {Kato}, {Kharbanda},
  {Kikunaga}, {Marmat}, {Susarla}, {Bagchi}, {Dhanda Batra}, {Choudhury},
  {Gopakumar}, {Gupta}, {Krishnakumar}, {Maan}, {Manoharan}, {Nobleson},
  {Pandian}, {Pathak}, \& {Takahashi}}]{singha_2021b}
{Singha}, J., {Surnis}, M.~P., {Joshi}, B.~C., {et~al.} 2021{\natexlab{b}},
  \mnras, 507, L57, \dodoi{10.1093/mnrasl/slab098}

\bibitem[{{Soglasnov} {et~al.}(2004){Soglasnov}, {Popov}, {Bartel}, {Cannon},
  {Novikov}, {Kondratiev}, \& {Altunin}}]{soglasnov_2004}
{Soglasnov}, V.~A., {Popov}, M.~V., {Bartel}, N., {et~al.} 2004, \apj, 616,
  439, \dodoi{10.1086/424908}

\bibitem[{{Speri} {et~al.}(2023){Speri}, {Porayko}, {Falxa}, {Chen}, {Gair},
  {Sesana}, \& {Taylor}}]{speri+2023}
{Speri}, L., {Porayko}, N.~K., {Falxa}, M., {et~al.} 2023, \mnras, 518, 1802,
  \dodoi{10.1093/mnras/stac3237}

\bibitem[{{Staelin} \& {Reifenstein}(1968)}]{staelin+1968}
{Staelin}, D.~H., \& {Reifenstein}, Edward~C., I. 1968, Science, 162, 1481,
  \dodoi{10.1126/science.162.3861.1481}

\bibitem[{Standish(1982)}]{standish82}
Standish, E.~M. 1982, \aap, 114, 297

\bibitem[{{Stinebring}(2013)}]{sti13}
{Stinebring}, D. 2013, Classical and Quantum Gravity, 30, 224006,
  \dodoi{10.1088/0264-9381/30/22/224006}

\bibitem[{{Stinebring} {et~al.}(1984){Stinebring}, {Cordes}, {Weisberg},
  {Rankin}, \& {Boriakoff}}]{stinebring+1984}
{Stinebring}, D.~R., {Cordes}, J.~M., {Weisberg}, J.~M., {Rankin}, J.~M., \&
  {Boriakoff}, V. 1984, \apjs, 55, 279, \dodoi{10.1086/190955}

\bibitem[{{Tarafdar} {et~al.}(2022){Tarafdar}, {Nobleson}, {Rana}, {Singha},
  {Krishnakumar}, {Joshi}, {Paladi}, {Kolhe}, {Batra}, {Agarwal}, {Bathula},
  {Dandapat}, {Desai}, {Dey}, {Hisano}, {Ingale}, {Kato}, {Kharbanda},
  {Kikunaga}, {Marmat}, {Pandian}, {Prabu}, {Srivastava}, {Surnis}, {Susarla},
  {Susobhanan}, {Takahashi}, {Arumugam}, {Bagchi}, {Banik}, {De}, {Girgaonkar},
  {Gopakumar}, {Gupta}, {Maan}, {Manoharan}, {Naidu}, \&
  {Pathak}}]{tarafdar2022}
{Tarafdar}, P., {Nobleson}, K., {Rana}, P., {et~al.} 2022, \pasa, 39, e053,
  \dodoi{10.1017/pasa.2022.46}

\bibitem[{{Taylor}(1992)}]{taylor1992}
{Taylor}, J.~H. 1992, Philosophical Transactions of the Royal Society of London
  Series A, 341, 117, \dodoi{10.1098/rsta.1992.0088}

\bibitem[{{Taylor}(2021)}]{tay21}
{Taylor}, S.~R. 2021, arXiv e-prints, arXiv:2105.13270,
  \dodoi{10.48550/arXiv.2105.13270}

\bibitem[{Taylor {et~al.}(2021)Taylor, Baker, Hazboun, Simon, \&
  Vigeland}]{e_e}
Taylor, S.~R., Baker, P.~T., Hazboun, J.~S., Simon, J., \& Vigeland, S.~J.
  2021, enterprise$\_$extensions.
\newblock \url{https://github.com/nanograv/enterprise_extensions}

\bibitem[{{Thorsett} {et~al.}(1999){Thorsett}, {Arzoumanian}, {Camilo}, \&
  {Lyne}}]{Thorsett+1999}
{Thorsett}, S.~E., {Arzoumanian}, Z., {Camilo}, F., \& {Lyne}, A.~G. 1999,
  \apj, 523, 763, \dodoi{10.1086/307771}

\bibitem[{{Thrane} \& {Romano}(2013)}]{thrane+2013}
{Thrane}, E., \& {Romano}, J.~D. 2013, \prd, 88, 124032,
  \dodoi{10.1103/PhysRevD.88.124032}

\bibitem[{{Tiburzi} {et~al.}(2016){Tiburzi}, {Hobbs}, {Kerr}, {Coles}, {Dai},
  {Manchester}, {Possenti}, {Shannon}, \& {You}}]{thk+2016}
{Tiburzi}, C., {Hobbs}, G., {Kerr}, M., {et~al.} 2016, \mnras, 455, 4339,
  \dodoi{10.1093/mnras/stv2143}

\bibitem[{{Tiburzi} {et~al.}(2019){Tiburzi}, {Verbiest}, {Shaifullah},
  {Janssen}, {Anderson}, {Horneffer}, {K{\"u}nsem{\"o}ller}, {Os{\l}owski},
  {Donner}, {Kramer}, {Kumari}, {Porayko}, {Zucca}, {Ciardi}, {Dettmar},
  {Grie{\ss}meier}, {Hoeft}, \& {Serylak}}]{tvs+19}
{Tiburzi}, C., {Verbiest}, J.~P.~W., {Shaifullah}, G.~M., {et~al.} 2019,
  \mnras, 487, 394, \dodoi{10.1093/mnras/stz1278}

\bibitem[{{Tiburzi} {et~al.}(2021){Tiburzi}, {Shaifullah}, {Bassa}, {Zucca},
  {Verbiest}, {Porayko}, {van der Wateren}, {Fallows}, {Main}, {Janssen},
  {Anderson}, {Bak Nielsen}, {Donner}, {Keane}, {K{\"u}nsem{\"o}ller},
  {Os{\l}owski}, {Grie{\ss}meier}, {Serylak}, {Br{\"u}ggen}, {Ciardi},
  {Dettmar}, {Hoeft}, {Kramer}, {Mann}, \& {Vocks}}]{tsb+21}
{Tiburzi}, C., {Shaifullah}, G.~M., {Bassa}, C.~G., {et~al.} 2021, \aap, 647,
  A84, \dodoi{10.1051/0004-6361/202039846}

\bibitem[{Turner {et~al.}(2023)Turner, Stinebring, McLaughlin, Archibald,
  Dolch, \& Lynch}]{turner23}
Turner, J.~E., Stinebring, D.~R., McLaughlin, M.~A., {et~al.} 2023, \apj, 944,
  191, \dodoi{10.3847/1538-4357/acb6fd}

\bibitem[{{Vallisneri}(2020)}]{libstempo}
{Vallisneri}, M. 2020, {libstempo: Python wrapper for Tempo2}.
\newblock \doeprint{2002.017}

\bibitem[{{Vallisneri} {et~al.}(2020){Vallisneri}, {Taylor}, {Simon},
  {Folkner}, {Park}, {Cutler}, {Ellis}, {Lazio}, {Vigeland}, {Aggarwal},
  {Arzoumanian}, {Baker}, {Brazier}, {Brook}, {Burke-Spolaor}, {Chatterjee},
  {Cordes}, {Cornish}, {Crawford}, {Cromartie}, {Crowter}, {DeCesar},
  {Demorest}, {Dolch}, {Ferdman}, {Ferrara}, {Fonseca}, {Garver-Daniels},
  {Gentile}, {Good}, {Hazboun}, {Holgado}, {Huerta}, {Islo}, {Jennings},
  {Jones}, {Jones}, {Kaplan}, {Kelley}, {Key}, {Lam}, {Levin}, {Lorimer},
  {Luo}, {Lynch}, {Madison}, {McLaughlin}, {McWilliams}, {Mingarelli}, {Ng},
  {Nice}, {Pennucci}, {Pol}, {Ransom}, {Ray}, {Siemens}, {Spiewak}, {Stairs},
  {Stinebring}, {Stovall}, {Swiggum}, {van Haasteren}, {Witt}, \&
  {Zhu}}]{vallisneri+20}
{Vallisneri}, M., {Taylor}, S.~R., {Simon}, J., {et~al.} 2020, \apj, 893, 112,
  \dodoi{10.3847/1538-4357/ab7b67}

\bibitem[{{van Eysden} \& {Link}(2018)}]{vEl2018}
{van Eysden}, C.~A., \& {Link}, B. 2018, \apj, 865, 60,
  \dodoi{10.3847/1538-4357/aacc24}

\bibitem[{{van Haasteren} \& {Levin}(2013)}]{vhaasteren+2013}
{van Haasteren}, R., \& {Levin}, Y. 2013, \mnras, 428, 1147,
  \dodoi{10.1093/mnras/sts097}

\bibitem[{{van Haasteren} {et~al.}(2009){van Haasteren}, {Levin}, {McDonald},
  \& {Lu}}]{vanhaasteren+2009}
{van Haasteren}, R., {Levin}, Y., {McDonald}, P., \& {Lu}, T. 2009, \mnras,
  395, 1005, \dodoi{10.1111/j.1365-2966.2009.14590.x}

\bibitem[{{van Haasteren} \& {Vallisneri}(2014)}]{vHv:2014}
{van Haasteren}, R., \& {Vallisneri}, M. 2014, \prd, 90, 104012,
  \dodoi{10.1103/PhysRevD.90.104012}

\bibitem[{{van Straten}(2006)}]{2006ApJ...642.1004V}
{van Straten}, W. 2006, \apj, 642, 1004, \dodoi{10.1086/501001}

\bibitem[{{Verbiest} \& {Shaifullah}(2018)}]{vs2018}
{Verbiest}, J. P.~W., \& {Shaifullah}, G.~M. 2018, Classical and Quantum
  Gravity, 35, 133001, \dodoi{10.1088/1361-6382/aac412}

\bibitem[{{Vigeland} \& {Vallisneri}(2014)}]{vigeland+2014}
{Vigeland}, S.~J., \& {Vallisneri}, M. 2014, \mnras, 440, 1446,
  \dodoi{10.1093/mnras/stu312}

\bibitem[{{Walker} {et~al.}(2013){Walker}, {Demorest}, \& {van
  Straten}}]{2013ApJ...779...99W}
{Walker}, M.~A., {Demorest}, P.~B., \& {van Straten}, W. 2013, \apj, 779, 99,
  \dodoi{10.1088/0004-637X/779/2/99}

\bibitem[{{Wang} {et~al.}(2021){Wang}, {Wang}, {Wang}, {Feng}, {Zhang}, {Lee},
  {Li}, {Lu}, {Xie}, {Zhou}, \& {Zhang}}]{wang_2021}
{Wang}, S.~Q., {Wang}, J.~B., {Wang}, N., {et~al.} 2021, \apj, 913, 67,
  \dodoi{10.3847/1538-4357/abf937}

\bibitem[{Williams \& Rasmussen(2006)}]{rw06}
Williams, C.~K., \& Rasmussen, C.~E. 2006, the MIT Press, 2, 4

\bibitem[{{Wolszczan} {et~al.}(1984){Wolszczan}, {Cordes}, \&
  {Stinebring}}]{wolszczan_1984}
{Wolszczan}, A., {Cordes}, J., \& {Stinebring}, D. 1984, in Birth and Evolution
  of Neutron Stars: Issues Raised by Millisecond Pulsars, ed. S.~P. {Reynolds}
  \& D.~R. {Stinebring}, 63

\bibitem[{{Xu} {et~al.}(2021){Xu}, {Huang}, {Burgay}, {Champion}, {Cognard},
  {Guillemot}, {Jang}, {Karuppusamy}, {Kramer}, {Lackeos}, {Lee}, {Liu},
  {Perrodin}, {Possenti}, {Stappers}, \& {Theureau}}]{xu_2021}
{Xu}, H., {Huang}, Y.~X., {Burgay}, M., {et~al.} 2021, The Astronomer's
  Telegram, 14642, 1

\end{thebibliography}

\appendix

\section{The Noise Covariance Matrix} \label{app_cov_matrix}
\subsection{Marginalizing the Timing Models} \label{app_marginalize}
In Equation \ref{eq:tm_marginalize}, we identify the full PTA marginalized timing model as 
\ba
\mathcal{P}(\resid|\mathbf{a}, \mathbf{E}) &=& \int \mathcal{P}(\resid|\delpar, \mathbf{a}, \mathbf{E})\mathcal{P}(\delpar) {\rm d} (\delpar)\\
&=& \frac{\exp\left(-\frac{1}{2}(\resid-F\mathbf{a})^\mathrm{T} D^{-1}(\resid-F\mathbf{a})\right)}{\sqrt{\det\left(2\pi D\right)}},
\ea
where
\be
D\equiv C+MXM^\mathrm{T}.
\label{eq:D_matrix}
\ee 
Using the Woodbury identity, $D^{-1}$ can be written as
\ba\label{eq:D_woodbury}
D^{-1}  &=& (C+MXM^\mathrm{T})^{-1}\nonumber\\
        &=& C^{-1}-C^{-1}M(X^{-1}+M^\mathrm{T} C^{-1} M)^{-1} M^\mathrm{T} C^{-1}.
\ea
One can now set the range on the priors for $\delpar$ to be infinite by setting $X={\rm diag}(\infty)$, hence $X^{-1}\rightarrow 0$ and the expression in \myeq{eq:D_woodbury} is further simplified to 
\be\label{eq:D_woodbury_infty}
D^{-1}= C^{-1}R=C^{-1}-C^{-1}M(M^\mathrm{T} C^{-1} M)^{-1} M^\mathrm{T} C^{-1},
\ee
which has completely removed any dependence on the timing model \emph{parameters} from the likelihood, though dependence on the timing model is preserved in the design matrix, $M$. Note we have defined $R=1-M(M^\mathrm{T} C^{-1} M)^{-1} M^\mathrm{T} C^{-1}$, where $R$ acts to project the effects of the timing model out of the covariance matrix. 

We use a similar marginalization to find the final form for a single pulsar likelihood (Equation \ref{eq:a_marginalize}): 
\ba
\mathcal{P}(\resid|\mathbf{E}) &=& \int \mathcal{P}(\resid|\mathbf{a}, \mathbf{E})\mathcal{P}(\mathbf{a}){\rm d} (\mathbf{a})\nonumber\\
&=& \frac{\exp\left(-\frac{1}{2}\resid^\mathrm{T} N^{-1}\resid\right)}{\sqrt{\det\left(2\pi N\right)}},
\ea
where
\be 
N\equiv D +F \varphi F^\mathrm{T}.
\label{eq:N_matrix}
\ee
The Woodbury lemma can again be used to get the following form for the inverse:
\ba\label{eq:N_woodbury}	
N^{-1}  &=& (D+F\varphi F^\mathrm{T})^{-1}\nonumber\\
        &=& D^{-1}-D^{-1}F(\varphi^{-1}+F^\mathrm{T} D^{-1} F)^{-1} F^\mathrm{T} D^{-1}.
\ea

The marginalizations discussed above and used to define the noise covariance matrix in Section \ref{subsec:phenommatrix} have been developed precisely for efficient calculation since $D^{-1}$ and $F$ are constant when the WN is fixed, and the only matrix inversion that needs to be calculated is the one in the parenthetical of \myeq{eq:N_woodbury}, a matrix that is only $(2N_{\rm freq}\times 2N_{\rm freq})$.

\subsection{Expanded Signal Covariance Matrix} \label{app_signal_matrix}
As described in \mysec{s:covmatrix}, we define the noise model of a particular observing epoch as the matrix containing information about the frequency dependence of template-fitting uncertainties $\sigma_{\rm S/N}$, pulse jitter $\sigma_{\rm J}$, and diffractive interstellar scattering $\sigma_{\rm DISS}$ (see Sections \ref{s:phenom_noise} and \ref{s:physical_noise}, respectively):

\begin{widetext}
\be C^\mathrm{epoch} =
\begin{bmatrix}
\sigma_{\rm S/N}^2(\nu_1) + \sigma_{\rm J}^2 + \sigma_{\rm DISS}^2(\nu_1, \nu_1) & \sigma_{\rm J}^2 + \sigma_{\rm DISS}^2(\nu_1, \nu_2) & \cdots & \sigma_{\rm J}^2 + \sigma_{\rm DISS}^2(\nu_1, \nu_N)\\ 
\sigma_{\rm J}^2 + \sigma_{\rm DISS}^2(\nu_2, \nu_1) & \sigma_{\rm S/N}^2(\nu_2) + \sigma_{\rm J}^2 + \sigma_{\rm DISS}^2(\nu_2, \nu_2) & \cdots & \sigma_{\rm J}^2 + \sigma_{\rm DISS}^2(\nu_2, \nu_N) \\ 
\vdots & \vdots & \ddots & \vdots \\
\sigma_{\rm J}^2 + \sigma_{\rm DISS}^2(\nu_N, \nu_1) & \sigma_{\rm J}^2 + \sigma_{\rm DISS}^2(\nu_N, \nu_2) & \cdots & \sigma_{\rm S/N}^2(\nu_N) + \sigma_{\rm J}^2 + \sigma_{\rm DISS}^2(\nu_N, \nu_N)
\end{bmatrix}\label{e:singlecovmatrix}%
\ee
\end{widetext}
Compare \myeq{e:singlecovmatrix} to the phenomenological covariance matrix presented in \myeq{covmatrix}. The diagonal elements describe the auto-covariance at each observing frequency, and hence include $\sigma_{\rm S/N}$, whereas the off-diagonal elements describe noise correlated across frequencies, here jitter and diffractive interstellar scattering.
Using this covariance matrix for each observing epoch, we can construct a full noise model following the structure of Equation \ref{eq:phenom_noise}:
\be
C =
 \begin{bmatrix}
C^\mathrm{epoch}(t_1) & 0 & \cdots & 0\\ 
0 & C^\mathrm{epoch}(t_2) & \cdots & 0 \\ 
\vdots & \vdots & \ddots & \vdots \\
0 & 0 & \cdots & C^\mathrm{epoch}(t_N)
\end{bmatrix} + C^{\rm observatory} + C^{\rm \delta DM} + C^{\rm spin}(|t-t'|),
\ee
where $C^{\rm spin}$ is modeled as a power-law spectral process, and $C^{\rm observatory}$ and $C^{\rm \delta DM}$ are derived directly from observation and encapsulate noise from the radio telescope and variations in the IISM respectively.

\section{Definitions}

Here we include an exhaustive list of the notation used throughout this work. A carat (\^{}) denotes an estimator and we have excluded certain subscripts on variables listed here (e.g., GWB).

\startlongtable
\begin{deluxetable*}{llc}
\centering
\tablecolumns{3}
\tablecaption{Symbols and Acronyms Used\label{t:notation}}
\tablehead{
\colhead{Symbols} & \colhead{Definition} & \colhead{Typical Units}\\
}
\startdata
$\mathbf{a}$ & Fourier amplitudes vector & \\
$A$ & Spectral amplitude & Strain (unitless)\\
$B$ & Observing bandwidth & MHz\\
$\mathcal{B}$ & Bayes factor & \\
$C$ & Generic covariance matrix  & \\
$C^\mathrm{epoch}$ & Covariance matrix of white noise on given epoch & \\
$C^{\rm observatory}$ & Covariance matrix for noise at the observatory & \\
$C^{\rm \delta DM}$ & Covariance matrix for DM noise & \\
$C^{\rm spin}$ & Covariance matrix for spin noise & \\
$C^{\rm RN}$ & Red-noise covariance matrix & \\
$\mathrm{d}f$ & Fourier frequency differential for power spectral density integrand & Hz \\
$D$ & $D\equiv C+MXM^\mathrm{T}$ & \\
DM & Dispersion Measure & pc cm$^{-3}$\\
$\mathbf{E}$ & Set of white noise parameters $(\ef, \eq, \ec)$ & \\
$f$ & Spectral/Fourier frequency & Hz\\
$f_{\rm Nyq}$ & Nyquist frequency & \\
$f_{\rm s}$ & Pulsar spin frequency & Hz\\
$f_{\rm yr}$ & Frequency of 1~yr$^{-1}$ & yr$^{-1}$ \\
$F$ & Fourier design matrix & \\
$G$ & [G matrix] & \\
$\ef$ & \efac, multiplicative error factor & \\
$h$ & Strain & \\
$h_c$ & Characteristic strain & \\\
$i$, $j$ & Time-of-arrival indices & \\
$\ec$ & \ecorr, error correlated across radio frequencies & $\mu$s\\
$M$ & Design matrix & \\
$\mathbf{n}$ & White noise (vector) & \\
$n_0$ & Electron number density at a distance of 1 au & cm$^{-3}$\\
$N$ & $N\equiv D +F \varphi F^\mathrm{T}$ & \\
$N_{\rm freq}$ & Number of Fourier frequencies & \\
$N_{\rm p}$ & Number of pulses & \\
$N_{\rm TOA}$ & Number of times of arrival & \\
$\mathcal{N}$ & Power spectral density of pulsar residuals, $\mathcal{N} = P_{\rm R}(f)$ & \\
 & Note $\mathcal{N}^{-1}$ is the noise-weighted transmission function & \\
$P(f)$ & Power spectral density & Hz$^{-1}$\\
$P_R(f)$ & Power spectral density of residuals & s$^2$ Hz$^{-1}$\\
$P^{\rm (WN)}_R$ & White-noise power spectral density of the residuals & \\
$P_{\rm s}$ & Pulsar spin period & ms \\
$\mathcal{P}$ & Probability density function & \\
$\mathbf{p}$ & Model parameters (vector) & \\
$\mathbf{p}_0$ & Best-fit model parameters (vector) & \\
$\eq$ & \equad, error added in quadrature & $\mu$s\\
$r$ & Radio-frequency ratio & \\
$\mathcal{R}(f)$ & Sky-averaged response function for the GWB & \\
$S$ & Strain power spectral density & \\
$\mathrm{S/N}$ & Signal-to-noise ratio & \\
$t$ & Time of arrival & \\
$\mathbf{t}$ & Times of arrival (vector) & \\
$\mathcal{T}(f)$ & Timing model transmission function & \\
$t_C$ & Non-dispersive chromatic delays & $\mu$s \\
$t_{\rm scatt}$ & Scattering delay timescale & \\
$t_{\rm sep}$ & Separation time between multi-frequency observations & \\
$T_{\rm obs}$ & Single observation length & min\\
$\Tspan$ & Total timespan of dataset & yrs \\
$\mathcal{U}$ & Block-diagonal matrix with values of 1 for TOAs from &\\
&  the same observation epoch and 0 otherwise & \\
$W$ & Pulse width & \\
$X$ & Covariance matrix of timing model parameters & \\
$\alpha$ & Spectral index of GWB in units of characteristic strain & \\
$\beta$ & Ecliptic latitude & deg\\ 
$\gamma$ & Spectral index & \\
$\delta_{ij}$ & Kronecker delta & \\
$\delta \mathbf{t}$ & Timing residuals (vector) & \\
$\delta \mathbf{t}_D$ & Deterministic non-gravitational-wave model times of arrival (vector) & \\
$\delta \mathbf{t}_M$ & Model/predicted times of arrival (vector) & \\
$\delta t_\infty$ & Perturbation on the infinite-frequency arrival time & $\mu$s \\
$\deltat$ & Diffractive scintillation timescale & s\\
$\Delta f_{\rm s}$ & Change in pulsar spin frequency & Hz\\
$\deltanu$ & Diffractive scintillation bandwidth & MHz \\
$\boldsymbol{\epsilon}$ & Perturbations to the model parameters  (vector) & \\
$\epsilon_{\nu}$ & Frequency-dependent timing errors & $\mu$s \\
$\varepsilon$ & Fractional gain error & \\
$\nu$ & Radio frequency & MHz, GHz\\
$\pi_{\rm V}$ & Degree of circular polarization & \\
$\sigma_{\rm DISS}$ & Diffractive interstellar scintillation noise & \\
$\sigma_{\rm J}$ & Jitter noise & \\
$\sigma_{\rm pol}$ & Polarization mis-calibration uncertainty & $\mu$s \\
$\sigma_{\rm S/N}$ & Signal-to-noise-ratio-dependent template-fitting uncertainty & $\mu$s\\
$\sigma_{\rm spin}$ & Spin noise & \\
$\sigma_{\delta \hat{t}_\infty}$ & Infinite-frequency arrival time uncertainty & \\
$\theta$ & Sun-pulsar separation angle & deg\\
$\varphi$ & Spectral components of RN model & $s^2$\\
\enddata

\end{deluxetable*}

\newpage

\begin{figure}
\centering
\includegraphics[width=0.9\linewidth]{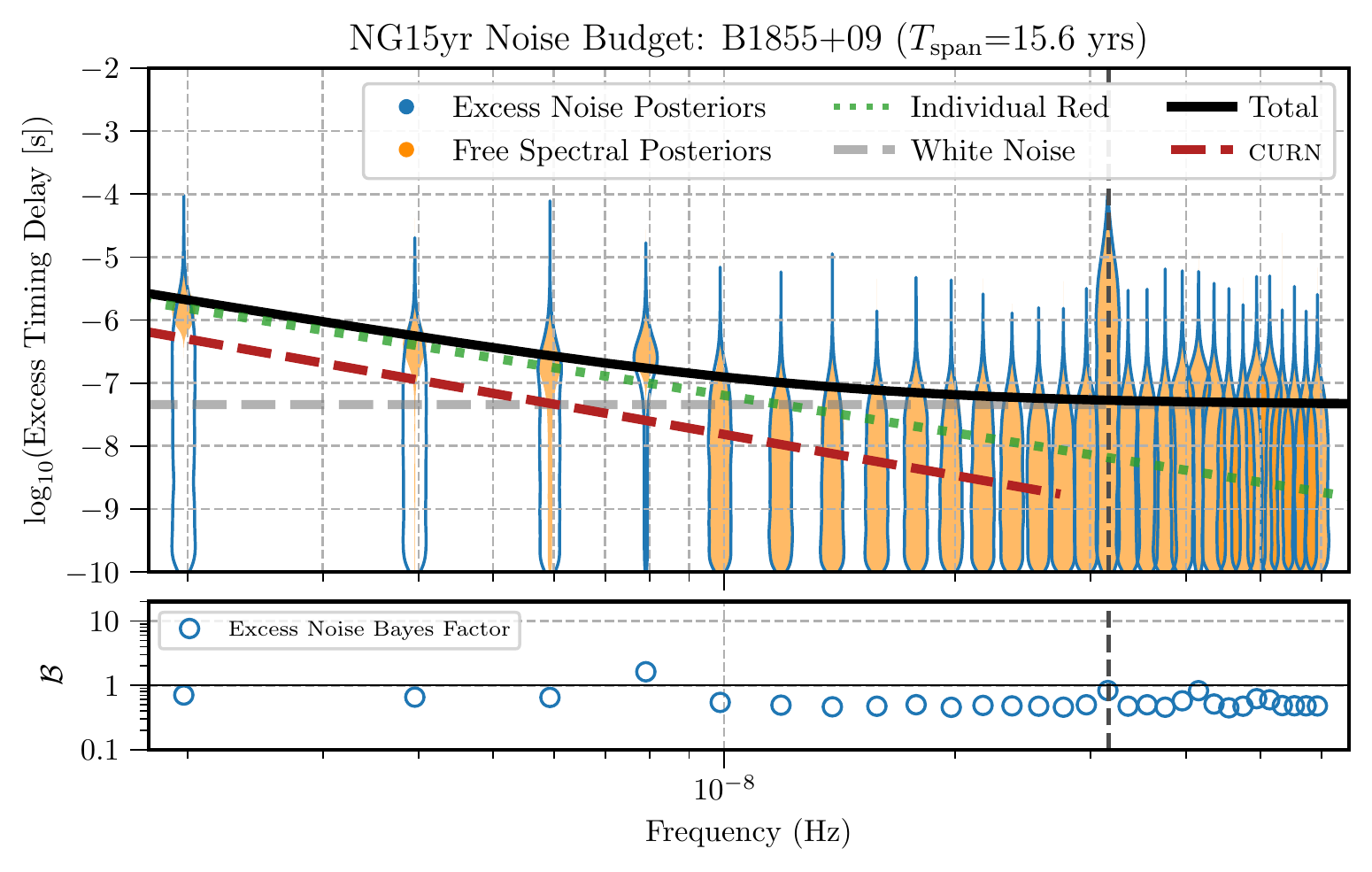}
\caption{The excess timing residual delay as a function of frequency for PSR~B1855+09. See \myfig{f:excess_j1909} for details.}
\label{f:budget_B1855+09}
\end{figure}

\begin{figure}
\centering
\includegraphics[width=0.9\linewidth]{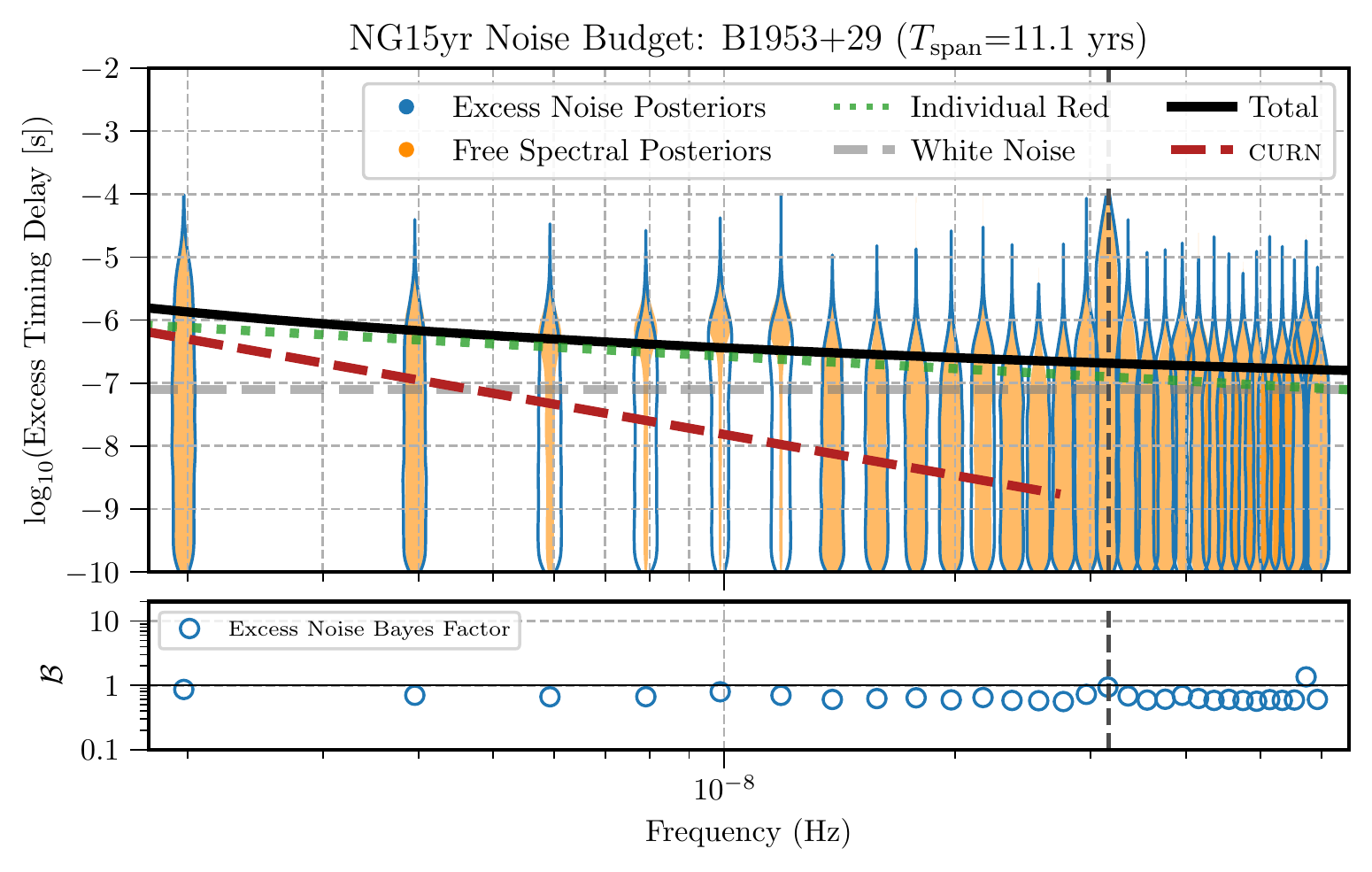}
\caption{The excess timing residual delay as a function of frequency for PSR~B1953+29. See \myfig{f:excess_j1909} for details.}
\label{f:budget_B1953+29}
\end{figure}

\begin{figure}
\centering
\includegraphics[width=0.9\linewidth]{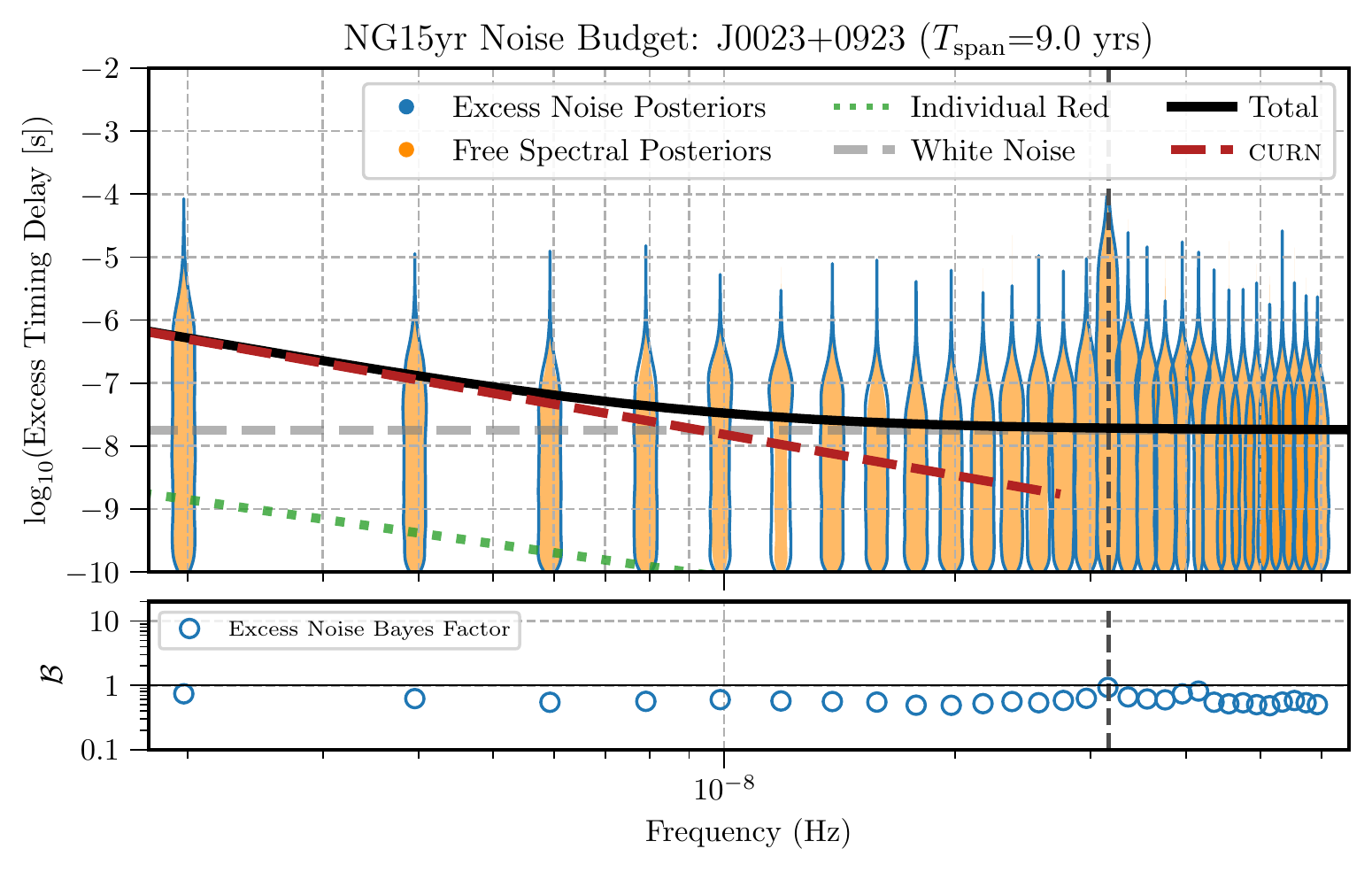}
\caption{The excess timing residual delay as a function of frequency for PSR~J0023+0923. See \myfig{f:excess_j1909} for details.}
\label{f:budget_J0023+0923}
\end{figure}

\begin{figure}
\centering
\includegraphics[width=0.9\linewidth]{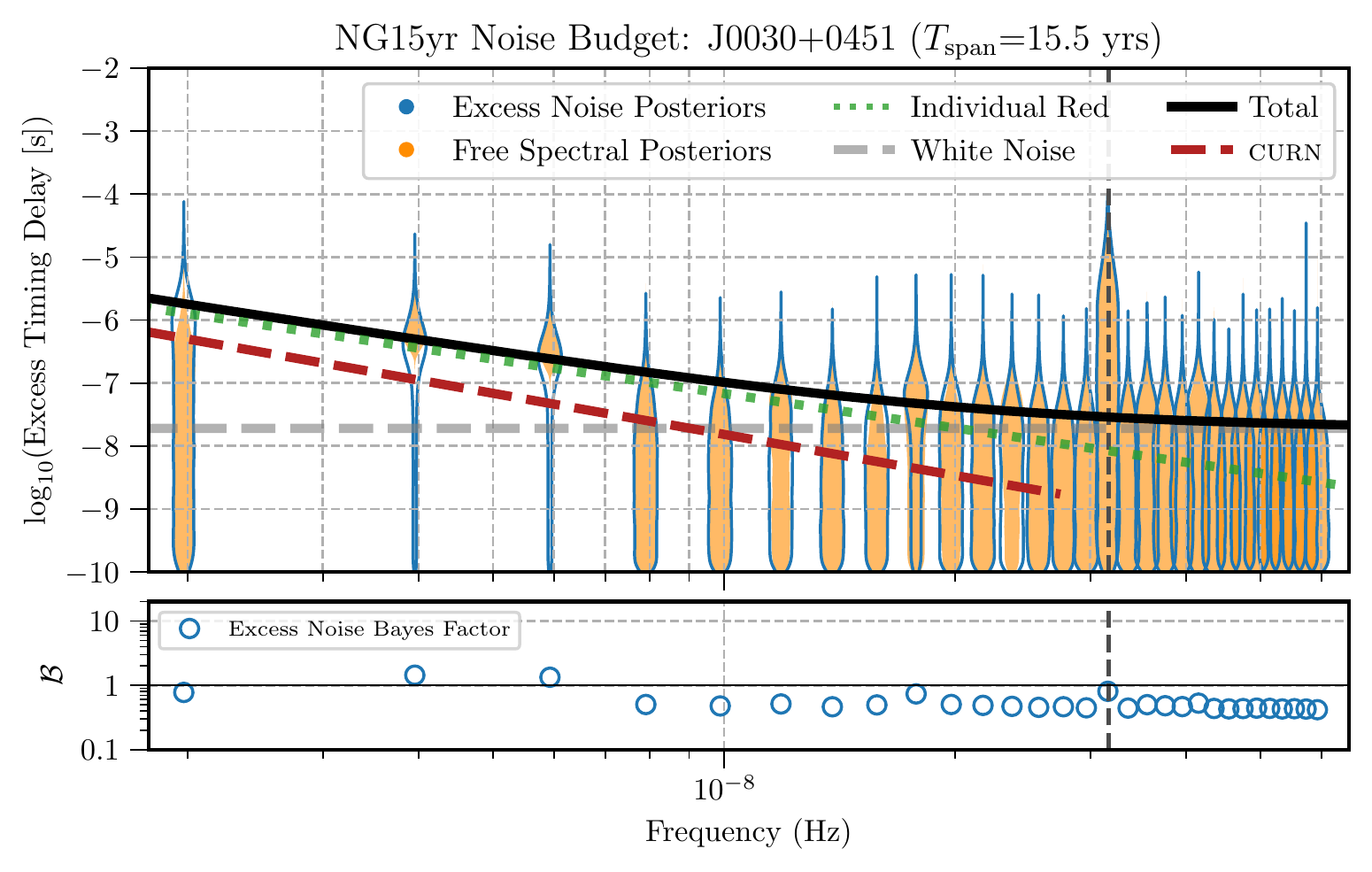}
\caption{The excess timing residual delay as a function of frequency for PSR~J0030+0451. See \myfig{f:excess_j1909} for details.}
\label{f:budget_J0030+0451}
\end{figure}

\begin{figure}
\centering
\includegraphics[width=0.9\linewidth]{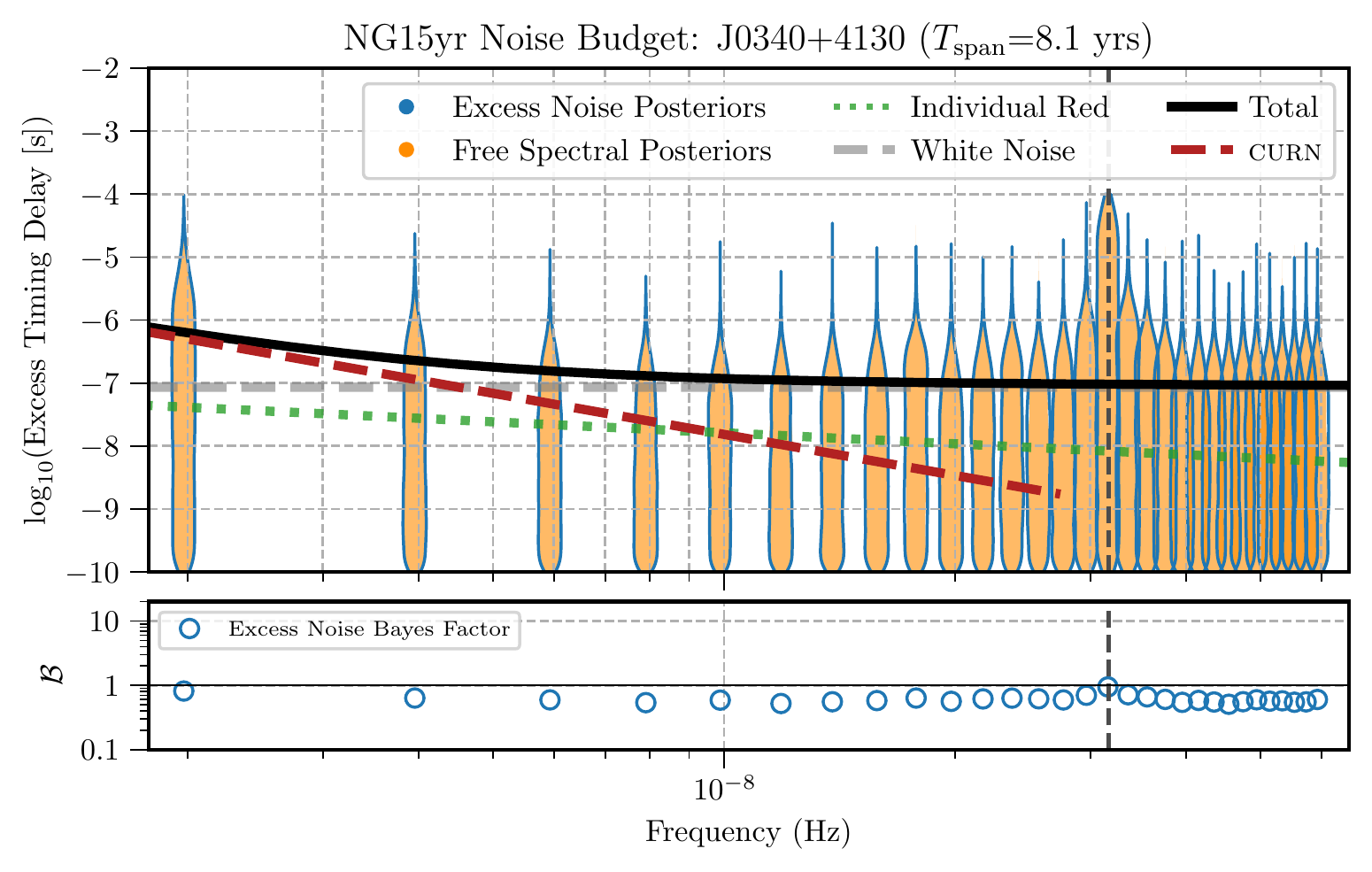}
\caption{The excess timing residual delay as a function of frequency for PSR~J0340+4130. See \myfig{f:excess_j1909} for details.}
\label{f:budget_J0340+4130}
\end{figure}

\begin{figure}
\centering
\includegraphics[width=0.9\linewidth]{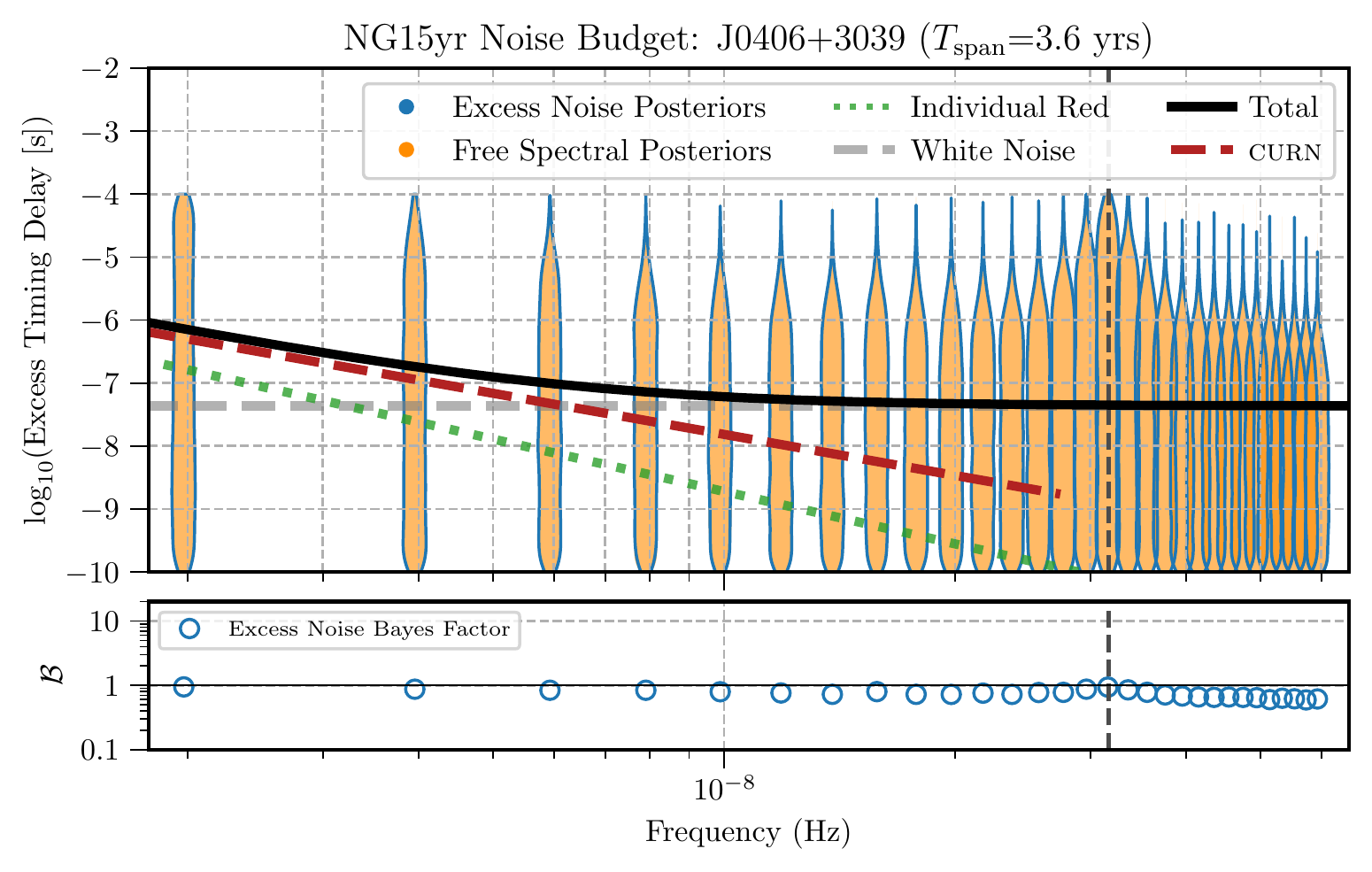}
\caption{The excess timing residual delay as a function of frequency for PSR~J0406+3039. See \myfig{f:excess_j1909} for details.}
\label{f:budget_J0406+3039}
\end{figure}

\begin{figure}
\centering
\includegraphics[width=0.9\linewidth]{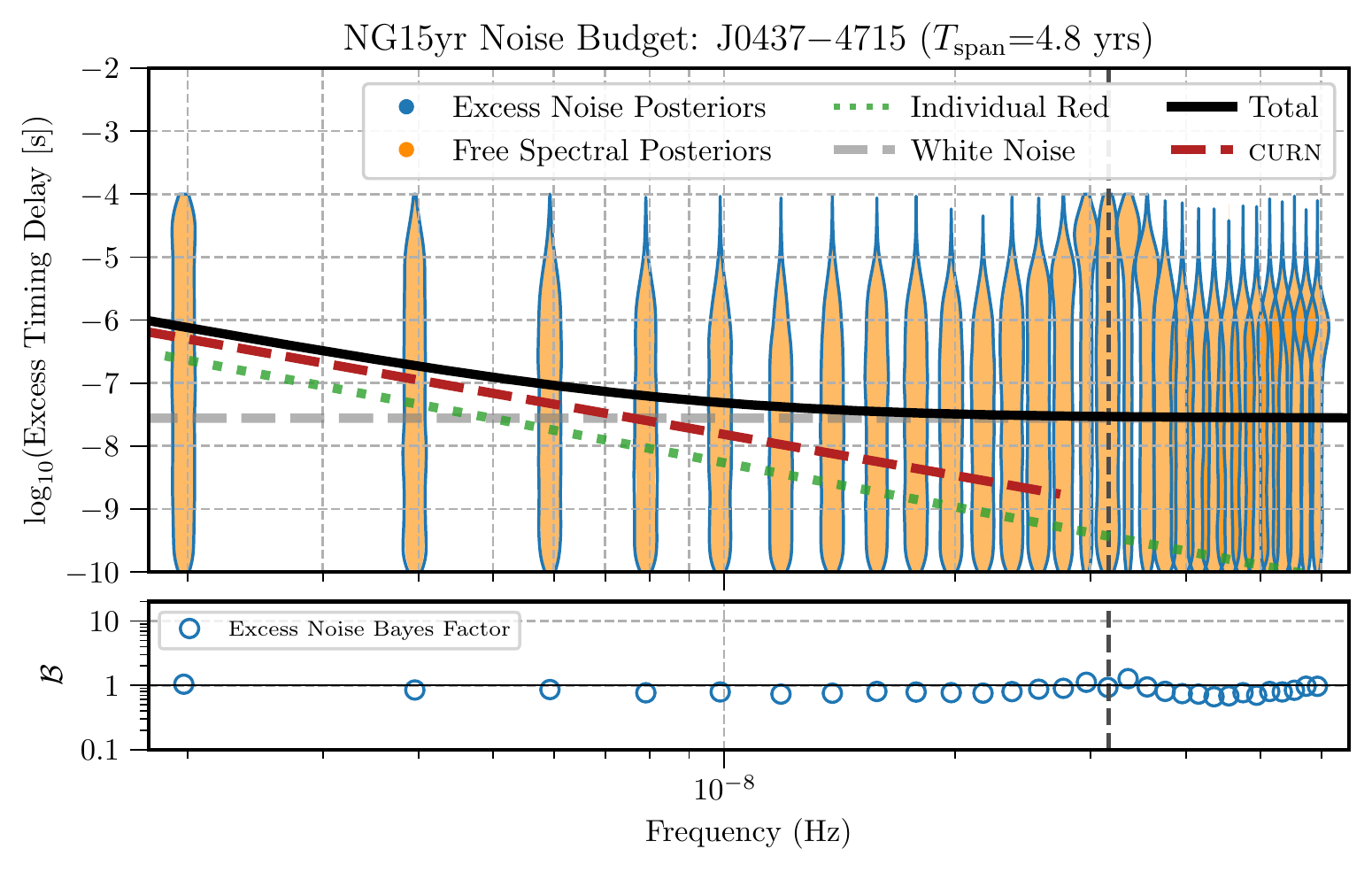}
\caption{The excess timing residual delay as a function of frequency for PSR~J0437$-$4715. See \myfig{f:excess_j1909} for details.}
\label{f:budget_J0437-4715}
\end{figure}

\begin{figure}
\centering
\includegraphics[width=0.9\linewidth]{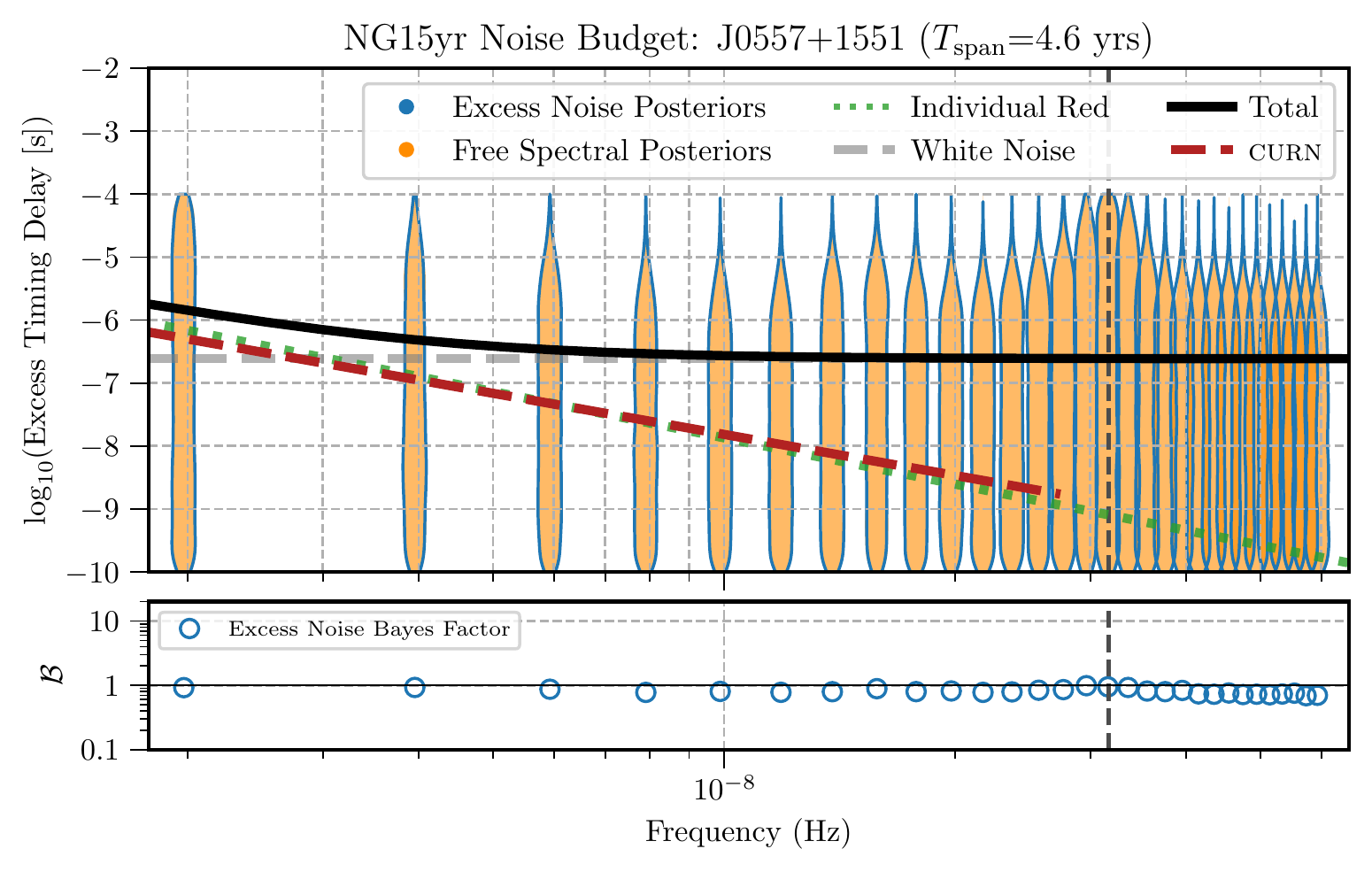}
\caption{The excess timing residual delay as a function of frequency for PSR~J0557+1551. See \myfig{f:excess_j1909} for details.}
\label{f:budget_J0557+1551}
\end{figure}

\begin{figure}
\centering
\includegraphics[width=0.9\linewidth]{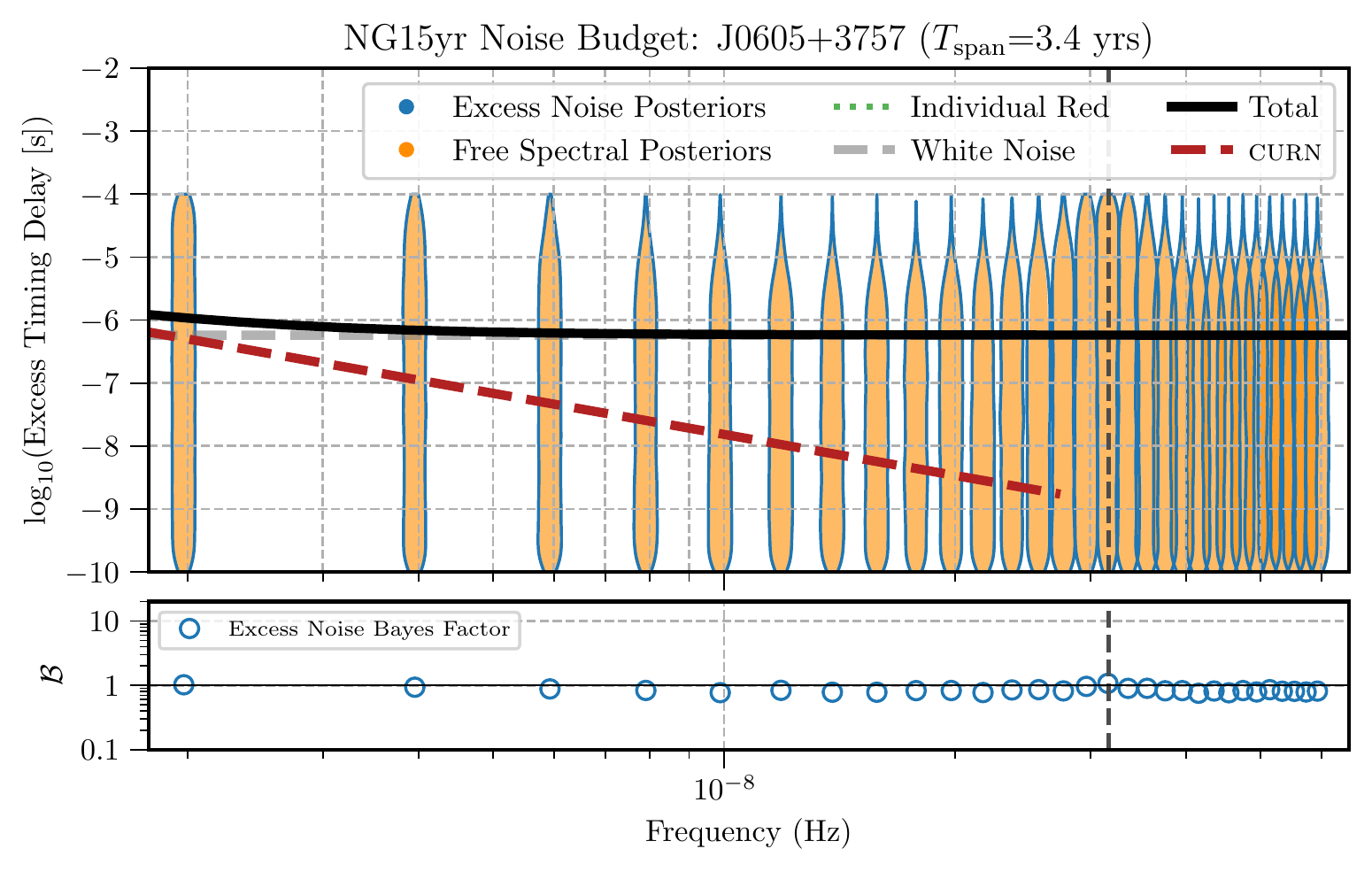}
\caption{The excess timing residual delay as a function of frequency for PSR~J0605+3757. See \myfig{f:excess_j1909} for details.}
\label{f:budget_J0605+3757}
\end{figure}

\begin{figure}
\centering
\includegraphics[width=0.9\linewidth]{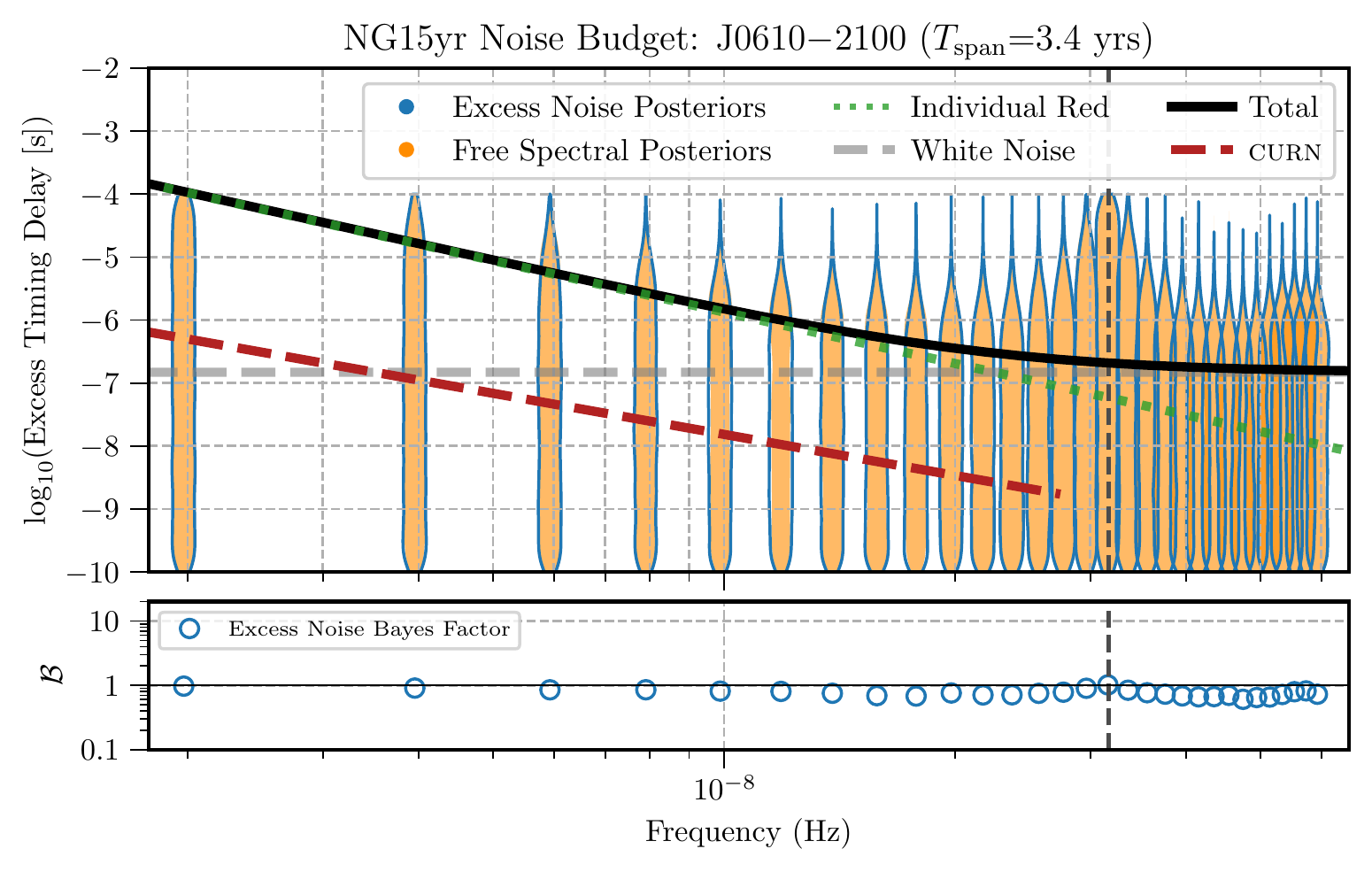}
\caption{The excess timing residual delay as a function of frequency for PSR~J0610$-$2100. See \myfig{f:excess_j1909} for details.}
\label{f:budget_J0610-2100}
\end{figure}

\begin{figure}
\centering
\includegraphics[width=0.9\linewidth]{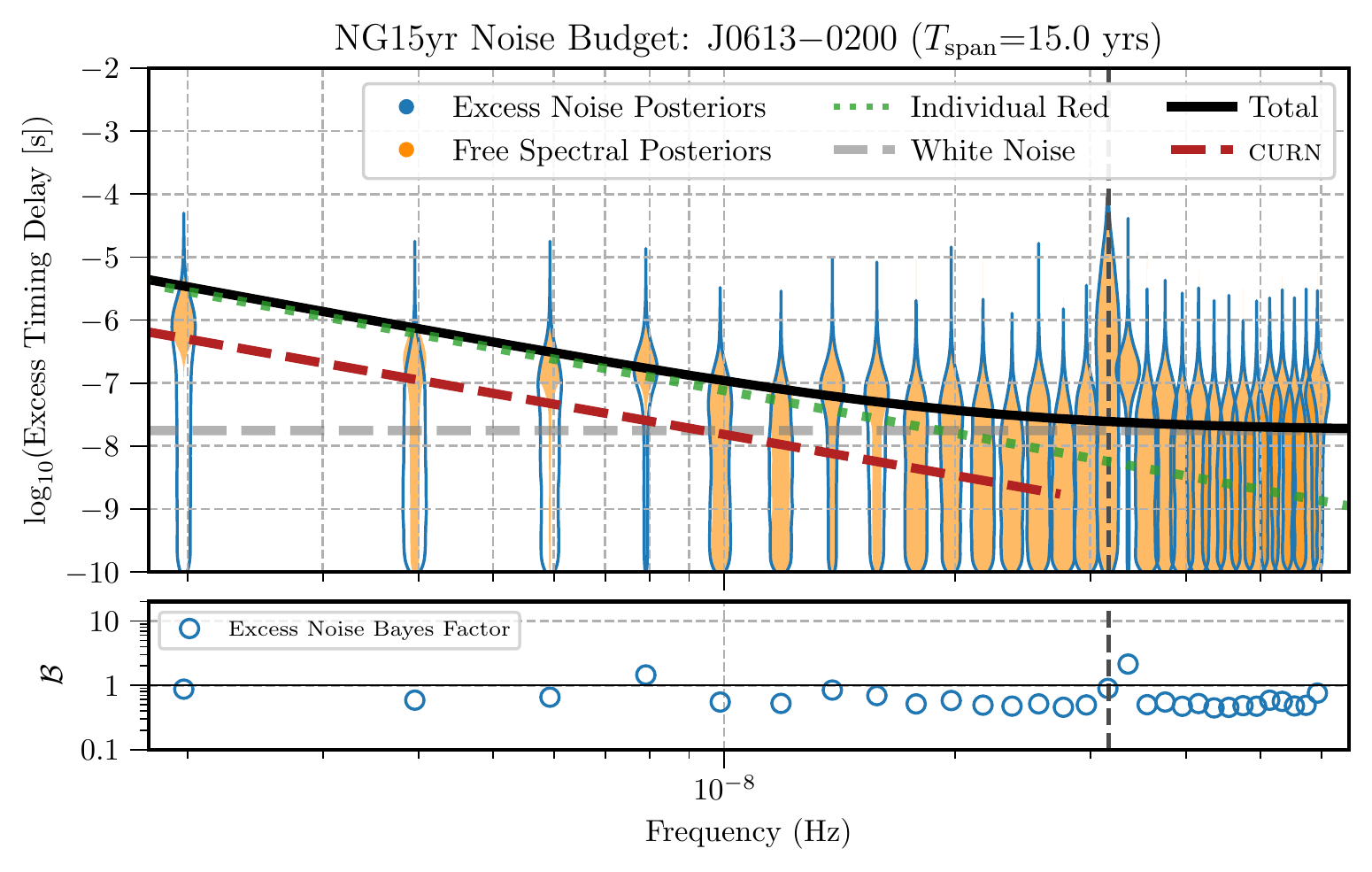}
\caption{The excess timing residual delay as a function of frequency for PSR~J0613$-$0200. See \myfig{f:excess_j1909} for details.}
\label{f:budget_J0613-0200}
\end{figure}

\begin{figure}
\centering
\includegraphics[width=0.9\linewidth]{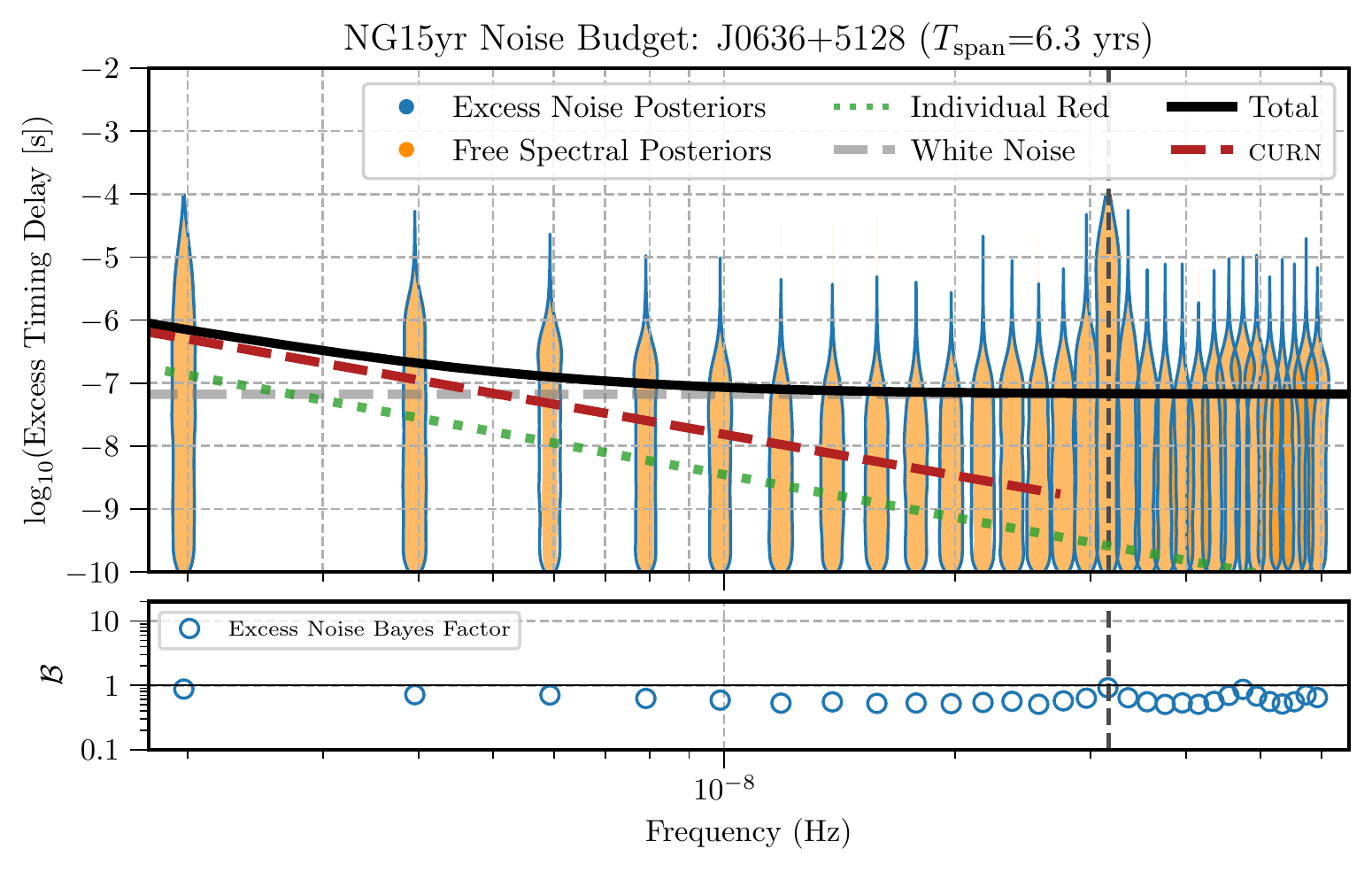}
\caption{The excess timing residual delay as a function of frequency for PSR~J0636+5128. See \myfig{f:excess_j1909} for details.}
\label{f:budget_J0636+5128}
\end{figure}

\begin{figure}
\centering
\includegraphics[width=0.9\linewidth]{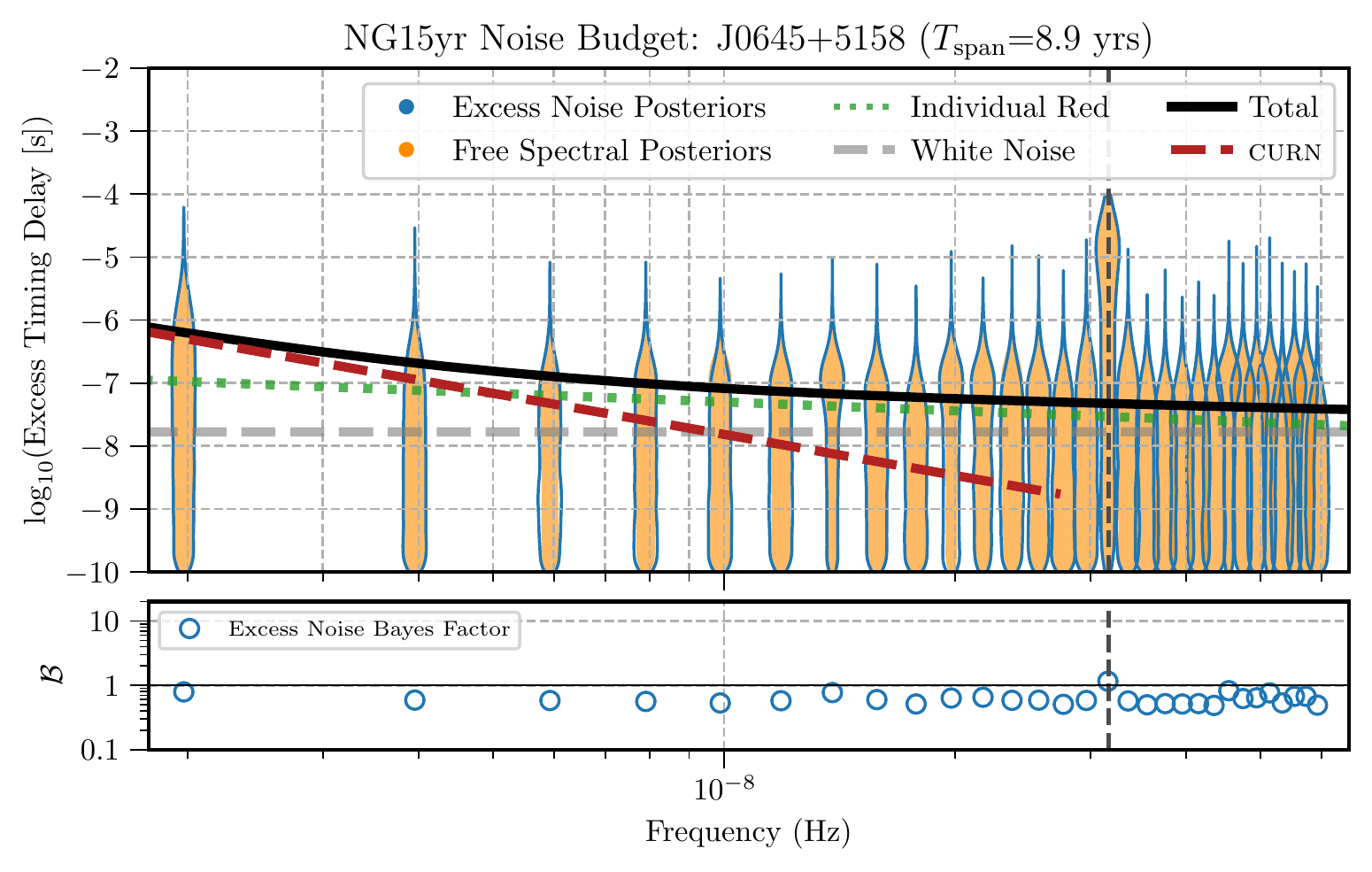}
\caption{The excess timing residual delay as a function of frequency for PSR~J0645+5158. See \myfig{f:excess_j1909} for details.}
\label{f:budget_J0645+5158}
\end{figure}

\begin{figure}
\centering
\includegraphics[width=0.9\linewidth]{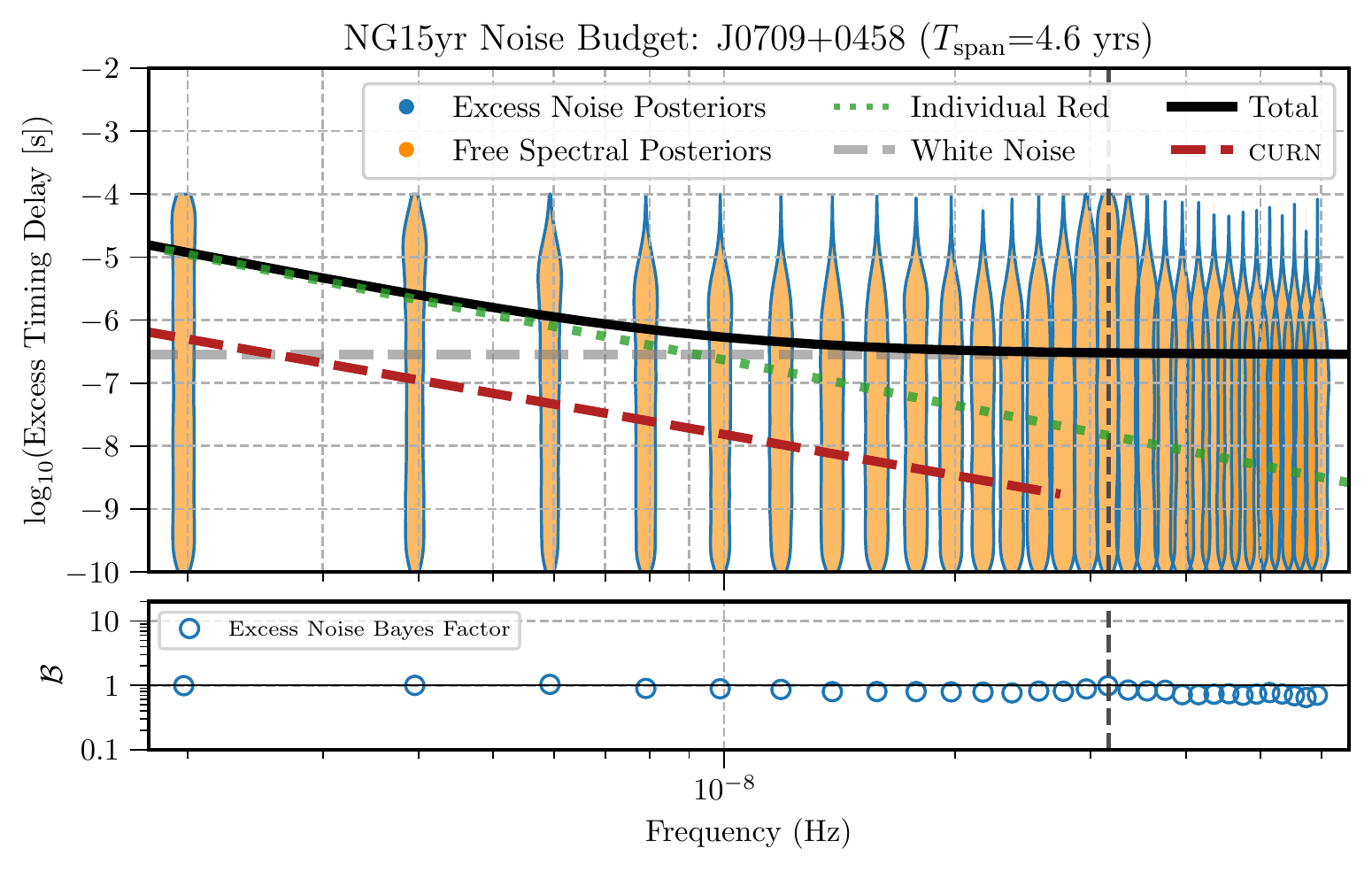}
\caption{The excess timing residual delay as a function of frequency for PSR~J0709+0458. See \myfig{f:excess_j1909} for details.}
\label{f:budget_J0709+0458}
\end{figure}

\begin{figure}
\centering
\includegraphics[width=0.9\linewidth]{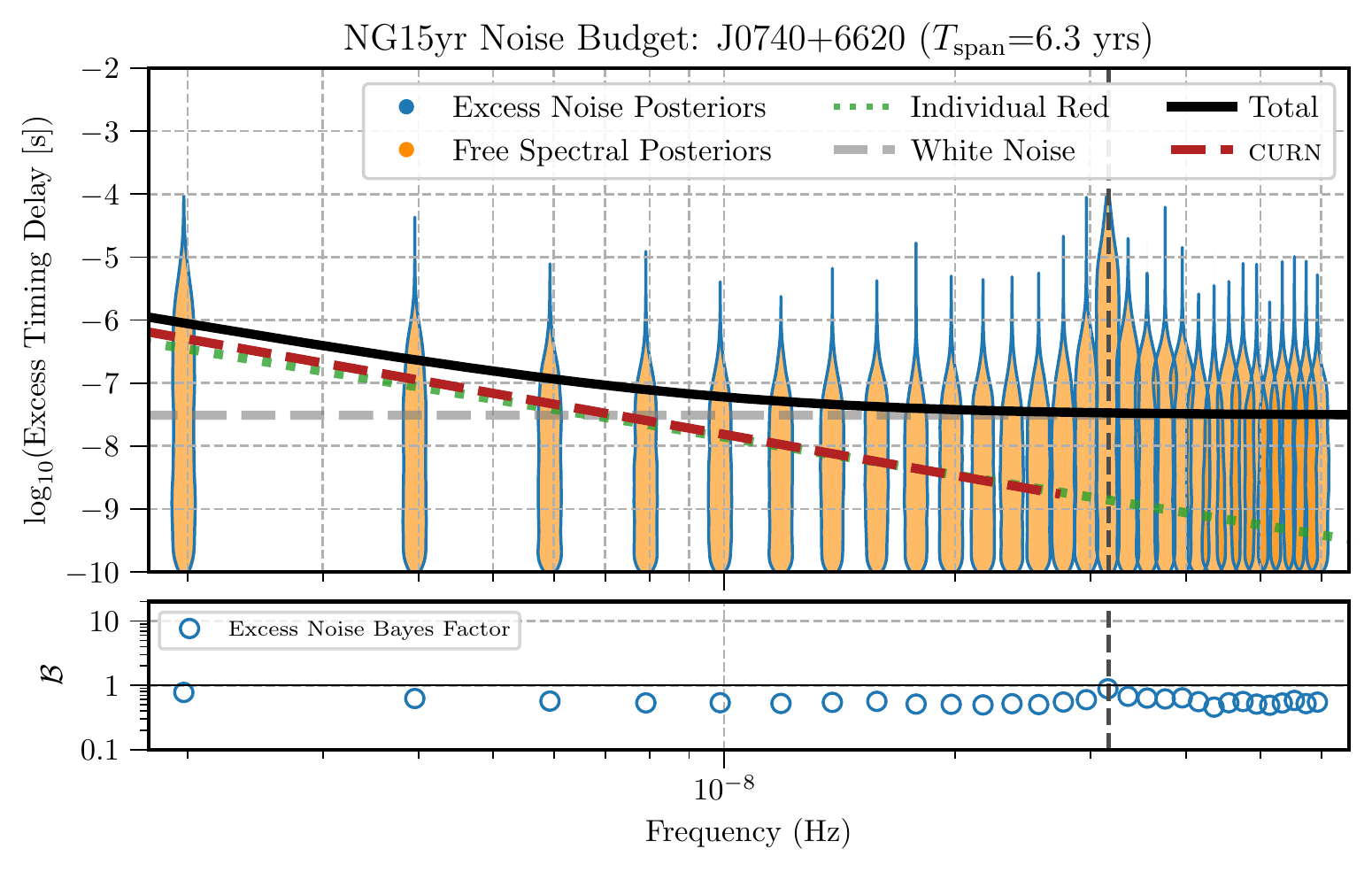}
\caption{The excess timing residual delay as a function of frequency for PSR~J0740+6620. See \myfig{f:excess_j1909} for details.}
\label{f:budget_J0740+6620}
\end{figure}

\begin{figure}
\centering
\includegraphics[width=0.9\linewidth]{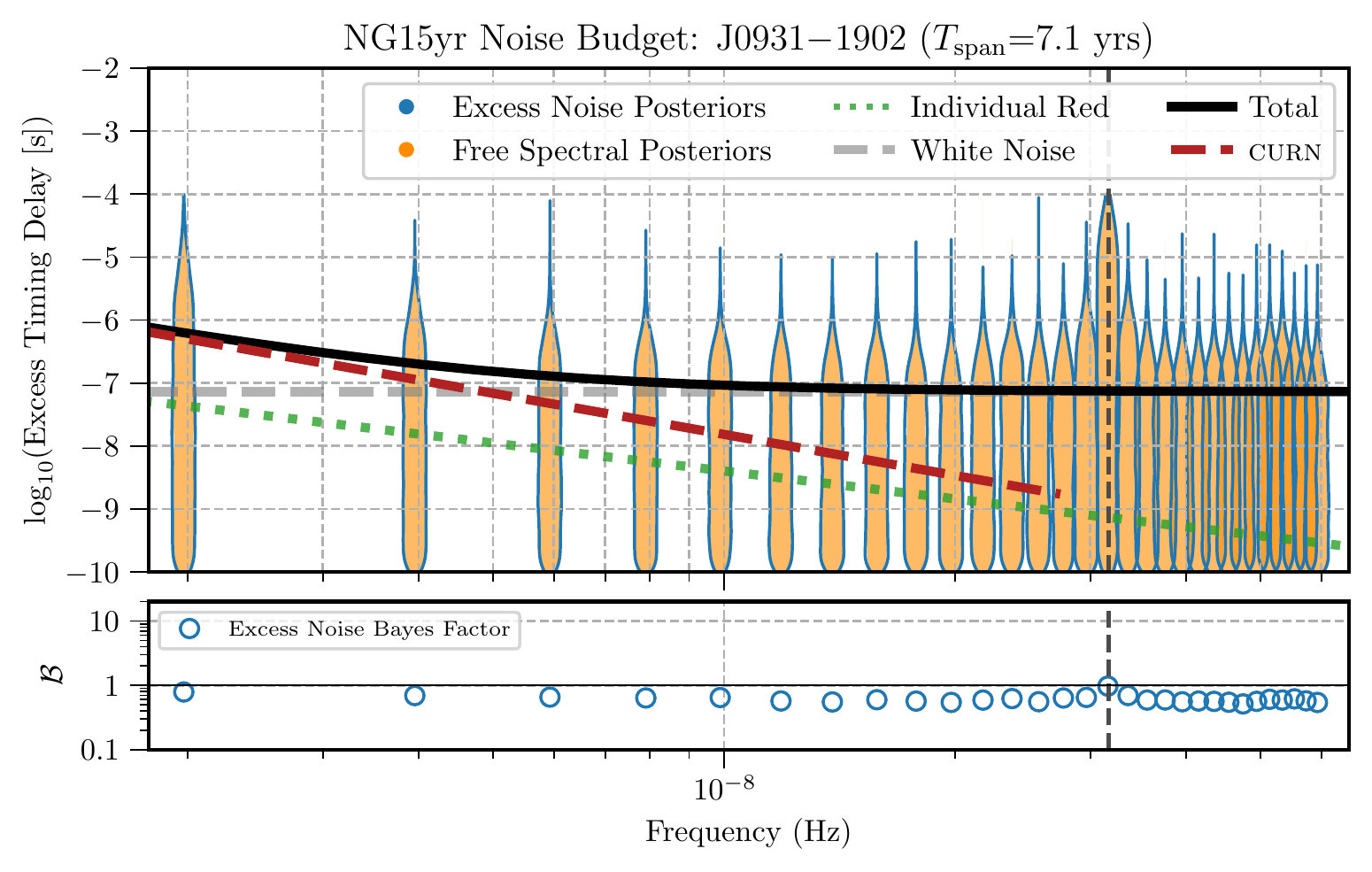}
\caption{The excess timing residual delay as a function of frequency for PSR~J0931$-$1902. See \myfig{f:excess_j1909} for details.}
\label{f:budget_J0931-1902}
\end{figure}

\begin{figure}
\centering
\includegraphics[width=0.9\linewidth]{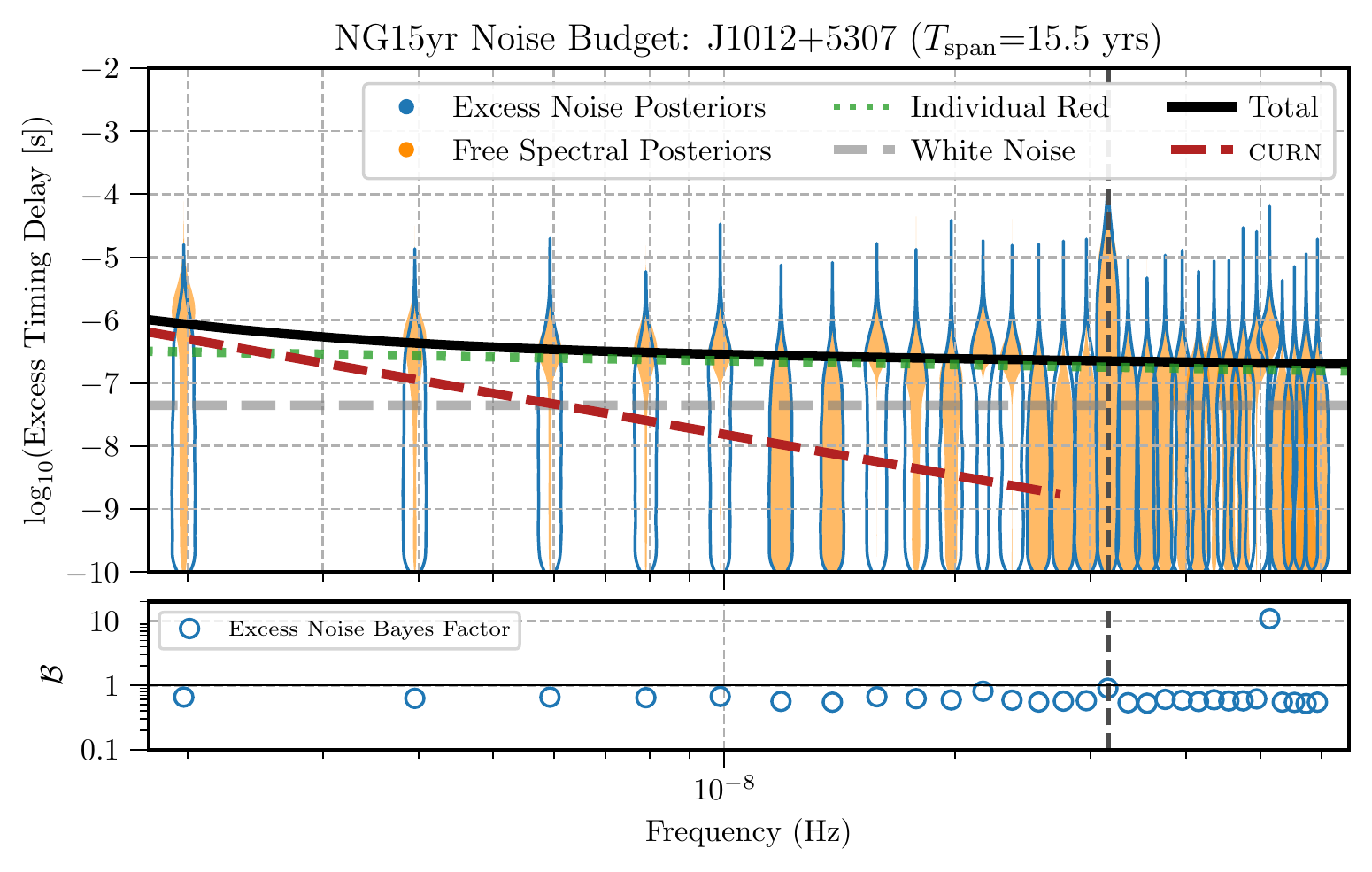}
\caption{The excess timing residual delay as a function of frequency for PSR~J1012+5307. See \myfig{f:excess_j1909} for details.}
\label{f:budget_J1012+5307}
\end{figure}

\begin{figure}
\centering
\includegraphics[width=0.9\linewidth]{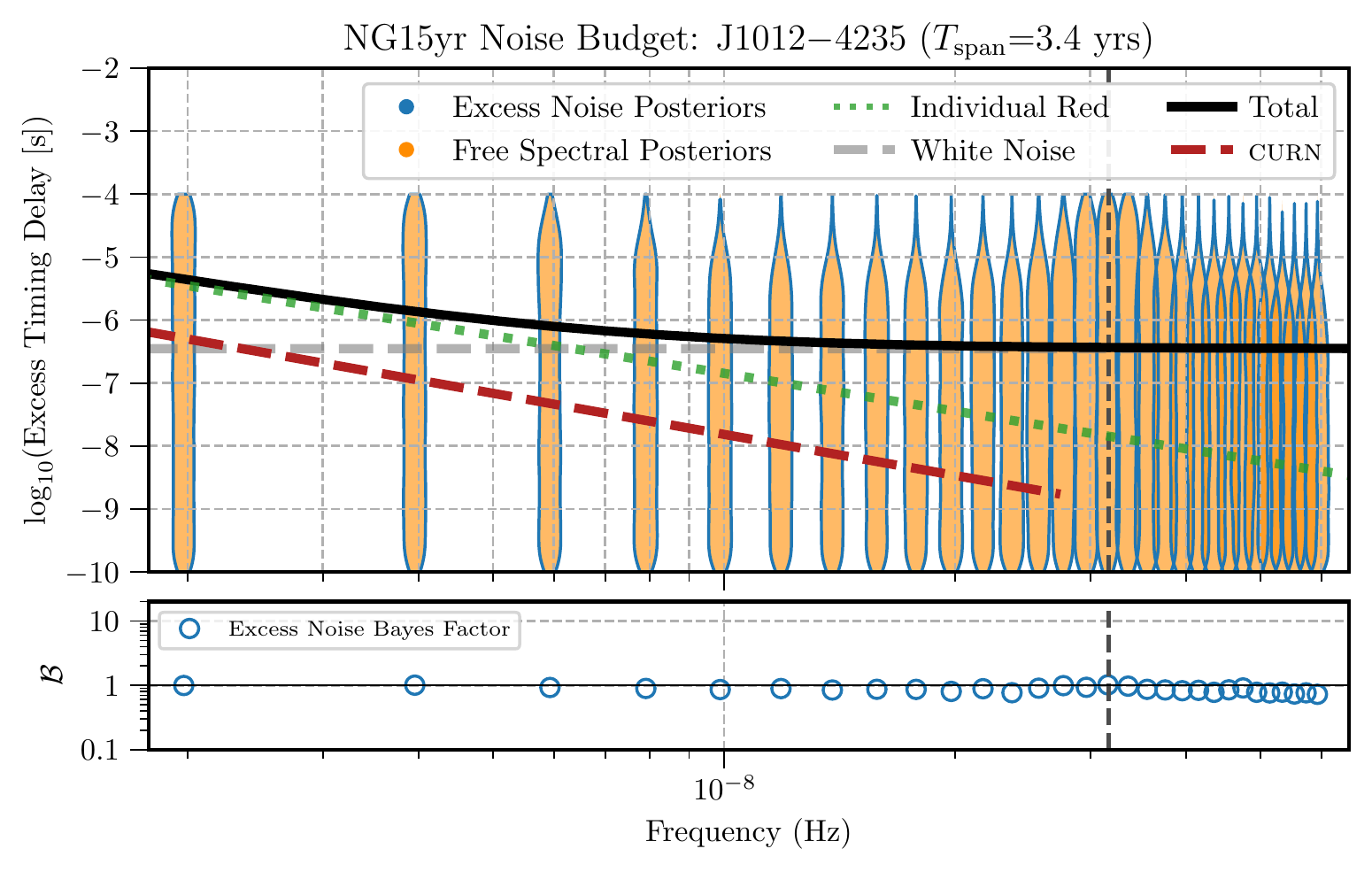}
\caption{The excess timing residual delay as a function of frequency for PSR~J1012$-$4235. See \myfig{f:excess_j1909} for details.}
\label{f:budget_J1012-4235}
\end{figure}

\begin{figure}
\centering
\includegraphics[width=0.9\linewidth]{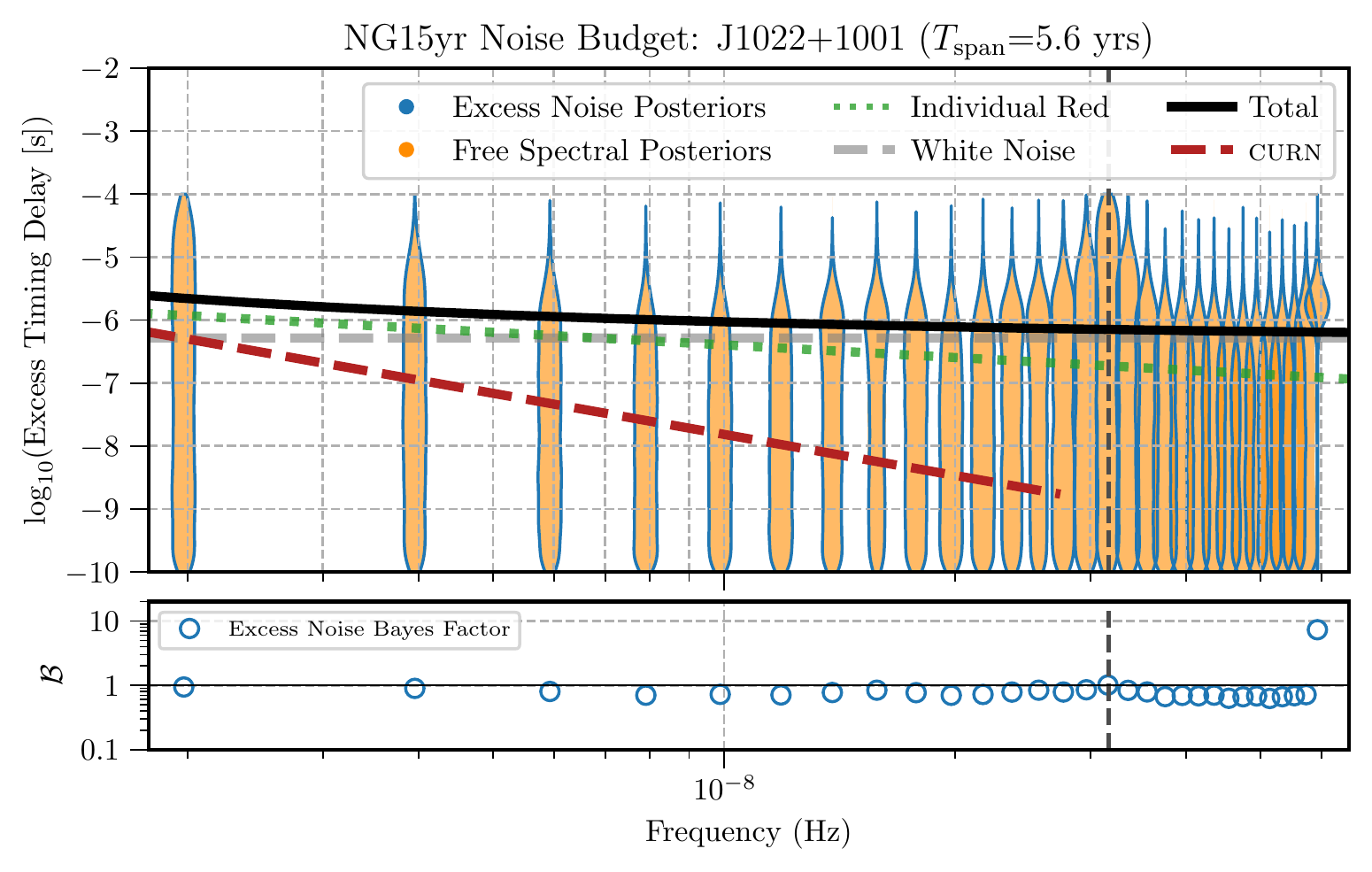}
\caption{The excess timing residual delay as a function of frequency for PSR~J1022+1001. See \myfig{f:excess_j1909} for details.}
\label{f:budget_J1022+1001}
\end{figure}

\begin{figure}
\centering
\includegraphics[width=0.9\linewidth]{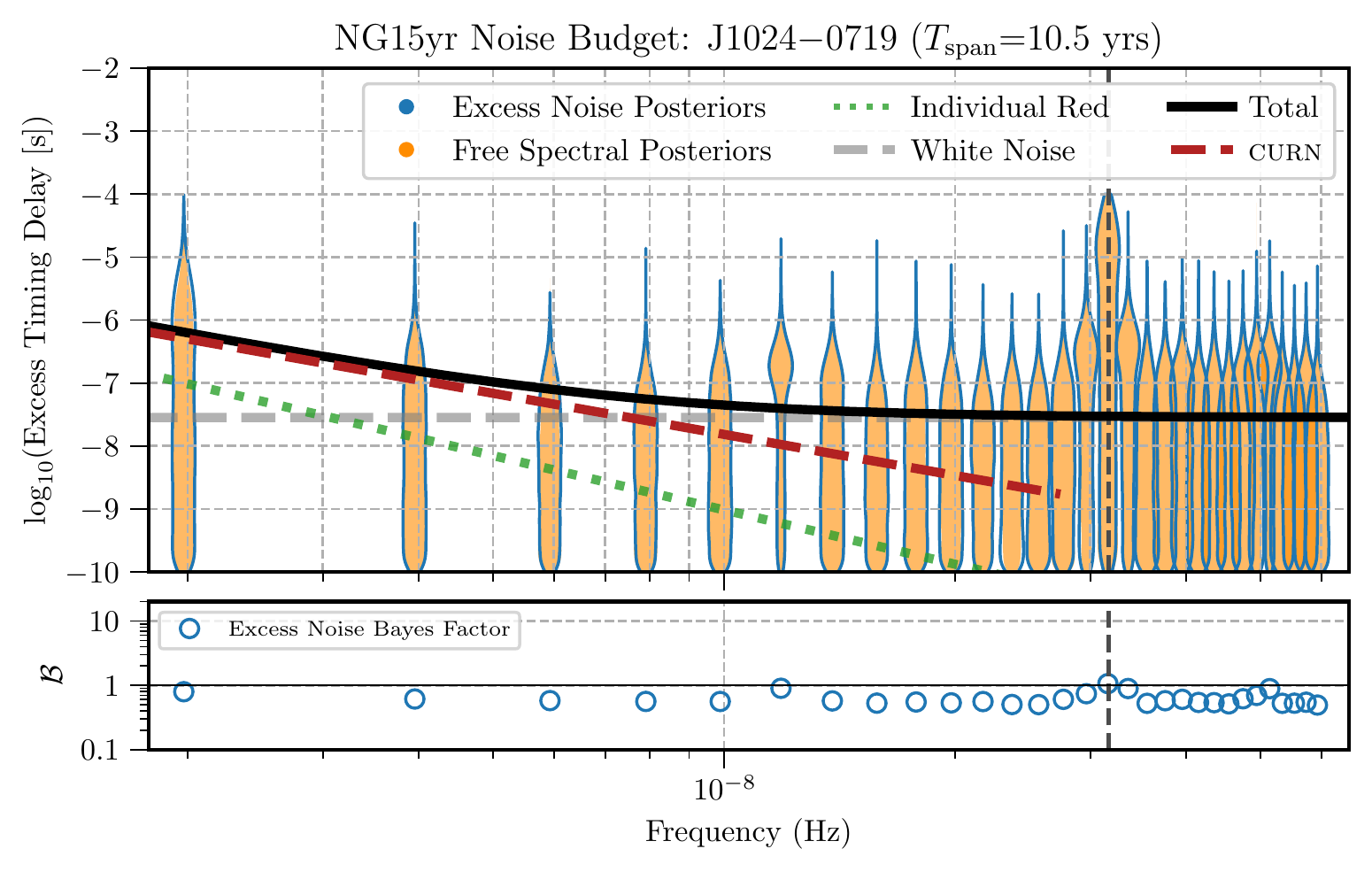}
\caption{The excess timing residual delay as a function of frequency for PSR~J1024$-$0719. See \myfig{f:excess_j1909} for details.}
\label{f:budget_J1024-0719}
\end{figure}

\begin{figure}
\centering
\includegraphics[width=0.9\linewidth]{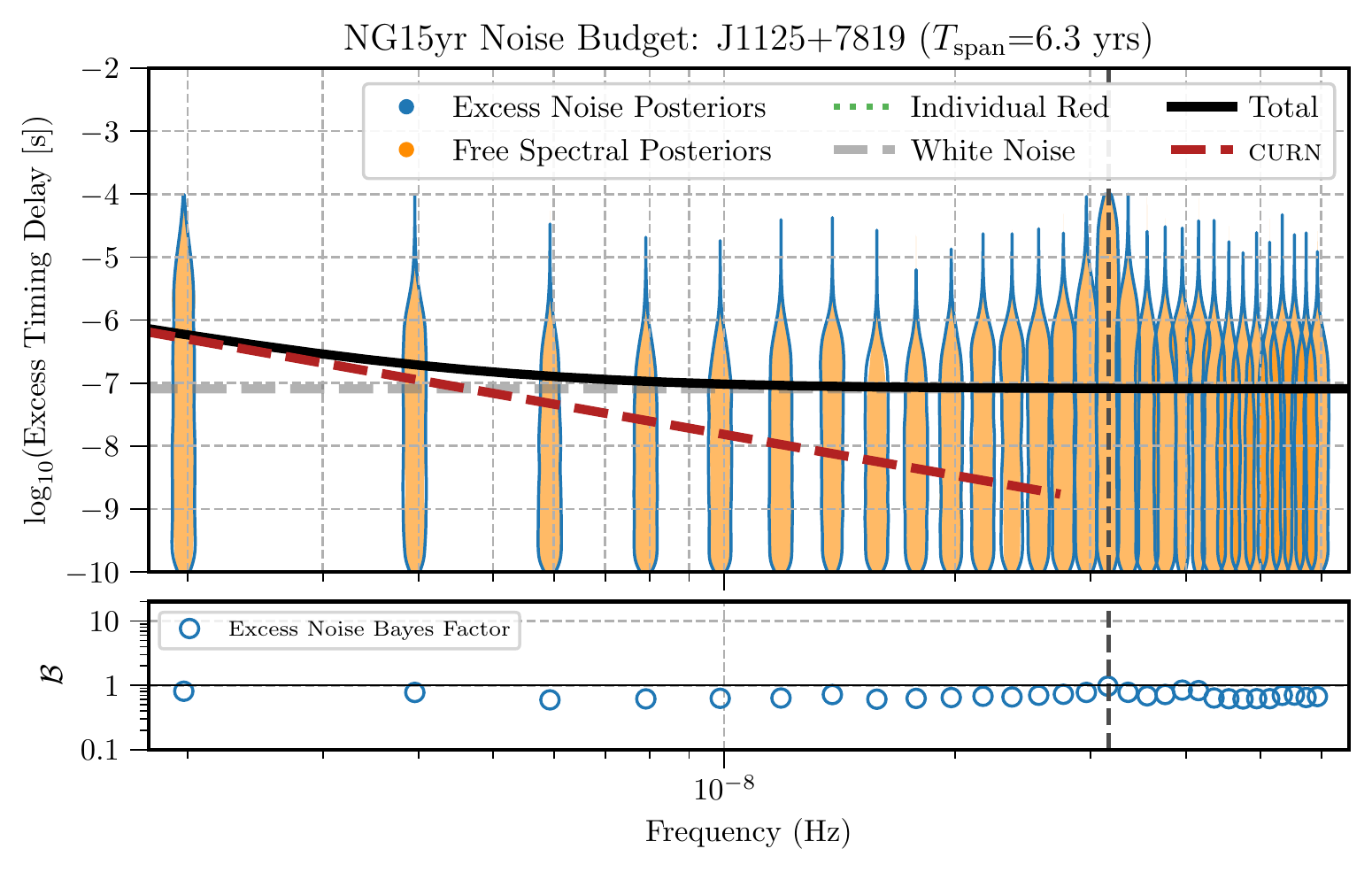}
\caption{The excess timing residual delay as a function of frequency for PSR~J1125+7819. See \myfig{f:excess_j1909} for details.}
\label{f:budget_J1125+7819}
\end{figure}

\begin{figure}
\centering
\includegraphics[width=0.9\linewidth]{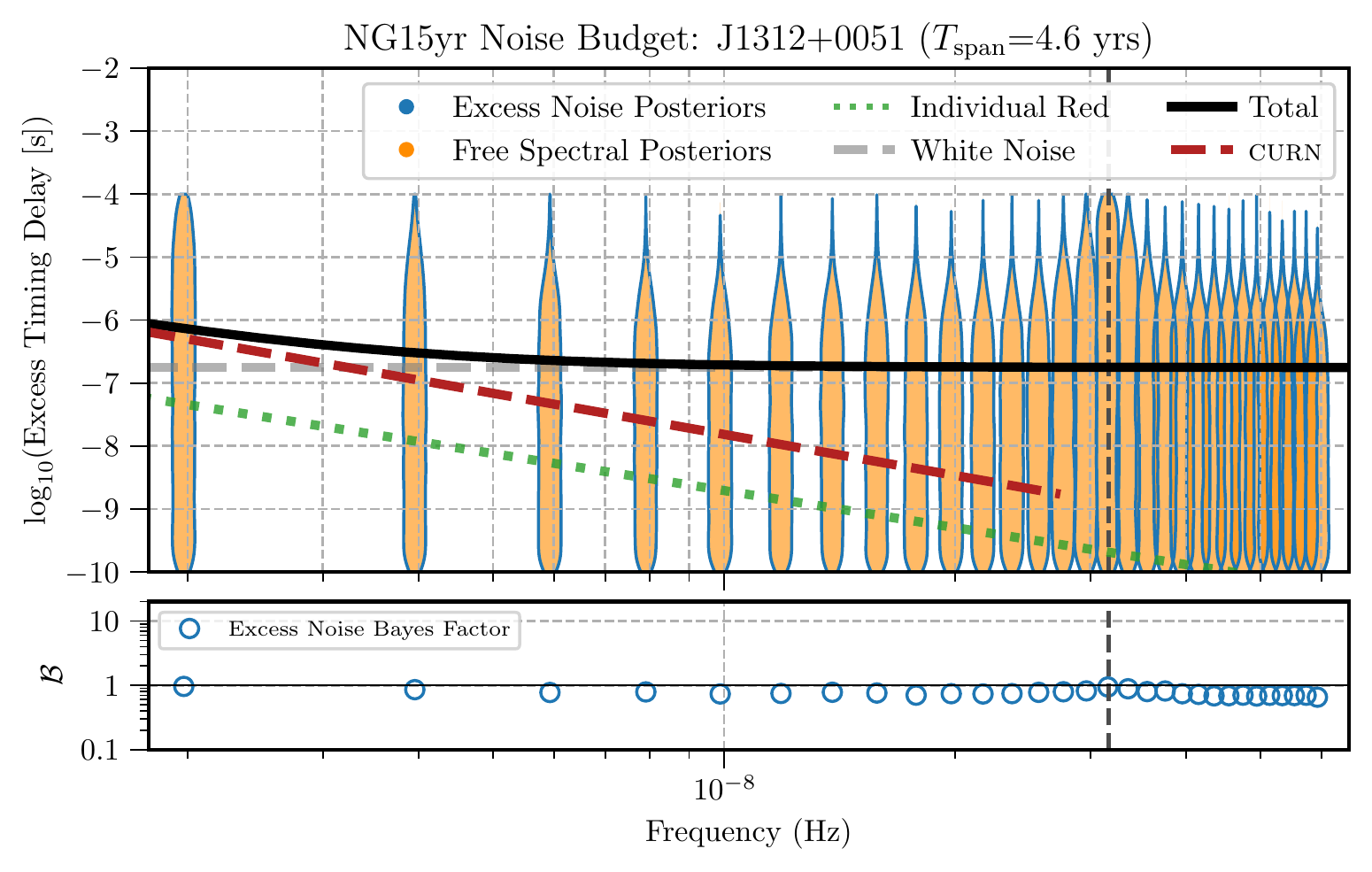}
\caption{The excess timing residual delay as a function of frequency for PSR~J1312+0051. See \myfig{f:excess_j1909} for details.}
\label{f:budget_J1312+0051}
\end{figure}

\begin{figure}
\centering
\includegraphics[width=0.9\linewidth]{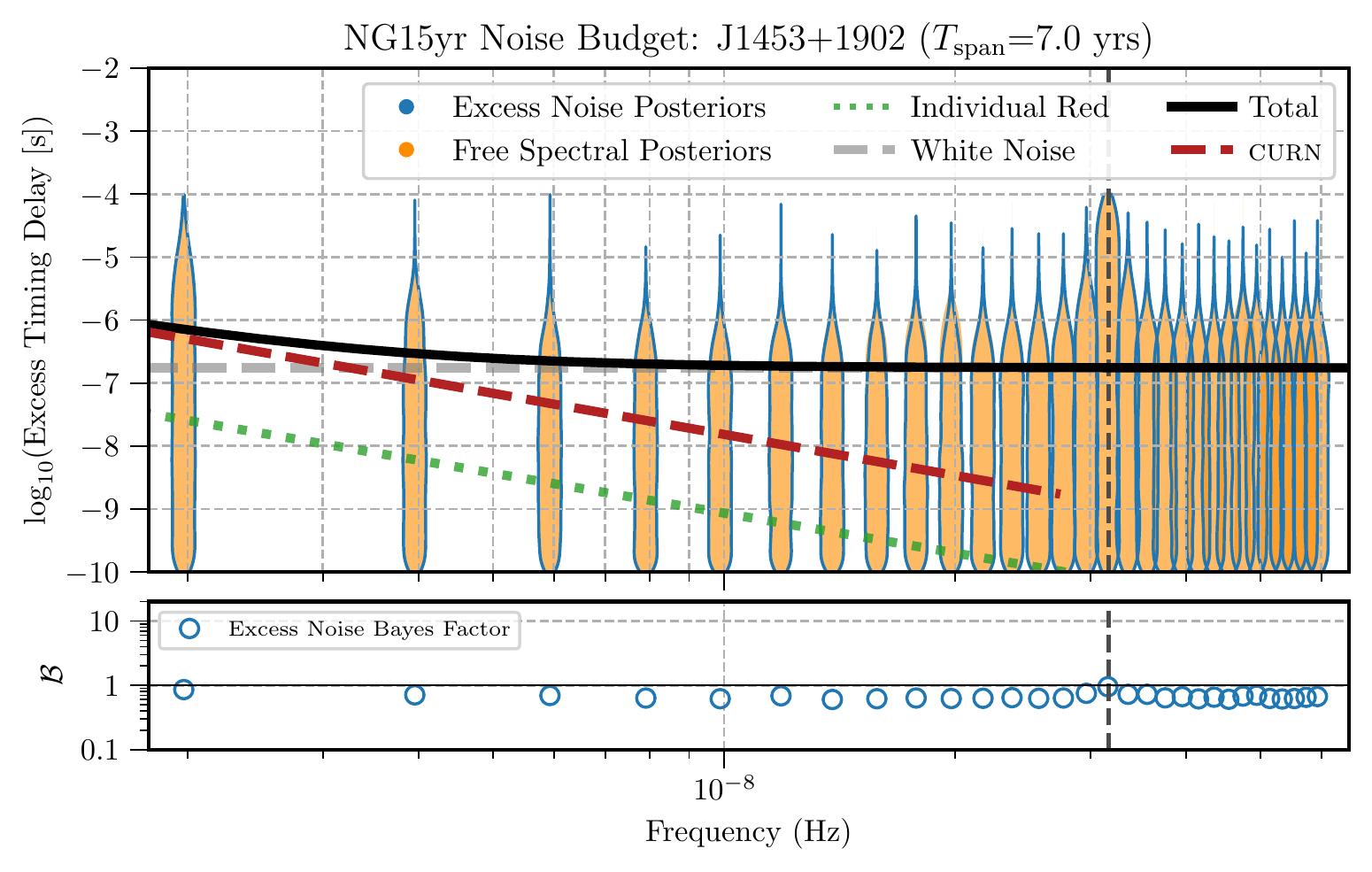}
\caption{The excess timing residual delay as a function of frequency for PSR~J1453+1902. See \myfig{f:excess_j1909} for details.}
\label{f:budget_J1453+1902}
\end{figure}

\begin{figure}
\centering
\includegraphics[width=0.9\linewidth]{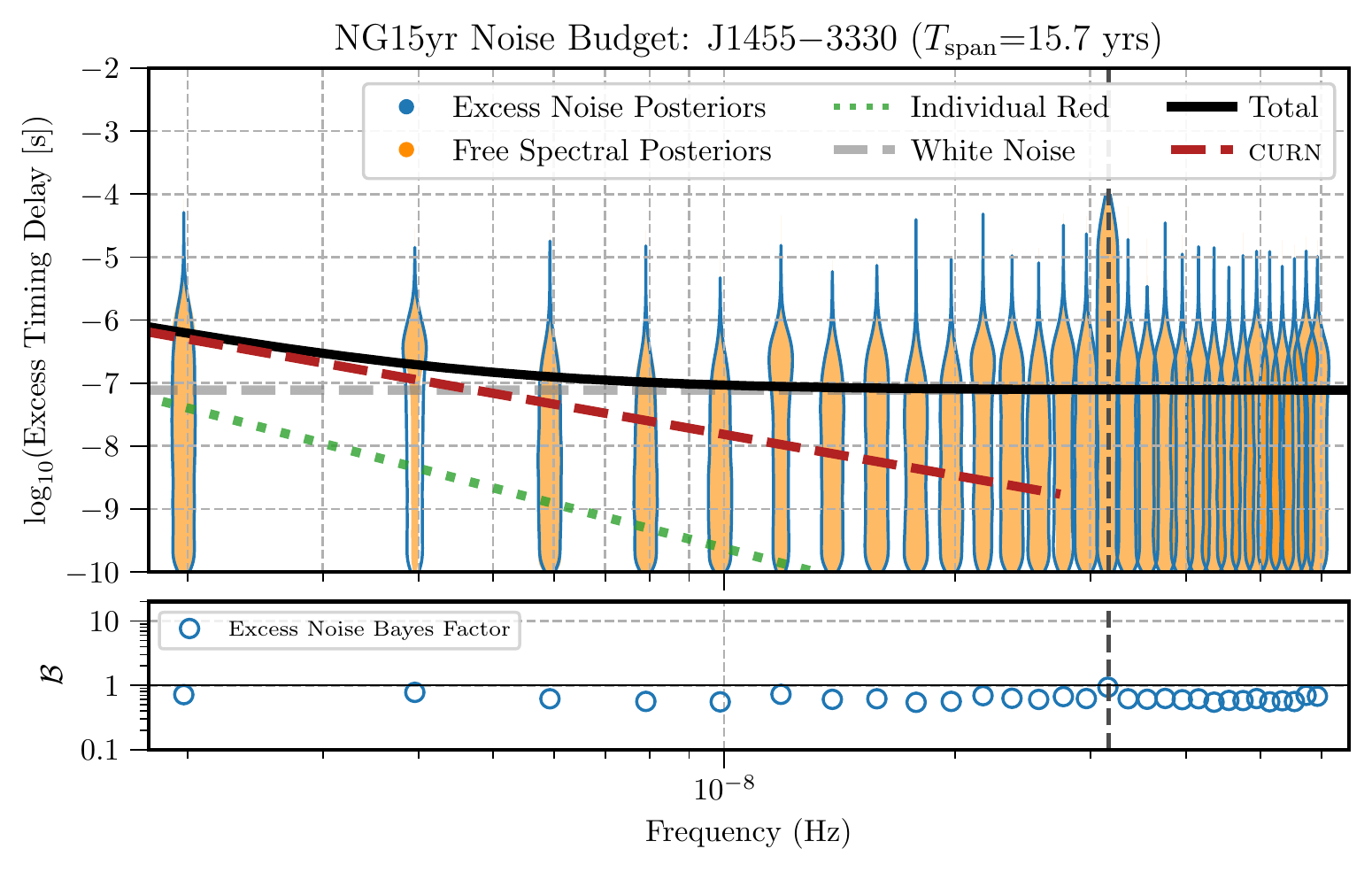}
\caption{The excess timing residual delay as a function of frequency for PSR~J1455$-$3330. See \myfig{f:excess_j1909} for details.}
\label{f:budget_J1455-3330}
\end{figure}

\begin{figure}
\centering
\includegraphics[width=0.9\linewidth]{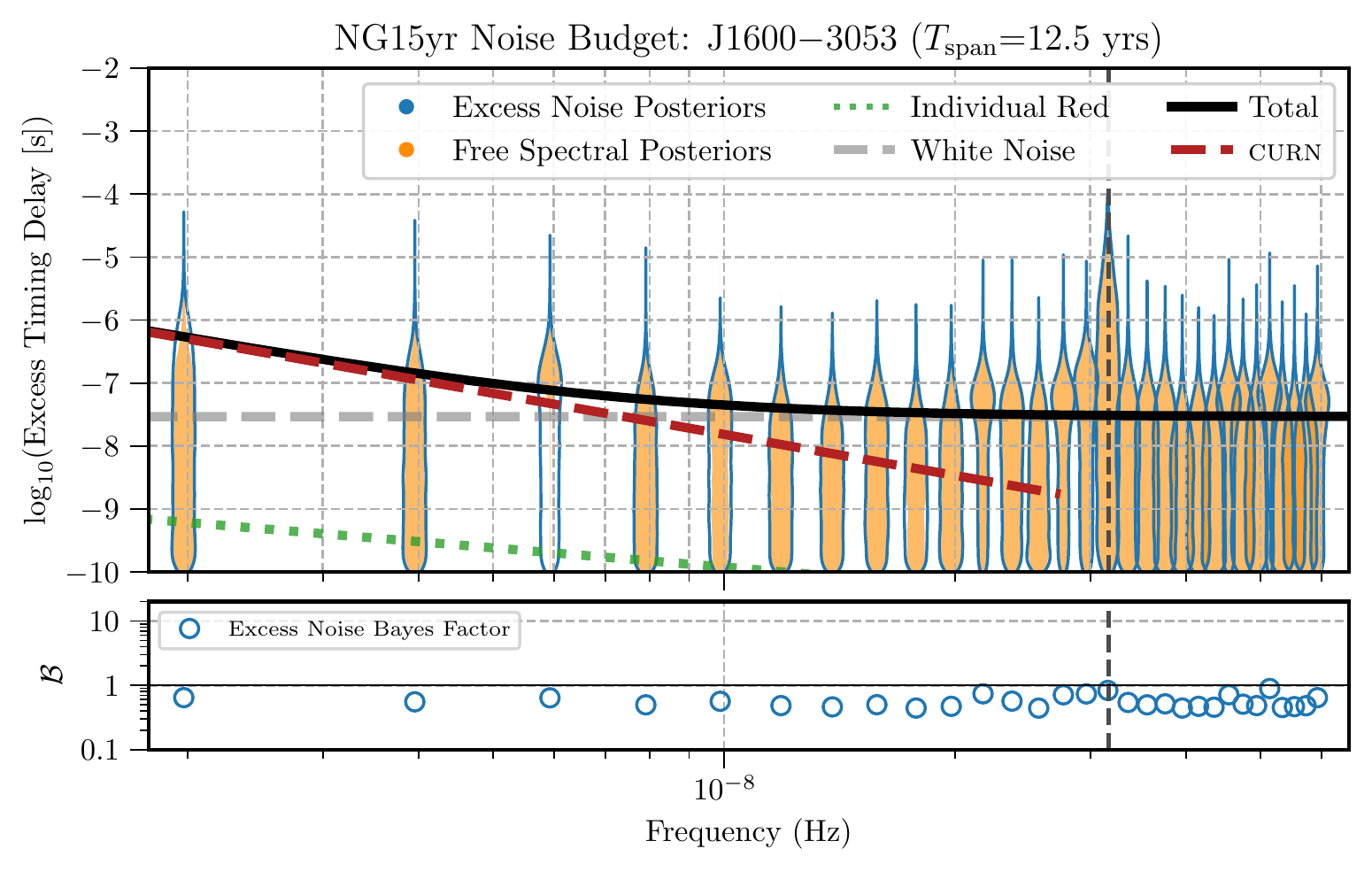}
\caption{The excess timing residual delay as a function of frequency for PSR~J1600$-$3053. See \myfig{f:excess_j1909} for details.}
\label{f:budget_J1600-3053}
\end{figure}

\begin{figure}
\centering
\includegraphics[width=0.9\linewidth]{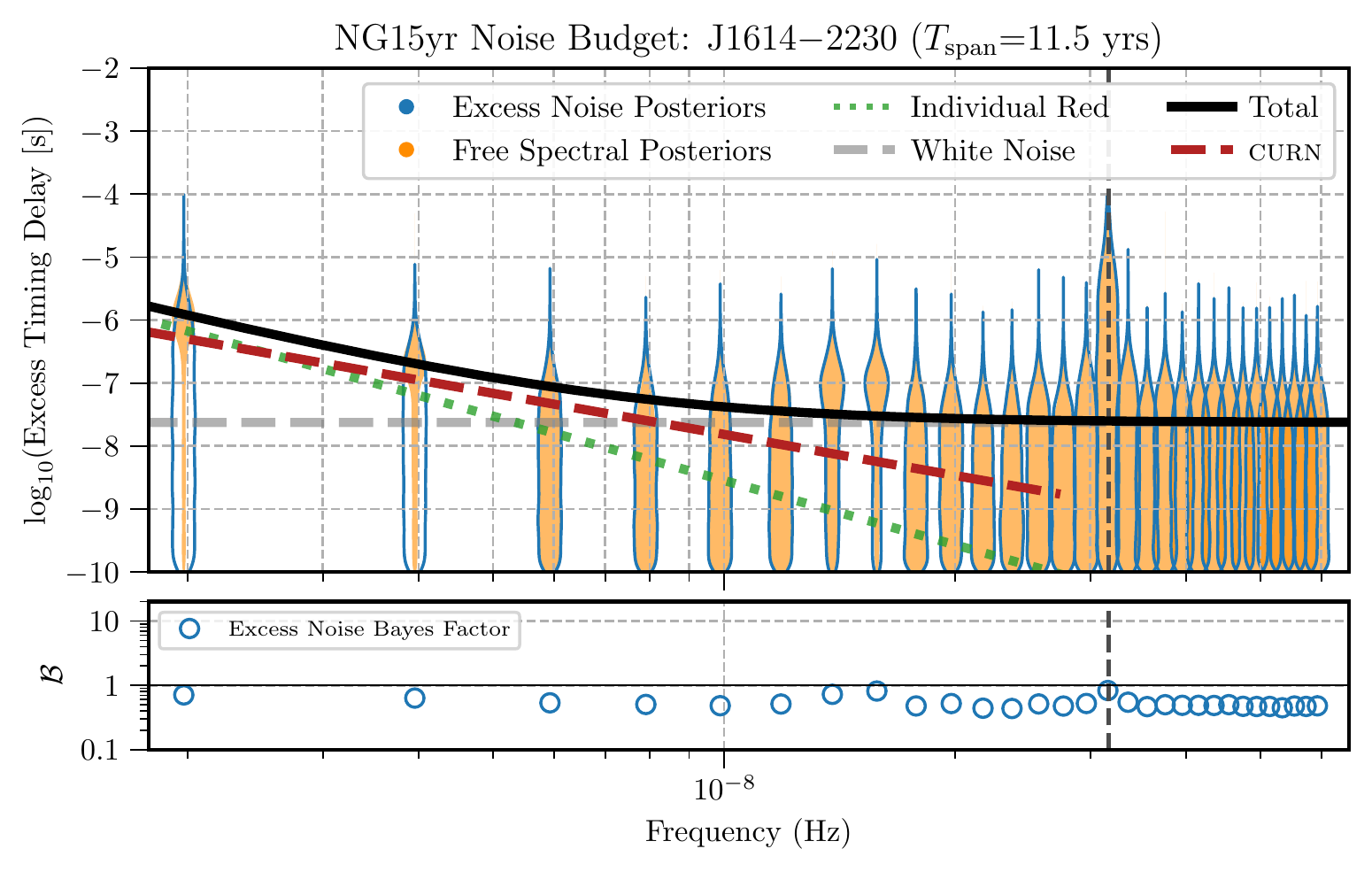}
\caption{The excess timing residual delay as a function of frequency for PSR~J1614$-$2230. See \myfig{f:excess_j1909} for details.}
\label{f:budget_J1614-2230}
\end{figure}

\begin{figure}
\centering
\includegraphics[width=0.9\linewidth]{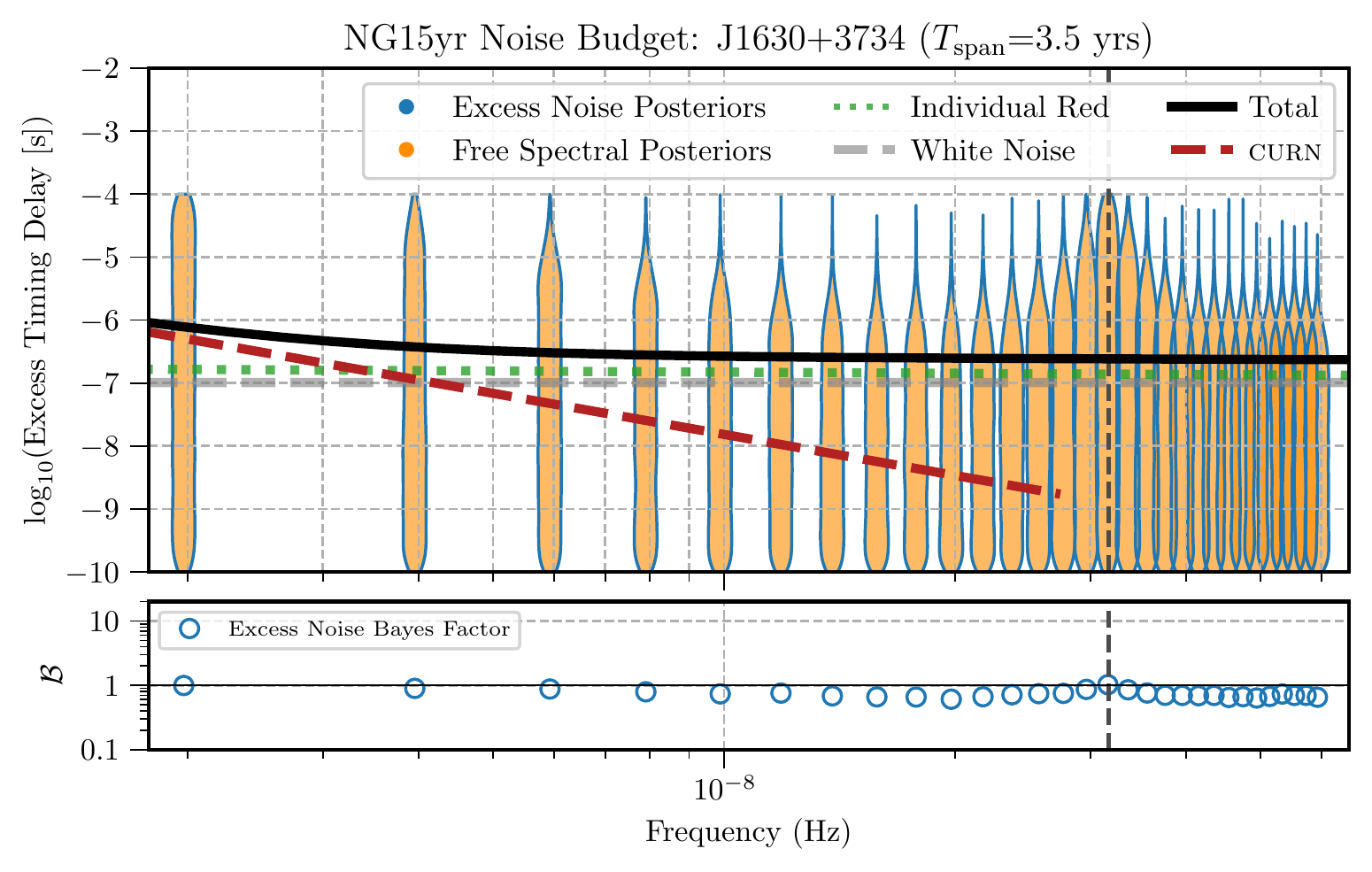}
\caption{The excess timing residual delay as a function of frequency for PSR~J1630+3734. See \myfig{f:excess_j1909} for details.}
\label{f:budget_J1630+3734}
\end{figure}

\begin{figure}
\centering
\includegraphics[width=0.9\linewidth]{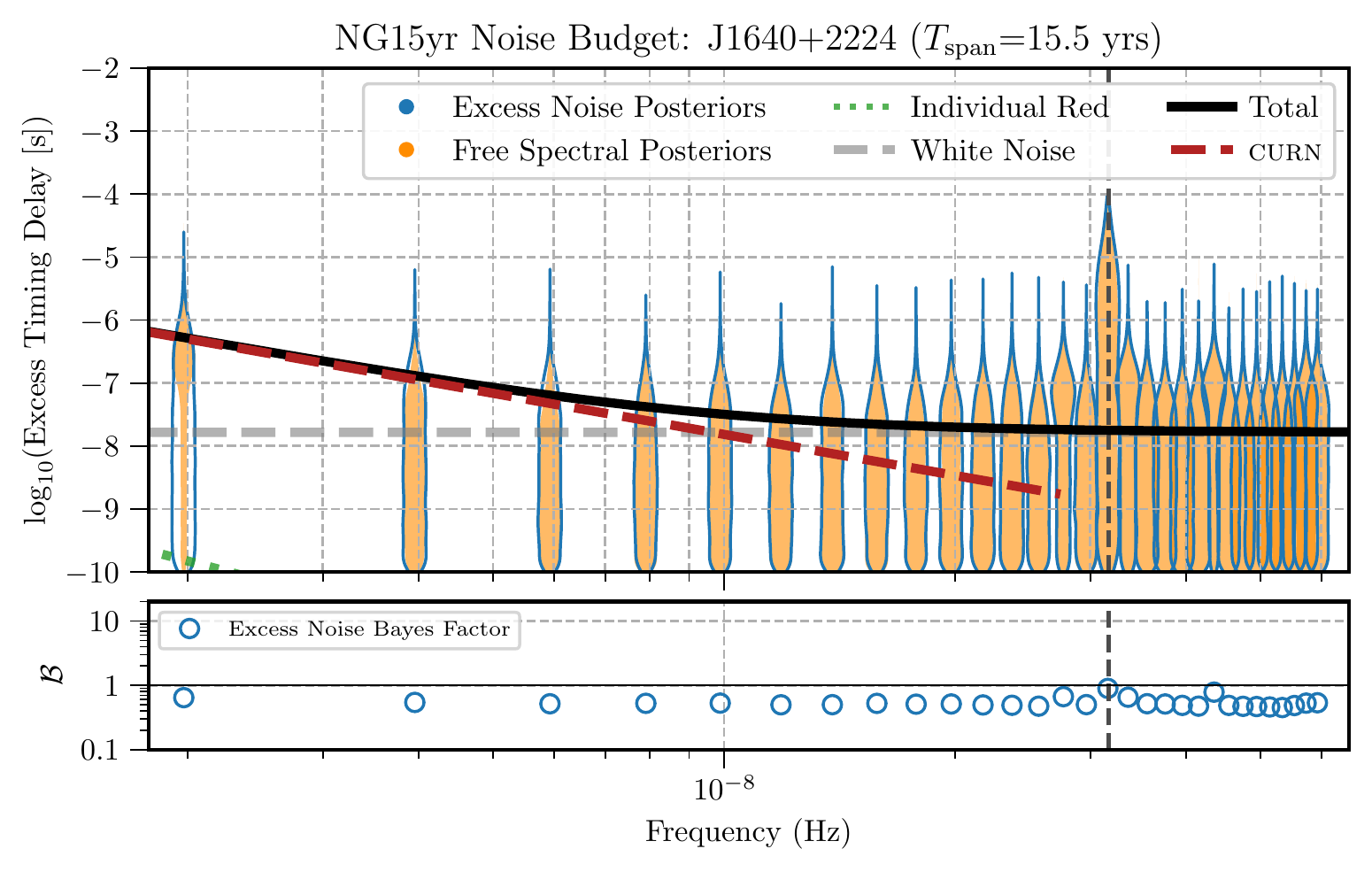}
\caption{The excess timing residual delay as a function of frequency for PSR~J1640+2224. See \myfig{f:excess_j1909} for details.}
\label{f:budget_J1640+2224}
\end{figure}

\begin{figure}
\centering
\includegraphics[width=0.9\linewidth]{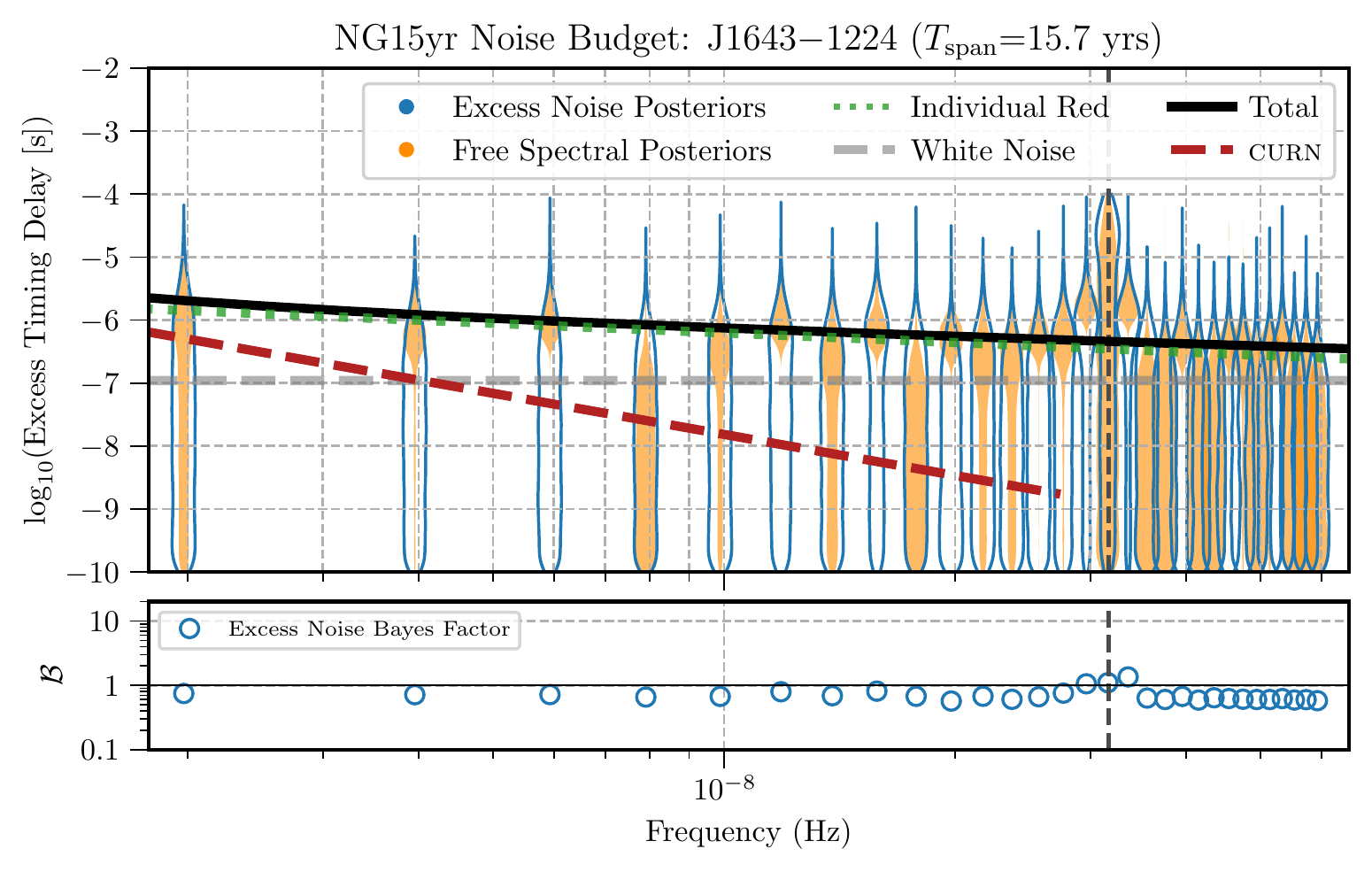}
\caption{The excess timing residual delay as a function of frequency for PSR~J1643$-$1224. See \myfig{f:excess_j1909} for details.}
\label{f:budget_J1643-1224}
\end{figure}

\begin{figure}
\centering
\includegraphics[width=0.9\linewidth]{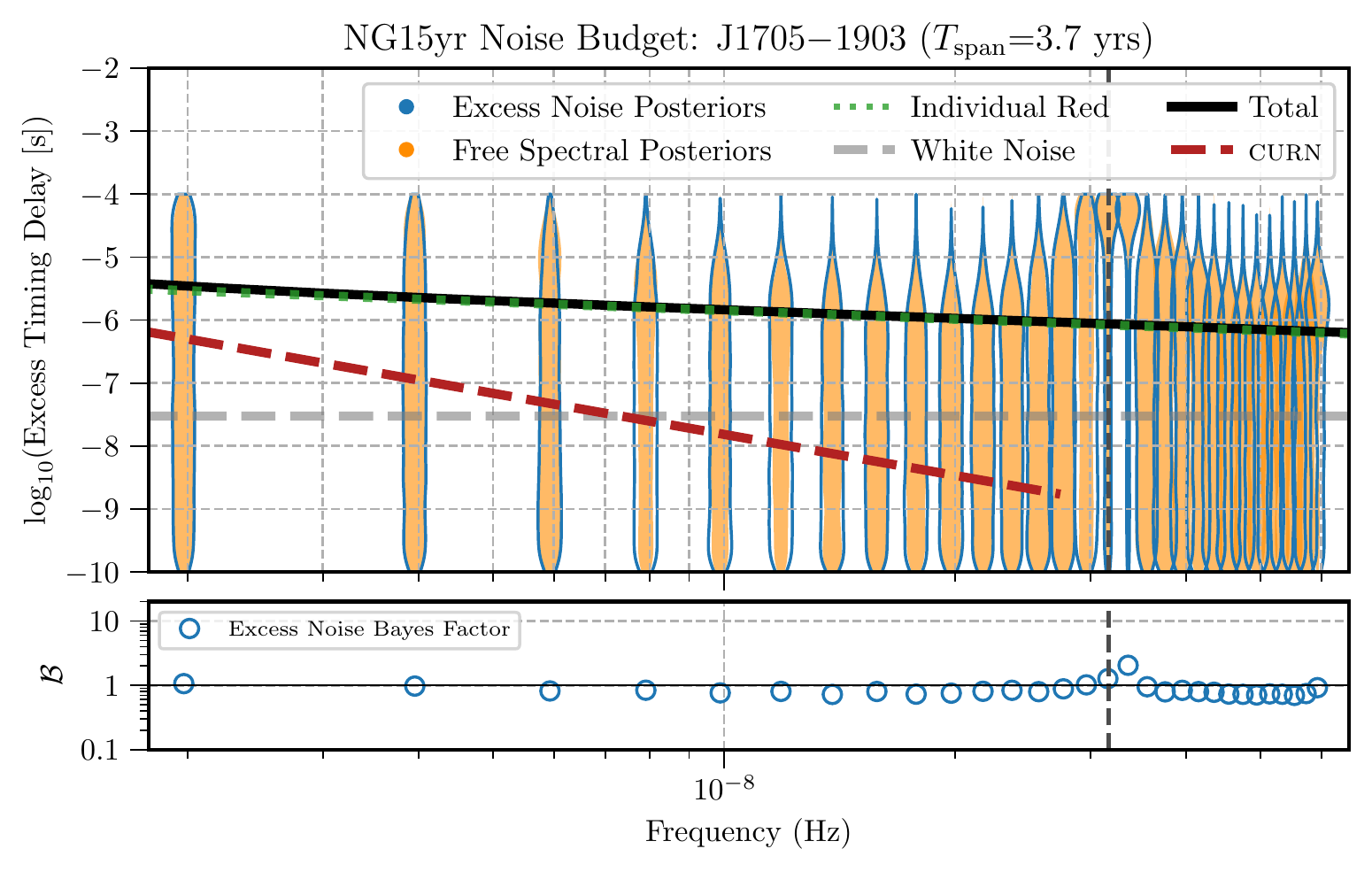}
\caption{The excess timing residual delay as a function of frequency for PSR~J1705$-$1903. See \myfig{f:excess_j1909} for details.}
\label{f:budget_J1705-1903}
\end{figure}

\begin{figure}
\centering
\includegraphics[width=0.9\linewidth]{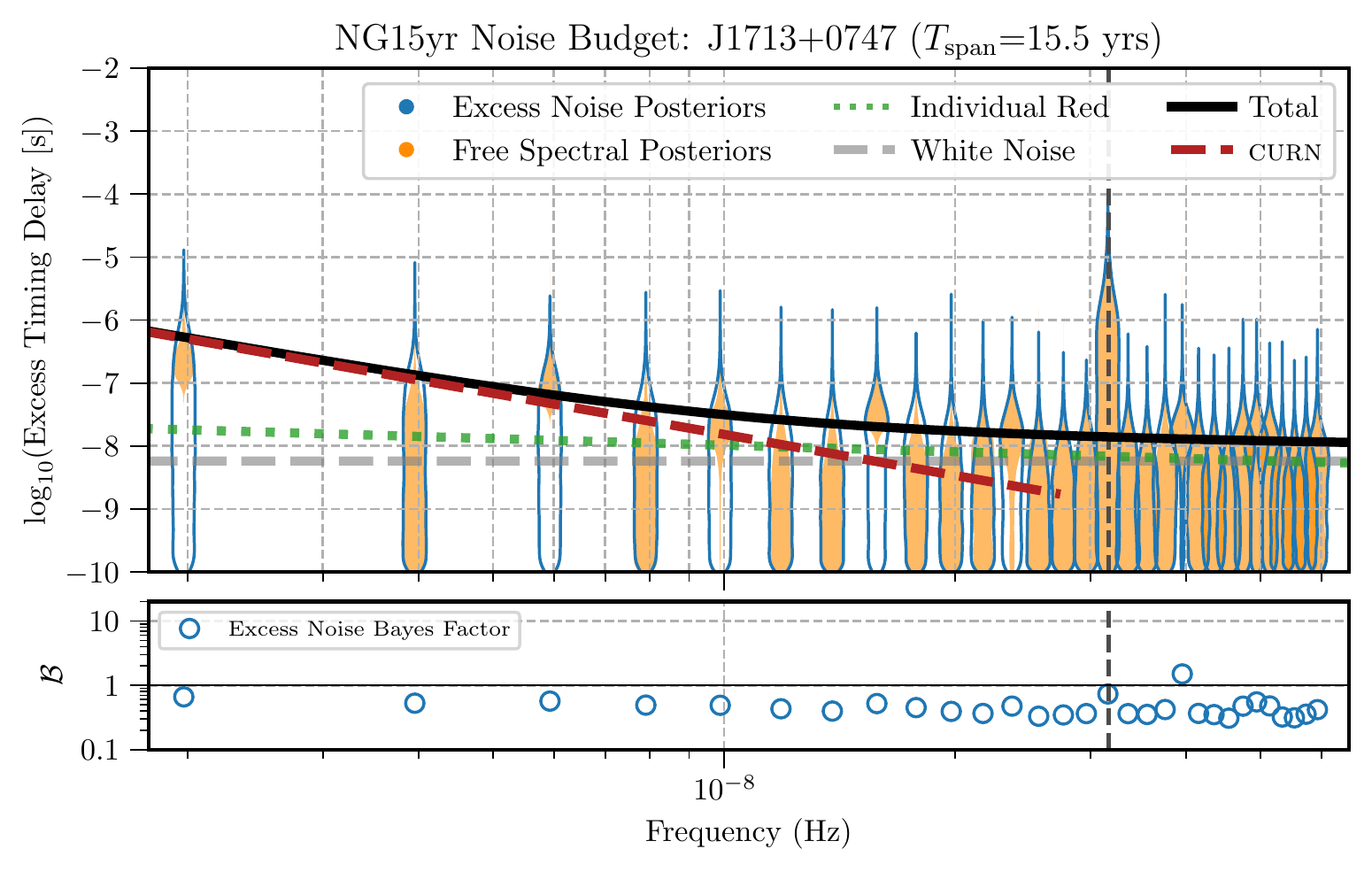}
\caption{The excess timing residual delay as a function of frequency for PSR~J1713+0747. See \myfig{f:excess_j1909} for details.}
\label{f:budget_J1713+0747}
\end{figure}

\begin{figure}
\centering
\includegraphics[width=0.9\linewidth]{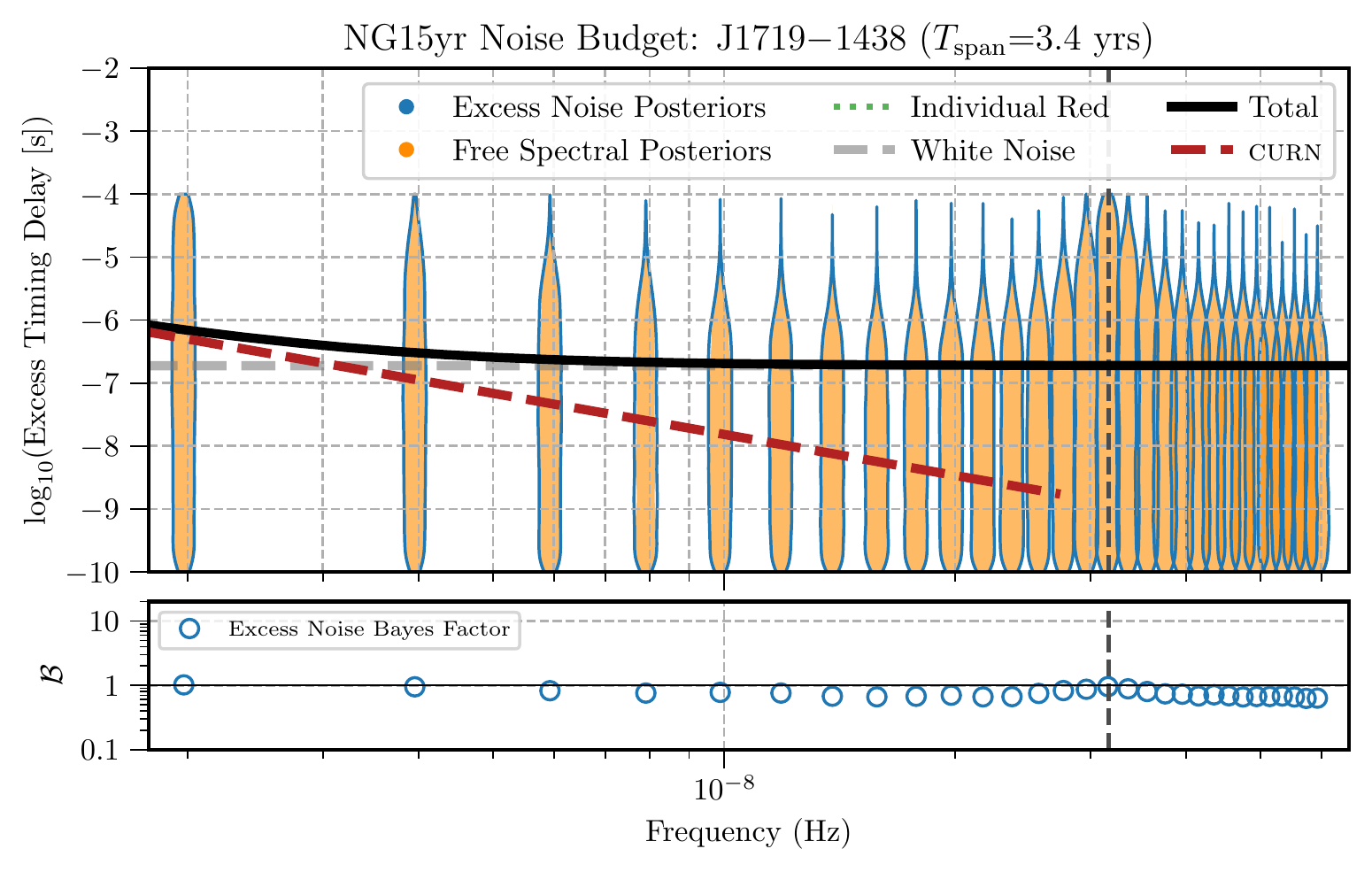}
\caption{The excess timing residual delay as a function of frequency for PSR~J1719$-$1438. See \myfig{f:excess_j1909} for details.}
\label{f:budget_J1719-1438}
\end{figure}

\begin{figure}
\centering
\includegraphics[width=0.9\linewidth]{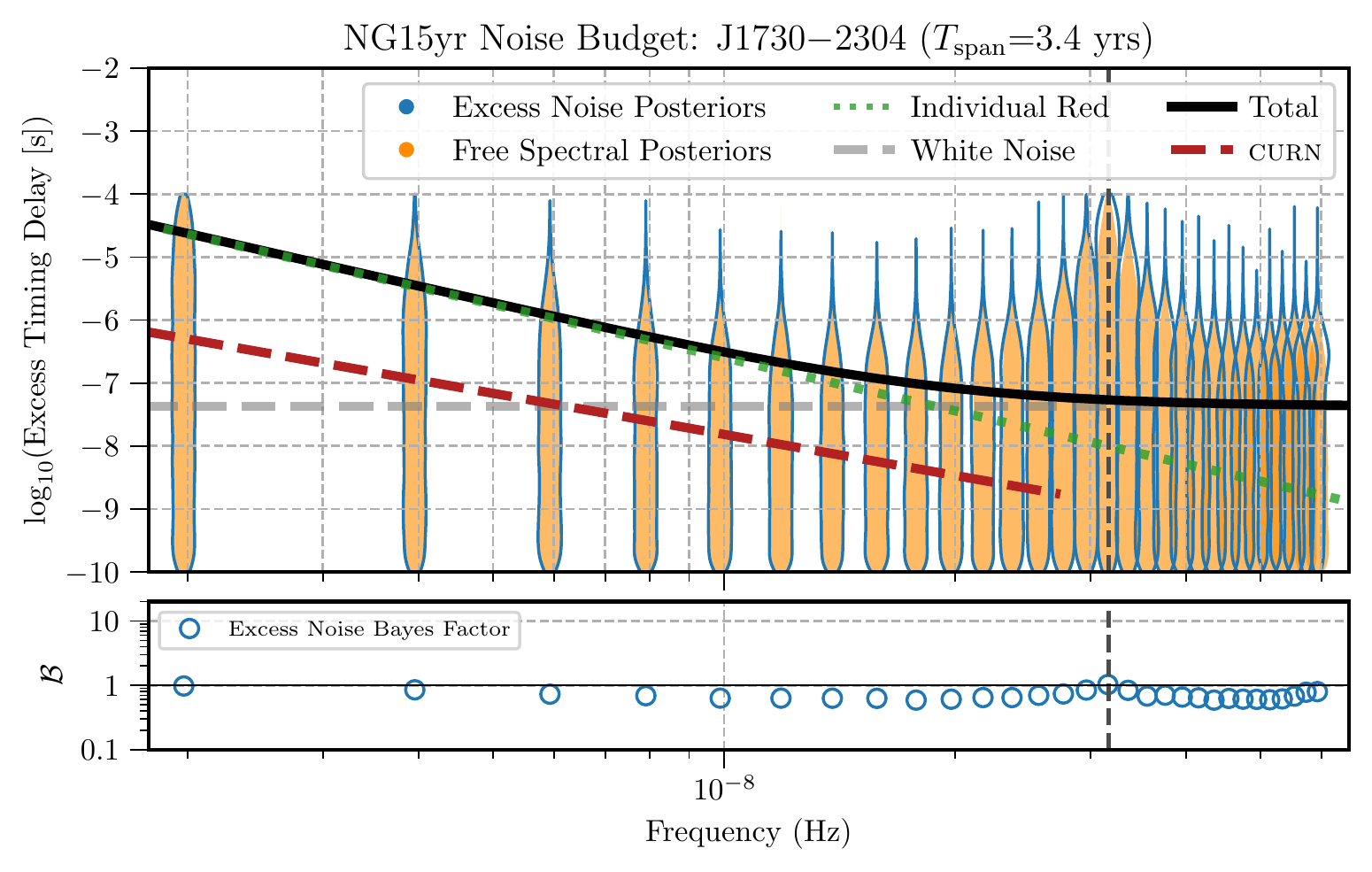}
\caption{The excess timing residual delay as a function of frequency for PSR~J1730$-$2304. See \myfig{f:excess_j1909} for details.}
\label{f:budget_J1730-2304}
\end{figure}

\begin{figure}
\centering
\includegraphics[width=0.9\linewidth]{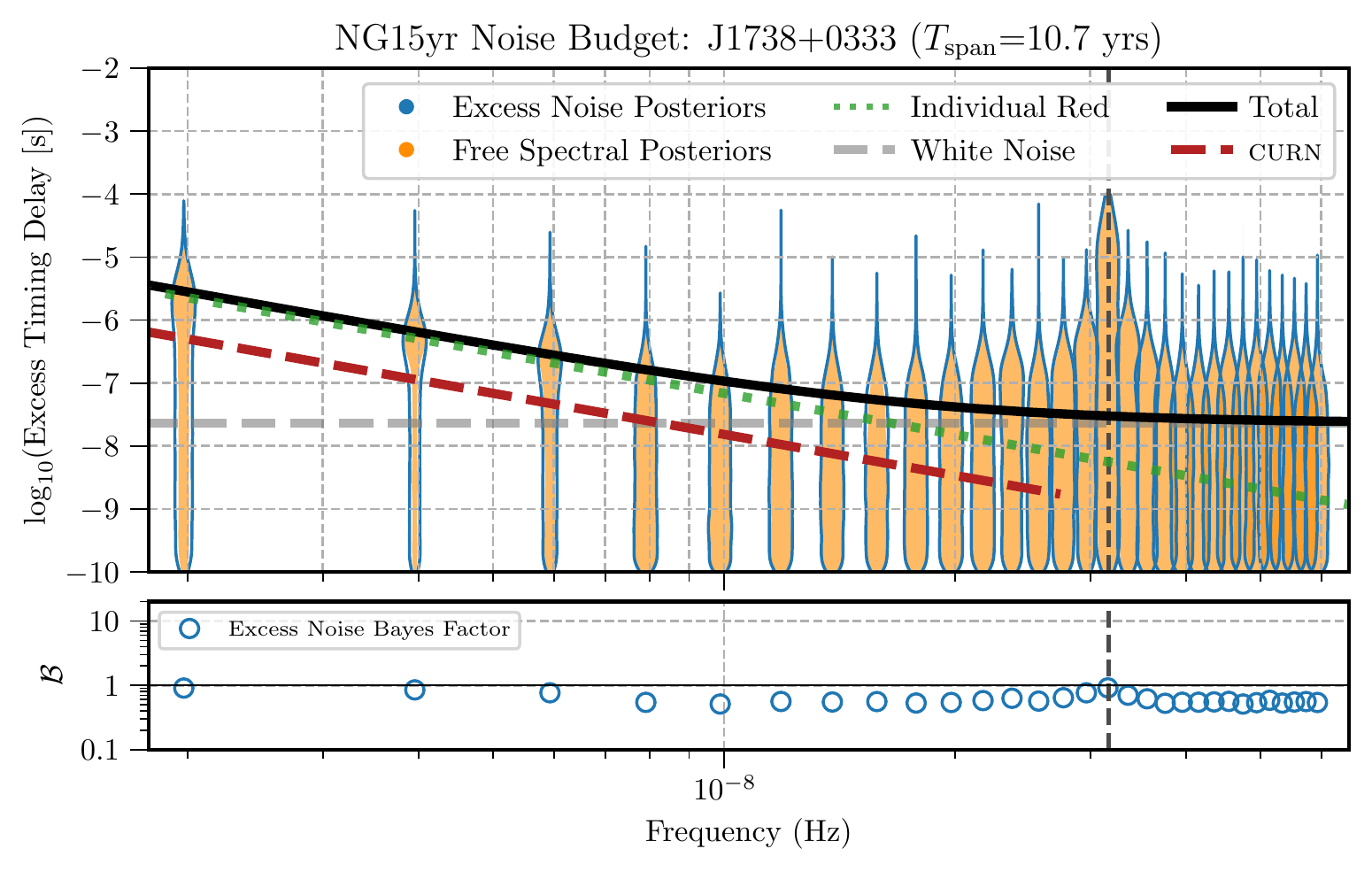}
\caption{The excess timing residual delay as a function of frequency for PSR~J1738+0333. See \myfig{f:excess_j1909} for details.}
\label{f:budget_J1738+0333}
\end{figure}

\begin{figure}
\centering
\includegraphics[width=0.9\linewidth]{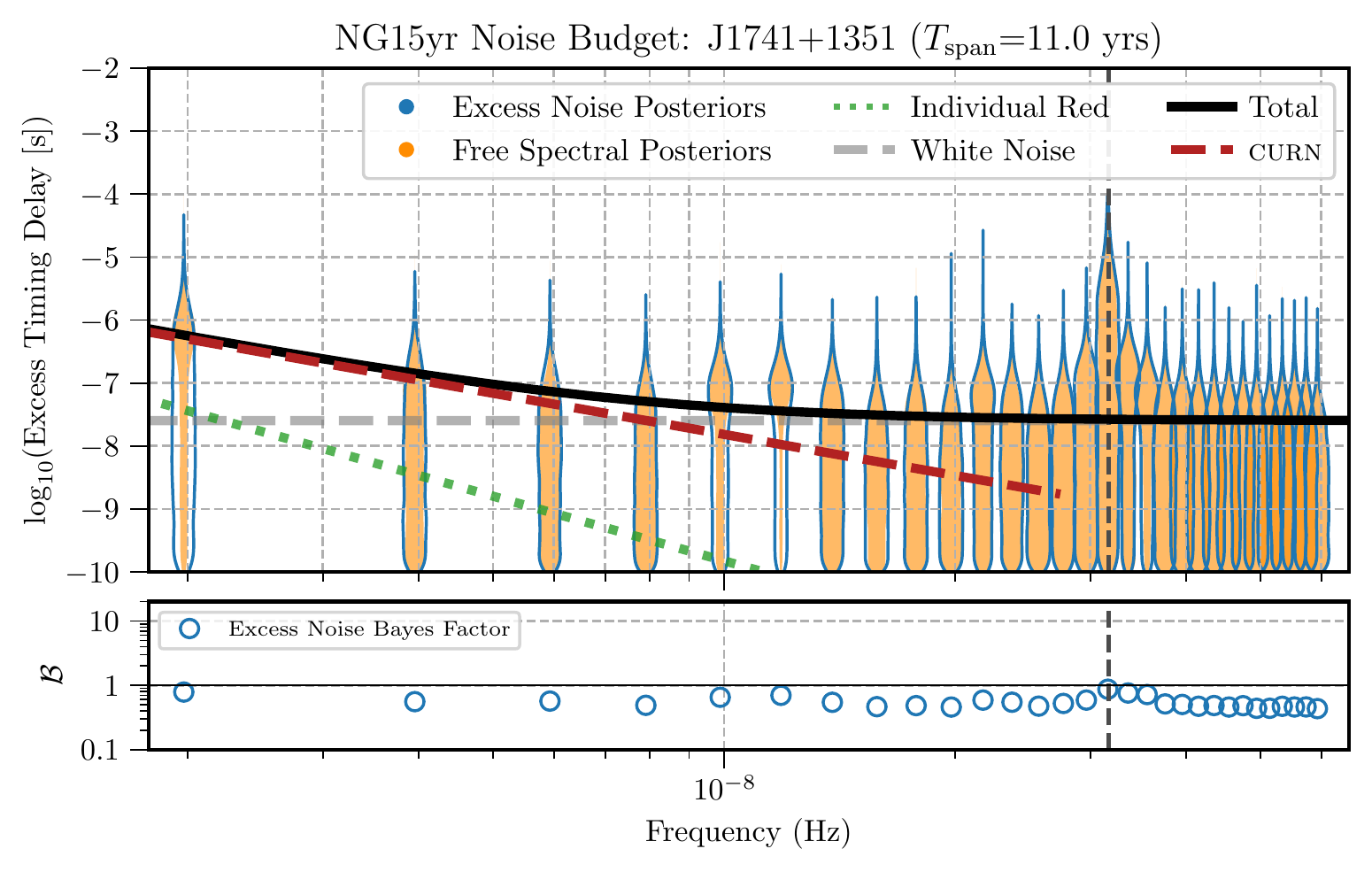}
\caption{The excess timing residual delay as a function of frequency for PSR~J1741+1351. See \myfig{f:excess_j1909} for details.}
\label{f:budget_J1741+1351}
\end{figure}

\begin{figure}
\centering
\includegraphics[width=0.9\linewidth]{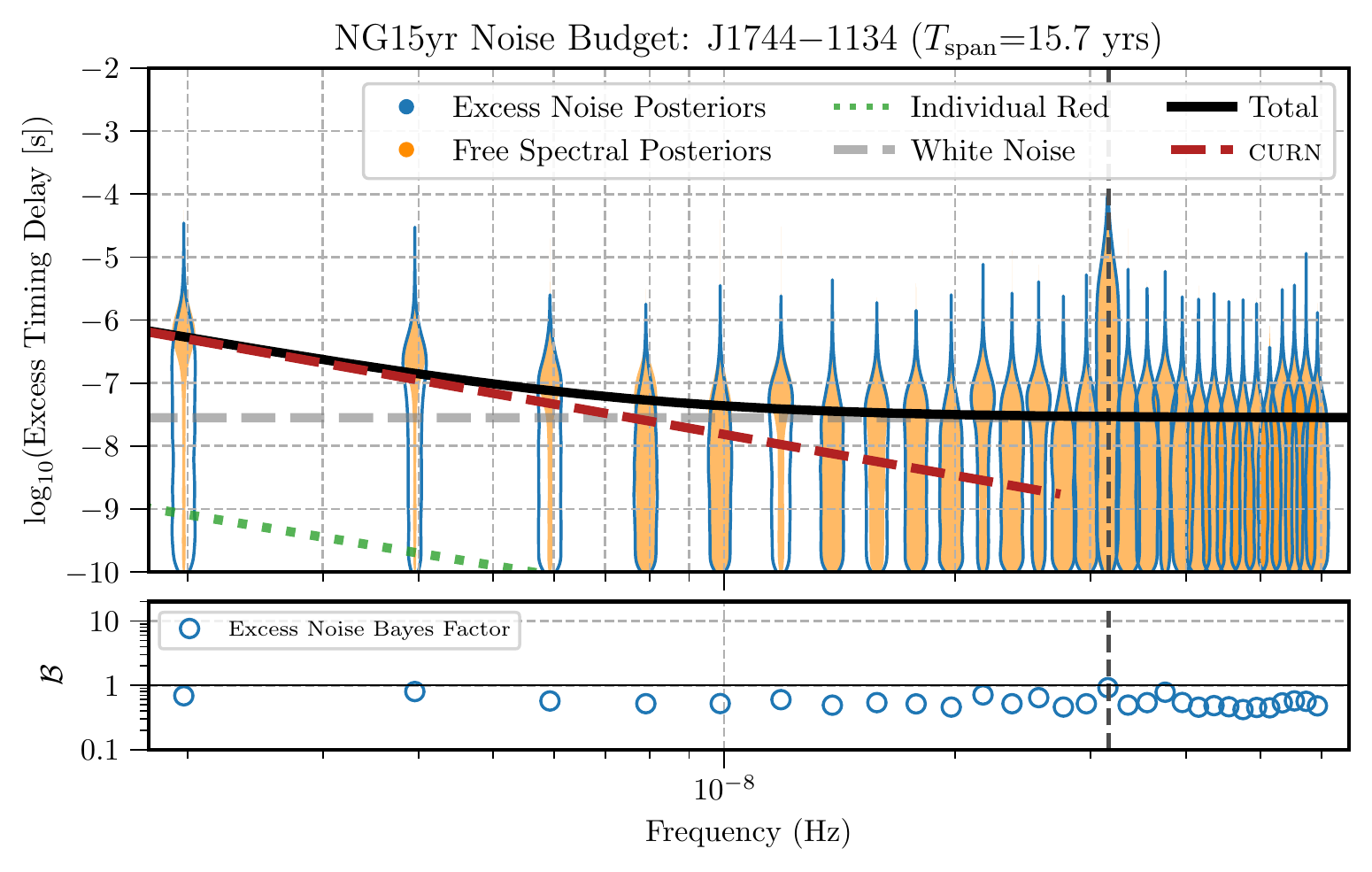}
\caption{The excess timing residual delay as a function of frequency for PSR~J1744$-$1134. See \myfig{f:excess_j1909} for details.}
\label{f:budget_J1744-1134}
\end{figure}

\begin{figure}
\centering
\includegraphics[width=0.9\linewidth]{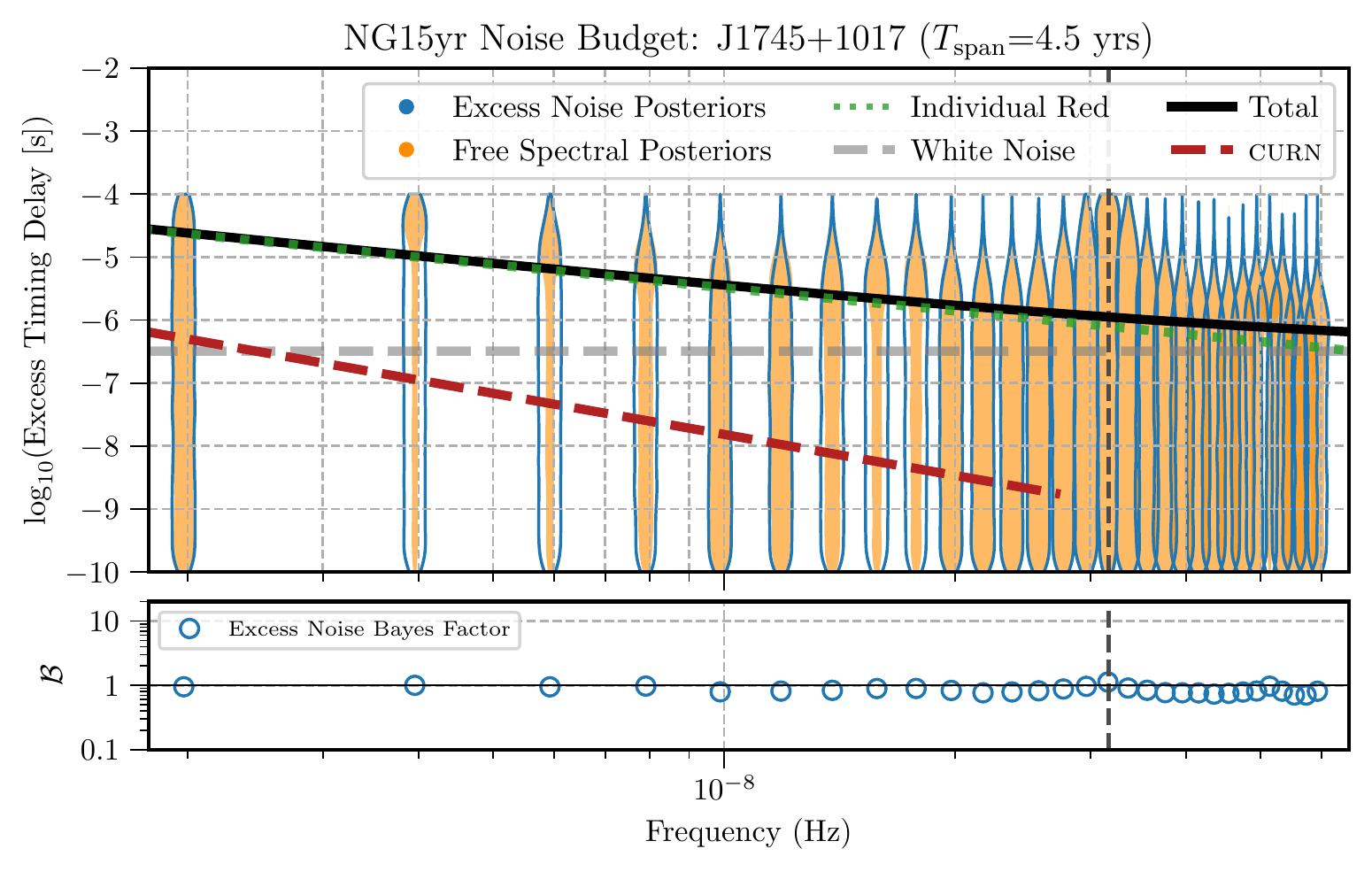}
\caption{The excess timing residual delay as a function of frequency for PSR~J1745+1017. See \myfig{f:excess_j1909} for details.}
\label{f:budget_J1745+1017}
\end{figure}

\begin{figure}
\centering
\includegraphics[width=0.9\linewidth]{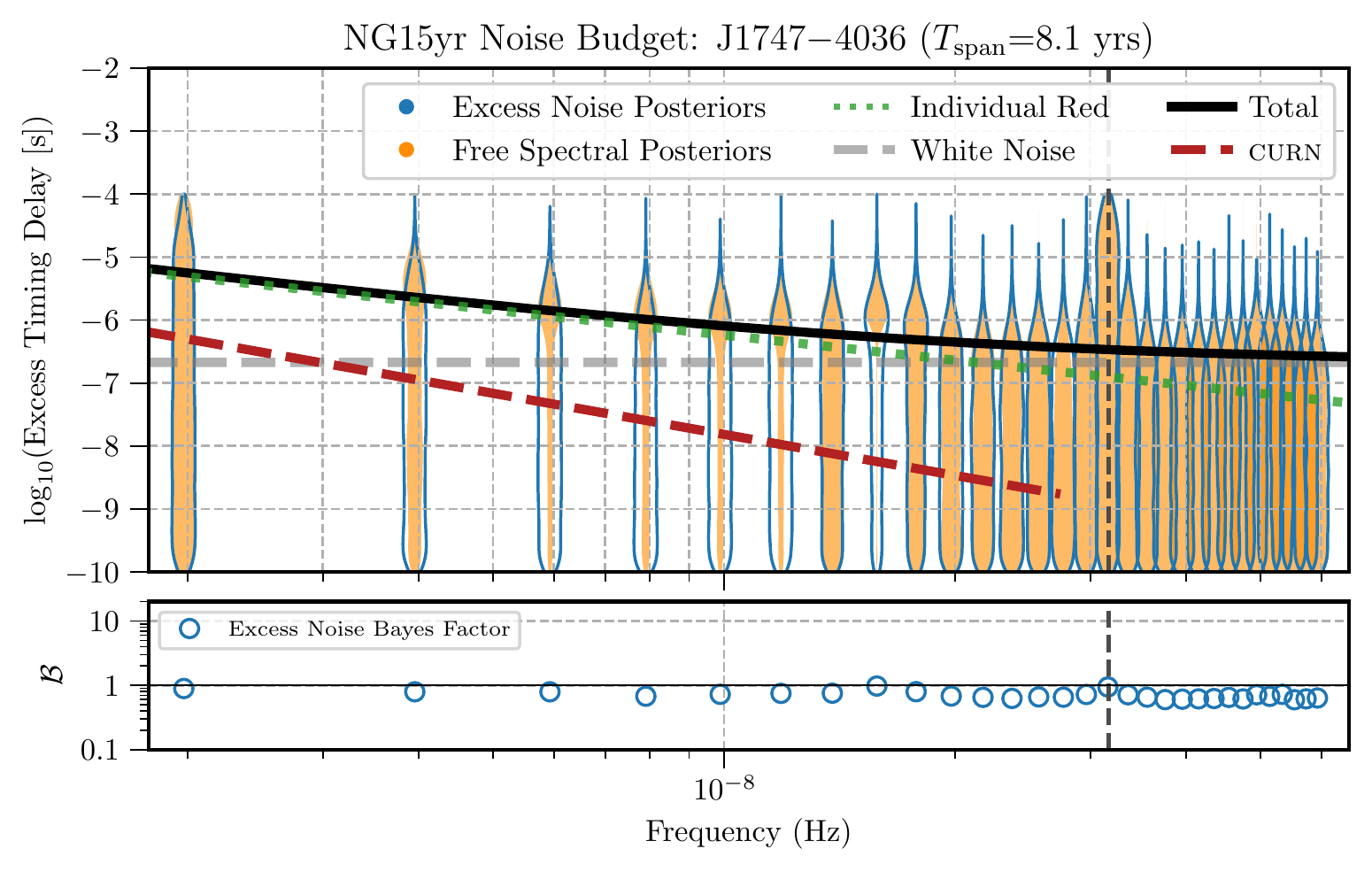}
\caption{The excess timing residual delay as a function of frequency for PSR~J1747$-$4036. See \myfig{f:excess_j1909} for details.}
\label{f:budget_J1747-4036}
\end{figure}

\begin{figure}
\centering
\includegraphics[width=0.9\linewidth]{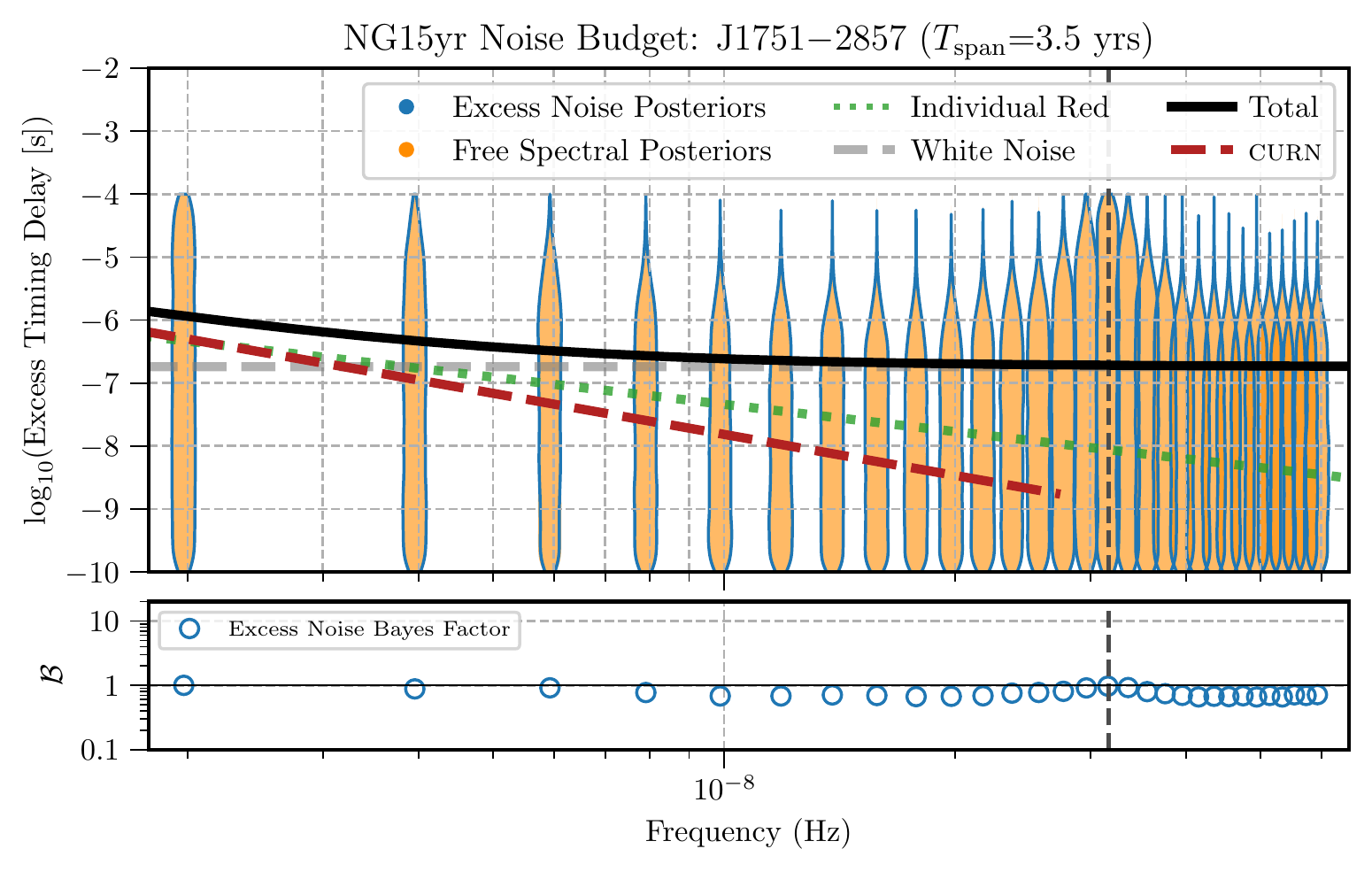}
\caption{The excess timing residual delay as a function of frequency for PSR~J1751$-$2857. See \myfig{f:excess_j1909} for details.}
\label{f:budget_J1751-2857}
\end{figure}

\begin{figure}
\centering
\includegraphics[width=0.9\linewidth]{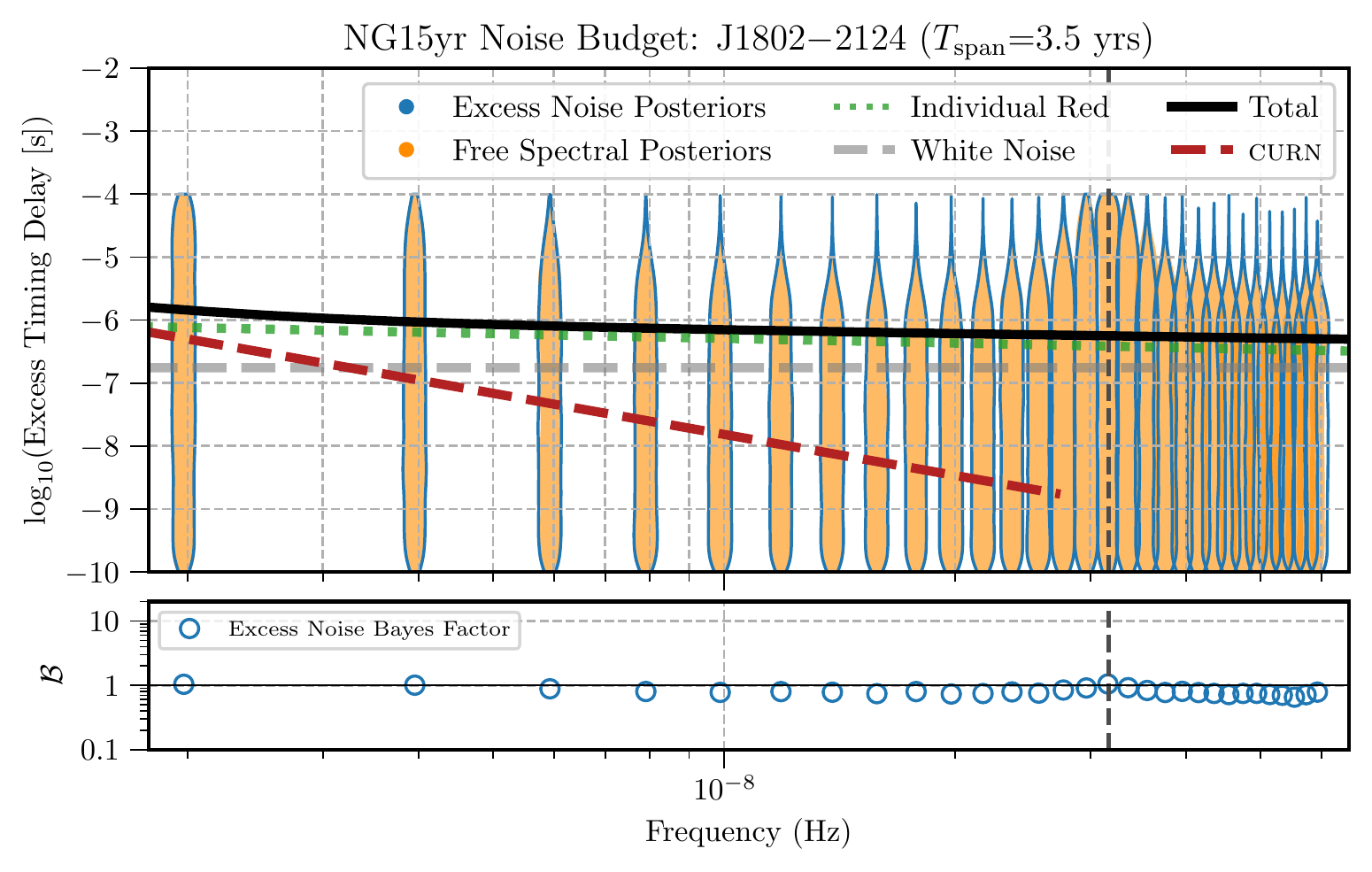}
\caption{The excess timing residual delay as a function of frequency for PSR~J1802$-$2124. See \myfig{f:excess_j1909} for details.}
\label{f:budget_J1802-2124}
\end{figure}

\begin{figure}
\centering
\includegraphics[width=0.9\linewidth]{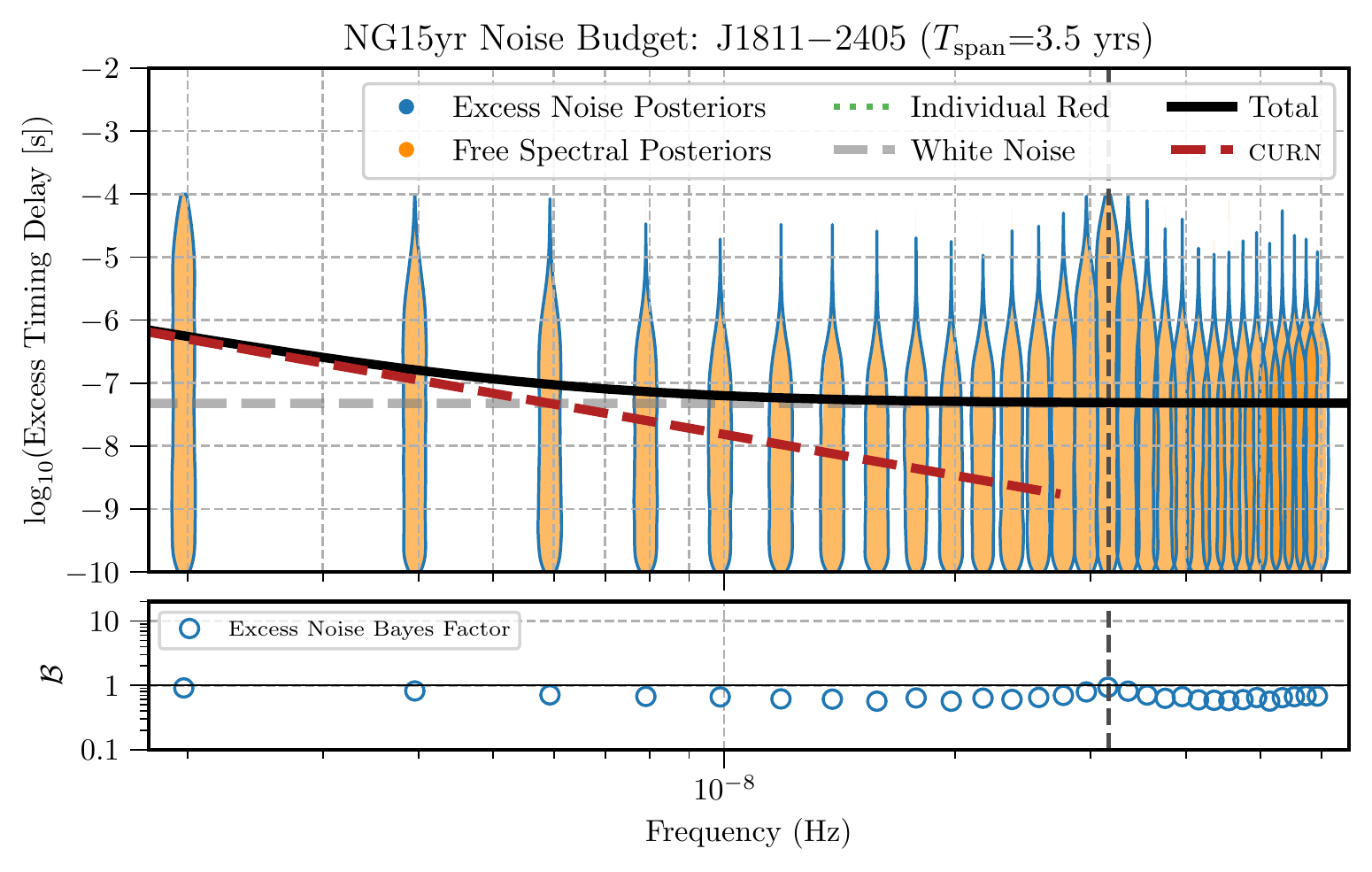}
\caption{The excess timing residual delay as a function of frequency for PSR~J1811$-$2405. See \myfig{f:excess_j1909} for details.}
\label{f:budget_J1811-2405}
\end{figure}

\begin{figure}
\centering
\includegraphics[width=0.9\linewidth]{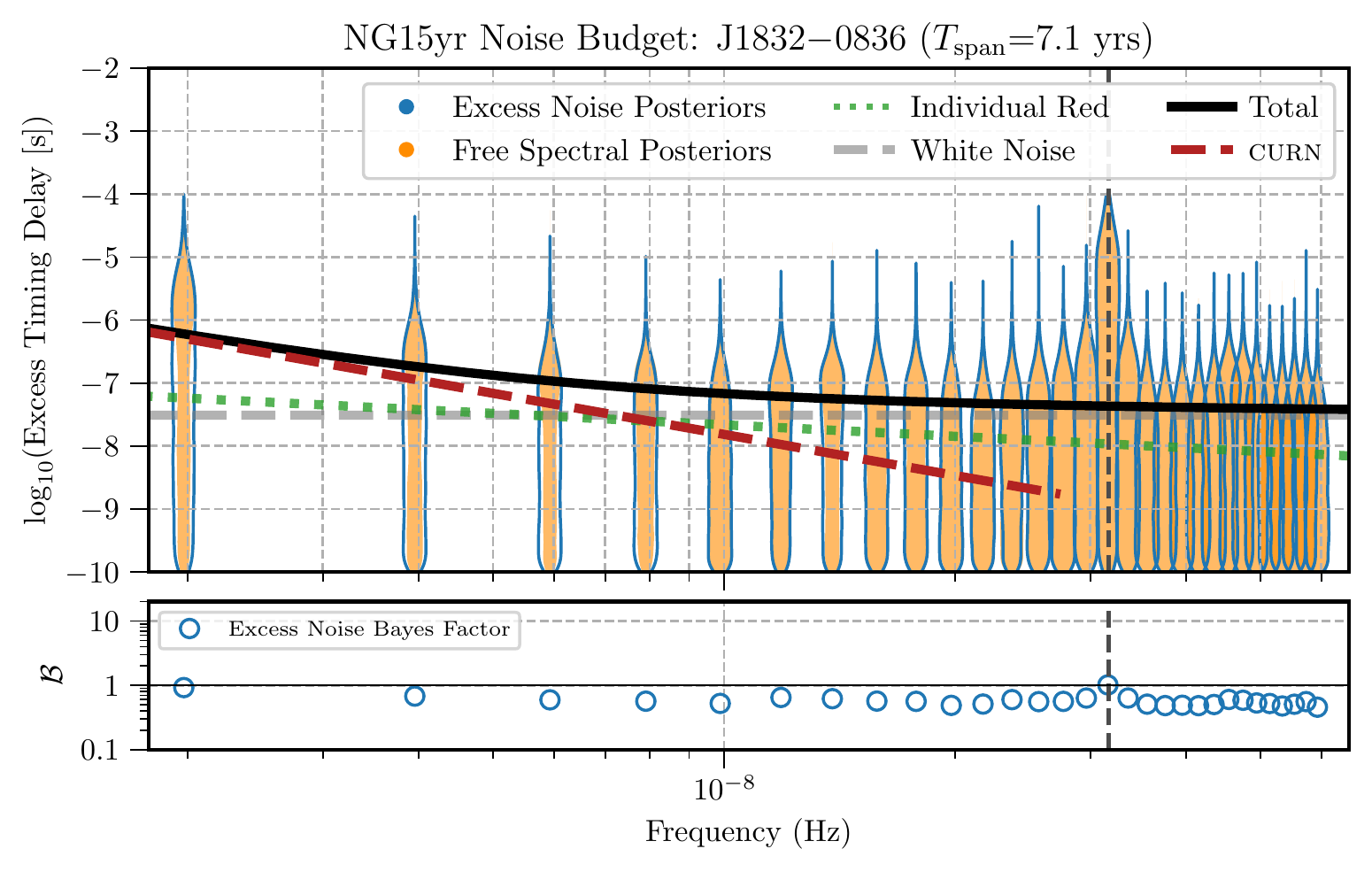}
\caption{The excess timing residual delay as a function of frequency for PSR~J1832$-$0836. See \myfig{f:excess_j1909} for details.}
\label{f:budget_J1832-0836}
\end{figure}

\begin{figure}
\centering
\includegraphics[width=0.9\linewidth]{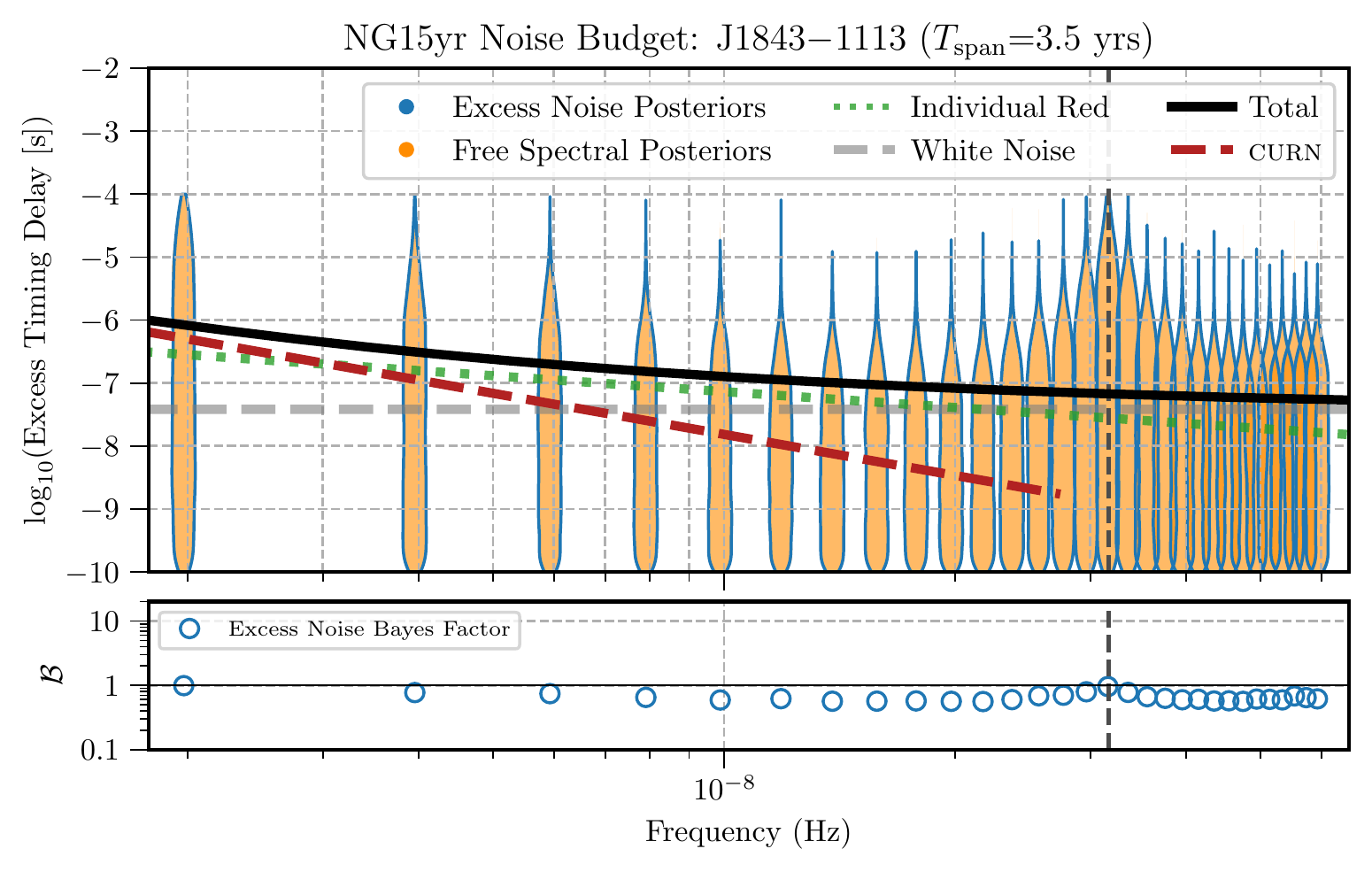}
\caption{The excess timing residual delay as a function of frequency for PSR~J1843$-$1113. See \myfig{f:excess_j1909} for details.}
\label{f:budget_J1843-1113}
\end{figure}

\begin{figure}
\centering
\includegraphics[width=0.9\linewidth]{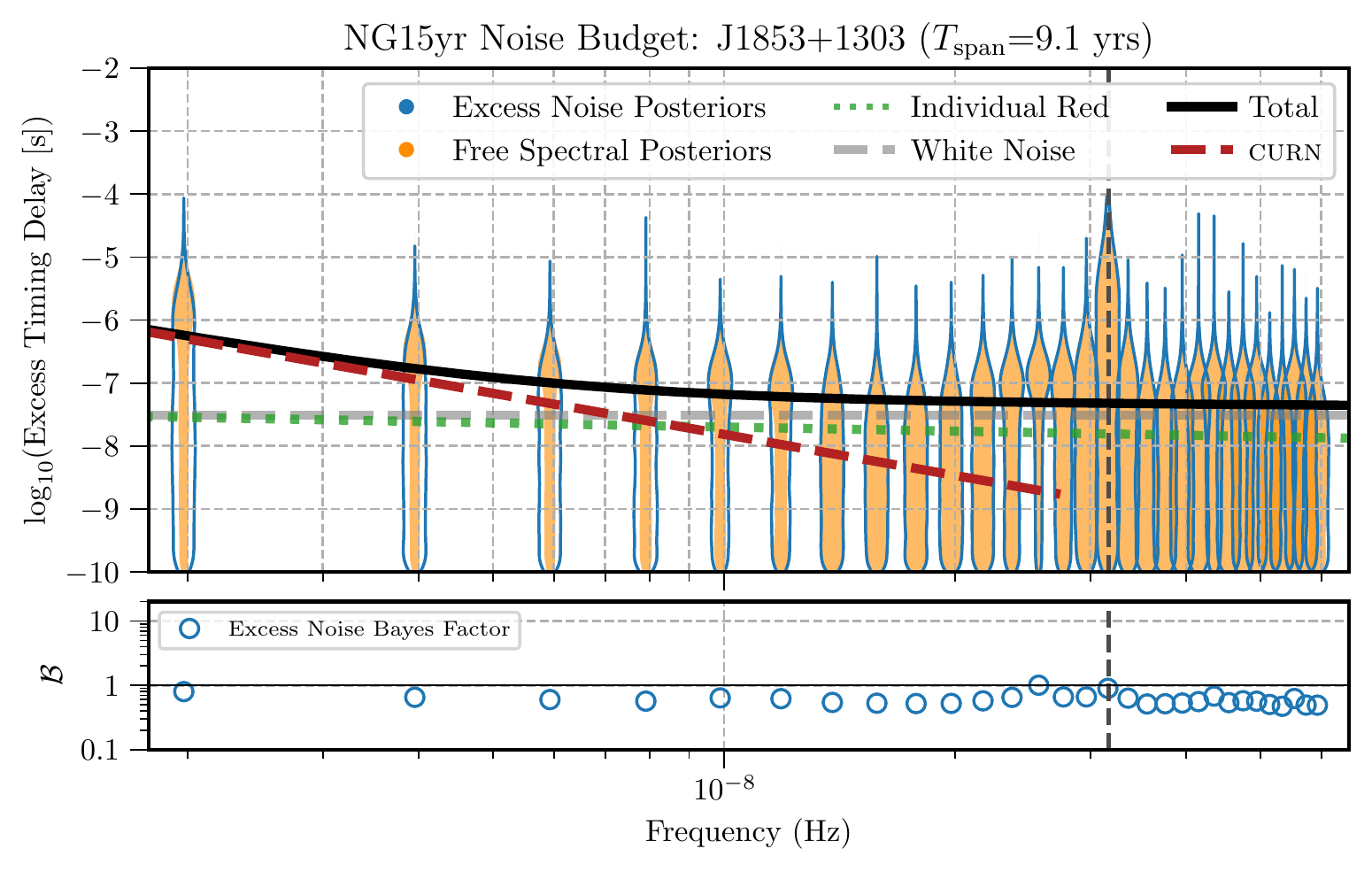}
\caption{The excess timing residual delay as a function of frequency for PSR~J1853+1303. See \myfig{f:excess_j1909} for details.}
\label{f:budget_J1853+1303}
\end{figure}

\begin{figure}
\centering
\includegraphics[width=0.9\linewidth]{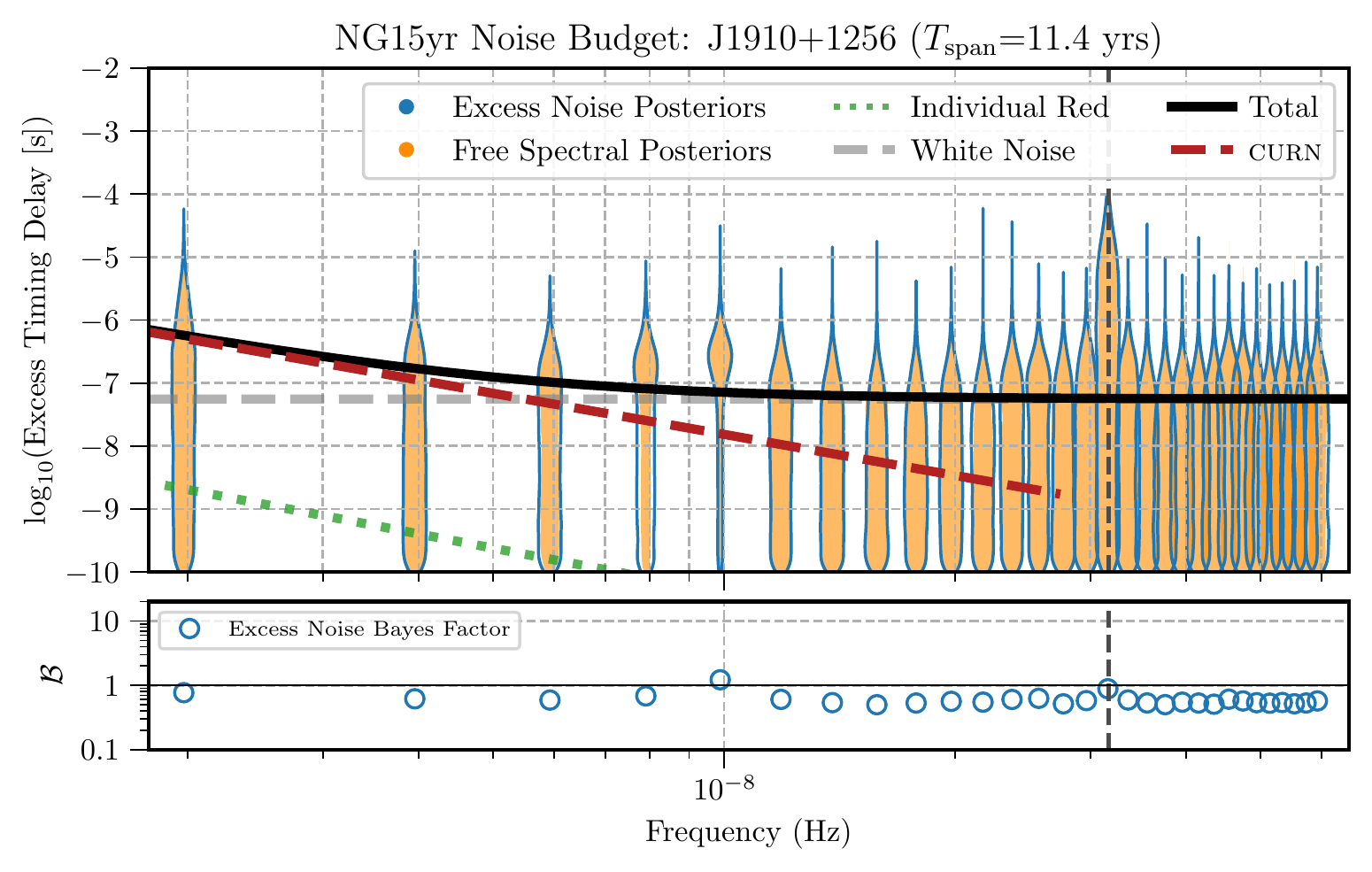}
\caption{The excess timing residual delay as a function of frequency for PSR~J1910+1256. See \myfig{f:excess_j1909} for details.}
\label{f:budget_J1910+1256}
\end{figure}

\begin{figure}
\centering
\includegraphics[width=0.9\linewidth]{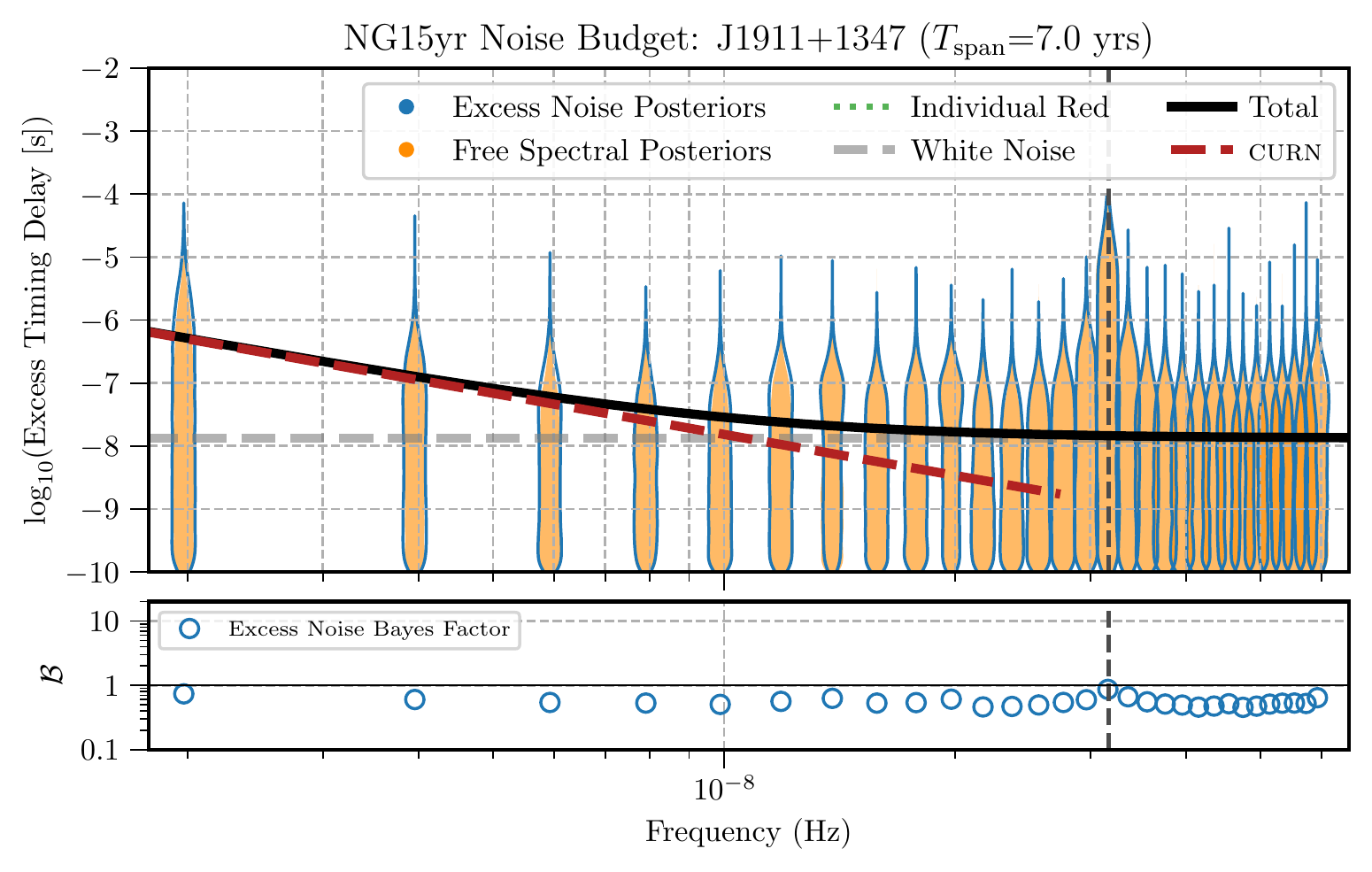}
\caption{The excess timing residual delay as a function of frequency for PSR~J1911+1347. See \myfig{f:excess_j1909} for details.}
\label{f:budget_J1911+1347}
\end{figure}

\begin{figure}
\centering
\includegraphics[width=0.9\linewidth]{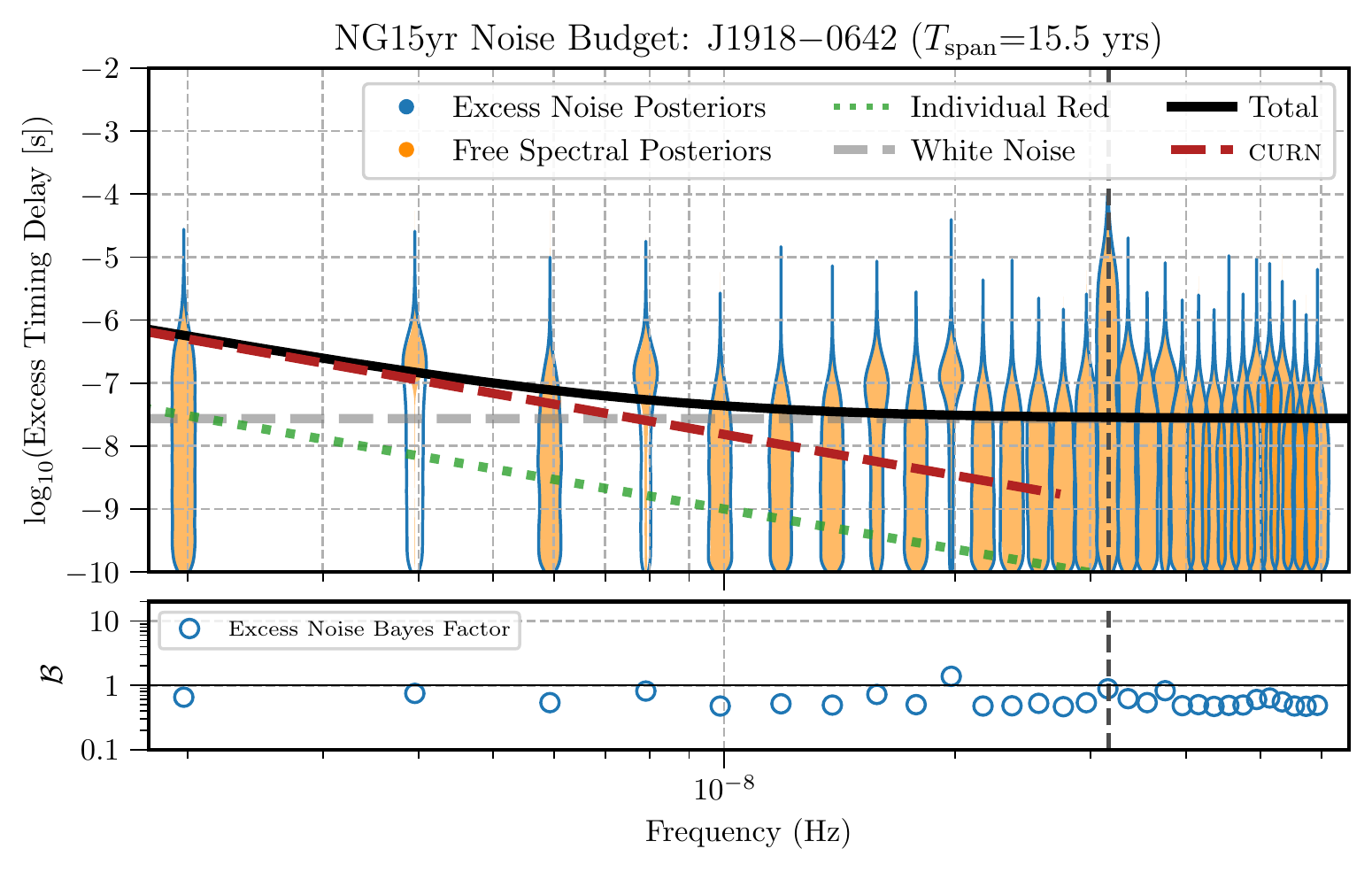}
\caption{The excess timing residual delay as a function of frequency for PSR~J1918$-$0642. See \myfig{f:excess_j1909} for details.}
\label{f:budget_J1918-0642}
\end{figure}

\begin{figure}
\centering
\includegraphics[width=0.9\linewidth]{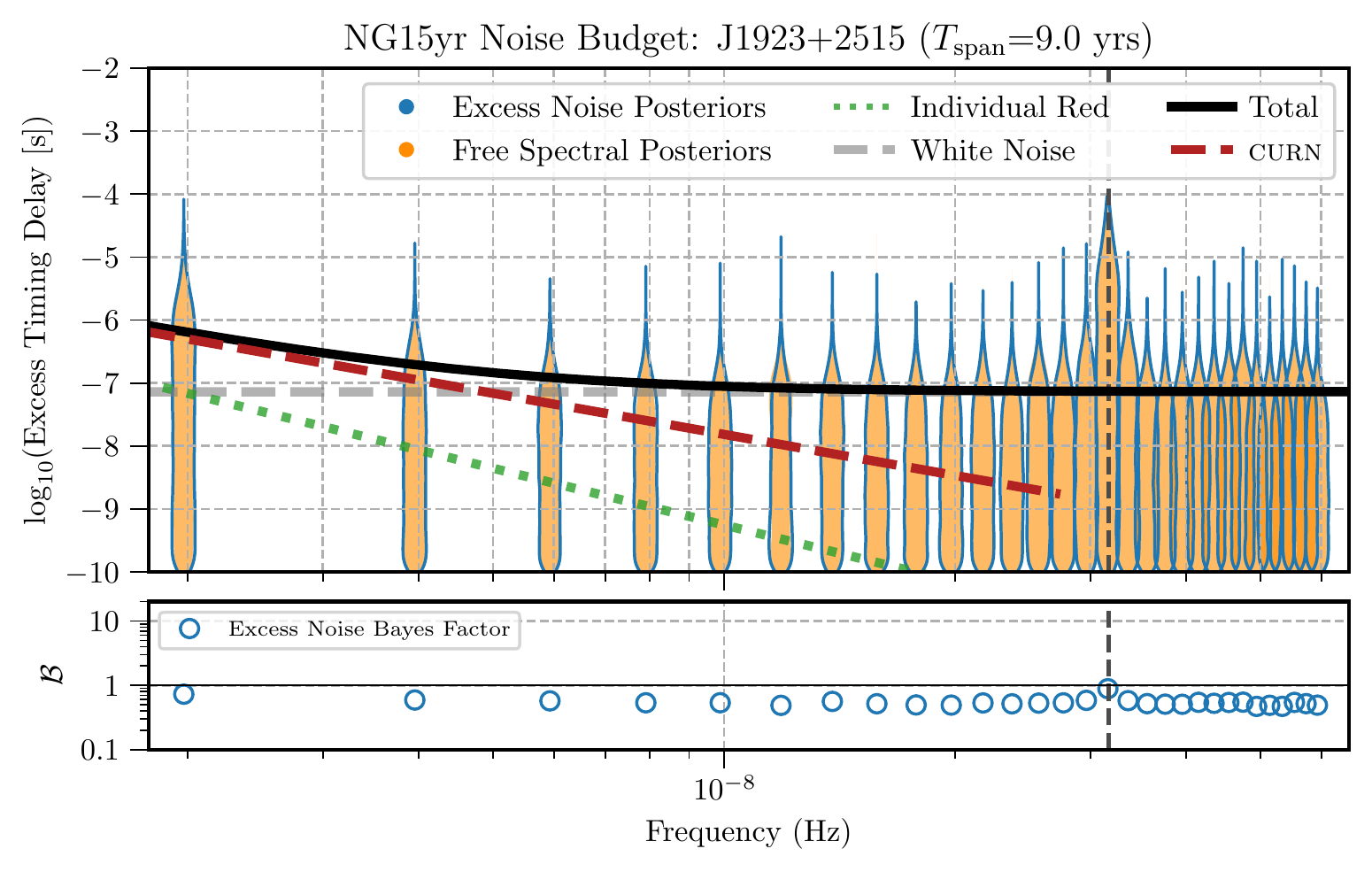}
\caption{The excess timing residual delay as a function of frequency for PSR~J1923+2515. See \myfig{f:excess_j1909} for details.}
\label{f:budget_J1923+2515}
\end{figure}

\begin{figure}
\centering
\includegraphics[width=0.9\linewidth]{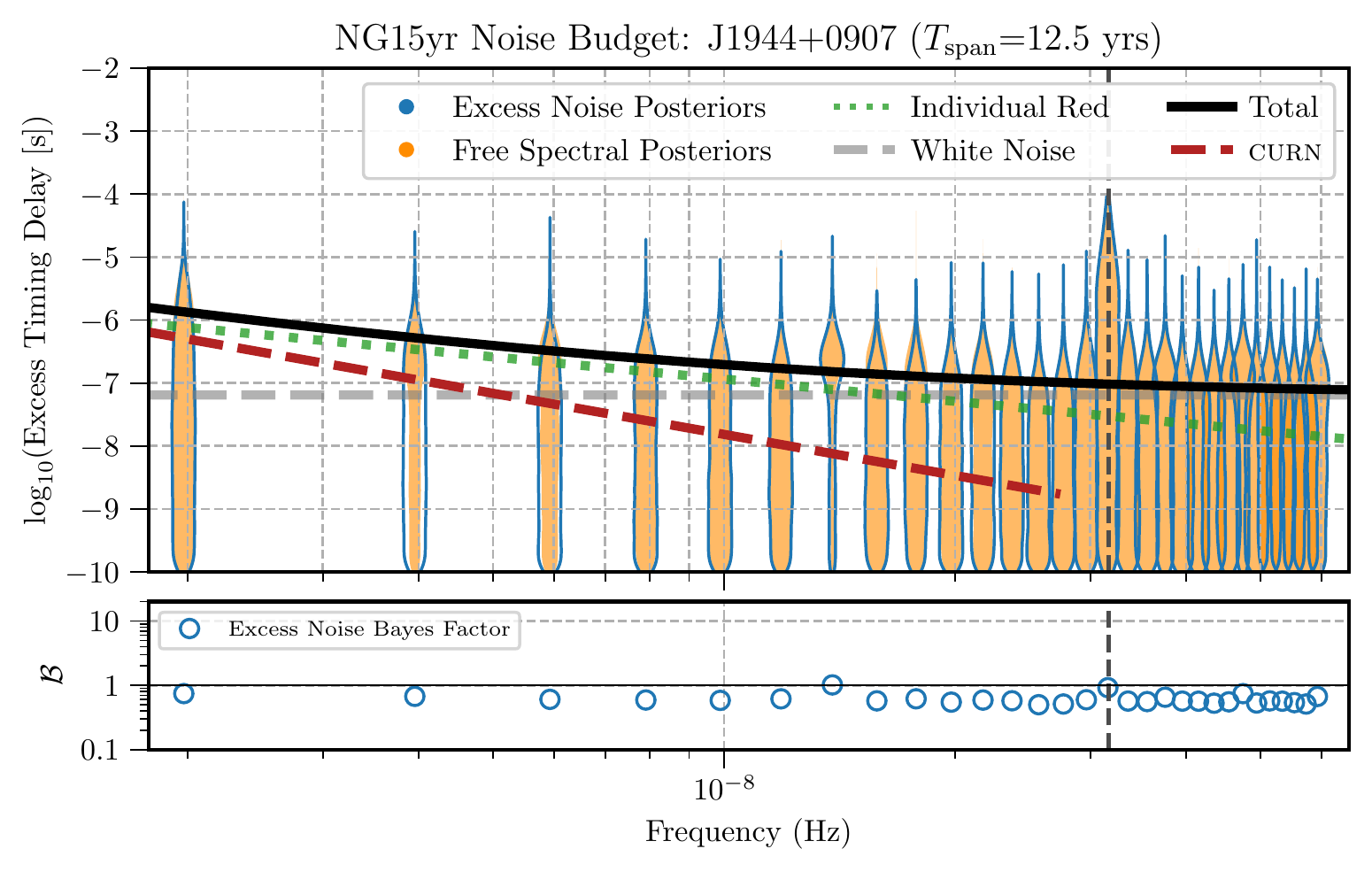}
\caption{The excess timing residual delay as a function of frequency for PSR~J1944+0907. See \myfig{f:excess_j1909} for details.}
\label{f:budget_J1944+0907}
\end{figure}

\begin{figure}
\centering
\includegraphics[width=0.9\linewidth]{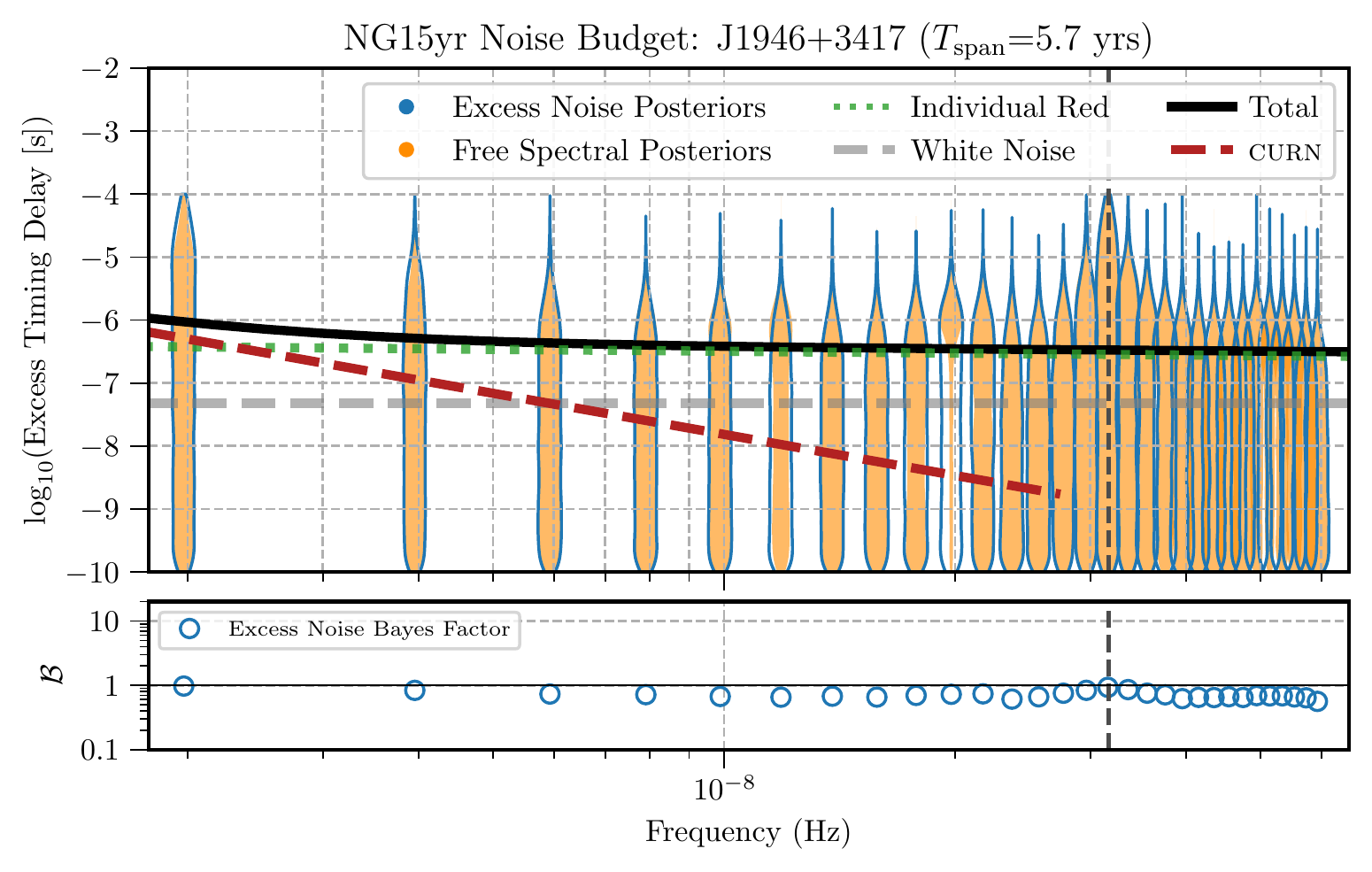}
\caption{The excess timing residual delay as a function of frequency for PSR~J1946+3417. See \myfig{f:excess_j1909} for details.}
\label{f:budget_J1946+3417}
\end{figure}

\begin{figure}
\centering
\includegraphics[width=0.9\linewidth]{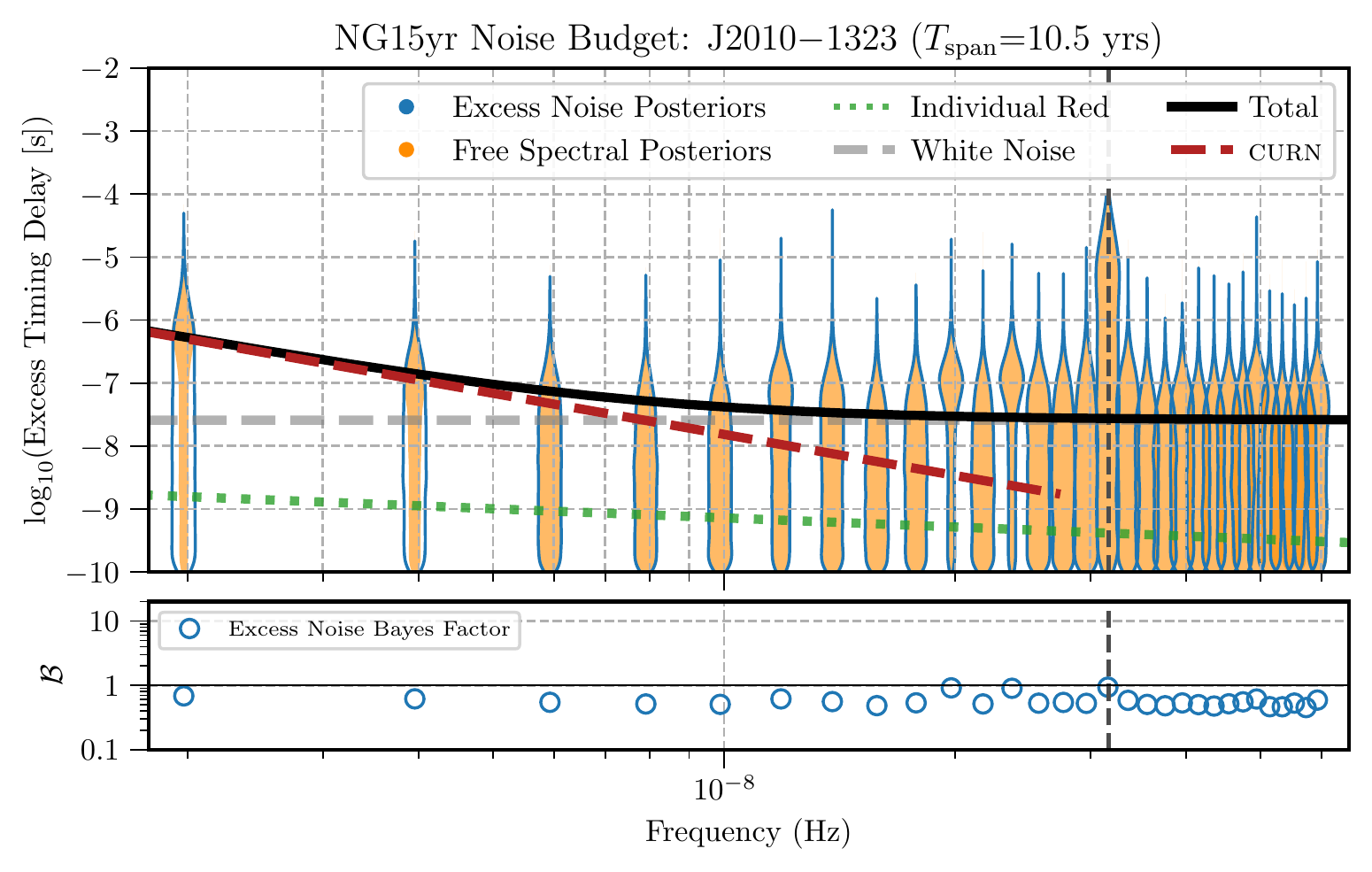}
\caption{The excess timing residual delay as a function of frequency for PSR~J2010$-$1323. See \myfig{f:excess_j1909} for details.}
\label{f:budget_J2010-1323}
\end{figure}

\begin{figure}
\centering
\includegraphics[width=0.9\linewidth]{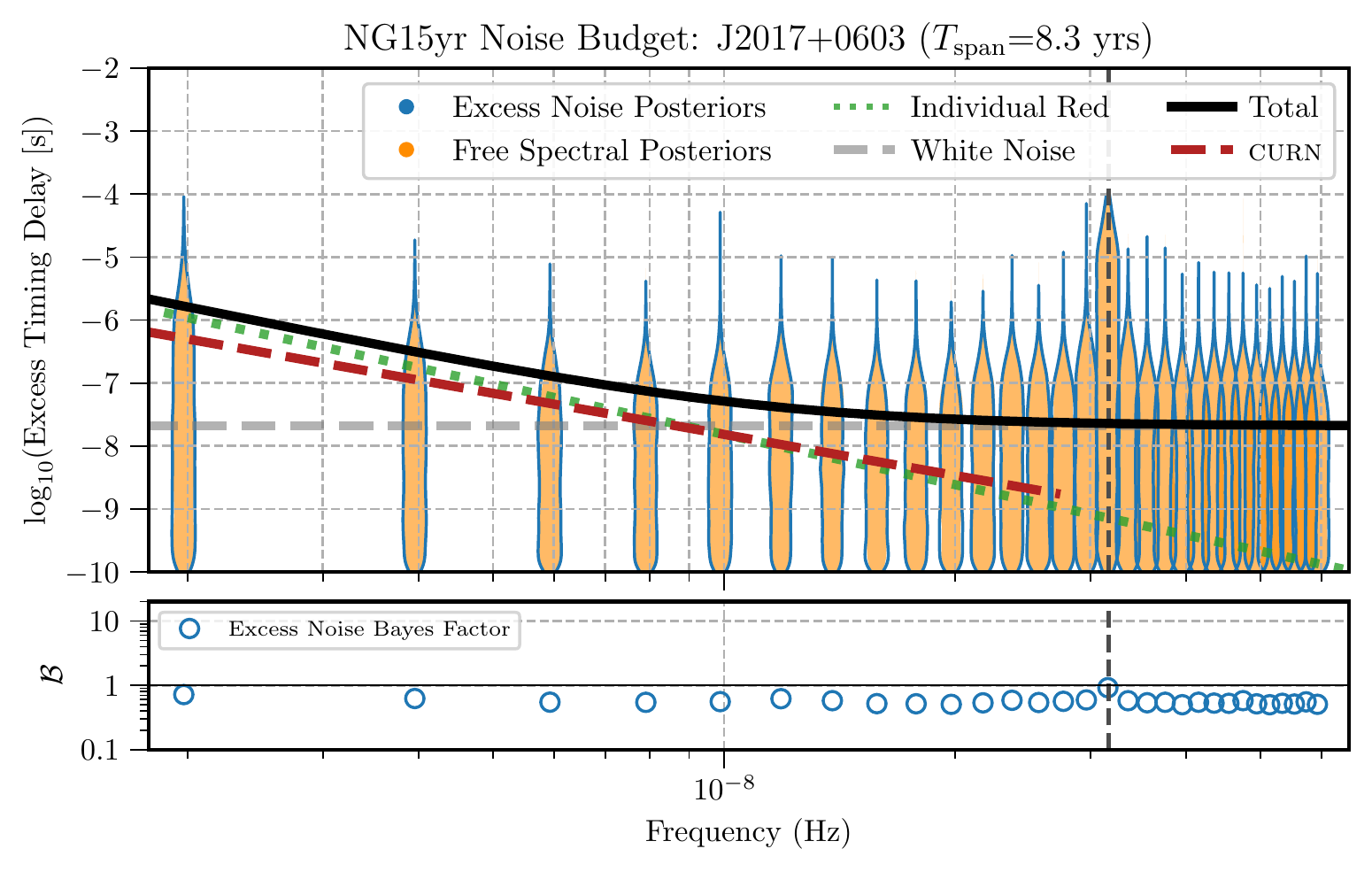}
\caption{The excess timing residual delay as a function of frequency for PSR~J2017+0603. See \myfig{f:excess_j1909} for details.}
\label{f:budget_J2017+0603}
\end{figure}

\begin{figure}
\centering
\includegraphics[width=0.9\linewidth]{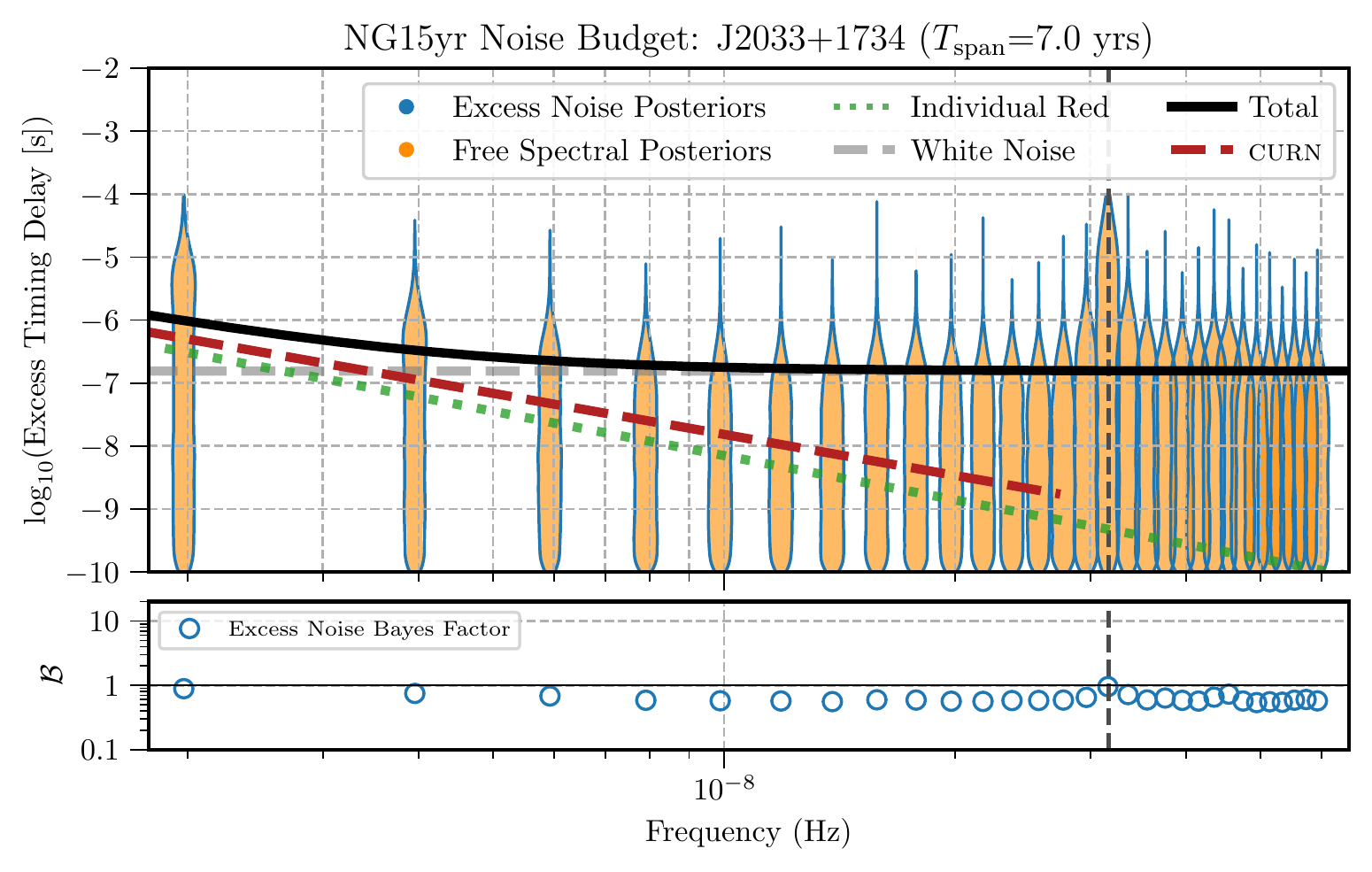}
\caption{The excess timing residual delay as a function of frequency for PSR~J2033+1734. See \myfig{f:excess_j1909} for details.}
\label{f:budget_J2033+1734}
\end{figure}

\begin{figure}
\centering
\includegraphics[width=0.9\linewidth]{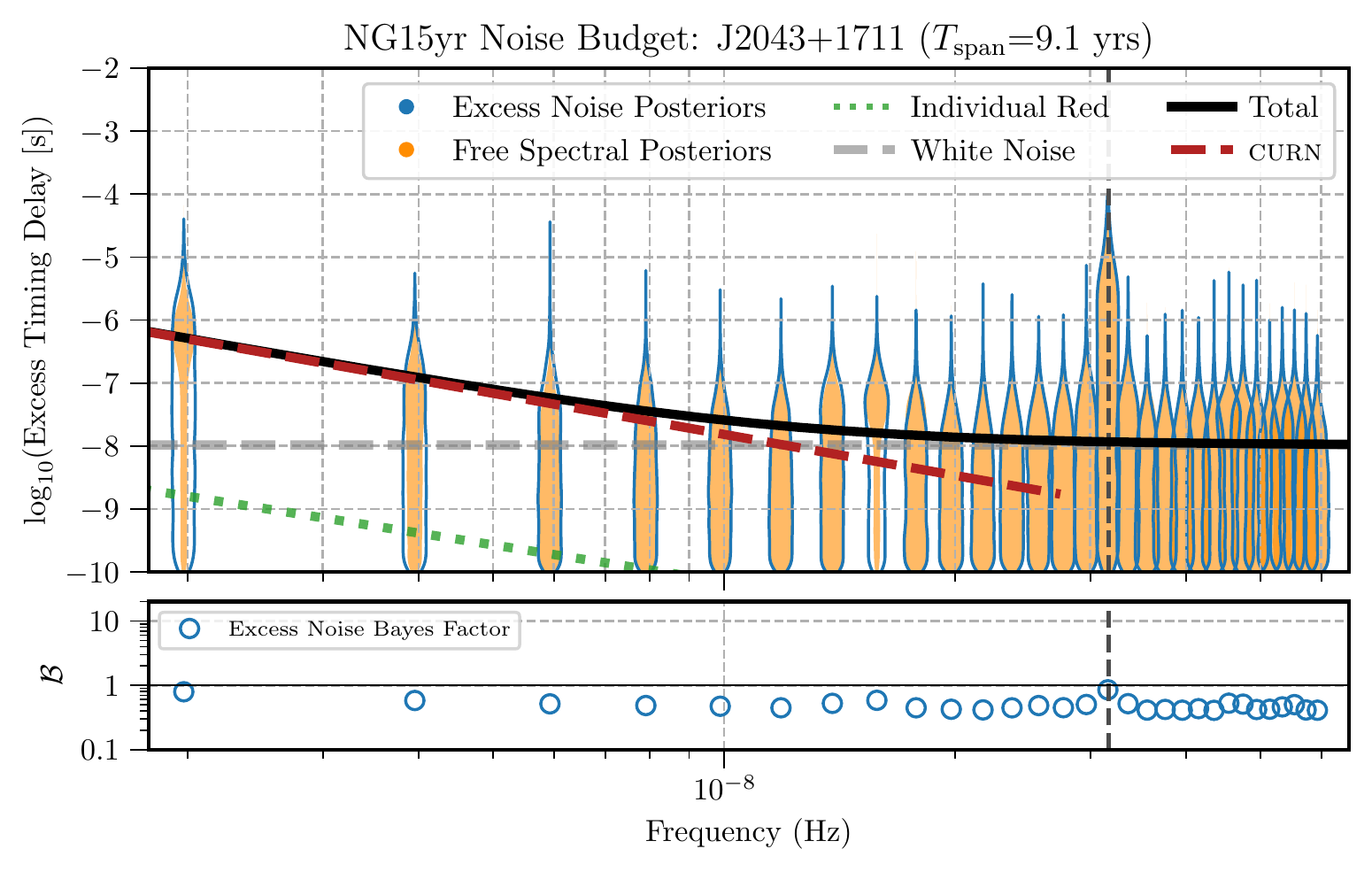}
\caption{The excess timing residual delay as a function of frequency for PSR~J2043+1711. See \myfig{f:excess_j1909} for details.}
\label{f:budget_J2043+1711}
\end{figure}

\begin{figure}
\centering
\includegraphics[width=0.9\linewidth]{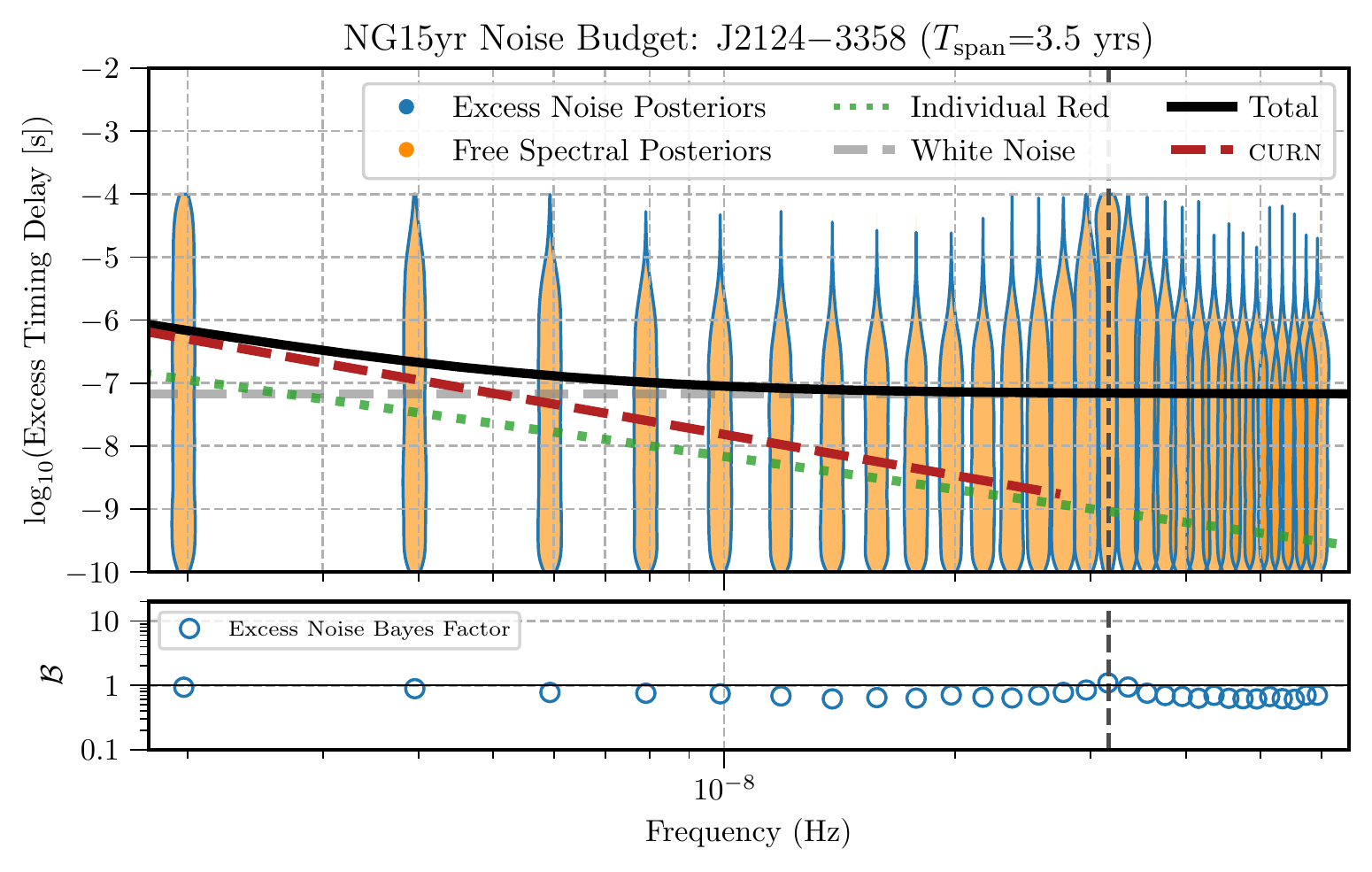}
\caption{The excess timing residual delay as a function of frequency for PSR~J2124$-$3358. See \myfig{f:excess_j1909} for details.}
\label{f:budget_J2124-3358}
\end{figure}

\begin{figure}
\centering
\includegraphics[width=0.9\linewidth]{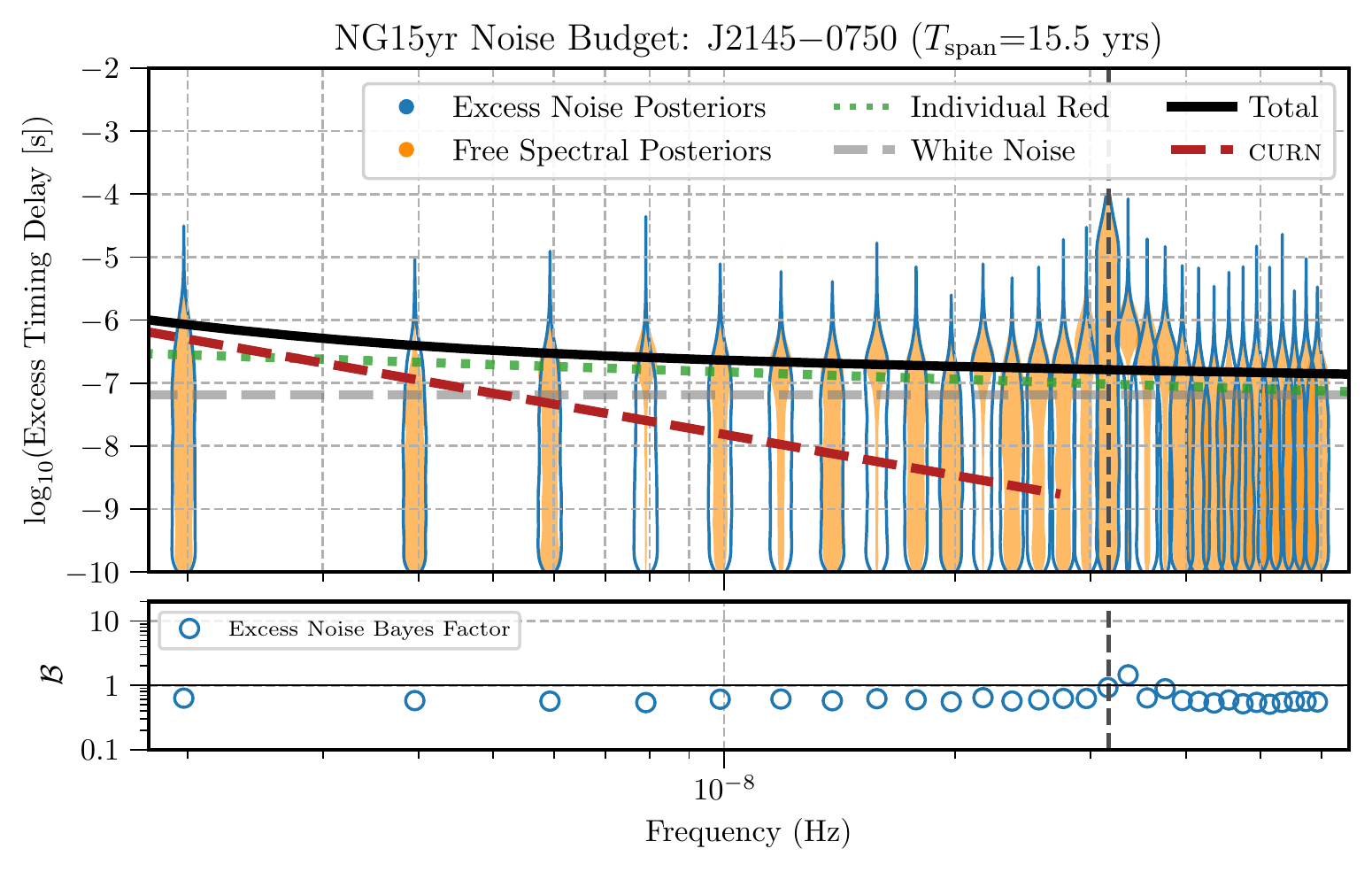}
\caption{The excess timing residual delay as a function of frequency for PSR~J2145$-$0750. See \myfig{f:excess_j1909} for details.}
\label{f:budget_J2145-0750}
\end{figure}

\begin{figure}
\centering
\includegraphics[width=0.9\linewidth]{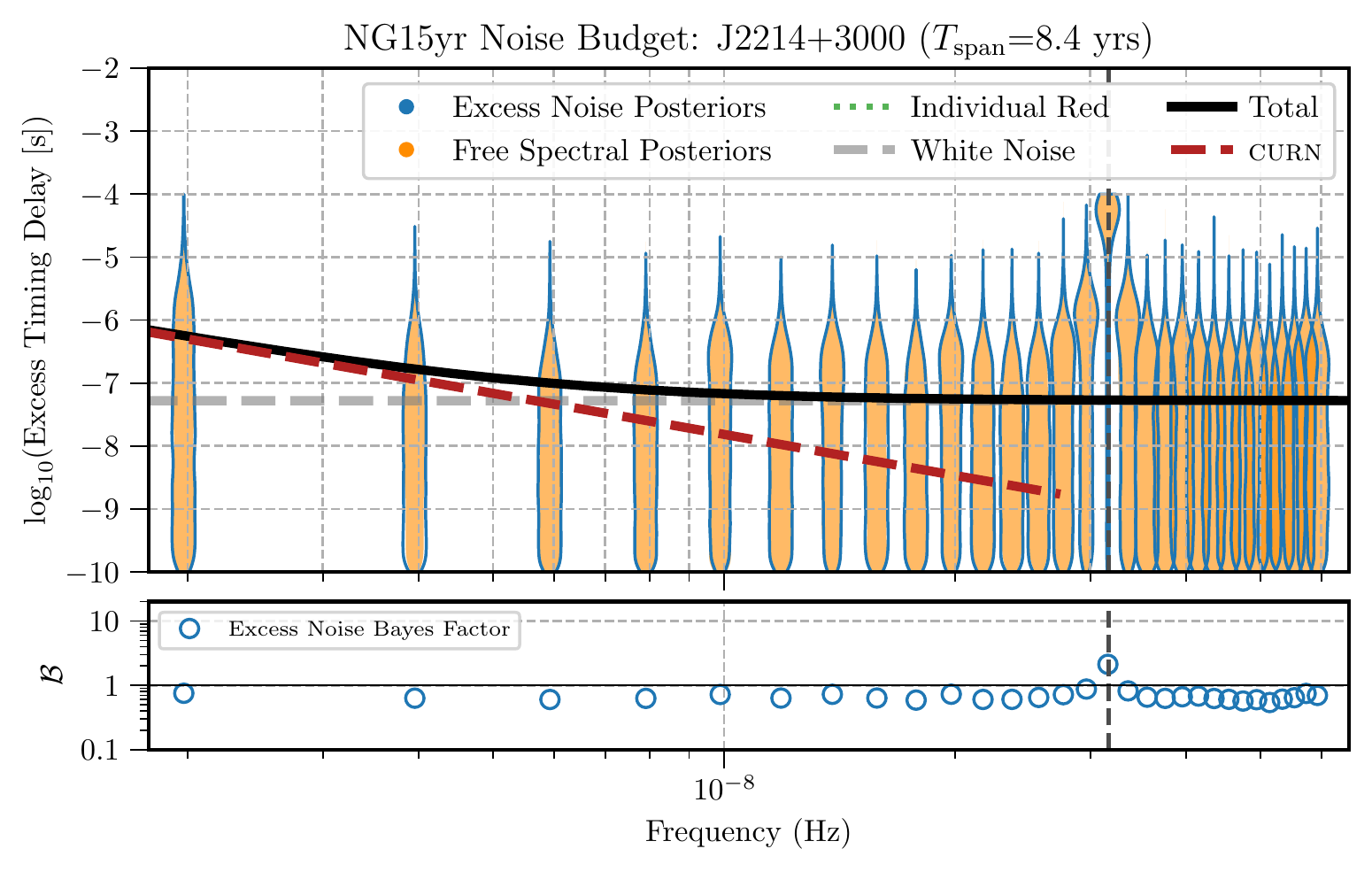}
\caption{The excess timing residual delay as a function of frequency for PSR~J2214+3000. See \myfig{f:excess_j1909} for details.}
\label{f:budget_J2214+3000}
\end{figure}

\begin{figure}
\centering
\includegraphics[width=0.9\linewidth]{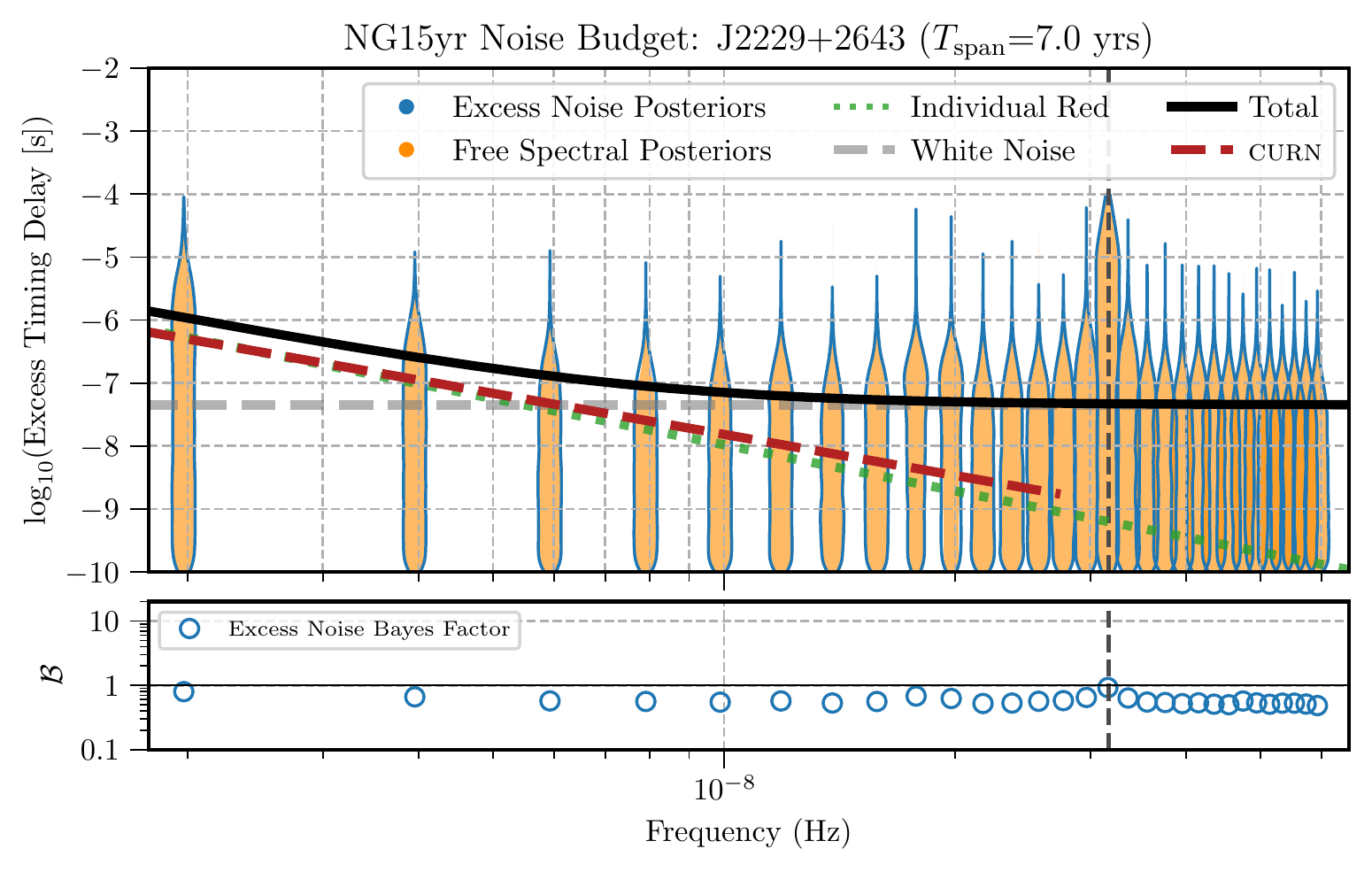}
\caption{The excess timing residual delay as a function of frequency for PSR~J2229+2643. See \myfig{f:excess_j1909} for details.}
\label{f:budget_J2229+2643}
\end{figure}

\begin{figure}
\centering
\includegraphics[width=0.9\linewidth]{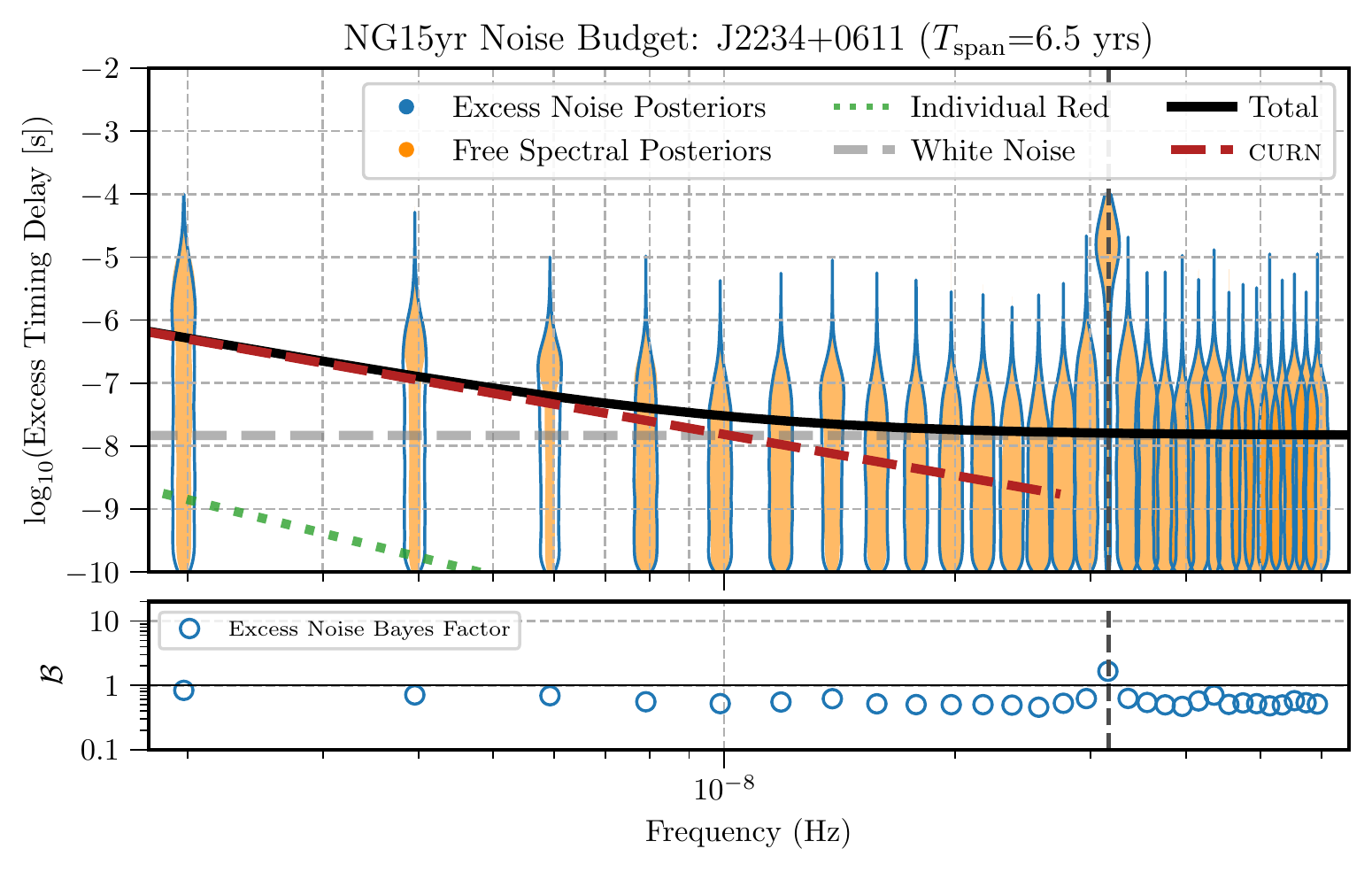}
\caption{The excess timing residual delay as a function of frequency for PSR~J2234+0611. See \myfig{f:excess_j1909} for details.}
\label{f:budget_J2234+0611}
\end{figure}

\begin{figure}
\centering
\includegraphics[width=0.9\linewidth]{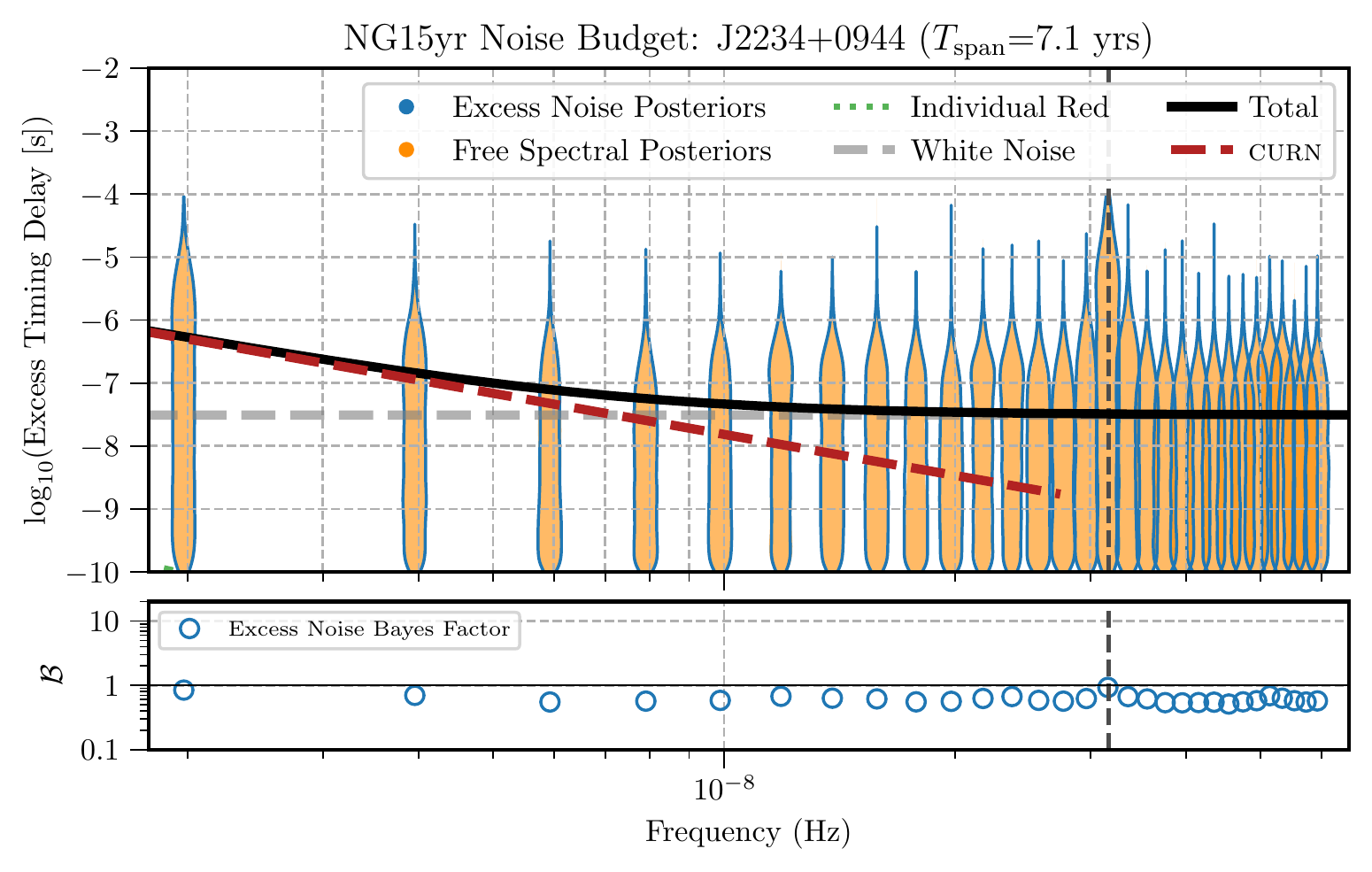}
\caption{The excess timing residual delay as a function of frequency for PSR~J2234+0944. See \myfig{f:excess_j1909} for details.}
\label{f:budget_J2234+0944}
\end{figure}

\begin{figure}
\centering
\includegraphics[width=0.9\linewidth]{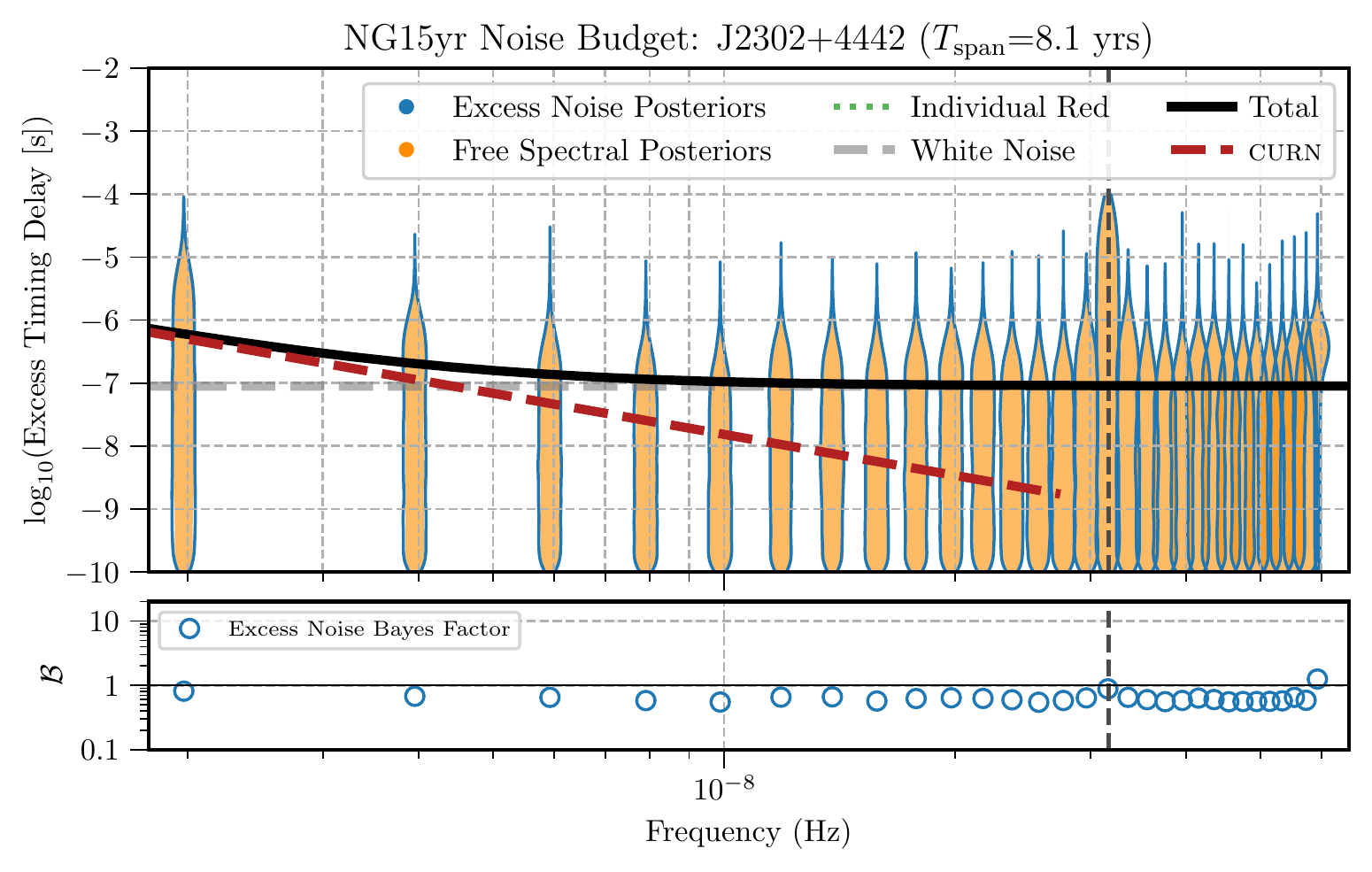}
\caption{The excess timing residual delay as a function of frequency for PSR~J2302+4442. See \myfig{f:excess_j1909} for details.}
\label{f:budget_J2302+4442}
\end{figure}

\begin{figure}
\centering
\includegraphics[width=0.9\linewidth]{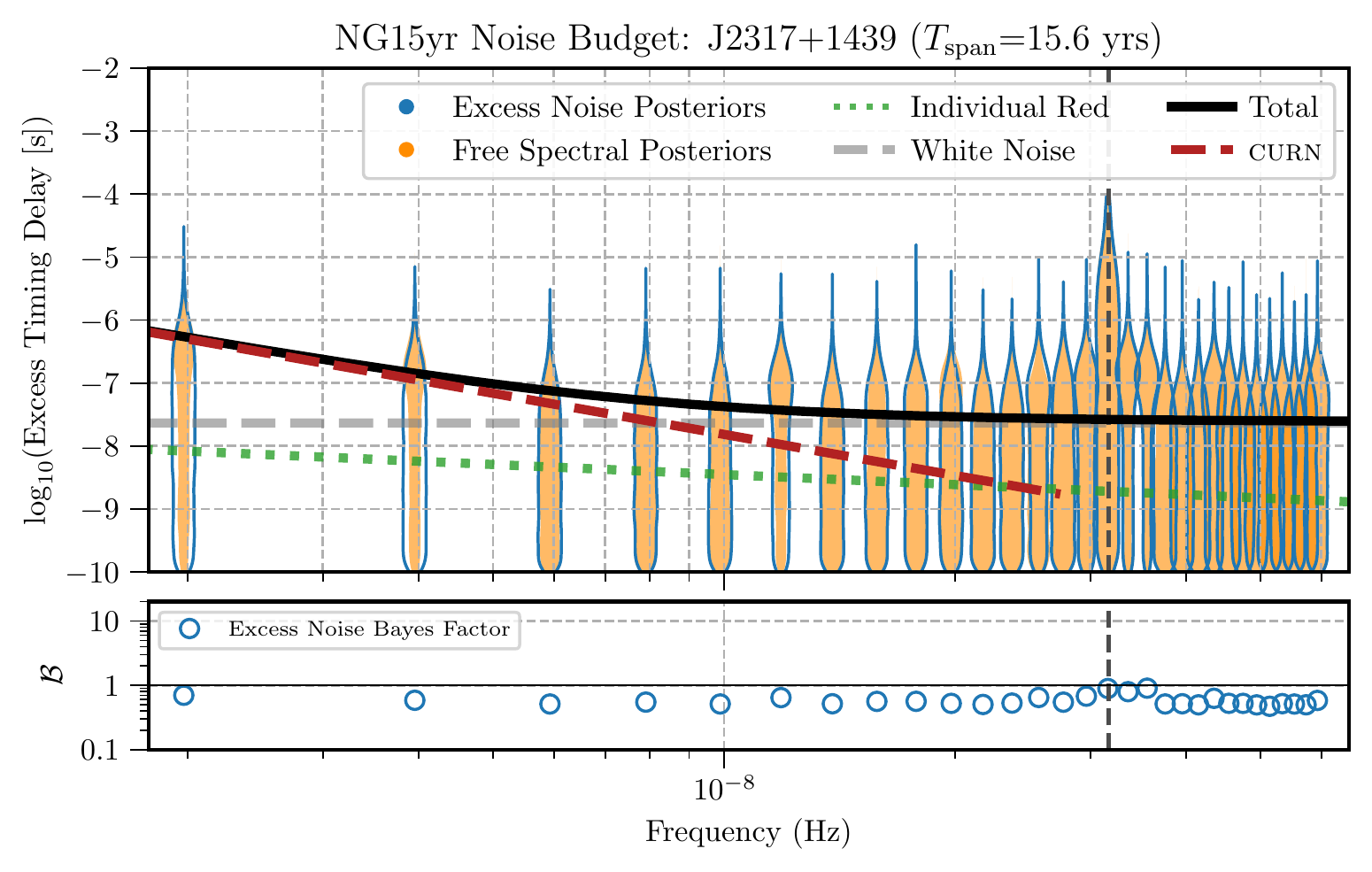}
\caption{The excess timing residual delay as a function of frequency for PSR~J2317+1439. See \myfig{f:excess_j1909} for details.}
\label{f:budget_J2317+1439}
\end{figure}

\begin{figure}
\centering
\includegraphics[width=0.9\linewidth]{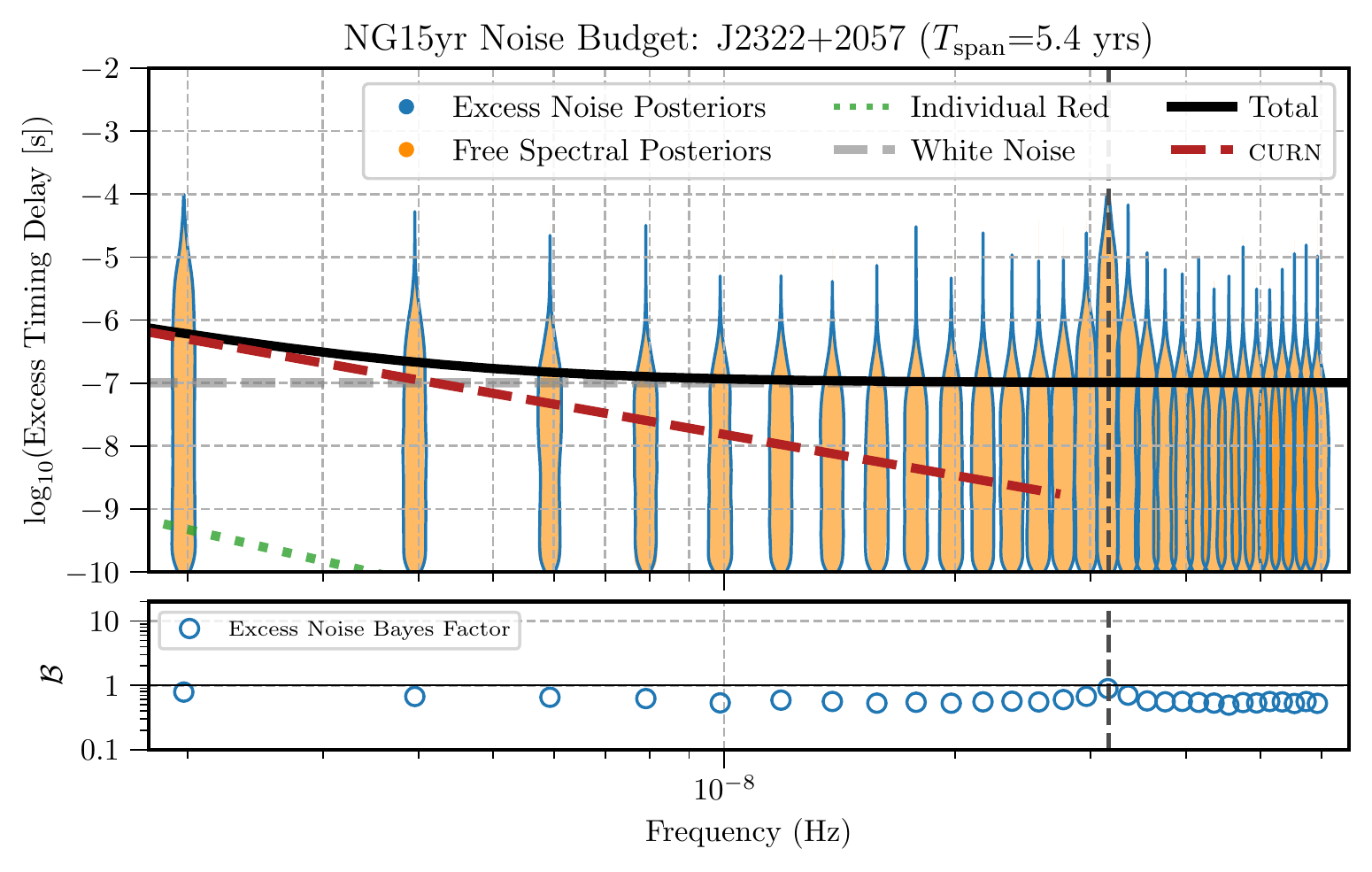}
\caption{The excess timing residual delay as a function of frequency for PSR~J2322+2057. See \myfig{f:excess_j1909} for details.}
\label{f:budget_J2322+2057}
\end{figure}

\end{document}